\documentclass[twoside,12pt]{article}
\usepackage{epsfig}

\def\Journal#1#2#3#4{{#1} {#2} (#4) #3 }

\def\NPA{{\em Nucl. Phys.} A}

\def\PRO{{\em Prog. Theor. Phys.}}
\def\NPB{{\em Nucl. Phys.} B}

\def\PLB{{\em Phys. Lett.} B}

\def\PRL{\em Phys. Rev. Lett.}

\def\PREP{\em Phys. Rep.}

\def\PRD{{\em Phys. Rev.} D}
\def\PRC{{\em Phys. Rev.} C}

\def\ZPC{{\em Z. Phys.} C}

\def\INT{{\em Int. J. Mod. Phys.} A}
\def\NPPS{{\em Nucl. Phys. Proc. Suppl.}}
\def\EPJC{{\em Eur. Phys. J.} C}
\def\EPJA{{\em Eur. Phys. J.} A}

\newcommand{\be}{\begin{equation}}
\newcommand{\ee}{\end{equation}}
\newcommand{\bea}{\begin{eqnarray}}
\newcommand{\eea}{\end{eqnarray}}
\newcommand{\nn}{\nonumber}

\newcommand{\bnP}{\bar {\cal P}}

\newcommand{\bnPd}{\bar {\cal P}^{\raisebox{0.8mm}{\scriptsize$\dagger$}} }
\newcommand{\cP}{{\cal P}}
\newcommand{\bn}{{\bar n}}

\def\Dslash{D\!\!\!\!\slash}

\def\bnslash{\bar n\!\!\!\slash}

\def\vma{{_{V-A}}}

\def\SppP{{\cal {P\!\!\!\!\hspace{0.04cm}\slash}}_T}
\def\Aslash{A\!\!\!\slash}
\def\la{\langle}
\def\ra{\rangle}
\def\etal{{\it et al }}
\begin{document}

\title{QCD Aspects of Exclusive $B$ Meson Decays}

\author{Hsinag-nan Li\\
Institute of Physics, Academia Sinica,\\
Nankang, Taipei, Taiwan 115, R.O.C.
}

\maketitle

\begin{abstract} 
I review recent progress on understanding QCD dynamics involved
in exclusive $B$ meson decays. Different frameworks, including light-cone
sum rules, QCD factorization, perturbative QCD,
soft-collinear effective theory, light-front QCD, are discussed.
Results from lattice QCD are quoted for comparison. I point out the
important issues in the above QCD methods, which require further
investigation. 
\end{abstract}

\section{Introduction}

We are now in the era of $B$ physics. $B$ fatories at KEK and SLAC have
collected about 80 $fb^{-1}$ data, based on which we are not only able 
to probe the origin of CP violation, but to explore rich QCD dynamics 
involved in exclusive $B$ meson decays. As announced in \cite{Nir}, the
Kobayashi-Maskawa (KM) ansatz \cite{KoMa} for CP violation is more
or less certain with the consistent measurements of $\sin 2\phi_1$
(or $\sin 2\beta$) from Belle and BaBar, $\phi_1$ being one of the
unitarity angles. The results are also in agreement with other 
indirect determination of the Cabibbo-Kobayashi-Maskawa (CKM) matrix 
elements. In this article I will forcus on the latter subject.
It will be realized that exclusive $B$ meson decays provide a unique
field, in which QCD theories with controllable theoretical uncertainty
can be developed. It turns out that these theories are simpler
than those for charm and kaon physics. This field has attracted wide
attention, and tremendous progress has been made recently.

Within the KM ansatz, the source of CP violation is organized in the form
of a unitarity triangle. On one hand, we overconstrain the unitarity
triangle as much as possible, and on the other hand, look for possible
discrepancies, which could reveal signals of new physics beyond the
Standard Model. The angle $\phi_1$ can be extracted from the CP asymmetry
in the $B\to J/\psi K_S$ decays \cite{BIS}, which arises from the 
$B$-$\bar B$ mixing. Through similar mechanism, the $B_d^0\to \pi^+\pi^-$ 
decays are appropriate for the extraction of the angle $\phi_2$ (or 
$\alpha$). However, these modes contain both tree and penguin 
contributions, such that the extraction suffers uncertainty. Strategies 
have been proposed to handle this penguin pollution, the best of which is 
known to be the isospin analysis \cite{GL}. Unfortunately, this strategy 
is difficult in practice, because of the small $B_d^0\to\pi^0\pi^0$ 
branching ratios. The angle $\phi_3$ (or $\gamma$) can be 
determined from the decays $B\to K\pi$ \cite{GRL,FM,NR,BRF}, which are
obviously also plagued by the similar penguin-tree interference.

We can move forward by learning how to estimate hadronic matrix 
elements involved in exclusive $B$ meson decays. For this purpose, 
symmetries of strong interaction
have been postulated to relate amplitudes among different modes.
For example, the penguin-over-tree ratio $|P/T|$ helps the extraction
of $\phi_2$ from the CP-violating
observables in $B_d^0\to\pi^+\pi^-$ \cite{GR-Bpipi1}. $SU(3)$ flavor 
symmetry and plausible dynamical assumptions were then employed to fix 
$|P|$ through the CP-averaged $B^\pm\to K\pi^\pm$ branching ratio 
\cite{GR-Bpipi2}. The information of $|T|$ can be obtained from the 
$B\to\pi l\nu$ data. Another strategy is to apply the $U$-spin flavor 
symmetry to the $B_d^0\to\pi^+\pi^-$ and $B_s\to K^+K^-$ modes 
\cite{RF-BsKK}, from which the penguin amplitudes are determined.
However, the above symmetries are in fact not exact, and it is not clear 
how to control theoretical uncertainties from symmetry breaking.
As an alternative, one searches for the special modes, in which relations
among decay amplitudes allow the elimination of hadronic uncertainties.
For example, $\phi_3$ can be extracted from the triangle 
relations for the $B^\pm_u\to K^\pm \{D^0,{\bar D}^0,D^0_+\}$ 
amplitudes \cite{gw,F03}, which receive only tree contributions, 
$D^0_+$ being
the CP-even eigenstate of the neutral $D$-meson system. However, 
this strategy, due to the squashed triangles, is experimentally difficult 
\cite{ADS}. The modes $B_d\to K^{\ast0}\{D^0,{\bar D}^0,D^0_+\}$ 
\cite{dun} and $B^\pm_c\to D_s^\pm \{D^0,{\bar D}^0,D^0_+\}$ 
\cite{FW}, providing more equilateral triangles, then deserve further 
feasibility studies.

The above discussion indicates that it is necessary to have deeper
understanding of QCD dynamics in exclusive $B$ meson decays
and control of hadronic uncertainties \cite{ZL}. The $b$ quark mass 
$m_b$, much larger than the QCD scale $\Lambda_{\rm QCD}$, renders such an
attempt possible: relevant hadronic matrix elements can be evaluated as 
an expansion in the strong coupling constant $\alpha_s(m_b)$ and in the 
ratio $\Lambda_{\rm QCD}/m_b$. The approaches based on this heavy quark
expansion include light-cone QCD sum rules (LCSR) 
\cite{Chernyak:1990ag,sum-rules,Khodja}, 
light-front QCD (LFQCD) \cite{CZL,CJ99}, QCD factorization (QCDF) 
\cite{BBNS1}, and perturbative QCD (PQCD) \cite{LY1,KLS,LUY}.
Soft-collinear effective theory (SCET) provides a systematic framework, 
in which the above expansion can be constructed in a
simple and formal way \cite{bfl,bfps}. Lattice QCD
is complementary to the above methods, whose results will be quoted 
for comparison. In this article I will explain the basic ideas behind the
various QCD theories, and review their applications to typical, such as 
semileptonic, radiative and nonleptonic, exclusive modes. 
That is, I emphasize the methodology, instead of the survey of all decay 
channels.

To be specific, I will not discuss the strategies to constrain the CKM 
matrix elements from experimental data. For nice reviews of
this topic, refer to \cite{RF,ABPS,MP03}. I will not explore analyses 
relying on symmetries of strong interaction, such as the $SU(3)$ flavor 
symmetry. Recent works along this vein, which have taken into account 
symmetry breaking effects, can be found in \cite{He,Wu,XZ}. The status 
of the important CKM global fitting have been presented in 
\cite{Hocker:2001xe,Ciuchini:2000de}. 
To demonstrate the applications of the QCD methods, 
I will consider only $B_{u,d}$ meson decays as an example. The subjects 
related to the $B_s$ and $B_c$ mesons and to heavy baryons, including 
their polarization effect, will be dropped. The perspectives for 
investigating $B_c$ mesons at LHCb have been surveyed in \cite{GKL}.
For a similar reason, I skip the applications to decays into baryons and
into tensor mesons \cite{KLO}. Studies of three-body $B$ meson decays
are still at the early stage \cite{CCY,CL2}, and will be reserved
for a future review. I will not touch
supersymmetric topics in $B$ physics either, which are too much for this
article. For a recent relevant review, refer to \cite{Ellis,K02}.

In Sec.~2 and Sec.~3 I briefly explain two types of factorization 
theorems, collinear and $k_T$ factorizations, which are the fundamental 
concepts of most of the QCD theories. The ideas and results derived
from the various QCD methods are reviewed in Sec.~4 for semileptonic
and radiative decays, and in Sec.~5 for two-body nonleptonic decays.
Charmed decays are discussed in Sec.~6. Other miscellaneous topics
are collected in Sec.~7. Section 8 is the conclusion.

\section{Collinear Factorization}

Most of QCD methods rely on some sorts of factorization theorems.
For example, QCDF is a generalization of collinear factorization theorem
to exclusive $B$ meson decays. In LCSR collinear factorization applies
to final-state hadron bound states, which are then expanded in terms of
parton Fock states characterized by different twists. $k_T$ factorization
theorem is the basis of the PQCD approach, which is more appropriate in 
the end-point region of parton momentum fractions. SCET for the kinematic
region with energetic final-state hadrons, is equivalent to collinear 
factorization theorem, but operated at the operator and Lagrange level. 
I will compare the factorization of high-energy QCD processes derived 
from perturbation theory and from SCET. I then discuss the
double logarithmic resummation, which is required for applying
collinear factorization theorem to semileptonic $B$ meson decays.

\subsection{\it Factorization Theorem \label{sec:all}}

I first review collinear factorization theorem for exclusive processes 
developed around 80's 
\cite{BL,BFL,ER,DM,CZS,CZ}. In this theorem nonperturbative dynamics of 
a high-energy QCD process, characterized by a large scale $Q$, either 
cancel or is absorbed into hadron distribution amplitudes. The remaining 
part, being infrared finite, is calculable in perturbation theory. A 
physical quantity is then expressed as the convolution of a 
hard-scattering kernel with the distribution amplitudes solely in 
parton momentum fractions. A distribution amplitude,
though not calculable, is universal, {\it i.e.}, process-independent.
With this universality, a distribution amplitude determined by
nonperturbative means, such as QCD sum rules
and lattice QCD, or extracted from experimental data, can be employed
to make predictions for other processes involving the same hadron.
Contributions of different orders in $\alpha_s$ and powers in $1/Q$ can be
included systematically.

Nonperturbative dynamics is reflected by infrared divergences in
radiative corrections, whose factorization leads to distribution
amplitudes at the parton level. Factorization of the above infrared 
divergences needs to be performed in momentum, spin, and color spaces. 
Factorization in momentum space means that a hard kernel depends on the 
loop momentum of a soft or collinear gluon, which has been absorbed into 
a distribution amplitude, only through the parton momentum fraction. 
Factorization in spin and color spaces means that there are separate 
fermion and color flows between a hard kernel and a distribution amplitude, 
respectively. I take the simple process $\pi\gamma^*\to\gamma$ as an 
example to demonstrate the proof of collinear factorization thheorem based 
on perturbation theory.
The collinear factorization of this process has been proved in \cite{BL},
but in the axial (light-cone) gauge $A^+=0$. In this gauge factorization
automatically holds and the analysis is straightforward, because
collinear divergences exist only in two-parton reducible diagrams. The
pion distribution amplitude has been constructed from
$\gamma^*\gamma\to\pi$ in the framework of covariant operator product
expansion \cite{BFL,ER}. The factorization of
$\pi\gamma^*\to\pi$ has been proved in \cite{DM} based on the
Zimmermann's ''$\Delta$-forest" prescription \cite{Z}, which involves
complicated diagram subtractions. 

Below I will adopt a simple proof proposed in \cite{NL}. 
To achieve factorization in momentum, spin, and color spaces,
one needs the eikonal approximation for loop integrals in leading
infrared regions, the insertion of the Fierz identity to separate fermion
flows, and the Ward identity to sum up diagrams with different color
structures. Under the eikonal approximation, a soft or collinear gluon is
detached from the lines in a hard kernel and in other distribution
amplitudes. The Fierz identity decomposes the  full amplitude into
contributions characterized by different twists. The Ward identity is
essential for proving factorization theorem in a nonabelian gauge theory.
The soft divergences exist in exclusive $B$ meson decays, which should 
be factorized into a $B$ meson distribution amplitude. The derivation in 
\cite{NL} is explicitly gauge-invariant, 
and appropriate for both the factorizations of soft and collinear
divergences, compared to those in the literature.

\begin{figure}[t!]
\begin{center}
\epsfig{file=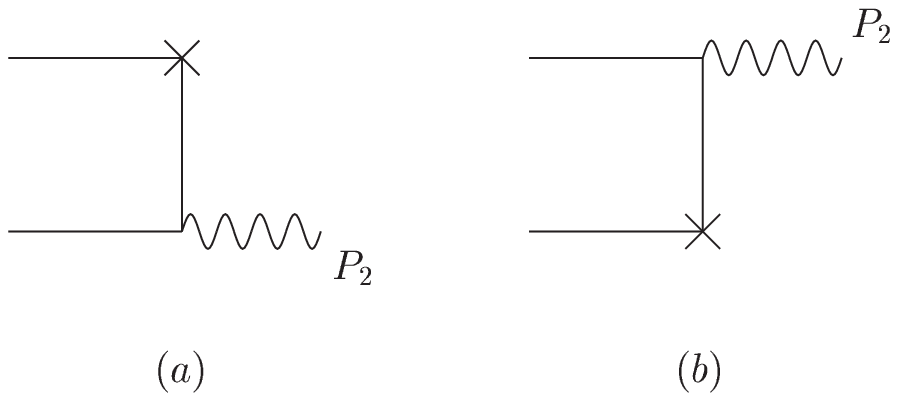, height=1.5in}
\caption{Lowest-order diagrams for $\pi\gamma^*\to\gamma$
($B\to\gamma l\bar\nu$), where the symbol $\times$ represents the
virtual photon (weak decay) vertex.\label{fig1}}
\end{center}
\end{figure}

The momentum $P_1$ of the pion and the momentum
$P_2$ of the outgoing on-shell photon are chosen, in light-cone
coordinates, as
\begin{eqnarray}
P_1=(P_1^+,0,{\bf 0}_T)=\frac{Q}{\sqrt{2}}(1,0,{\bf 0}_T)\;,\;\;\;\;
P_2=(0,P_2^-,{\bf 0}_T)=\frac{Q}{\sqrt{2}}(0,1,{\bf 0}_T)\;.
\label{mpp}
\end{eqnarray}
Let $\epsilon$ denote the polarization vector of the outgoing photon,
which contains only the transverse components. Consider the kinematic
region with large $Q^2=-q^2$, $q=P_2-P_1$ being the virtual photon
momentum, where perturbative expansion is reliable.
The lowest-order diagrams are displayed in Fig.~\ref{fig1}. Assume that 
the on-shell quark and antiquark carry the fractional momenta 
${\bar x}P_1$ and $xP_1$, respectively, with ${\bar x}\equiv 1-x$.  
Figure~\ref{fig1}(a) gives the 
parton-level amplitude,
\begin{eqnarray}
{\cal G}^{(0)}(x)= -ie^2 {\bar q}(xP_1)\not \epsilon 
\frac{\not P_2-x\not P_1}{(P_2-xP_1)^2}\gamma_\mu q({\bar x}P_1)\;.
\label{ga0} 
\end{eqnarray}
The analysis for Fig.~\ref{fig1}(b) is the same. The internal quarks are 
regarded as being hard, {\it i.e.}, being off-shell by $O(Q^2)$. 

The factorization in the fermion flow is achieved by inserting the 
Fierz identity,
\begin{eqnarray}
I_{ij}I_{lk}=\frac{1}{4}I_{ik}I_{lj}
+\frac{1}{4}(\gamma_5)_{ik}(\gamma_5)_{lj}
+\frac{1}{4}(\gamma_\alpha)_{ik}(\gamma^\alpha)_{lj}
+\frac{1}{4}(\gamma_5\gamma^\alpha)_{ik}(\gamma_\alpha\gamma_5)_{lj}
+\frac{1}{8}(\sigma^{\alpha\beta})_{ik}(\sigma_{\alpha\beta})_{lj}\;,
\label{fierz}
\end{eqnarray}
where $I$ represents the identity matrix, and $\sigma_{\alpha\beta}$ is
defined as $\sigma_{\alpha\beta}\equiv i[\gamma_\alpha,\gamma_\beta]/2$.
Different terms in the above identity lead to contributions of different
twists. Equation~(\ref{ga0}) is then factorized into
\begin{eqnarray}
{\cal G}^{(0)}(x)=\int d\xi\phi^{(0)}(x,\xi){\cal H}^{(0)}(\xi)\;,
\label{gl0}
\end{eqnarray}
where the functions,
\begin{eqnarray}
\phi^{(0)}(x,\xi)&=&\phi^{(0)}(x)\delta(x-\xi)\;,\;\;\;\;
\phi^{(0)}(x)=\frac{1}{4P_1^+}{\bar q}(xP_1)\gamma_5\not{\bar n}
q({\bar x}P_1)\;,
\nonumber\\
{\cal H}^{(0)}(x)&=& ie^2
\frac{tr(\not \epsilon \not P_2 \gamma_\mu
\not P_1\gamma_5)}{2x P_1\cdot P_2}\;,
\label{h0}
\end{eqnarray}
with the dimensionless vector $\bar n=(0,1,{\bf 0}_T)$  on the light cone,
define the lowest-order distribution amplitude and hard kernel 
in perturbation theory, respectively. For the momenta 
chosen in Eq.~(\ref{mpp}), only the pseudo-vector structure 
$\gamma_\alpha\gamma_5$ with $\alpha=+$ survives, as it is contracted
with the hard kernel to form the factor $\not P_1\gamma_5$ in 
Eq.~(\ref{h0}). This piece of contribution is of leading 
twist (twist 2). For other processes, such as the pion form factor, 
higher-twist structures survive, but the analysis is the same \cite{NL}.

\begin{figure}[tb]
\begin{center}
\epsfig{file=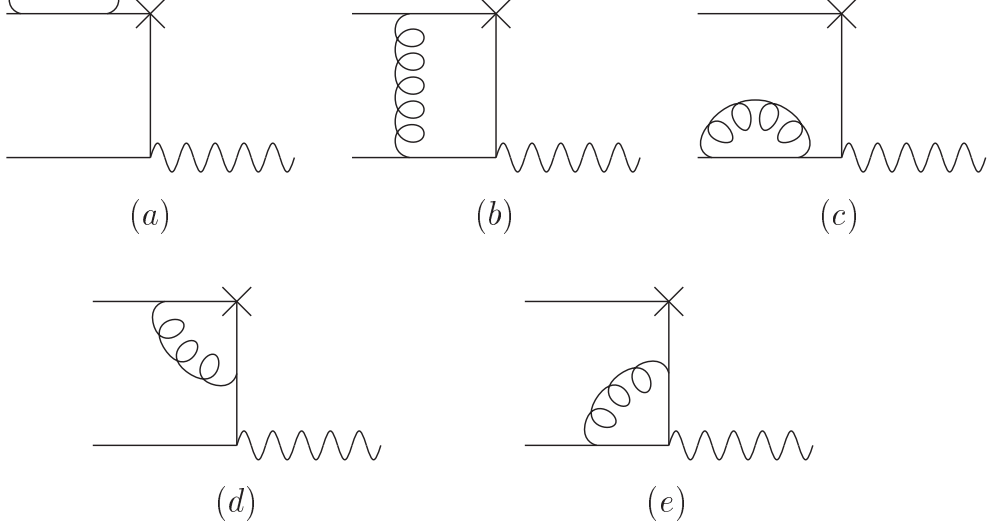, height=3.0in}
\caption{$O(\alpha_s)$ radiative corrections to 
Fig.~\ref{fig1}(a).\label{fig2}}
\end{center}
\end{figure}

There are two types of infrared divergences in radiative corrections, 
soft and collinear. In the soft region and in the collinear region 
associated with the pion momentum $P_1$, the components of a loop 
momentum $l$ behave like
\begin{eqnarray}
l^\mu=(l^+,l^-,{\bf l}_T)\sim Q(\lambda,\lambda,\lambda)\;,\;\;\;\;
l^\mu\sim Q(1,\lambda^2,\lambda)\;,
\label{sog2}
\end{eqnarray}
respectively, where $\lambda$ is a small parameter. In both regions the 
invariant mass of the radiated gluon diminishes as $\lambda^2$, and 
the corresponding loop integrand may diverge as $1/\lambda^4$. As the 
phase space for loop integration vanishes like $d^4 l\sim \lambda^4$, 
logarithmic divergences are generated.

We identify the infrared divergences in the $O(\alpha_s)$
corrections \cite{AC,B,KMR} to Fig.~\ref{fig1}(a), which are shown in 
Fig.~\ref{fig2}. Self-energy correction to the internal quark, giving a
next-to-leading-order hard kernel, is not included. The collinear 
factorization formula is written as the convolution over the momentum 
fraction $\xi$,
\begin{eqnarray}
{\cal G}^{(1)}(x)&=& \sum_{i=a}^{e}{\cal G}^{(1)}_i(x)\;,
\nonumber\\
{\cal G}^{(1)}_i(x)&=&\int d\xi\phi_i^{(1)}(x,\xi){\cal H}^{(0)}(\xi)
+\phi^{(0)}(x){\cal H}_i^{(1)}(x)\;, 
\label{cf}
\end{eqnarray}
The above expression, with the $O(\alpha_s)$ distribution amplitudes 
$\phi_i^{(1)}(x,\xi)$ specified, defines the 
$O(\alpha_s)$ hard kernels ${\cal H}_i^{(1)}(x)$, which do not contain
collinear divergences. This procedure is referred to as matching the
effective theory to the full thoery
in the determination of Wislon coefficients in SCET.
It is now obvious why an arbitrary $x$ is
considered for the parton-level diagrams in Figs.~\ref{fig1} and 
\ref{fig2}: one can obtain the functional form of ${\cal H}_i^{(1)}(x)$ 
in $x$. 

Figures~\ref{fig2}(a)-\ref{fig2}(c) are the two-particle reducible 
diagrams with the additional gluon attaching the two valence quarks of 
the pion. It has been known that soft divergences cancel among these 
diagrams \cite{NL}. The 
reason for this cancellation is that soft gluons, being huge in 
space-time, do not resolve the color structure of the pion. Collinear 
divergences in Figs.~\ref{fig2}(a)-\ref{fig2}(c) do not cancel, since 
the loop momentum flows into the internal quark line in 
Fig.~\ref{fig2}(b), but not in Figs.~\ref{fig2}(a) and \ref{fig2}(c). 
To absorb the collinear divergences, one introduces a nonperturbative 
distribution amplitude. The factorization of the above diagrams is 
achieved by inserting the Fierz identity.
For example, one obtains, from Fig.~\ref{fig2}(b), the $O(\alpha_s)$
distribution amplitude,
\begin{eqnarray}
\phi^{(1)}_b(x,\xi)=\frac{ig^2C_F}{4P_1^+}
\int\frac{d^4l}{(2\pi)^4}{\bar q}(xP_1)
\frac{\gamma^\nu(x\not P_1-\not l)\gamma^5\not{\bar n}
({\bar x}\not P_1+\not l)\gamma_\nu} 
{(xP_1-l)^2({\bar x}P_1+l)^2l^2}q({\bar x}P_1)
\delta\left(\xi-x+\frac{l^+}{P_1^+}\right)\;,
\label{c2b}
\end{eqnarray}
with $C_F=4/3$ being a color factor. 
$\phi^{(1)}_b$ contains the collinear divergence, because
the integrand in Eq.~(\ref{c2b}) diverges as $1/\lambda^4$. The
dependences on $l^-$ and on $l_T$ in ${\cal H}^{(0)}$, being subleading
according to Eq.~(\ref{sog2}), have been neglected.

In the collinear region of Fig.~\ref{fig2}(d), the following 
approximation for part of the loop integrand holds,
\begin{eqnarray}
(\not P_2-x\not P_1)\gamma^\nu(\not P_2-x\not P_1+\not l)
&\approx& 2P_2^\nu\not P_2\;,
\label{ap00}
\end{eqnarray}
where the $l^-$ and $l_T$ terms, being power-suppressed compared to 
$P_2^-$, have been dropped.
The factorization of the collinear divergence from Figs.~\ref{fig2}(d) 
requires the further approximation for the product of the two internal 
quark propagators \cite{NL},
\begin{eqnarray}
\frac{2P_2^\nu}{(P_2-xP_1)^2(P_2-xP_1+l)^2}
\approx \frac{\bar n^\nu}{\bar n\cdot l}\biggl[\frac{1}{(P_2-xP_1)^2}
-\frac{1}{(P_2-xP_1+l)^2}\biggr]\;,
\label{pi}
\end{eqnarray}
which is an example of the Ward identity. Similarly, the power-suppressed 
terms, such as $l^2$ and $xP_1\cdot l$, have been neglected .
The numerator $2P_2^\nu$ comes from Eq.~(\ref{ap00}), and the factor
$\bar n^\nu/\bar n\cdot l$ is exactly the Feynman rule associated with a
Wilson line. Therefore, the appearence of the
Wilson line is a consequence of the Ward identity.

The first (second) term on the right-hand side of Eq.~(\ref{pi})
corresponds to the case without (with) the loop momentum $l$ flowing
through the hard kernel. Hence, the extracted $O(\alpha_s)$
distribution amplitude is written as 
\begin{eqnarray}
\phi^{(1)}_{d}(x,\xi)&=&
\frac{-ig^2C_F}{4P_1^+}\int\frac{d^4l}{(2\pi)^4}
{\bar q}(xP_1)\gamma^5\not{\bar n}
\frac{ {\bar x}\not P_1+\not l}{({\bar x}P_1+l)^2}
\gamma_\nu q({\bar x}P_1)
\frac{1}{l^2}\frac{\bar n^\nu}{\bar n\cdot l}
\nonumber\\
& &\times\left[\delta(\xi-x)-\delta\left(\xi-x+\frac{l^+}{P_1^+}\right)
\right]\;,
\label{c2e}
\end{eqnarray}
where the first (second) term in the bracket is associated with the 
first (second) term on the right-hand side of Eq.~(\ref{pi}). The
factorization of the distribution amplitude from Figs.~\ref{fig2}(e) is
performed in a similar way.

The above $O(\alpha_s)$ distribution amplitudes can be reproduced by the 
$O(\alpha_s)$ terms of the following nonlocal matrix element with the 
structure $\gamma_5\not{\bar n}$ sandwiched,
\begin{eqnarray}
\phi(x,\xi)=i\int\frac{dy^-}{2\pi }e^{-i\xi P_1^+y^-}
\langle 0|{\bar q}(y^-)\gamma_5\not{\bar n}
{\cal P}\exp\left[-ig\int_0^{y^-}dz \bar n\cdot A(z\bar n)\right]q(0)
|{\bar q}(xP_1) q({\bar x}P_1)\rangle\;.
\label{cw1}
\end{eqnarray}
The notation $\cal P$ means the path ordering for the Wilson line. The 
integral over $z$ contains two pieces: to generate the first term in 
Eq.~(\ref{c2e}), $z$ runs from 0 to $\infty$; to generate the second term, 
$z$ runs from $\infty$ back to $y^-$. The light-cone coordinate 
$y^-\not =0$ corresponds to the fact that the collinear divergences in 
Fig.~\ref{fig2} do not cancel. The Wilson line along the light 
cone collects collinear gluons in irreducible diagrams. By expanding the 
quark field ${\bar q}(y^-)$ and the Wilson line into powers of $y^-$, 
the above matrix element can be expressed as a series of covariant 
derivatives $(D^+)^n{\bar q}(0)$, $D=\partial-igA$, implying that 
Eq.~(\ref{cw1}) is gauge-invariant.

I then review the all-order proof of leading-twist collinear
factorization theorem for the process $\pi\gamma^*\to \gamma$, and
justify the definition of the parton-level distribution amplitude in
Eq.~(\ref{cw1}). The proof is performed in the covariant gauge, in which
collinear divergences also exist in two-particle irreducible diagrams.
It has been mentioned that factorization of a QCD
process in momentum, spin, and color spaces requires summation of many
diagrams, especially at higher orders. Hence, the diagram summation
must be handled in an elegant way. For this purpose, one employs the Ward
identity,
\begin{eqnarray}
l_\mu G^\mu(l,k_1,k_2,\cdots, k_n)=0\;,
\label{war}
\end{eqnarray}
where $G^\mu$ represents a physical amplitude with an external gluon
carrying the momentum $l$ and with $n$ external quarks carrying the
momenta $k_1$, $k_2$, $\cdots$, $k_n$. All these external particles are
on mass shell. The Ward identity can be easily derived by means of the
Becchi-Rouet-Stora (BRS) transformation \cite{BRS}.

Factorization theorem can be proved by induction. The factorization of the
$O(\alpha_s)$ collinear divergences associated with the pion has been
worked out in Eq.~(\ref{cf}).
Assume that factorization theorem holds up to $O(\alpha_s^N)$,
\begin{eqnarray}
{\cal G}^{(j)}(x)=\sum_{i=0}^{j}\int d\xi 
\phi^{(i)}(x,\xi){\cal H}^{(j-i)}(\xi)\;,\;\;\;\;
j=1,\cdots, N\;,
\label{cbn}
\end{eqnarray} 
where $\phi^{(i)}(x,\xi)$ is given by the $O(\alpha_s^{i})$ terms in the
perturbative expansion of Eq.~(\ref{cw1}), and ${\cal H}^{(j-i)}(\xi)$
stands for the $O(\alpha_s^{j-i})$ infrared-finite hard kernel. It will be
shown that the $O(\alpha_s^{N+1})$ diagrams ${\cal G}^{(N+1)}$
is written as the convolution of the $O(\alpha_s^N)$ diagrams
${\cal G}^{(N)}$ with the $O(\alpha_s)$ distribution amplitude by
employing the Ward identity in Eq.~(\ref{war}).

Look for the gluon in a complete set of $O(\alpha_s^{N+1})$ diagrams 
${\cal G}^{(N+1)}$, one of whose ends attaches the outer most vertex on 
the upper quark line in the pion. Let $\alpha$ denote the outer most
vertex, and $\beta$ denote the attachments of the other end of the 
identified gluon inside the rest of the diagrams. There are two 
types of collinear configurations associated with this gluon, depending on 
whether the vertex $\beta$ is located on an internal line with a momentum 
along $P_1$. The quark spinor adjacent to the vertex $\alpha$ is 
$q({\bar x}P_1)$. If $\beta$ is not located on a collinear line along 
$P_1$, the component $\gamma^+$ in $\gamma^\alpha$ and the minus 
component of the vertex $\beta$ give the leading contribution. If 
$\beta$ is located on a collinear line along $P_1$, $\beta$ can not be
minus, and both $\alpha$ and $\beta$ label the transverse components. 
This configuration is the same as of the self-energy correction to an 
on-shell particle.

According to the above classification, one decomposes the tensor
$g_{\alpha\beta}$ appearing in the propagator of the identified gluon as
\begin{eqnarray}
g_{\alpha\beta}=\frac{\bar n_{\alpha} l_\beta}{\bar n\cdot l}
-\delta_{\alpha T}\delta_{\beta T}
+\left(g_{\alpha\beta}-\frac{\bar n_{\alpha} l_\beta}{\bar n\cdot l}
+\delta_{\alpha T}\delta_{\beta T}\right)\;.
\label{dec}
\end{eqnarray}
The first term on the right-hand side extracts the first type of
collinear enhancements, since the light-like vector $\bar n_{\alpha}$
selects the plus component of $\gamma^\alpha$, and the dominant component
$l_{\beta=-}$ in the collinear region selects the minus component of the 
vertex $\beta$. The components $l_{\beta=+,T}$ do not change the 
collinear structure, since they are negligible in the numerators compared 
to the leading terms proportional to $P_1^+$ or $P_2^-$. This can be 
confirmed by contracting $l_\beta$ to Fig.~\ref{fig2}(d),
from which Eq.~(\ref{pi}) is obtained. The second term
extracts the second type of collinear enhancements. The last term does
not contribute a collinear enhancement due to the equation of motion for
the valence quark. We shall concentrate on the factorization of
${\cal G}_\parallel^{(N+1)}$ corresponding to the first term on the
right-hand side of Eq.~(\ref{dec}), and the factorization associated
with the second term can be included following the procedure
in \cite{NL}.

\begin{figure}[t!]
\begin{center}
\epsfig{file=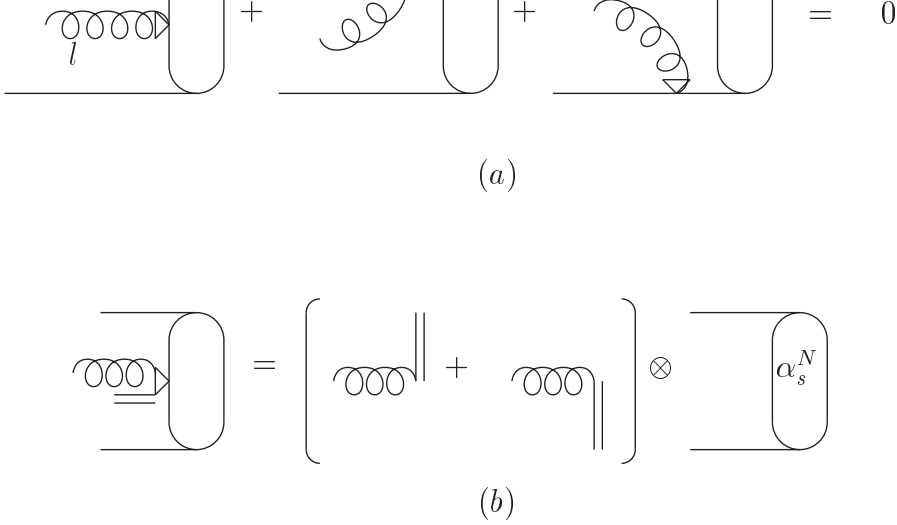, height=3.0in}
\caption{(a) Ward identity. (b) Factorization of ${\cal G}^{(N+1)}$.
\label{fig3}}
\end{center}
\end{figure}

The contraction of $l_\beta$ hints the application of the Ward identity
in Eq.~(\ref{war}) to the case with two external on-shell quarks.
Those diagrams with Figs.~\ref{fig2}(a) and \ref{fig2}(b) as the
$O(\alpha_s)$ subdiagrams are excluded from the set of
${\cal G}_\parallel^{(N+1)}$ as discussing the first type of collinear
configurations, since the identified gluon does not attach a line
parallel to $P_1$. Consider the physical amplitude, in which the two
on-shell quarks and the on-shell gluon carry the momenta
${\bar \xi} P_1$, $x P_1$ and $l$, respectively. Figure~\ref{fig3}(a),
describing the Ward identity, contains a complete set of contractions of 
$l_\beta$, since the second and third diagrams have been added back. The
second and third diagrams in Fig.~\ref{fig3}(a) lead to
\begin{eqnarray}
& &l_\beta \frac{1}{{\bar \xi}\not P_1-\not l}\gamma^\beta
q({\bar \xi} P_1)
=\frac{1}{{\bar \xi}\not P_1-\not l}(\not l-{\bar \xi}\not P_1 +
{\bar \xi}\not P_1)q({\bar \xi} P_1)
=-q({\bar \xi} P_1)\;,
\nonumber\\
& &l_\beta{\bar q}(xP_1)\gamma^\beta\frac{1}{x\not P_1-\not l}
=-{\bar q}(xP_1)\;,
\label{ide2}
\end{eqnarray}
respectively. The terms $q({\bar \xi} P_1)$ and ${\bar q}(xP_1)$ at the 
ends of the above expressions correspond to the $O(\alpha_s^N)$ diagrams.

Figure~\ref{fig3}(b) shows that the diagrams
${\cal G}_{\parallel}^{(N+1)}$ associated with the first term in
Eq.~(\ref{dec}) are factorized into the convolution of the parton-level
$O(\alpha_s^N)$ diagrams ${\cal G}^{(N)}$ with the $O(\alpha_s)$
collinear piece extracted from Fig.~\ref{fig2}(d). The factor
$\bar n_{\alpha}/\bar n\cdot l$ from the collinear replacement in
Eq.~(\ref{dec}) is exactly the Feynman rule associated with the Wilson
line in the direction of $\bar n$, represented by the double line.
The first diagram means that the gluon
momentum does not flow into ${\cal G}^{(N)}$, while in the second diagram
the gluon momentum does. The similar reasoning applies to the identified
gluon, one of whose ends attaches the outer most vertex of the lower
antiquark line. Substituting Eq.~(\ref{cbn}) into
${\cal G}^{(N)}(\xi)$ on the right-hand side of Fig.~\ref{fig3}(b), and
following the procedure in \cite{NL}, one arrives at
\begin{eqnarray}
{\cal G}^{(N+1)}(x) =\sum_{i=0}^{N+1} \int d\xi 
\phi^{(i)}(x,\xi) {\cal H}^{(N+1-i)}(\xi)\;,
\label{cf2}
\end{eqnarray}
with the infrared-finite $O(\alpha_s^{N+1})$ hard kernel
${\cal H}^{(N+1)}$. Equation (\ref{cf2}) implies that all the collinear
enhancements in the process $\pi\gamma^* \to \gamma$ can be factorized
into the distribution amplitude in Eq.~(\ref{cw1}) order by order. 

Equation~(\ref{cw1}) plays the role of an infrared regulator for
the parton-level diagrams. A hard kernel ${\cal H}(x,Q^2,\mu)$ can then 
be regarded as the regularized parton-level diagrams, where
the dependences on the momentum transfer $Q$ and on the factorization 
scale $\mu$ have been made explicit. Different choices of $\mu$ 
correspond to different factorization schemes. Since the parton-level 
diagrams with on-shell external particles and Eq.~(\ref{cw1}) are both
gauge-invariant, the hard kernel is gauge-invariant. After determining
the gauge-invariant infrared-finite hard kernel, one
convolutes it with the physical two-parton pion distribution amplitude,
whose all-order gauge-invariant definition is given by
\begin{eqnarray}
\phi_\pi(x,\mu)=i\int\frac{dy^-}{2\pi }e^{-ixP_1^+y^-}
\langle 0|{\bar q}(y^-)\gamma_5\not{\bar n}
{\cal P}\exp\left[-ig\int_0^{y^-}dz\bar n\cdot\cdot A(z\bar n)\right]
q(0)|\pi(P_1)\rangle\;.
\label{cw1p}
\end{eqnarray}
The valence-quark state $|{\bar q}(xP_1) q({\bar x}P_1)\rangle$
has been replaced by the pion state $|\pi(P_1)\rangle$, and the pion decay 
constant $f_\pi$ has been omitted. Equation (\ref{cw1p}) can also be
derived in SCET as argued in the next subsection. The 
$\pi\gamma^\ast \to\gamma$ scattering amplitude is then expressed as
the convolution over the parton momentum fraction $x$,
\begin{eqnarray}
{\cal M}(Q^2)=\int_0^1 dx\phi_\pi(x,\mu){\cal H}(x,Q^2,\mu)\;.
\end{eqnarray}
Hence, predictions derived from collinear factorization theorem
are gauge-invariant and infrared-finite.

\subsection{\it Soft-Collinear Effective Theory \label{sec:sof}}

Final-state hadrons in exclusive $B$ meson decays may carry energy $E$ of
$O(m_B)$, $m_B$ being the $B$ meson mass, which is much larger than 
$\Lambda_{\rm QCD}$. These processes can be analyzed in the collinear 
factorization framework discussed in the previous subsection. To study 
the collinear factorization at the operator and Lagrange level, SCET has 
been developed \cite{bfl,bfps,cbis,bpssoft}. After integrating out 
short-distance fluctuations characterized by the invariant mass 
$p^2\gg(E\lambda)^2$, which appear in Wilson coefficients, long-distance 
fluctuations are then described by new effective degrees of freedom. 
SCET then exhibits symmetries
in the large energy limit, such as the reduction of spin structures,
helicity constraints, and collinear gauge invariance, which apply to the
new effective fields. Power corrections in SCET are included in terms of
the small parameter $\lambda=\Lambda_{\rm QCD}/E$ (or
$\lambda=\sqrt{\Lambda_{\rm QCD}/E}$) 
\cite{Chay:2002vy,bpspc,Beneke:2002ph}.
For a recent review, refer to \cite{S02}.


The effective fields contain collinear quarks and gluons
($\xi_{n,p}$, $A_{n,q}^\mu$), massless soft quarks and gluons ($q_{s}$,
$A_{s}^\mu$), and massless ultrasoft (usoft) quarks and gluons ($q_{us}$, 
$A_{us}^\mu$). The collinear fields, labelled by the light cone direction
$n=(1,0,{\bf 0}_T)$ and their momentum $p$,
come from the phase redefinitions,
\begin{eqnarray}
\phi_n(x) =\sum_p e^{-ip\cdot x}\phi_{n,p}(x)\;.
\end{eqnarray}
Derivatives on the collinear fields, $\partial^\mu\phi_{n,p}(x)
\sim (E\lambda^2) \phi_{n,p}(x)$, pick up only the small scale. The
large momenta are picked up by introducing the label operator,
$\bnP \xi_{n,p}=(\bn\!\cdot\!  p)\xi_{n,p}$. Similarly, the operators
$\cP_T^\mu$ and $\cP^\mu$ are defined to pick up the $O(\lambda)$ labels
of the collinear and soft fields, respectively. In the discussion 
in Sec.~\ref{sec:all} the
usoft fields can be regarded as the leftover pieces with the collinear
or soft dynamics being factorized out. That is, the usoft fields,
due to their slow variation, appear as the background fields to
the collinear or soft ones.

The collinear Wilson line $W[\bn\cdot A_{n,q}]$ and
soft Wilson line $S_n[n\cdot A_s]$ are induced to preserve gauge
invariance \cite{bfps,bpssoft}. For the former, the explicit expression
is given by
\begin{eqnarray}
W &=& \sum_{\rm perms}
\exp\Big[-\frac{g}{\bnP}\: \bn\cdot A_{n,q}(x) \Big] \;,
\end{eqnarray}
which is equivalent to that in Eq.~(\ref{cw1p}) derived from perturbation
theory. The meaning of a collinear Wilson line in the definition of the
pion distribution amplitude has been emphasized in Sec.~\ref{sec:all}.
The usoft Wilson line $Y_n[n\cdot A_{us}]$ is introduced by
the further field redefinitions $\xi_{n,p}=Y_n \xi^{(0)}_{n,p}$ and
$A_{n,q}=Y_n A_{n,q}^{(0)} Y_n^\dagger$. 

Assuming the action for the
kinetic terms in SCET to be of $O(\lambda^0)$, the scaling of each
effective field in $\lambda$ can be defined straightforwardly. The power
counting rules for the momenta, fields, momentum label operators, and
collinear, soft and usoft Wilson lines are summarized in Table~\ref{pc}
\cite{bfl,bfps}. It is found that the scaling of the collinear and soft
momenta in SCET is the same as that defined in Eq.~(\ref{sog2}). The usoft
momentum in the framework based on perturbation theory will appear as
discussing the factorization of soft divergence from exclusive $B$
meson decays in Sec.~\ref{sec:b}.

\begin{table}[t!]
\begin{center}
\begin{tabular}{|c|l|c|clc||}
\hline
  Type & Momenta $(+,-,T)$\hspace{0.4cm} 
   & \hspace{0.2cm}Field Scaling  \hspace{0.cm} 
   & \hspace{0.2cm}Operators\hspace{0.2cm} \\ 
   \hline
  collinear & $p^\mu\sim (1,\lambda^2,\lambda)$ \hspace{0.2cm}
& \hspace{0.2cm} $\xi_{n,p}\sim \lambda$ & $\bnP$, $W$ $\sim\lambda^0$ \\
  && ($A_{n,p}^+$, $A_{n,p}^-$, $A_{n,p}^T$) $\sim$ 
(1,$\lambda^2$,$\lambda$) & \hspace{0.4cm} $\cP_T^\mu\sim \lambda$ \\
   \hline
  soft &  $p^\mu\sim (\lambda,\lambda,\lambda)$ \hspace{0.25cm}
   & \hspace{0.8cm} $q_{s,p}\sim \lambda^{3/2}$ & \hspace{0.5cm} 
   $S_n \sim \lambda^0$ \\
  & & \hspace{0.2cm} $A_{s,p}^\mu\sim \lambda$ & \hspace{0.3cm} 
    $\cP^\mu\sim \lambda$ \\ \hline 
  usoft &  $k^\mu\sim (\lambda^2,\lambda^2,\lambda^2)$
   & \hspace{0.5cm} $q_{us}\sim \lambda^3$ & \hspace{0.4cm}
   $Y_n\sim\lambda^0$ \\
  & & \hspace{0.4cm} $A_{us}^\mu\sim \lambda^2$  \\
\hline
\end{tabular}
\end{center}
\caption{\setlength\baselineskip{12pt} Power counting rules for momenta,
fields, momentum label operators ($\bnP,\cP_T^\mu$, $\cP^\mu$),
and collinear, soft and usoft Wilson lines ($W$, $S_n$, $Y_n$) in SCET.
\label{pc}
\setlength\baselineskip{18pt}}
\end{table}

The leading-order Lagrangians for (u)soft light quarks and gluons are the
same as in QCD. For heavy quarks $h_v$ labelled by the velocity $v$, we
have the heavy quark effective theory (HQET) Lagrangian \cite{HQET},
\begin{equation}
{\cal L}=\bar h_v\,iv\cdot D\, h_v+\ldots\;.
\label{lhe}
\end{equation}
The collinear quark Lagrangian can be expanded as
\begin{eqnarray}
 {\cal L}_c &=&  {\cal L}_c^{(0)} + {\cal L}_c^{(1)} + {\cal L}_c^{(2)} 
   + \ldots \,.
\label{lco}
\end{eqnarray}
The first three terms are 
\begin{eqnarray}
{\cal L}_{c}^{(0)} 
 &=&   \bar\xi_{n,p'}\:  \bigg\{  i\, n\cdot  D\!+\! g n\cdot A_{n,q} 
  + \Big( \SppP\! + g \Aslash_{n,q}^T\Big)\, W \frac{1}{\bnP} 
W^\dagger\,
   \Big( \SppP\!  + g \Aslash_{n,q'}^T\Big) \bigg\}
  \frac{\bnslash}{2}\, \xi_{n,p}  \,,\nn \\
{\cal L}_{c}^{(1)}
 &=&  \bar\xi_{n,p'}\bigg\{ i\Dslash_T W \frac{1}{\bnP} W^\dagger\,
   \big(\SppP\!+g\Aslash_{n,q'}^T\big) 
  + \Big( \SppP\! + g \Aslash_{n,q}^T\Big)\, W \frac{1}{\bnP} 
W^\dagger\,
   i\Dslash_T \bigg\} \frac{\bnslash}{2}\, \xi_{n,p} \,, \nn \\
{\cal L}_c^{(2)} 
 &=& \bar \xi_n \left\{ i\Dslash_T W\frac{1}{\bnP} W^\dagger 
i\Dslash_T 
 -\Big( \SppP\! + g \Aslash_{n,q}^T\Big)
W \frac{1}{\bnP} W^\dagger
(\bn\cdot iD) W \frac{1}{\bnP} W^\dagger 
\Big( \SppP\! + g \Aslash_{n,q}^T\Big)
\right\}
\frac{\bnslash}{2}\xi_n \,,
\label{Lc}
\end{eqnarray}
with $D^\mu = \partial^\mu -i g A^\mu_{us}$, where ${\cal L}_c^{(0)}$ 
gives the order $O(\lambda^0)$ interactions \cite{bfps,cbis}, and the 
expressions of ${\cal L}_c^{(1)}$ and of ${\cal L}_c^{(2)}$
were derived in \cite{Chay:2002vy} and in \cite{mmps}, respectively.
For the mixed effects, which are power-suppressed, the  
usoft-collinear Lagrangian has been derived up to
$O(\lambda^2)$ \cite{PS02}. Note that the results in \cite{PS02} represent
an expansion of SCET in the hybrid momentum-position space. A manifestly
gauge-invariant expansion in the position space has been derived
\cite{BF02}, in which each operator has a homogeneous power counting in
$\lambda$.

\begin{figure}[t!]
\begin{center}
 \epsfig{file=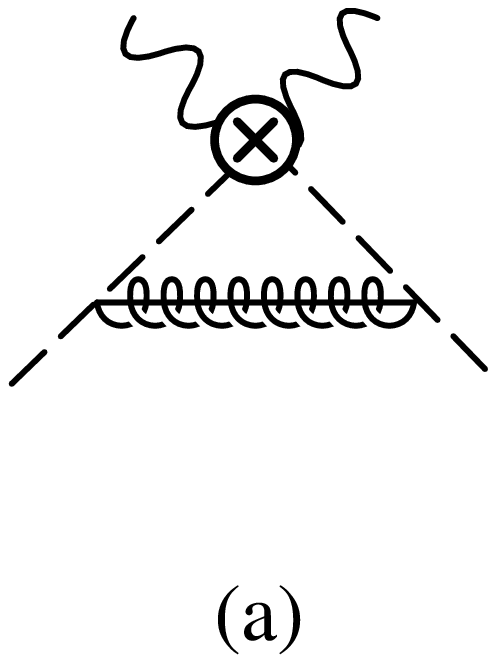,height=2.0in} 
 \hspace{2.5cm} 
 \epsfig{file=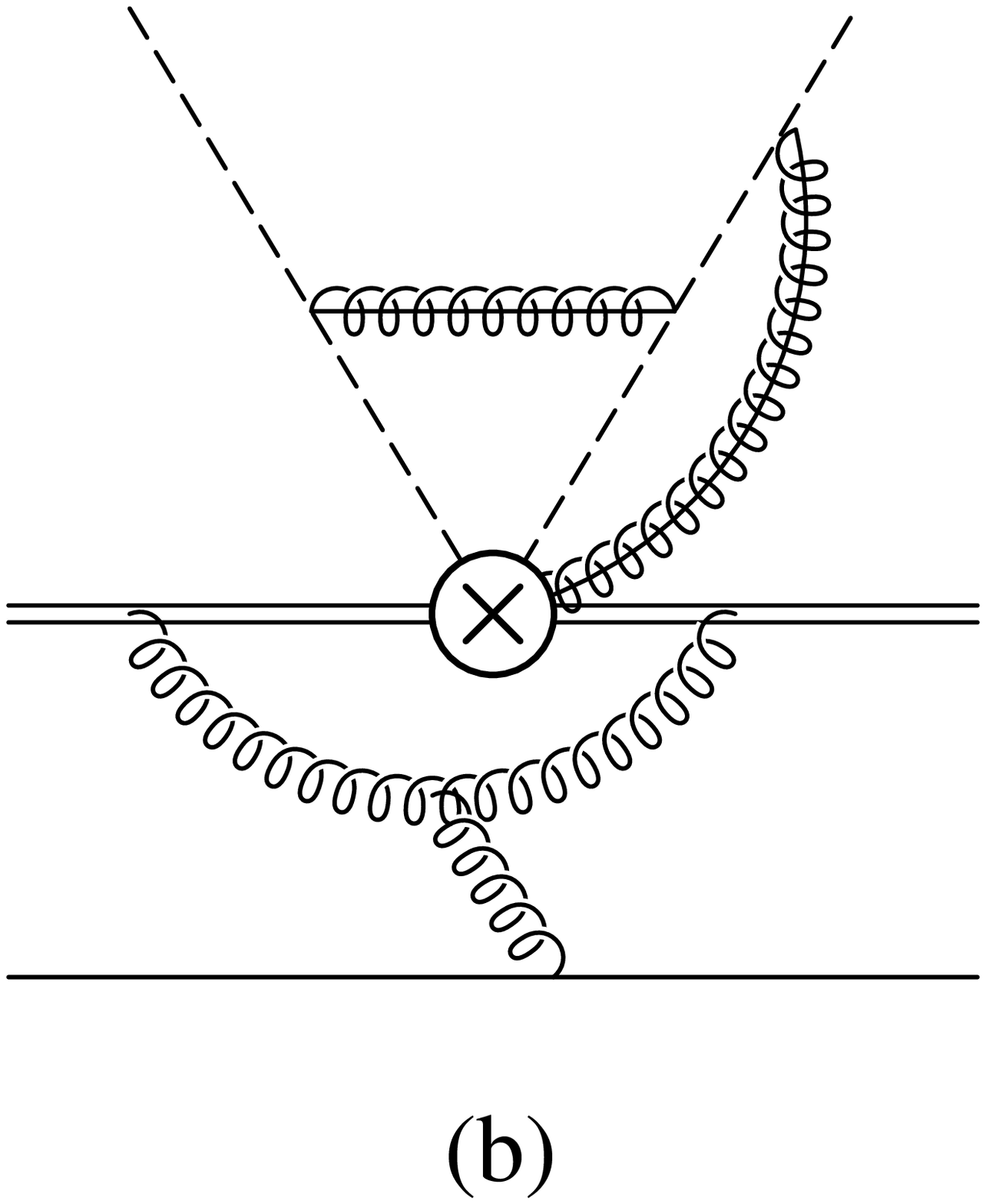,height=2.0in} 
\caption{Diagrams for comparing the power counting in SCET and in
perturbation theory for the processes (a) DIS
and (b) $B\to D\pi$, respectively. Dashed lines are
collinear quarks, double solid lines are usoft or soft heavy quarks, and
the single solid lines are soft light quarks.  Gluons with a line
through them are collinear, while those without a line are soft or usoft.
\label{fig4}}
\end{center}
\end{figure}

After defining the power counting rules, I explain how to construct
collinear factorization theorem at the operator and Lagrange level.
The idea is to start with an operator relevant for a high-energy QCD
process, which is characterized by some power of $\lambda$. Draw the
diagrams based on this operator and the effective Lagrangians in
Eqs.~(\ref{lhe}) and (\ref{lco}). The power of a diagram is the sum of 
those for the loop measures, propagators, vertices, and external lines. 
Note that powers in $\lambda$ shift between the propagators and vertices in
different gauges. I will adopt the Feynman gauge the same as in
Sec.~\ref{sec:all} in order to compare the formalism in SCET and in
perturbation theory. Those diagrams, whose contributions scale like the
power the same as of the specified operator, contribute to the 
corresponding matrix element.

Consider the diagrams in Fig.~\ref{fig4}. The one-loop diagram in
Fig.~\ref{fig4}(a) involves the leading-power operator for deeply
inelastic scattering (DIS) \cite{bfprs},
\begin{eqnarray} \label{DIS}
 O_{\rm DIS} = \frac{1}{Q}\: \bar\xi_{n,p'} W\: \frac{\bnslash}{2}\: 
  C(\bnP_+,\bnP_-,Q,\mu)\: W^\dagger\, \xi_{n,p}\,,
\label{diso}
\end{eqnarray}
with $\bnP_\pm=\bnP^\dagger \pm \bnP$, and $Q$ representing the momentum
transfer from the virtual photon. The Wilson coefficient $C$,
equivalent to the hard kernel, absorbs short-distance dynamics.
When taking the proton matrix element of $O_{\rm DIS}$, the $\bnP_+$
dependence of the Wilson coefficient leads to a convolution with 
parton distribution functions, which is the collinear factorization
formula for DIS. Because of $\xi\sim\lambda$ and
$W\sim\lambda^0$, the operator in Eq.~(\ref{diso}) scales as $\lambda^2$.

The two collinear gluon vertices come from the leading-power interactions
of ${\cal L}_c^{(0)}$ in Eq.~(\ref{Lc}).
Hence, Fig.~\ref{fig4}(a) is characterized by the power,
\begin{eqnarray} \label{dpc1}
 (\lambda)^2 \Big[  \lambda^4 \times \Big(\frac{1}{\lambda^2}\Big)^3 
   \times \lambda^2 \Big] = \lambda^2 \;.
\end{eqnarray}
The factor outside the square brackets is for the external fields,
the first term in the square brackets counts the collinear loop measure,
and the second factor counts the three collinear propagators following
Eq.~(\ref{sog2}) or Table~\ref{pc}. The last factor in the bracket is the
power of momentum in the quark-quark-gluon vertices in
${\cal L}_c^{(0)}$, which are either $(T, T) \sim
(\lambda, \lambda)$ or $(+, -) \sim (\lambda^0, \lambda^2)$ in Feynman
gauge (see the all-order proof of the collinear factorization theorem 
in Sec.~\ref{sec:all}). Equation~(\ref{dpc1}) indicates that 
Fig.~\ref{fig4}(a) is of the
same power as the operator $O_{\rm DIS}$, and that nonperturbative
collinear gluon exchanges of this type contribute to the leading-twist
parton distribution function defined by $\langle\bar\xi_{n,p'} W
(\bnslash/2) W^\dagger\xi_{n,p}\rangle$.

It is easy to see that the above power counting is similar to
that for Fig.~\ref{fig2}(b) in Sec.~\ref{sec:all}: if one drops the power
associated with the external collinear quark fields, Eq.~(\ref{dpc1}),
being of $O(\lambda^0)$, corresponds to a logarithmic divergence, which 
should be absorbed into the distribution function. The strategies of the 
two approaches are compared below. In perturbation theory one starts with
Feynman diagrams in full QCD. Look for the leading region of the
loop momentum defined by Eq.~(\ref{sog2}), in which one makes the power 
counting of the Feynman diagrams. It can be found that the approximate
loop integral in the leading region is represented by a diagram
of the type of Fig.~\ref{fig4}(a) at $O(\alpha_s)$. The distribution 
amplitude in Eq.~(\ref{cw1}) then collects this type of diagrams to all
orders. In SCET one first constructs the various effective degrees
of freedom describing infrared dynamics and the interactions
in Eqs.~(\ref{lhe}) and (\ref{lco}), and defines their powers. 
Draw the diagrams based on the effective theory and then
make the power counting. It can be shown that the diagrams of the type of
Fig.~\ref{fig4}(a) build up the leading-twist distribution amplitude in 
Eq.~(\ref{cw1}). That is, one arrives at Fig.~\ref{fig4}(a) through 
approximating loop integrals in the full theory in the former, but does
at the operator and Lagrange level in the latter.
Therefore, the derivations of collinear factorization 
theorem from both approaches are equivalent.


At leading power, the external operator for the nonleptonic decay
$B\to D\pi$ in SCET is given by \cite{cbis,bps}
\begin{eqnarray}\label{Q08fact}
 O_{\bf \{0,8\}} &=&  \Big( \bar h_{v'}^{(c)}S \Gamma_h \:\{{\bf 1,T^A}\} 
  S^\dagger h_v^{(b)} \Big) \Big( \bar \xi_{n,p'}^{(d)} W 
  C_{\bf \{0,8\}}(\bnP,\bnPd) 
  \, \Gamma_\ell\:\{{\bf 1,T^A}\} \, W^\dagger\,
  \xi_{n,p}^{(u)} \Big) \,,
\label{bdp}
\end{eqnarray}
where $\Gamma_{h,\ell}$ are the spin structures from the
Fierz identity in Eq.~(\ref{fierz}). From Table~\ref{pc}, 
$O_{\bf \{0,8\}}\sim\lambda^5$ is the base $\lambda$-dimension for this
process. In the three-loop diagram in Fig.~\ref{fig4}(b) all interactions
are taken from the lowest-power Lagrangian ${\cal L}_c^{(0)}$. The direct
power counting gives $\lambda^5$:
\begin{eqnarray} \label{dpc3}
  \lambda^5\: 
 \Big\{ \frac{(\lambda^{3/2})^2}{\lambda^2}\Big\} 
 \ \Big[\lambda^4\times\frac{1}{(\lambda)^2\,(\lambda^2)^2}
 \times\lambda\Big]
 \ \Big[ (\lambda^4)^2 \times \Big(\frac{1}{\lambda^2}\Big)^5
 \times \lambda^2
  \Big] = \lambda^5 \,.
\end{eqnarray}
Here the first term counts the dimension of the external heavy quark
fields and collinear quark fields. The term in curly brackets counts
powers of $\lambda$ from the light soft spectator quark lines and the
soft gluon propagator that does not participate in a loop. The factors
in the first square bracket are the measure, propagators, and vertices
for the soft loop. In the final square bracket the $\lambda$ factors
are given for the measures, propagators, and vertices in the two collinear
loops. The above power counting implies that Fig.~\ref{fig4}(b)
contributes to the leading collinear factorization formula of the
$B\to D\pi$ decay.

Note that collinear gluons do not attach the heavy quarks, which
can not remain on-shell after emitting or absorbing collinear gluons
\cite{bfl}. Equivalently, no collinear divergence is associated with a 
massive particle in perturbation theory \cite{NL}. Collinear
gluons are not emitted by the soft spectator quarks in the $B$ and $D$
mesons either, since they do not produce pinched singularities \cite{NL}.
The collinear divergences associated with the pion have been collected by
the Wilson line $W$ in Eq.~(\ref{bdp}). Nonfactorizable soft gluons
decouple from the pion due to the argument of color transparency
\cite{transparency} mentioned in Sec~\ref{sec:all}.
They contribute only at the subleading power.

\begin{figure}[t!]
\begin{center}
 \epsfig{file=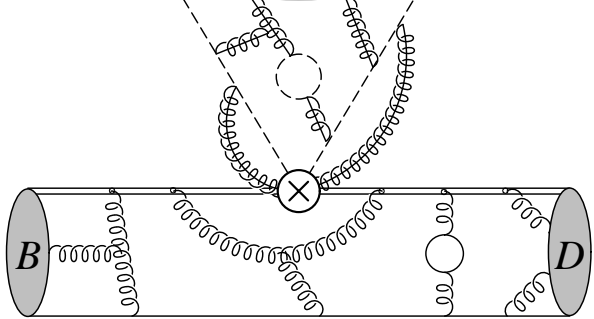,height=2.0in} 
\caption{Collinear factorization of the $B\to D\pi$ decay in SCET.}
\label{fig:bdpi}
\end{center}
\end{figure}

Following the above explanation, the proof of the collinear factorization
theorem for the decay $B\to D\pi$ in SCET is trivial \cite{bps}:
the leading-power diagrams
involve soft gluons exchanged among the quarks in the $B$ and $D$ mesons,
and collinear gluons exchanged between the quarks in the pion.
The former give the $B\to D$ form factor $F^{BD}$, defined as the matrix 
element of the frist piece of $O_{\bf \{0,8\}}$. The latter lead to the 
pion distribution amplitude $\phi_\pi(x)$, defined as the matrix
element of the second piece with the Wilson coefficient $C_{\bf \{0,8\}}$
being excluded. The above discussion indicates that the pion distribution 
amplitude in Eq.~(\ref{cw1p}) can be constructed in the framework of SCET. 
One simply identifies the correspondence of the quark fields $q$ in 
Eq.~(\ref{cw1p}) and the collinear effective fields $\xi_{n,p}$ in 
Eq.~(\ref{bdp}), which are equivalent in the collinear region, and choose 
$\Gamma_\ell$ as $\gamma_5\not{\bar n}$. The two collinear Wilson
lines $W$ correspond to the two pieces of path-ordered exponential
in Eq.~(\ref{cw1p}). The contributions from the diagrams in 
Fig.~\ref{fig2} are of the same power in $\lambda$ as of the 
effective current $\bar \xi_{n,p'}^{(d)} W \Gamma_\ell W^\dagger
\xi_{n,p}^{(u)}$ in Eq.~(\ref{bdp}).

Hence, only the diagrams of the type 
shown in Fig.~\ref{fig:bdpi} exist at leading power in SCET.
The corresponding collinear factorization formula is then written as
\begin{eqnarray} \label{fact1}
\langle D^{(*)} \pi | H_w  | B\rangle = N { F^{B\to D^{(*)}}(0)}\ 
 \int_0^1 d x\ { T(x,E,\mu)}\ 
 {\phi_\pi(x,\mu)} + \cdots \,,
\label{bdpi}
\end{eqnarray}
where $H_w$ is the weak effective Hamiltonian, $T(x,E,\mu)$ a calculable 
hard kernel, and the ellipses denote 
terms that vanish faster than the leading term as the pion energy 
$E\to\infty$. The dependences on the renormalization scale $\mu$ cancel
between $T$ and $\phi_\pi$. The convolution in $x$ is a consequence of
the non-commutative nature of the Wilson coefficients and the effective
fields. The hard kernel can be determined by matching the effective
theory to the full theory. Equation~(\ref{fact1}) was proposed in
\cite{PW}, proven to two-loops in \cite{BBNS2}, and proven to all orders
in $\alpha_s$ in \cite{bps}.

\subsection{\it Jet Function \label{sec:thr}}

The application of collinear factorization theorem to exclusive $B$
meson decays has encountered a difficulty: the evaluation of the
$B\to\pi$ transition form factors suffers the singularities
from the end point of a momentum fraction $x\to 0$ \cite{SHB,BD,BF}.
These singularities are logarithmic and linear in the leading-twist
(twist-2) and next-to-leading-twist (twist-3) contributions, respectively.
On the other hand, the double logarithms $\alpha_s\ln^2 x$ from radiative
corrections were observed in the semileptonic decay
$B\to\pi l\nu$ \cite{ASY} and in the radiative decay $B\to\gamma l\nu$
\cite{KPY}. It has been argued that when the end-point region is
important, these double logarithms are not small expansion parameters,
and should be organized into a quark jet function systematically
in order to improve perturbative expansion. The procedure is referred
to as threshold resummation \cite{L5}. The resultant Sudakov factor is
found to vanish quickly as $x\to 0$. It turns out that in a
self-consistent perturbative evaluation of the $B\to\pi$ form factors,
where the original factorization formula is convoluted with the jet
function, the end-point singularities do not exist \cite{L5}.

I take the radiative decay $B\to\gamma l\nu$ as an example.
The momentum $P_1$ of the $B$ meson and the
momentum $P_2$ of the outgoing on-shell photon are parametrized as
\begin{eqnarray}
P_1=\frac{m_B}{\sqrt{2}}(1,1,{\bf 0}_T)\;,\;\;\;\;
P_2=\frac{m_B}{\sqrt{2}}(0,\eta,{\bf 0}_T)\;,
\label{bmpp}
\end{eqnarray}
where $\eta$ denotes the energy fraction carried by the photon.
Assume that the light spectator quark in the $B$ meson carries the
momentum $k$. Consider the kinematic region with small $q^2$,
$q=P_1-P_2$ being the lepton pair momentum, {\it i.e.}, with large
$\eta$, where perturbative expansion is reliable.
The four components of the spectator quark
momentum $k$ are of the same order as $\Lambda$.
Here $\Lambda$ represents a hadronic scale, such
as the $B$ meson and $b$ quark mass diffetrence, $m_B-m_b$.
In collinear factorization, only the plus
comonent $k^+$ is relevant through the inner product $k\cdot P_2$.

The lowest-order diagrams for the $B\to \gamma l\nu$ decay are
displayed in Fig.~\ref{fig1}, but with the upper quark (virtual photon)
replaced by a $b$ quark ($W$ boson). It is easy to observe that
Figs.~\ref{fig1}(a) and \ref{fig1}(b) scale like $1/(\Lambda m_B)$
and $1/m_B^2$, respectively, implying that Fig.~\ref{fig1}(b) is
power-suppressed. Below I will concentrate on Fig.~\ref{fig1}(a).
According to the leading-twist collinear factorization theorem discussed
in Sec.~\ref{sec:all}, the $B\to\gamma l\nu$ decay amplitude is written 
as the convolution of a hard kernel $H(x,\eta,m_B,\mu)$ with the $B$ 
meson distribution amplitude $\phi_+(x,\mu)$ over the parton momentum 
fraction $x=k^+/P_1^+$ \cite{DS},
\begin{eqnarray}
A(\eta,m_B)=\int dx \phi_+(x,\mu)H(x,\eta,m_B,\mu)\;,
\label{bglv}
\end{eqnarray}
This expression has been derived in the framework of SCET 
\cite{LPW,BHL}, in which the hard kernel was further factorized 
into $H=H_hH_i$ with the function $H_h$ and $H_i$ being characterized
by the scale $m_b$ and $\sqrt{\Lambda m_b}$, respectively. Note that
the jet function in \cite{LPW,BHL}, referred to $H_i$, differs from
that in \cite{L5}.

Equation (\ref{bglv}) is appropriate for the region with 
$k^+\sim O(\Lambda)$, in which the only infrared divergences are
the soft ones absorbed into the $B$ meson distribution amplitude
\cite{NL}. Near the end point $k^+\sim O(\Lambda^2/m_B)$, the internal 
quark in Fig.~\ref{fig1}(a) carries a 
large momentum $P_2-k$ with its invariant mass vanishing like 
$(P_2-k)^2=-2xP_1\cdot P_2\sim O(\Lambda^2)$. This kinematics is 
similar to the threshold region of DIS with
the Bjorken variable $x_B\to 1$, where the scattered quark also carries a 
large momentum and possesses a small invariant mass 
$(1-x_B)s$, $s$ being the center-of-mass energy. In this region the 
scattered quark produces a jet of particles, to which the radiative
corrections contain additional collinear divergences. Hence, a jet function
needs to be introduced into the collinear factorization formula for DIS
\cite{G}. Similarly, a jet function has been incorporated into the
factorization of direct photon production at a large photon transverse
momentum (threshold) \cite{LSV}. Here a jet function is associated with
the internal quark near the end point of the momentum
fraction involved in the decay $B\to\gamma l\nu$.

An additional collinear divergence from the loop momentum parallel to 
$P_2$ appears in the higher-order correction to the weak decay vertex
shown in Fig.~\ref{fig2}(d). This divergence can be extracted by
replacing the $b$ quark line by an eikonal line in the direction of $n$.
The factorization of the fermion flow is achieved by inserting the Fierz
identity in Eq.~(\ref{fierz}), in which the first and last terms
contribute in the combined structure
$I_{ij}I_{lk}\to I_{ik}(\not n\not{\bar n})_{lj}/4$.
Assigning the identity matrix $I$ to the trace for the
hard kernel, one obtains Fig.~\ref{fig1}(a). The matrix 
$\not n\not{\bar n}/4$ then leads to the loop integral \cite{L5}, 
\begin{eqnarray}
J_{\parallel}^{(1)}&=&-ig^2C_F\int\frac{d^4 l}{(2\pi)^4}
\frac{1}{4}tr\left[\not n\not{\bar n}
\gamma_\beta\frac{\not P_2-\not k+\not l}
{(P_2-k+l)^2}\right]\frac{n^\beta}{n\cdot l l^2}\;,
\nonumber\\
&=&-\frac{\alpha_s}{4\pi}C_F\ln^2 x+\cdots\;
\label{j1c}
\end{eqnarray}
which are the same as those derived in \cite{KPY,DS,BHL}. The correction to 
the photon vertex in Fig.~\ref{fig2}(e) contains only the single logarithm
$\alpha_s\ln x$, since the phase space of the loop momentum is 
restricted to $0<l^+<k^+\sim O(\Lambda^2/m_B)$. The self-energy correction 
to the virtual light quark deos also. For the explicit expressions for
the $O(\alpha_s)$ corrections from Fig.~\ref{fig2}, refer to \cite{DS}.

The all-order factorization of the jet function from the decay 
$B\to\gamma l\nu$ has been proved following the procedure in 
Sec.~\ref{sec:all}, which provides a solid theoretical 
ground for the modified formalism appropriate for the end-point region. 
The jet function $J(x)$ is defined via
\begin{eqnarray}
J(x)\bar q(P_2-k)\equiv
\langle q(P_2-k)|{\bar q}(0)\frac{1}{4}\not n\not{\bar n}
\exp\left[-ig\int_{-\infty}^0dzn \cdot A(zn)\right]|0\rangle\;.
\label{pwj}
\end{eqnarray}
The spinor $q(P_2-k)$ is associated with the internal quark, through 
which the momeutm $P_2-k$ flows. It is then understood from 
Eq.~(\ref{pwj}) that the jet function is universal.

I then discuss threshold resummation of the double logarithms
$\alpha_s\ln^2 x$ in the covariant gauge $\partial\cdot A=0$, which have
been collected into the jet function to all orders. Threshold resummation
for inclusive QCD processes has been studied intensively \cite{S0,CT}.
Here I will adopt the framework developed in \cite{L1,L2}, which has
been shown to lead to the same results as in \cite{S0,CT}. First, allow
the vector $n$ to contain a (small) minus component $n^-$. This
modification, regularizing the collinear pole, extracts the double
logarithm as shown in Eq.~(\ref{j1c}). The definition in
Eq.~(\ref{pwj}) involves three variable vectors: the Wilson line
direction $n$, the large momentum $P_2$, and the spectator momentum $k$.
The scale invariance in $n$, as indicated by the Feynman rule
associated with the eikonal line along $n$, implies that the jet function
must depend on $k$ through the ratio $n\cdot k/n\cdot P_2$.

The next step is to derive the evolution of the jet function in $x$,
{\it i.e.}, in $k^+=xP_1^+$ by considering the derivative,
\begin{eqnarray}
k^+\frac{dJ}{dk^+}=\frac{n\cdot k}{P_2\cdot k}P_2^\alpha
\frac{dJ}{dn^\alpha}\;,
\end{eqnarray}
where the chain rule has been applied to relate the derivatives with
respect to $k$ and to $n$. The differentiation $d/dn^\alpha$ operates 
on the eikonal line along $n$, giving
\begin{eqnarray}
\frac{n\cdot k}{P_2\cdot k}P_2^\alpha
\frac{d}{dn^\alpha}\frac{n_\mu}{n\cdot l}=\frac{{\hat n}_\mu}
{n\cdot l}\;,\;\;\;\;
{\hat n}_\mu=-\frac{n\cdot k}{P_2\cdot k}\frac{P_2\cdot l}
{n\cdot l}n_\mu\;.
\label{dp}
\end{eqnarray}

The loop momentum $l$ flowing through the special vertex does not
generate a collinear divergence due to vanishing of the numerator
$P_2\cdot l$ in the special vertex ${\hat n}_\mu$. It is easy to 
confirm that the ultraviolet 
region of $l$ does not produce $\ln x$ either. Therefore, one
concentrates on the factorization of the soft gluon emitted from the
special vertex, which can be achieved by applying the eikonal
approximation to internal quark propagators, leading to
${\bar n}_{\nu}/{\bar n}\cdot l$. Following the reasoning in \cite{L2},
the derivative of the jet function is written as
\begin{eqnarray}
x\frac{dJ(x)}{dx}=-ig^2C_F\int\frac{d^{4} l}
{(2\pi)^{4}}\frac{{\hat n}_\mu}{n\cdot l}
\frac{g^{\mu\nu}}{l^2}\frac{{\bar n}_{\nu}}{{\bar n}\cdot l}
J(x-l^+/P_1^+)\;,
\label{dij}
\end{eqnarray}
where the argument of $J$ in the integral arises from the invariant
mass of the internal quark,
$(P_2-k+l)^2\approx -2(x-l^+/P_1^+)P_1\cdot P_2$. The integrand
corresponds to the diagram with the soft gluon attaching the eikonal
lines along $n$ and along $\bar n$.
Performing the integration over $l^-$ and $l_T$, one derives the evolution
equation,
\begin{eqnarray}
x\frac{dJ(x)}{dx}=\frac{\alpha_s}{2\pi}C_F
\int_x^1\frac{d\xi}{(\xi-x)_+}J(\xi)\;,
\label{dj}
\end{eqnarray} 
where the variable change from $l^+$ to $\xi$ has been made. 
The plus distribution is defined such that, when $1/(\xi-x)_+$ is 
integrated with a function $f(\xi)$, one must replace it by
$f(\xi)-f(x)$ in the integral. It has been
shown that the above evolution equation is similar to
that for unintegrated parton distribution functions involved in inclusive
QCD processes \cite{KZ}, which resums the same double logarithm
$\alpha_s\ln^2 x$.

The analytical solution is a Sudakov factor,
%
%
\begin{eqnarray}
J(x)=-\exp\left(\frac{\pi^2}{4}\gamma_K\right)
\int_{-\infty}^{\infty}\frac{dt}{\pi}(1-x)^{\exp(t)}
\sin\left(\frac{\pi}{2}\gamma_Kt\right)
\exp\left(-\frac{1}{4}\gamma_K t^2\right)\;,
\label{mjx}
\end{eqnarray}
with the anomalous dimension $\gamma_K=\alpha_sC_F/\pi$.
It is trivial to check that $J(x)$ is normalized to unity,
$\int J(x)dx={\tilde J}(1)=1$ \cite{LL2}.
Obviously, Eq.~(\ref{mjx}) vanishes at $x\to 0$, because the integrand
is an odd function in $t$, and at $x\to 1$ due to the factor
$(1-x)^{\exp(t)}$.
Moreover, Eq.~(\ref{mjx}) provides suppression near the end point $x\to 0$, 
which is stronger than any power of $x$. This is understood from 
vanishing of all the derivatives of Eq.~(\ref{mjx}) with respect to $x$ 
at $x\to 0$ \cite{L5}. To the accuracy of the next-to-leading
logarithms, the running of the coupling constant $\alpha_s$ should 
be taken into account, and Eq.~(\ref{mjx}) will be modified. However, 
the above features remain.

I emphasize the differences among the Sudakov resummations for the
$B\to\gamma l\nu$ decay in the literature.
In \cite{KPY} it is the double logarithm $\ln^2(k_T/m_B)$ that was
resummed. In \cite{DS} it is the double logarithm $\ln^2(E_\gamma/m_B)$,
$E_\gamma$ being the photon energy, that was resummed.
In \cite{BHL} the evolution from the scale of $O(\sqrt{\Lambda m_b})$ 
to the scale of $O(m_b)$ was derived by 
solving the renormalization-group equations,
\begin{eqnarray}\label{Hevol}
   \frac{d}{d\ln\mu}\,H_h(\mu)
   &=& \left[ - \Gamma_{\rm cusp}(\alpha_s)\,\ln\frac{\mu}{2E_\gamma}
    + \gamma(\alpha_s) - \gamma'(\alpha_s) \right] H_h(\mu) \,, \\
   \frac{d}{d\ln\mu}\,H_i(l_+,\mu)
   &=& \left[ \Gamma_{\rm cusp}(\alpha_s)\,
    \ln\frac{\mu^2}{2E_\gamma\,l_+} + \gamma'(\alpha_s) \right]
    H_i(l_+,\mu) + \int_0^\infty\!d\omega\,\Gamma(\omega,l_+,\alpha_s)\,
    H_i(\omega,\mu) \,.
\end{eqnarray}
$\Gamma_{\rm cusp}$ is the universal cusp anomalous dimension 
familiar from the theory of the renormalization of Wilson loops 
\cite{Korchemsky:wg}. The anomalous dimensions $\gamma$ and $\gamma'$
are given by \cite{BHL}
\begin{equation}
   \gamma(\alpha_s) = -2C_F\,\frac{\alpha_s}{4\pi} + O(\alpha_s^2) \,,
    \qquad
   \gamma'(\alpha_s) = O(\alpha_s^2) \,.
\end{equation}
The function $\Gamma$ obeys 
$\int d\omega\,\Gamma(\omega,\omega',\alpha_s)=0$, whose one-loop 
expression is written as
\begin{equation}\label{G1l}
   \Gamma^{\rm 1-loop}(\omega,\omega',\alpha_s)
   = - \Gamma_{\rm cusp}^{\rm 1-loop}(\alpha_s)
   \left[ \frac{\omega'}{\omega}\,
   \frac{\theta(\omega-\omega')}{\omega-\omega'}
   + \frac{\theta(\omega'-\omega)}{\omega'-\omega} \right]_+ \;.
\end{equation}
It was found that the resummation effect 
decreases the magnitude of the radiative corrections, i.e., the 
renormalization-group improved  
kernel is closer to the tree-level value than the one-loop 
result \cite{BHL}. 

The formalism for threshold resummation has been extended to the 
semileptonic decay
$B\to\pi l\nu$ in the fast recoil region of the pion. The $B$
meson momentum $P_1$ is the same as in the decay $B\to\gamma l\nu$,
and the pion momentum $P_2$ is the same as the photon momentum.
Leading-twist factorization theorem for the $B\to\pi$ form factor 
$F(q^2)$ has been proved in \cite{NL}, 
\begin{eqnarray}
F(q^2)= \sum_{m=+,-}\int_0^1 dx_1dx_2\phi_m(x_1)
{\cal H}_m(x_1,x_2,\eta) \phi_\pi(x_2)\;,
\end{eqnarray}
which holds in the region with $x_1\sim O(\Lambda/m_B)$ and 
with $x_2\sim O(1)$. The light-cone $B$ meson distribution amplitudes
$\phi_m$ will be defined in the next section.

\begin{figure}[t!]
\begin{center}
\epsfig{file=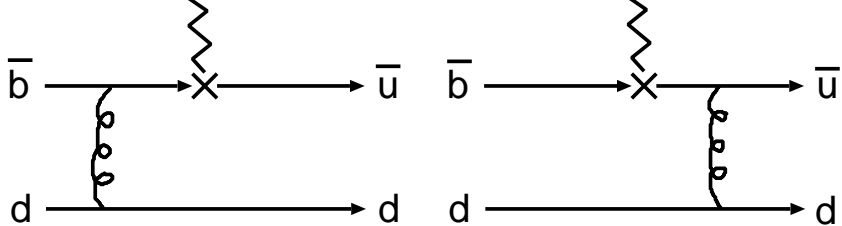,height=1.5in}
\caption{Leading-order contribution to $F^{B\pi}$.}
\label{aa2}
\end{center}
\end{figure}

Since Fig.~\ref{aa2}(a), proportional to $1/(x_1x_2^2)$, is more
singular at small $x_2$, one considers the end-point region with
$x_2\sim O(\Lambda/m_B)$, where the internal $b$ quark propagator
scales like $1/({\Lambda} m_B)$. The loop correction to the weak
vertex, where the radiative gluon attaches the virtual $b$ quark
and upper valence quark in the pion, generates the double logarithm
$\alpha_s\ln^2 x_2$ from the collinear region with the loop momentum
parallel to $P_2$. This double logarithm, similar to that in
Eq.~(\ref{j1c}), is grouped into a jet funciton. It is easy to show that
this jet function obeys the evolution equation in Eq.~(\ref{dj}). Hence,
the threshold resummation leads to a result the same as Eq.~(\ref{mjx}).
That is, the Sudakov factor is universal. The analysis for
Fig.~\ref{aa2}(b) is similar to that for the decay $B\to\gamma l\nu$. In
the end-point region with $x_1\sim O(\Lambda^2/m_B^2)$, additional
collinear divergences associated with the internal light quark are
produced. The loop correction to the weak vertex, where the radiative
gluon attaches the $b$ quark and the virtual light quark,
gives the double logarithm $\alpha_s\ln^2 x_1$, whose
factorization is the same as of Fig.~\ref{fig2}(d).

The modified collinear factorization formula appropriate for the
end-point region is then written as
\begin{eqnarray}
F(q^2)= \sum_{m=+,-}\sum_{i=1,2}\int_0^1 dx_1dx_2\phi_m(x_1)
H_m^{(i)}(x_1,x_2,\eta) J(x_i) \phi_\pi(x_2)\;,
\end{eqnarray}
with the index $i=1$ (2) corresponding to Fig.~\ref{aa2}(b) [(a)].
If $J(x)$ is excluded, the above expression is divergent because
of $H_m^{(2)}\propto 1/x_2^2$ and $\phi_\pi\propto x_2$ at $x_2\to 0$. 
Including the threshold resummation, the $B\to\pi$ form factor is 
calculable without introducing any infrared cutoffs \cite{SHB,BF}.
The numerical effect from the jet function on the $B\to\pi $ form
factor has been examined in \cite{L5}.

In a recent work based on SCET, a jet function has also been
defined in the analysis of the decay $B\to\pi l\nu$ \cite{CPS}.
It was also concluded that the end-point singularity does not exist in
the $B\to\pi$ transiton form factors in
the convolution with the jet function. I stress that the jet function
in \cite{CPS} differs from the one considered here, and that the
smearing mechanism of the end-point singularity is also different: it
is not attributed to the Sudakov mechanism discussed above. 
The jet function in \cite{CPS}, absorbing dynamics 
characterized by $O(\sqrt{\Lambda m_B})$, more or less corresponds to 
the finite piece of the hard kernels in collinear factorization theorem
without threshold resummation. In the case of the $B\to\pi$ transiton
form factors, it can be identified as the piece from
Fig.~\ref{aa2}(a), which is proportional to $1/x_2$. This piece is free
of the end-point singularity, nothing to do with the Sudakov effect. This
point will be elucidated in detail in Sec.~\ref{sec:pqcd}.

\section{$k_T$ Factorization}

Both collinear and $k_T$ factorizations are the fundamental tools of
QCD perturbation theory, where $k_T$ denotes parton transverse momenta.
For inclusive processes, consider DIS of a hadron, carrying a momentum
$p$, by a virtual photon, carrying a momentum $q$. Collinear factorization
\cite{Ste} and $k_T$ factorization \cite{CCH,CE,LRS} apply, when DIS is
measured at a large and small Bjorken variable
$x_B\equiv -q^2/(2p\cdot q)$, respectively. The cross section is
written as the convolution of a hard subprocess with a hadron distribution 
function in a parton momentum fraction $x$ in the former, and in both $x$ 
and $k_T$ in the latter. When $x_B$ is small, $x \ge x_B$ can reach a small 
value, at which $k_T$ is of the same order of magnitude as the 
longitudinal momentum $x p$, and not negligible. For exclusive processes, 
such as hadron form factors, collinear factorization was developed in 
\cite{BL,BFL,ER,DM,CZS,CZ} as stated in the previous section. The range of 
a parton momentum fraction $x$, contrary to that in the inclusive case, is
not experimentally controllable, and must be integrated over between 0
and 1. Hence, the end-point region with a small $x$ is not avoidable. If
no end-point singularity is developed, collinear factorization works. If 
such a singularity occurs, indicating the breakdown of collinear 
factorization, $k_T$ factorization should be employed. In fact,
the recent observation $QF_2(Q^2)/F_1(Q^2)\sim$ const. \cite{JLAB},
$F_1$ and $F_2$ being the proton Dirac and Pauli form factors,
respectively, indicates that $k_T$ factorization is the appropriate
tool for studying exclusive processes \cite{RJ03}.
Since $k_T$ factorization theorem 
was proposed \cite{BS,LS}, there had been wide applications to various 
processes \cite{LFF}.

In this section I review $k_T$ factorization theorem for exclusive
processes. It is more convenient to perform $k_T$ factorization in the
impact parameter $b$ space, in which infrared divergences in radiative 
corrections can be extracted from parton-level diagrams explicitly. 
The procedure is basically similar to that for collinear factorization in
Sec.~\ref{sec:all}, if the proof is performed in the impact parameter $b$
space. It has been observed that collinear factorization is
the $b\to 0$ limit of $k_T$ factorization. I explain how to construct a 
gauge-invariant $b$-dependent meson wave function defined as a nonlocal 
matrix element with a special path for the Wilson line. The application
of $k_T$ factorization theorem to exclusive $B$ meson decays, and
the behavior of $b$-dependent $B$ meson wave functions are discussed. 
Retaining the parton transverse degrees of freedom, the double
logarithms $\alpha_s\ln^2 k_T$ appear, which should be organized to
all orders. The basic idea for $k_T$ resummation of these double
logarithms into a Sudakov factor \cite{LY1} is given.
The end-point singularity in the heavy-to-light transition form factors 
can also be smeared \cite{LY1,KLS,LUY,TLS,WY} by including this Sudakov 
factor.

\subsection{\it Gauge Invariance \label{sec:gau}}

I again start with the process $\pi\gamma^\ast \to\gamma$ \cite{MR}. This 
process, though containing no end-point singularity, is simple and 
appropriate for a demonstration. The momentum $P_1\;(P_2)$ of the 
initial-state pion
(final-state photon) is chosen as in Eq.~(\ref{mpp}). I explain how to
perform the factorization of the collinear enhancement from $l$ parallel 
to $P_1$ without integrating out the transverse components $l_T$.
The lowest-order diagrams are displayed in Fig.~\ref{fig1}, and
the $O(\alpha_s^0)$ $k_T$ factorization formula is the same as the
collinear factorization formula in Eq.~(\ref{gl0}). That is, none of
${\cal G}^{(0)}$, $\phi^{(0)}$, and ${\cal H}^{(0)}$ depends on
a transverse momentum. The wave function and the hard kernel become 
$l_T$-dependent through collinear gluon exchanges at higher orders.

The $O(\alpha_s)$ $k_T$ factorization formula is a sum over
the diagrams in Fig.~\ref{fig2}, the same as Eq.~(\ref{cf}), but with
each term being written as the convolution in the momentum fraction $\xi$ 
and in the impact parameter $b$,
\begin{eqnarray}
{\cal G}^{(1)}_i(x)&=&\int d\xi\frac{d^2b}{(2\pi)^2}
\Phi_i^{(1)}(x,\xi,b)H^{(0)}(\xi,b)
+\phi^{(0)}(x){\cal H}_i^{(1)}(x)\;.
\label{h11}
\end{eqnarray}
The above expression, with the $O(\alpha_s)$ wave functions 
$\Phi_i^{(1)}(x,\xi,b)$ and $H^{(0)}(\xi,b)$ specified, defines the 
$O(\alpha_s)$ hard kernels ${\cal H}_i^{(1)}(x)$, which do not contain
collinear divergences. Equation (\ref{h11}) is
a consequence of the assertion that partons acquire transverse
degrees of freedom through collinear gluon exchanges:
${\cal H}^{(1)}$, convoluted with the lowest-order $l_T$-independent
$\phi^{(0)}$, is then identical to that in
collinear factorization. As shown later, this consequence is
crucial for constructing gauge-invariant hard kernels.

The $O(\alpha_s)$ wave functions obtained from Figs.~\ref{fig2}(a) and
\ref{fig2}(c) are the same as in collinear factorization.
The $k_T$ factorization of Fig.~\ref{fig2}(b) leads to the wave function,
\begin{eqnarray}
\Phi^{(1)}_b(x,\xi,b)&=&\frac{ig^2C_F}{4P_1^+}
\int\frac{d^4l}{(2\pi)^4}{\bar q}(xP_1)
\frac{\gamma^\nu(x\not P_1-\not l)\gamma^5\not {\bar n}
({\bar x}\not P_1+\not l)\gamma_\nu} 
{(xP_1-l)^2({\bar x}P_1+l)^2l^2}q({\bar x}P_1)
\nn\\
& &\times
\delta\left(\xi-x+\frac{l^+}{P_1^+}\right)e^{-i{\bf l}_T\cdot {\bf b}}\;.
\label{p2b}
\end{eqnarray}
The Fourier transformation introduces the additional factor 
$\exp(-i{\bf l}_T\cdot {\bf b})$ into the wave function $\Phi^{(1)}_b$ 
compared to the result in collinear factorization in Eq.~(\ref{c2b}), 
since the hard kernel depends on $l_T$ in this case.
The $O(\alpha_s)$ wave function extracted from Fig.~\ref{fig2}(d)
is written as
\begin{eqnarray}
\Phi^{(1)}_{d}(x,\xi,b)&=&
\frac{-ig^2C_F}{4P_1^+}\int\frac{d^4l}{(2\pi)^4}
{\bar q}(xP_1)\gamma^5\not {\bar n}
\frac{ {\bar x}\not P_1+\not l}{({\bar x}P_1+l)^2}
\gamma_\nu q({\bar x}P_1)
\frac{1}{l^2}\frac{{\bar n}^\nu}{{\bar n}\cdot l}
\nonumber\\
& &\times\left[\delta(\xi-x)-\delta\left(\xi-x+\frac{l^+}{P_1^+}\right)
e^{-i{\bf l}_T\cdot {\bf b}}\right]\;.
\label{p2d}
\end{eqnarray}
The second term acquires the additional factor
$\exp(-i{\bf l}_T\cdot {\bf b})$ from the Fourier transformation,
because it corresponds to the case with the loop momentum $l$ flowing
through the hard kernel. It is easy to observe that the soft divergences
cancel among the $O(\alpha_s)$ radiative corrections: in the soft 
region of $l$ we have $\exp(-i{\bf l}_T\cdot {\bf b})\approx 1$ and 
$l^+\approx 0$, and the two terms in Eq.~(\ref{p2d})
cancel. Similarly, the soft divergences cancel among 
Figs.~2(a)-2(c).

\begin{figure}[t!]
\begin{center}
\epsfig{file=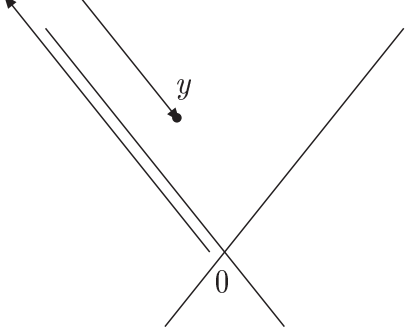,height=2.0in}
\caption{The special path for the Wilson line in a $b$-dependent wave
function.}
\label{path}
\end{center}
\end{figure}

One constructs the parton-level wave function as the 
nonlocal matrix element in the $b$ space,
\begin{eqnarray}
\Phi(x,\xi,b)=i\int\frac{dy^-}{2\pi }e^{-i\xi P_1^+y^-}
\langle 0|{\bar q}(y)\gamma_5\not {\bar n}
P\exp\left[-ig\int_0^{y}ds\cdot A(s)\right]q(0)
|{\bar q}(xP_1) q({\bar x}P_1)\rangle\;,
\label{pw1}
\end{eqnarray}
with the coordinate $y=(0,y^-,{\bf b})$. The path for the Wilson line is
composed of three pieces: from 0 to $\infty$ along the direction of
${\bar n}$, from $\infty$ to $\infty+{\bf b}$, and from $\infty+{\bf b}$
back to $y$ along the direction of $-{\bar n}$ as displayed in
Fig.~\ref{path}. The first (third) piece
corresponds to the eikonal line associated with the first (second) term in 
Eq.~(\ref{p2d}).


For the evaluation of the lowest-order hard kernel, one
neglects only the minus component $l^-$ in the denominator [see the second
term on the right-hand side of Eq.~(\ref{pi})], 
\begin{eqnarray}
(P_2-xP_1+l)^2 \approx -(2\xi P_1\cdot P_2 + l_T^2) \;.
\label{appI}
\end{eqnarray}
Note that in collinear factorization both $l^-$ and $l_T$ are dropped. 
The $b$-dependent hard amplitude is then given by,
\begin{eqnarray}
H^{(0)}(\xi,b)&=&\int d^2l_T
{\cal H}^{(0)}(\xi,l_T)\exp(i{\bf l}_T\cdot {\bf b})\;,
\nonumber\\
{\cal H}^{(0)}(\xi,l_T)&=& ie^2
\frac{tr(\not \epsilon \not P_2\gamma_\mu
\not P_1\gamma^5)}{2\xi P_1\cdot P_2+l_T^2}\;.
\label{psi0}
\end{eqnarray}
Equivalently, the above ${\cal H}^{(0)}(\xi,l_T)$ is derived by
considering an off-shell $\bar q$ quark, which carries the momentum
$\xi P_1-{\bf l}_T$, and the leading structure $\not P_1\gamma_5$
associated with the pion, which is the same as in collinear 
factorization.

I now demonstrate the gauge invariance of $k_T$ factorization
theorem. Equation~(\ref{pw1}) is explicitly gauge-invariant because 
of the presence of the Wilson link from $0$ to $y$ \cite{CE,CS2}. 
${\cal H}^{(1)}(x)$, the same as in collinear
factorization, is gauge-invariant. From Eq.~(\ref{h11}), the
gauge invariance of $\Phi^{(1)}(x,\xi,b)$ stated above, together with the
gauge invariance of ${\cal G}^{(1)}(x)$ and ${\cal H}^{(1)}(x)$, then
imply the gauge invariance of $H^{(0)}(\xi,b)$. Similarly, the $k_T$
factorization formula of $O(\alpha_s^2)$,
\begin{eqnarray}
{\cal G}^{(2)}(x)=\int d\xi\frac{d^2b}{(2\pi)^2}
\left[\Phi^{(2)}(x,\xi,b)H^{(0)}(\xi,b)
+\Phi^{(1)}(x,\xi,b)H^{(1)}(\xi,b) \right]
+\phi^{(0)}(x){\cal H}^{(2)}(x)\;, 
\end{eqnarray}
leads to the gauge invariance of $H^{(1)}(\xi,b)$.
Therefore, the hard kernels in $k_T$ 
factorization are gauge-invariant at all orders.

After determining the gauge-invariant infrared-finite hard kernel
$H$, one convolutes it with the physical two-parton pion wave
function, whose all-order gauge-invariant definition
is written as
\begin{eqnarray}
\Phi_\pi(x,b,Q,\mu)=i\int\frac{dy^-}{2\pi }e^{-ixP_1^+y^-}
\langle 0|{\bar q}(y)\gamma_5\not{\bar n}
{\cal P}\exp\left[-ig\int_0^{y}ds\cdot A(s)\right]q(0)|\pi(P_1)\rangle\;.
\label{pwt}
\end{eqnarray}
The relevant form factor $F(Q^2)$ for the process $\pi\gamma^* \to\gamma$
is then expressed as
\begin{eqnarray}
F(Q^2)=\int_0^1 dx \int\frac{d^2b}{(2\pi)^2}
\Phi_\pi(x,b,Q,\mu)H(x,b,Q,\mu)\;.
\end{eqnarray}
where both the dependences on $Q$ and on the factorization scale $\mu$
have been made explicit. It has been concluded that predictions derived 
from $k_T$ factorization theorem are gauge-invariant and infrared-finite
\cite{NLK}.

In summary, a two-parton $b$-dependent wave function is factorized
from parton-level diagrams in a way the same as in collinear 
factorization (for example, under the same eikonal approximation), but the
loop integrand is associated with an additional Fourier factor
$\exp(-i{\bf l}_T\cdot {\bf b})$, when the loop momentum $l$ flows through 
a hard kernel. A $k_T$-dependent hard kernel is obtained in a way
the same as in collinear factorization, but considering off-shell external
partons, which carry the fractional momenta $k=xP-{\bf k}_T$ 
($k^2=-k_T^2$), $P$ being the external meson momenta. Then Fourier 
transform this hard kernel into the $b$ space.
The insertion of the Fierz identity to separate the fermion
flow between a wave function and a hard amplitude is the same as
in collinear factorization. For inclusive
processes in small $x_B$ physics, the gauge invariance of the 
unintegrated gluon distribution function and of the hard subprocess 
of reggeized gluons, being also off-shell by $-k_T^2$, is ensured in a 
similar way. The distinction is that the structures of $\gamma$-matrices 
from the Fierz identity are replaced by eikonal vertices, which contain 
only the longitudinal components \cite{CE}.

\subsection{\it $B$ Meson Wave Functions \label{sec:b}}

In this subsection I review the $k_T$ factorization theorem for exclusive
$B$ meson decays by considering the radiative decay $B\to \gamma l\nu$.
It has been shown that in heavy quark limit a gauge-invariant
$b$-dependent $B$ meson wave function can be defined, which absorbs soft
divergences in the decay process, differing from the collinear
divergences in the pion wave function. As explained below, exclusive $B$
meson decays are characterized by the scale $\sqrt{\Lambda m_B}$.
In terms of the power counting in SCET, the soft dynamics discussed here
is referred to as the usoft one, since the typical momentum behaves like
\cite{NL}
\be
l^\mu\sim (\Lambda,\Lambda,\Lambda)\sim
m_B(\lambda^2,\lambda^2,\lambda^2)\;,
\label{usoft}
\ee
for the expansion parameter $\lambda\sim \sqrt{\Lambda/m_B}$. It is
possible to construct a light-cone $B$ meson wave function, if
an appropriate frame with the photon moving along the light cone
is chosen.

Figure~\ref{fig1}(a) gives the parton-level amplitude,
\begin{eqnarray}
{\cal G}^{(0)}(x)&=&
e{\bar q}(k)\not \epsilon \frac{\not P_2-\not k}{(P_2-k)^2}
\gamma_\mu(1-\gamma_5) b(P_1-k)\;,
\label{b1a}
\end{eqnarray}
which does not depend on a transverse momentum.
Inserting the Fierz identity in Eq.~(\ref{fierz}) into Eq.~(\ref{b1a}),
one obtains Eq.~(\ref{gl0}) with
\begin{eqnarray}
\phi^{(0)}(x)&=&\frac{1}{4P_1^+}{\bar q}(k)\gamma_5\not{\bar n}
b(P_1-k)\;,
\nonumber\\
{\cal H}^{(0)}(x)
&=&-e\frac{tr[\not \epsilon \not P_2 \gamma_\mu(1-\gamma_5)
(\not P_1+m_B)(\not n/\sqrt{2})\gamma^5]}{2x P_1\cdot P_2}\;,
\label{h0B}
\end{eqnarray}
where the higher-power term $\not k$ in the numerator has been dropped.
For the $B$ meson, there are two leading-twist wave functions
associated with the structures $\gamma_5\gamma^\pm$. For the
$B\to\gamma l\nu$ decay, only the structure
$\gamma_5\gamma^+=\gamma_5\not{\bar n}$ contributes:
since $\not \epsilon$ in Eq.~(\ref{h0B}) involves $\gamma_T$,
only the structure $\gamma^-\gamma_5=\not n\gamma_5$ contributes to
the hard kernel.

Next one considers the $O(\alpha_s)$ radiative corrections to
Fig.~\ref{fig1}(a) shown in Fig.~\ref{fig2}, and discuss the
factorization of the soft divergence from the region of the loop
momentum in Eq.~(\ref{usoft}).
The dependence of the $B$ meson wave function on the transverse momentum
is generated by soft gluon exchanges. The analysis is similar to that in
Sec.~\ref{sec:gau}, and one derives Eq.~(\ref{h11}). The factorization of
the two-particle reducible diagrams in Fig.~\ref{fig2}(a)-\ref{fig2}(c)
is straightforward. Take Fig.~\ref{fig2}(b) as an example.
Employing the eikonal approximation in the heavy quark limit, one has
\begin{eqnarray}
\frac{\not P_1-\not k+\not l+m_b}{(P_1-k+l)^2-m_b^2}
\gamma_\nu b(P_1-k) \approx
\frac{v_\nu}{v\cdot l}b(P_1-k)\;,
\label{appII}
\end{eqnarray}
with the velocity $v=P_1/m_B$.
The $O(\alpha_s)$ wave function
extracted from Fig.~\ref{fig2}(b) is then written as
\begin{eqnarray}
\Phi^{(1)}_{b}(x,\xi,b)&=&\frac{ig^2C_F}{4P_1^+}
\int\frac{d^4l}{(2\pi)^4}{\bar q}(k)
\frac{\gamma^\nu(\not k-\not l)}{(k-l)^2 l^2}
\gamma_5\not{\bar n}b(P_1-k)\frac{v_\nu}{v\cdot l}
\delta\left(\xi-x+\frac{l^+}{P_1^+}\right)
e^{-i{\bf l}_T\cdot {\bf b}}\;.
\label{p2bB}
\end{eqnarray}
Performing the contour integration over, say, $l^-$, one observes that the
integral is singular only when the component $l^+$ is of $O(\Lambda)$. This 
observation implies that the infrared divergence associated with the $B$ 
meson is of the soft type, and that $l^+$, being of the same order of
magnitude as $k^+=xP_1^+$, is not negligible in the $\delta$-function. 
Therefore, the soft divergences in Figs.~\ref{fig2}(a)-\ref{fig2}(c) do 
not cancel in $B$ meson decays \cite{LY1}. The explanation is simple: the 
light spectator quark, carrying a small amount of momenta, forms a color 
cloud around the $b$ quark. This cloud is also huge in space-time, such 
that soft gluons resolve the color structure of the $B$ meson.

Diagrams with the radiative gluon attaching the internal quark in
Figs.~\ref{fig2}(d) and \ref{fig2}(e) also contain soft divergences,
because the internal quark is off-shell by $O(\Lambda m_B)$, which
defines the characteristic scale of the decay $B\to\gamma l\nu$. Note
that the internal quark in the process $\pi\gamma^*\to\gamma$ is
off-shell by $O(Q^2)$. One extracts the $O(\alpha_s)$ wave function
from Fig.~\ref{fig2}(d),
\begin{eqnarray}
\Phi^{(1)}_{d}(x,\xi,b)&=&
\frac{-ig^2C_F}{4P_1^+}\int\frac{d^4l}{(2\pi)^4}
{\bar q}(xP_1)\gamma_5\not{\bar n} b(P_1-k)
\frac{1}{l^2}\frac{{\bar n}\cdot v}{{\bar n}\cdot l v\cdot l}
\nonumber\\
& &\times\left[\delta(\xi-x)-\delta\left(\xi-x+\frac{l^+}{P_1^+}\right)
e^{-i{\bf l}_T\cdot {\bf b}}\right]\;.
\label{p2dB}
\end{eqnarray}
The eikonal approximation in Eq.~(\ref{appII}) has been applied.
The above expression implies that the infrared divergences in the 
irreducible diagrams can also be collected by the eikonal line along the 
light cone. This is attributed to the choice of the frame, in which the 
photon moves in the minus direction.

Following the procedure in Sec.~\ref{sec:gau}, one constructs a
gauge-invariant light-cone $B$ meson wave function,
\begin{eqnarray}
\Phi_+(x,b,m_B,\mu)&=&i\int\frac{dy^-}{2\pi }e^{-ix P_1^+y^-}
\langle 0|{\bar q}(y)\gamma_5\gamma^+
{\cal P}\exp\left[-ig\int_0^{y}ds\cdot A(s)\right]b_v(0)
|B(P_1)\rangle\;,
\label{bw}
\end{eqnarray}
where $b_v$ is the rescaled $b$ quark field, and the decay constant
$f_{B}$ has been omitted. The Feynman rules associated with $b_v$ are
those for an eikonal line in the direction of $v$ defined in
Eq.~(\ref{appII}). The lowest-order hard kernel in the $b$ space is given
by Eq.~(\ref{psi0}) with
\begin{eqnarray}
{\cal H}^{(0)}(\xi,l_T)&=&-e
\frac{tr[\not \epsilon \not P_2 \gamma_\mu(1-\gamma_5)
(\not P_1+m_B)(\not n/\sqrt{2})\gamma^5]}{2\xi P_1\cdot P_2+l_T^2}\;,
\end{eqnarray}
$\xi= (k^+-l^+)/P_1^+$ being the momentum fraction.
The above expression can be derived by
considering an off-shell $\bar q$ quark of the momentum
$(\xi P_1^+,0,-{\bf l}_T)$, and by contracting the parton-level diagram
with the leading structure $(\not P_1+m_B)(\not n/\sqrt{2})\gamma^5$,
which is the same as in collinear factorization.

The semileptonic decay $B\to\pi l\nu$, because of the end-point
singularity, demands $k_T$ factorization, whose all-order proof can be
performed in a similar way. For this mode, both the leading-twsit $B$
meson wave functions $\phi_\pm$, associated with the structures
$\gamma_5\gamma^\pm$, contribute \cite{NL}. Moreover, contributions
from the pseudo-scalar and pseudo-tensor two-parton twist-3 pion
distribution amplitudes are also leading-power \cite{TLS}. The
factorization of the corresponding collinear divergences has been
proved \cite{NL}. The point is to replace the Dirac structure
$\gamma_5\gamma^\alpha$ by the corresponding ones $\gamma_5$ and
$\gamma_5\sigma^{\alpha\beta}$ in Eq.~(\ref{fierz}). 

I then discuss the behavior of the $B$ meson wave functions constructed
in Eq.~(\ref{bw}). In the heavy quark limit the two-parton light-cone
wave functions $\tilde{\Phi}_{\pm}(t,z^2)$ are defined in terms of the
nonlocal matrix element \cite{GN,GS}:
\bea
\langle 0 | \bar{q}(y) \Gamma b_{v}(0) |\bar{B}(P_1) \rangle
 = - \frac{i f_{B} m_B}{2} tr
 \left[ \gamma_{5}\Gamma \frac{1 + \not{v}}{2}
\left\{ \tilde{\Phi}_{+}(t,y^2) - \not{y}
\frac{\tilde{\Phi}_{+}(t,y^2)
 -\tilde{\Phi}_{-}(t,y^2)}{2t}\right\} \right]\;,
 \label{phi}
\eea
with $t=v\cdot y$, $y^2=-b^2$, and $\Gamma$ being a Dirac matrix .

Consider the light-cone distribution amplitudes in terms of the
variable $\omega=xm_B$ \cite{GN},
\be
\phi_{\pm}(\omega)=\lim_{y^2\to 0}\Phi_{\pm}(\omega,y^2)\;,
\ee
where the wave functions $\Phi_{\pm}(\omega,y^2)$, defined in 
Eq.~(\ref{bw}), come from the Fourier transformation of
$\tilde{\Phi}_{\pm}(t, y^{2})$.
The differential equations are written as \cite{KKQT}
\bea
 \omega \frac{d \phi_{-}(\omega)}{d \omega}
  &+& \phi_{+}(\omega) = I(\omega)\ ,
  \nn \\
  \left(\omega - 2 \Lambda\right)\phi_{+}(\omega)
 &+& \omega \phi_{-}(\omega) = J(\omega) \ , \label{mde2}
\eea
where 
$I(\omega)$ and $J(\omega)$ denote the source terms
due to three-parton wave functions $\Psi_{A}$, $\Psi_{V}$ and
$X_{A}$:
\bea
I(\omega)&=& 2\frac{d}{d\omega}
\int_{0}^{\omega}d\rho \int_{\omega - \rho}^{\infty} \frac{d\xi}{\xi}  
\frac{\partial}{\partial \xi}\left[ \Psi_{A}(\rho, \xi)
   - \Psi_{V}(\rho, \xi)\right]\;,
\nonumber\\
J(\omega) &=& -2\frac{d}{d\omega}
\int_{0}^{\omega}d\rho \int_{\omega - \rho}^{\infty} \frac{d\xi}{\xi}
\left[ \Psi_{A}(\rho, \xi) + X_{A}(\rho, \xi)\right]
-4 \int_{0}^{\omega}d\rho \int_{\omega - \rho}^{\infty}\frac{d\xi}{\xi}
\frac{\partial \Psi_{V}(\rho, \xi)}{\partial \xi} \;.
\label{sj}
\eea

The solution can be decomposed into two pieces:
\be
  \phi_{\pm}(\omega) = \phi_{\pm}^{(W)}(\omega) 
  + \phi_{\pm}^{(g)}(\omega) \;.
\label{decomp}
\ee
The functions $\phi_{\pm}^{(W)}(\omega)$ are the solution with 
$I(\omega)=J(\omega)=0$, corresponding to the ``Wandzura-Wilczek 
approximation'' \cite{PB1,Braun:1990iv} $\Psi_{V}=\Psi_{A}=X_{A}=0$.
The functions $\phi_{\pm}^{(g)}(\omega)$ are induced by the source 
terms $I(\omega)$ and $J(\omega)$.
The analytic expressions for the Wandzura-Wilczek part are given by
\be
 \phi_{\pm}^{(W)}(\omega) = \frac{\Lambda \pm
 (\omega-\Lambda)}{2 \Lambda^{2}}
 \theta(\omega)\theta(2 \Lambda - \omega)\;.
\label{solm}
\ee
The expressions for $\phi_{\pm}^{(g)}$, in terms of
$\Psi_{A}$, $\Psi_{V}$ and $X_{A}$, can be found in \cite{KKQT2}.
Equation~(\ref{solm}) is quite different from the model
distribution amplitudes appearing in the literature. One
example of such models is \cite{GN}
\bea
\phi^{GN}_{+}(\omega)&=&\frac{\omega}{\omega_{0}^{2}}
\exp\left(- \frac{\omega}{\omega_{0}}\right)\;,
\nn\\
\phi^{GN}_{-}(\omega)& =& \frac{1}{\omega_{0}}
\exp\left(- \frac{\omega}{\omega_{0}}\right)\;,
\eea
with $\omega_{0} = 2\Lambda/3$, which are inspired by the QCD sum
rule estimates \cite{GN}. Note that, however, the behavior
$\phi_{+}^{GN}(\omega) \sim \omega$ and
$\phi_{-}^{GN}(\omega) \sim {\rm constant}$
at $\omega \rightarrow 0$ is consistent with Eq.~(\ref{solm}).
Another example comes from solving an integro-differential
equation \cite{LN03}: evolution effects generate a radiative tail,
which falls off slower than $1/\omega$.

Including the $k_T$ (or $y^2=-b^2$) dependence, one has the differential 
equations \cite{KKQT2} 
\bea
&&\omega \frac{\partial \Phi_{-}}{\partial \omega}+ \Phi_{+}
+ z^2\frac{\partial}{\partial z^2}
\left(\Phi_{+}-\Phi_{-}\right) 
= 0\; , \nn\\
&&\left(\omega 
\frac{\partial}{\partial \omega} + 2\right)\left(\Phi_{+} -
\Phi_{-}\right)
+ 4  \frac{\partial^{2}}{\partial \omega^{2}}\frac{\partial
\Phi_{+}}{\partial z^2}
= 0\; ,\nn \\
&&\left[(\omega - \Lambda) \frac{\partial}
{\partial \omega}\!+\!\frac{3}{2}\right]\Phi_{+}
\!-\!\frac{1}{2}\Phi_{-}
\! + 2  \frac{\partial^{2}}{\partial \omega^{2}}
\frac{\partial \Phi_{+}}{\partial z^2}
\!= 0\;,\nn \\
\label{eq:4}
&&\left[ (\omega - \Lambda)\frac{\partial}
{\partial \omega}+ 2\right] \left(\Phi_{+}
- \Phi_{-}\right)+2  \frac{\partial^{2}}{\partial \omega^{2}}
\left(\frac{\partial \Phi_{+}}{\partial z^2} -
\frac{\partial \Phi_{-}}{\partial z^2} \right)
= 0\;.
\eea
The solution in the Wandzura-Wilczek approximation is given by
\be
\label{eq:12}
\Phi^{(W)}_{\pm}(\omega, b)
\sim \frac{1}{\sqrt{b}}\cos\left(
\sqrt{\omega(2\Lambda-\omega)}\;b
-\frac{\pi}{4}\right)\;,
\ee
where the $b$ dependence corresponds to 
$\delta(k_{T}^{2} - \omega (2 \Lambda -\omega))$ in $k_T$ space.
It is observed that the longitudinal and transverse momentum dependences
do not seperate (factorize) in Eq.~(\ref{eq:12}), contrary to the
assumption in many models \cite{KLS,GS,BW}. The Gaussian
distribution for the $k_{T}$-dependence has been adopted in 
\cite{KLS}, which exhibits strong damping at large
$b$ as $\exp\left(-\omega_B^2b^{2}/2\right)$ (see Sec.~\ref{sec:pqcd}).
In contrast, the results in Eq.~(\ref{eq:12}) show slow-damping with 
oscillatory behavior. I mention that, despite of the different functional 
forms, the numerical results of the $B\to\pi$ form factor derived from 
the $B$ meson wave function in \cite{KLS} and from Eq.~(\ref{eq:12}) are
very similar \cite{Kuri}.

\subsection{\it $k_T$ Resummation \label{sec:kt}}

The inclusion of parton transverse degrees of freedom introduces
a soft logarithm $\alpha_s\ln b$. Its overlap with the original
collinear logarithm leads to a double logarithm $\alpha_s\ln^2(Qb)$. This 
large logarithm must be organized in order not to spoil perturbative 
expansion. I explain the idea of $k_T$ resummation by taking
the pion wave function as an example. It is known that single 
logarithms can be summed to all orders using renormalization group 
methods, while double logarithms are organized by the technique 
developed in \cite{CS,LL}. I choose the axial gauge $n\cdot A=0$, in 
which the two-particle reducible diagrams, like 
Figs.~\ref{fig2}(a)-\ref{fig2}(c), contain 
the double logarithms, while the two-particle irreducible corrections, 
like Figs.~\ref{fig2}(d) and \ref{fig2}(e), contain only single soft
logarithms. If the double logarithms appear in an exponential form
$\Phi_\pi\sim \exp[-{\rm const.}\times \ln^2(Qb)]$,
the task will be simplified by studying the derivative of $\Phi_\pi$,
$d\Phi_\pi/d\ln Q=C\Phi_\pi$. It is obvious that the coefficient $C$
contains only large single logarithms, and can be treated by
renormalization group methods. Therefore, working with $C$ one reduces
the double-logarithm problem to a single-logarithm one.

Consider the pion wave function $\Phi_\pi(x,b,Q,\mu)$ defined in
Eq.~(\ref{pwt}). The two invariants appearing in $\Phi_\pi$ are
$P_1\cdot \bar n$ and $\bar n^2$, where $\bar n$ is allowed to
vary away from the light cone. By the scale invariance of $\bar n$ in
the gluon propagator,
\begin{equation}
N^{\mu\nu}(l)=\frac{-i}{l^2}\left[g^{\mu\nu}-\frac{\bar n^{\mu}l^{\nu}+
l^{\mu}\bar n^{\nu}}{\bar n\cdot l}
+\bar n^2\frac{l^{\mu}l^{\nu}}{(\bar n\cdot l)^2}\right]\;,
\end{equation}
$\Phi_\pi$ depends only on a single large scale
$\nu^2=(P_1\cdot \bar n)^2/\bar n^2$.
It is then easy to show that the differential operator $d/d\ln Q$ can
be replaced by $d/d \bar n$:
\begin{equation}
\frac{d}{d \ln Q}\Phi_\pi=-\frac{\bar n^2}{P_1\cdot \bar n}P_1^{\alpha}
\frac{d}{d \bar n^{\alpha}}\Phi_\pi\;.
\label{qn}
\end{equation}
The motivation for this replacement is that the momentum $P_1$ flows
through both quark and gluon lines, but $\bar n$ appears only on gluon
lines. The analysis then becomes simpler by studying the $\bar n$, instead
of $P_1$, dependence.

Applying $d/d \bar n_{\alpha}$ to the gluon propagator, one obtains
\begin{equation}
\frac{d}{d \bar n_{\alpha}}N^{\mu\nu}=-\frac{1}{\bar n\cdot l}
(N^{\mu\alpha}l^{\nu}+N^{\nu\alpha}l^{\mu})\;.
\label{dn}
\end{equation}
The momentum $l$ that appears at both ends of a gluon line is contracted
with the vertex, where the gluon attaches. After adding together all
diagrams with different differentiated gluon lines and
using the Ward identity in Eq.~(\ref{war}), one arrives at the 
differential equation of $\Phi_\pi$, and the result $\bar \Phi_\pi$
contains the special vertex \cite{LY1},
\begin{eqnarray}
g\frac{\bar n^2}{P_1\cdot \bar n \bar n\cdot l}P_{1\alpha}\;.
\end{eqnarray}
An important feature of this special vertex is that the gluon momentum $l$
does not lead to collinear enhancements because of the nonvanishing
$\bar n^2$. The leading regions of $l$ are then soft and ultraviolet,
in which the subdiagram containing the special vertex can be
factorized from the new function $\bar\Phi_\pi$.
The left-over part is exactly $\Phi_\pi$, 
and the subdiagram is assigned to the coefficient $C$
introduced before.

Therefore, one needs a function $K$ to organize the soft
divergences and $G$ to organize the ultraviolet divergences in
the subdiagrams. The differential equation of $\Phi_\pi$ is then written
as,
\begin{eqnarray}
\frac{d}{d \ln Q}\Phi_\pi(x,b,Q,\mu)=\left[\,2K
(b\mu)+\frac{1}{2}\;G(x\nu/\mu)+\frac{1}{2}\;G(
(1-x)\nu/\mu)\right]\Phi_\pi(x,b,Q,\mu)\; .
\label{qp}
\end{eqnarray}
The functions $K$ and $G$ have been calculated
to one loop, and the single logarithms have been organized to give
their evolutions in $b$ and $Q$, respectively \cite{BS}. They possess
individual ultraviolet poles, but their sum $K+G/2$ is
finite, such that Sudakov logarithms are renormalization-group invariant.
Substituting the expressions for $K$ and $G$ into
Eq.~(\ref{qp}), one derives the solution,
\begin{equation}
\Phi_\pi(x,b,Q,\mu)=\exp\left[-\sum_{\xi=x,\;\bar x}\;s(\xi,b,Q)\right]
\Phi_\pi(x,b,\mu)\;,
\label{sp}
\end{equation}
where the initial condition of the Sudakov evolution, $\Phi_\pi(x,b,\mu)$,
contains the single-logarithm evolution in $\mu$, and the intrinsic
dependence on $b$ \cite{JK93}. The distribution
amplitude $\phi_\pi(x,\mu)$, defined in Eq.~(\ref{cw1p}), is the
$b\to 0$ limit of $\Phi_\pi(x,b,\mu)$.  
The explicit expression for the exponent $s$,
grouping the double logarithms, is referred to \cite{KLS}.

Note that the vector $\bar n$ has been varied away from the light cone
in the above technique. The leading-logarithm resummation, being 
independent of the $\bar n$, is still gauge invariant. The
$\bar n$ dependence indeed appears in the next-to-leading-logarithm
resummation for the wave function \cite{BS,GS}, such that this piece 
becomes gauge dependent. However, this $\bar n$ dependence will be 
cancelled by that from the resummation of nonfactorizable soft gluons,
which is also next-to-leading-logarithm \cite{L1,L98}.
That is, in a complete next-to-leading-logarithm resummation,
the Sudakov factor is gauge invariant.

\begin{figure}[t!]
\begin{center}
 \epsfig{file=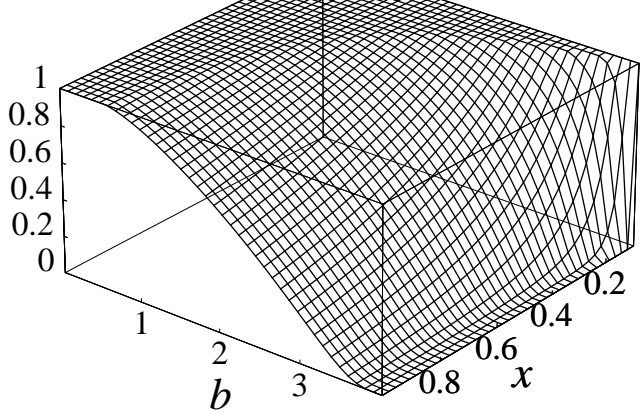,height=2.5in} 
\caption{The Sudakov factor $\exp[-s(x,b,Q)]$ for
$\Lambda_{\rm QCD}=250$ MeV.}
\label{fig6}
\end{center}
\end{figure}

Variation of $\exp(-s)$ with $b$ and $x$ is displayed in Fig.~\ref{fig6}, 
which shows a strong falloff in the large $b$ and large $x$ region, and
vanishes for $b > 1/\Lambda_{\rm QCD}$. Hence, Sudakov suppression
selects components of the pion wave functions with
small spatial extent $b$, and makes the hard scattering more perturbative.
Once the main contributions to the factorization formula
come from the small $b$, or short-distance, region,
perturbation theory becomes relatively self-consistent.

The above formalism has been generalized to the $B$ meson wave function. 
In the axial gauge only the two-particle reducible diagrams generate the 
double logarithms. Figure~\ref{fig2}(a), giving the self-energy correction 
to the massive $b$ quark, produces only soft enhancement, and is 
subleading. If the component $k^+$ of the spectator momentum is as small as 
$O(\Lambda)$, collinear
divergences in Figs.~\ref{fig2}(b) and \ref{fig2}(c), which arise from
the loop momentum with a large component parallel to $k$,
will not be pinched, and they also give only soft enhancements.
This is consistent with the physical picture that the soft light quark
can not interact with the heavy quark through a fast moving gluon. If
there is nonvanishing probability of finding the light spectator with
$k^+$ being of $O(m_B)$, such as in the model with a power-law decrease
in $k^+$ \cite{Schl}, Figs.~\ref{fig2}(b) and \ref{fig2}(c) contribute
large double logarithms. Most of the models for the $B$ meson wave
function in the literature favor $k^+\sim O(\Lambda)$. That is, the $k_T$
resummation for the $B$ meson is not important. However, I will discuss
this resummation, and allow the behavior of the $B$ meson wave function 
to determine whether its effect is essential.

The major difficulity in summing up the double logarithms in
Figs.~\ref{fig2}(b) and \ref{fig2}(c) arises from the many invariants
that can be constructed from $P_1$, $k$ and $\bar n$, such as $P_1^2$,
$P_1\cdot k$, $P_1\cdot \bar n$, $k\cdot \bar n$ and $\bar n^2$.
In the pion case the invariants are only $P_1\cdot \bar n$ and $\bar n^2$.
The fact that the $B$ meson wave function contains many invariants fails 
the replacement of $d/d k^+$ by $d/d \bar n$, because 
some large scales like $P_1^2$ can not be related to $\bar n$.
Fortunately, this difficulity can be overcome by applying
the heavy quark approximation in Eq.~(\ref{appII}). 
This approximation also holds for collinear gluons with momenta
parallel to $k$, since collinear divergences are independent of the 
direction of the eikonal line that collects the collinear gluons.
Different directions correspond to different shifts of
finite contributions between the wave function and hard kernels,
i.e., to different factorization schemes \cite{L98}. However, it was
argued \cite{GS} that the approximation in Eq.~(\ref{appII}) is not
suitable for collecting collinear gluons.
 
Substituting the eikonal line along $v$ for the $b$ quark line,
self-energy diagrams like Fig.~\ref{fig2}(a) are excluded by
definition \cite{C}. The eikonal approximation also reduces the number
of large invariants involved in the $B$ meson wave function. We have the 
scale invariance in $P_1$ in addition to the scale invariance in $\bar n$. 
Hence, $P_1$ does not lead to a large scale, and the only large scale is 
$k^+$, which must appear through the ratios $(k\cdot \bar n)^2/\bar n^2$ 
and $(k\cdot v)^2/v^2$. At leading-logarithm accuracy,
the second scale does not appear. This observation can be verified
by evaluating the soft function $K$
and the hard function $G$ for $\Phi_+$ \cite{LY1}. However, the above
argument for the survival of a single large scale has been questioned
\cite{GS}. The Sudakov effect associated with the $B$ meson is not
important, and the dispute does not affect the numerics discussed in the
following sections.

Since $\Phi_+$ depends only on the single large scale
$(k\cdot \bar n)^2/\bar n^2$, the derivation reduces to the one in
analogy with the pion case. One obtains
\begin{equation}
\Phi_+(x,b,m_B,\mu)=\exp\left[-s(x,b,m_B)\right]\Phi_+(x,b,\mu)\;,
\label{sb}
\end{equation}
with the same exponent $s$. The intrinsic $b$ dependence, which is
more important for a heavy meson, has been included into the initial
condition of the Sudakov evolution, $\Phi_+(x,b,\mu)$.
The behavior of $\Phi_+(x,b)$, ignoring the single-logarithm evolution
in $\mu$, has been discussed in the previous subsection.

\section{Semileptonic and Radiative Decays}

The $B$ meson decay constant and transition form factors, involving
the hadronic effects in semileptonic and radiative decays,
provide the nonperturbative inputs of many QCD methods. In this section
I review the recent studies of these topics in LCSR, lattice QCD, PQCD
QCDF, SCET, and LFQCD. I will skip the topics on
the $b$ quark mass and on the $B-\bar{B}$ mixing parameter. On one hand,
the heavy quark mass can be determined by means of a two-point 
correlation function similar to that for the heavy meson decay constant.
On the other hand, the above quantities are not very relevant to the
leading-power formalism of nonleptonic $B$ meson decays discussed in
Sec.~5. The results for corresponding $D$ meson decays will be quoted
for comparison.

\subsection{\it Light-Cone Sum Rules \label{sec:sum}}

QCD sum rules \cite{SVZ,ISNR} have been applied to various problems in 
heavy flavor physics. The idea is to calculate a quark-current
correlation function and to relate it to hadronic parameters
via dispersion relations.
Take the $B$ meson decay constant $f_B$ as an example
\cite{6auth,RRY,fB}, which is defined via the matrix element
$\langle 0| m_B\bar{q}i\gamma_5 b| \bar B\rangle=f_Bm_B^2$, $q=u,d$.
Consider the correlation function of two heavy-light currents,
\begin{eqnarray}
\Pi(q^2)=i \int d^4ye^{iq\cdot y}\langle 0| T[m_B\bar{q}i\gamma_5 b(y),
m_B\bar{b}i\gamma_5 q(0)]| 0\rangle\;.
\label{fBcorr}
\end{eqnarray}
The amplitude $\Pi(q^2)$ can be treated by operator product expansion
(OPE) at the quark level, if
$q^2$ is far below $m_b^2$, or parametrized as a sum over hadronic states
including the ground-state $B$ meson for $q^2 \geq m_B^2$.
Assuming the quark-hadron duality, the expressions in the above two
regions are related. Therefore, on the left-hand side of the sum rule,
one has
\begin{eqnarray}
\Pi(q^2)=
\frac{f_B^2m_B^4}{m_B^2-q^2}+\cdots\;,
\label{dispfB}
\end{eqnarray}
where the contribution of the ground-state $B$ meson has been singled 
out, and $\cdots$ represents those from the excited resonances and from
the continuum of hadronic states with the $B$ meson quantum numbers.
On the right-hand side of the sum rule, we have the expansion including 
the perturbative series in $\alpha_s$ and the quark, gluon and
quark-gluon condensates. A simple explanation of the quark-hadron
duality has been given in \cite{CRS93,KR}. Inserting the values of
$\alpha_s$, $m_b$ and the condensates $\langle G^2\rangle$ and
$\langle \bar{q}q \rangle$ into the above sum rule, one estimates $f_B$.

LCSR \cite{lcsr}, employed frequently for studying
exclusive $B$ meson decays, is a simplified version of QCD sum rules.
Consider the $B\to \pi $ transition form factors 
\cite{KR,KRWY,Bagan,BKR,BBKR,PB3}, for which
the correlation function is chosen as
\begin{eqnarray}
i \int \!d^4ye^{iq\cdot y}\langle \pi^+(P_2)|
T[\bar{u}\gamma_\lambda b(y),
m_b\bar{b}i\gamma_5 d(0)]| 0\rangle
\nonumber\\
= F((P_2+q)^2,q^2)P_{2\mu}+\tilde{F}((P_2+q)^2,q^2)q_\mu\;.
\label{lcsrcorr}
\end{eqnarray}
Compared to Eq.~(\ref{fBcorr}), the final state has been specified as a
pion, and the twist expansion has been applied. The presence of
the heavy quark mass justifies the twist expansion.

At large virtuality $\mid (P_2+q)^2-m_b^2 \mid \gg \Lambda_{\rm QCD}^2$
and $q^2\ll m_b^2$, the correlation function is treated by OPE near the
light-cone $y^2=0$. The perturbative part involves a convolution with the
pion distribution amplitude $\phi_\pi(x)$ according to
collinear factorization theorem in Sec.~\ref{sec:all}. The evaluation
becomes simpler: it contains an integral only over the
one-dimensional momentum fraction $x$, instead of over the
four-dimensional loop momentum. The price to pay is that higher-twist
contributions need to be included in terms of inverse powers of
$(P_2+q)^2-m_b^2$. On the hadron side, one has
\begin{eqnarray}
F((P_2+q)^2,q^2)=\frac{2f_B F_+(q^2)m_B^2}{m_B^2-(P_2+q)^2}+\cdots\;,
\label{displcsr}
\end{eqnarray}
where the ground-state contribution from the $B$ meson contains a product
of $f_B$ and the $B\to \pi$ form factor $F_+(q^2)$. The form factor $F_+$,
along with another one $F_0$, are defined via
\begin{eqnarray}
\langle\pi^+(P_2)|{\bar q}\gamma_\mu b|\bar B(P_1)\rangle
=F_+(q^2)\left[(P_1+P_2)_\mu-\frac{m_B^2-m_\pi^2}{q^2}q_\mu\right]
+F_0(q^2)\frac{m_B^2-m_\pi^2}{q^2}q_\mu\;.
\label{f+0}
\end{eqnarray}
The quark-hadron duality then gives the information of $F_+(q^2)$
with $f_B$ being extracted from Eq.~(\ref{dispfB}). 

The resulting sum rule is written as
\bea\label{fplus}
F_+(q^2)&=&\frac{1}{2m_B^2f_B}
\exp \left( \frac{m_B^2}{M^2}\right)
\Bigg [ F_0^{(2)}(q^2,M^2,m_b^2,s_0^B,\mu)
\nonumber\\
&&+\frac{\alpha_s(\mu)}{3\pi} F_1^{(2)}(q^2,M^2,m_b^2,s_0^B,\mu)
+ F_0^{(3,4)}(q^2,M^2,m_b^2,s_0^B,\mu) \Bigg ]\;.
\eea
The mass scale $M$ is associated with a Borel transformation usually
performed in sum rule calculations. The scale $\mu$ is the
factorization scale separating soft and hard dynamics.
The effective threshold parameter $s_0^B$ sets the lower bound of
the $B$ meson invariant, $(P_2+q)^2 \ge s_0^B$, above which
the quark-hadron duality is assumed to hold. Long-distance
dynamics characterized by scales lower than $\mu$ is absorbed into the
universal nonperturbative pion distribution amplitudes. The first two
terms on the right-hand side of Eq.~(\ref{fplus}) represent the twist-2
contributions up to next-to-leading order. The third term represents
the leading-order twist-3 and twist-4 contributions.

For illustration, the leading term $F_0^{(2)}$ is given by
\bea
\label{twist2}
F_0^{(2)}(q^2,M^2,m_b^2,s_0^B,\mu)  = 
m_b^2 f_{\pi} \int\limits_{\Delta}^1\frac{dx}{x} 
\exp\left(-\frac{m_b^2-q^2(1-x)}{xM^2} \right)\phi_\pi(x,\mu)\;.
\eea
The lower integration boundary $\Delta=(m_b^2-q^2)/(s_0^B-q^2)$
originates from the subtraction of excited resonances and continuum
states from both sides of the sum rule, which contribute to the
dispersion integral in the $B$ meson channel. These contributions are
identical on both sides of the sum rule because of the
quark-hadron duality assumed above.
The explicit expressions for the remaining terms $F_1^{(2)}$ and
$F_0^{(3,4)}$ can be found in \cite{KRWY,Bagan} and in
\cite{BKR,BBKR}, respectively. The radiative correction to the
twist-3 cobtribution has been available \cite{BZ01}.

The $D$-meson decay constant $f_D$ can be derived by a simple $b\to c$
($\bar{B}\to D$) replacement in the sum rule for $f_B$ in
Eq.~(\ref{dispfB}), and by the necessary adjustment of the renormalization
scale. One can also predict the ratios $f_{B_s}/f_{B}$ and
$f_{D_s}/f_{D}$ by setting the quark field $q=s$ in Eq.~(\ref{fBcorr}).
The values of $f_{B}$ and $f_D$
are sensitive to the $b$ and $c$ quark pole masses, $m_b$ and $m_c$.
Varying these masses in the intervals,
\begin{eqnarray}
m_b=4.8 \pm 0.1~ \mbox{GeV}, ~~m_c= 1.3 \pm 0.1 ~\mbox{GeV},
\label{masses}
\end{eqnarray}
one obtains \cite{CK}
\bea
f_B= 170 \mp 30~\mbox{MeV},~~   f_D=180 \mp 30~\mbox{MeV}\,, 
\label{fBD}
\nonumber
\\
f_{Bs}/f_B=1.16\pm 0.09 \,, ~~f_{D_s}/f_D=1.19\pm 0.08\,.  
\eea
Within uncertainties, the predictions in Eq.~(\ref{fBD})
agree with the lattice determinations of the heavy meson decay constants
quoted in the next subsection.

\begin{figure}[t!]
\begin{center}
 \epsfig{file=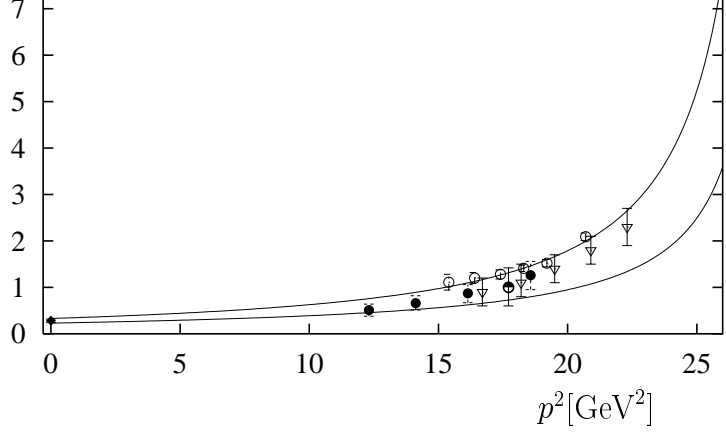,height=2.5in} 
\caption{The LCSR prediction for the $B\to \pi$ form factor $F_+(p^2)$
\cite{KRWWY00}. The full curves 
indicate the theoretical uncertainty, and the points
represent various lattice QCD calculations
from FNAL \cite{Flynnk} (full circles),
UKQCD \cite{UKQCDk} (triangles), APE \cite{APEk} (full square), 
JLQCD \cite{JLQCDk} (open circles), and ELC \cite{Flynnk}
(semi-full circle).}
\label{fig7}
\end{center}
\end{figure}

The LCSR predictions for $F_+$ \cite{KRWWY00} are presented in
Fig.~\ref{fig7}. This calculation includes twist-2 
(leading-order and next-to-leading-order) and twist-3,4 effects. The
twist-2 and twist-3 contributions are roughly equal. The twist-4
contribution is less than 10\% in the fast recoil region. The results are
insensitive to the nonasymptotic behavior of the pion distribution
amplitudes. At the maximal recoil, one finds \cite{KRWWY00}
\begin{eqnarray}
F_+^{B\pi}(0)=0.28 \pm 0.05\;,\;\;\;\;
F_+^{D\pi}(0)=0.65 \pm 0.11\;.
\end{eqnarray}

Note that QCD sum rules have a limited accuracy due to the truncation in
OPE, to the duality approximation, to the variation of the corresponding
auxiliary parameters, such as the Borel mass $M$, and to the contributions
of excited states. A detailed discussion on the uncertainty from
the above sources can be found in \cite{KRWWY00}. Moreover, large 
radiative correction to
the $B$ meson vertex, which reaches 35\% of the full contribution,
or about half of the leading-order contribution, has been noticed
in the correlation function in Eq.~(\ref{lcsrcorr}). 
This $O(\alpha_s)$ correction renders the sum rule for $f_BF_+$
quite unstable relative to the variation of input parameters
\cite{KRWY,PB3}. This is the reason one considers the sum rule for $f_B$
at the same time in order to stabilize the sum rule for $f_BF_+$:
the sum rule for $f_B$ also receives large radiative correction to the
$B$ meson vertex, such that the two large vertex corrections cancel in 
the ratio $f_BF_+/f_B$ \cite{WZH}.
However, the radiative correction to $f_B$ then becomes large. Therefore,
an evaluation of $O(\alpha_s^2)$ corrections to both the sum rules is
necessary. Progress has been made in the calculation of the three-loop
radiative corrections to the heavy-to-light correlator
\cite{Chetyrkin2001}.


Replacing the pion with the kaon (including the distribution amplitudes
and the decay constants) in the correlation function in
Eq.~(\ref{lcsrcorr}), one obtains LCSR for the $B\to K$ form factor,
and the ratios \cite{KRWWY00},
\begin{eqnarray}
F_+^{BK}(0)/F_+^{B\pi}=1.28^{+0.18}_{-0.10} \;,\;\;\;\;
F_+^{DK}(0)/F_+^{D\pi}=1.20\;,
\end{eqnarray}
where the uncertainty of the first ratio arises from the strange quark
mass $m_s( 1 \mbox{GeV}) = 150 \mp 50$ MeV. This result indicates
that $SU(3)$ breaking effects might be significant.

\begin{figure}[t!]
\begin{center}
 \epsfig{file=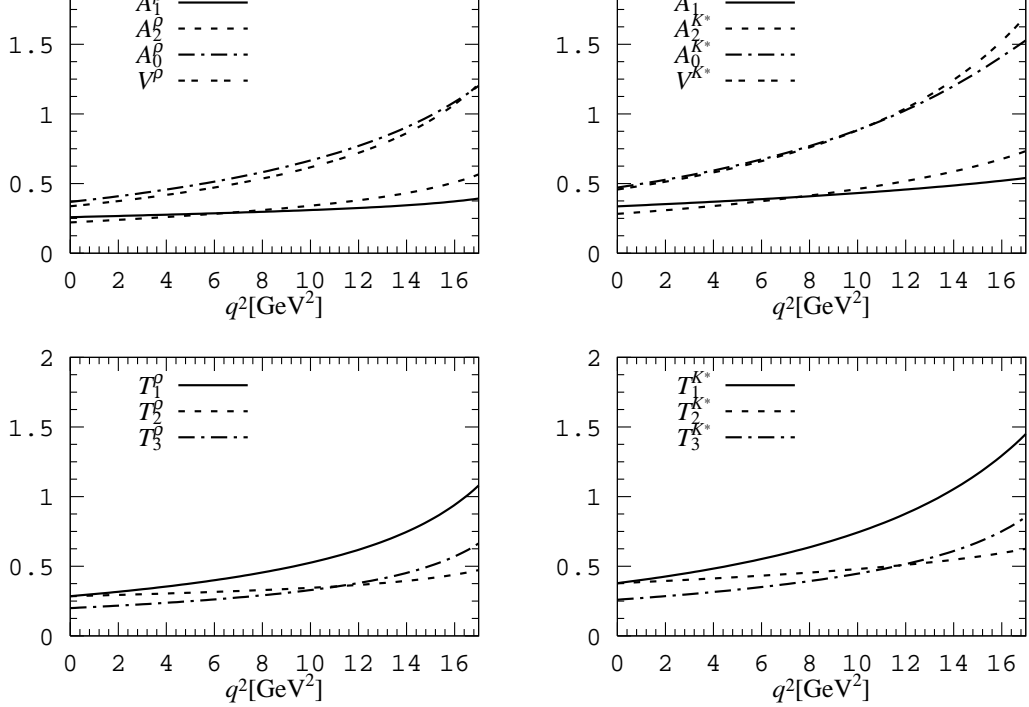,height=4.0in} 
\caption{LCSR results for $B\to V$ transition form factors.}
\label{tBud}
\end{center}
\end{figure}

The $B\to V$ form factors, $V=K^*,\rho,\phi$, associated with the
semileptonic decays $B\to V l \nu $ and with the radiative decays
$B\to V \gamma$, can be analyzed in a similar way. The semileptonic form
factors and {\em penguin} form factors are defined via the matrix
elements,
\begin{eqnarray}
\langle V(P_2,\epsilon) |{\bar q}\gamma_\mu(1-\gamma_5)b |
\bar B(P_1)\rangle & = &
-i\epsilon^*_\mu (m_B+m_V)A_1^V(q^2) + i (P_1+P_2)_\mu
(\epsilon^*\cdot P_1)\,\frac{A_2^V(q^2)}{m_B+m_V}
\nonumber\\
\lefteqn{+\: i q_\mu (\epsilon^* \cdot P_1) \,\frac{2m_V}{q^2}\,
\left[A_3^V(q^2)-A_0^V(q^2)\right] +
\epsilon_{\mu\nu\rho\sigma}\epsilon^{*\nu} P_1^\rho P_2^\sigma\,
\frac{2V^V(q^2)}{m_B+m_V}\,,}\hspace*{2cm}
\label{eq:SL}
\end{eqnarray}
and
\begin{eqnarray}
\langle V (P_2,\epsilon)| \bar q \sigma_{\mu\nu} q^\nu (1+\gamma_5)b|
\bar B(P_1)\rangle & = & i\epsilon_{\mu\nu\rho\sigma} \epsilon^{*\nu}
P_1^\rho P_2^\sigma \, 2 T_1(q^2)\nonumber\\
& & {} + T_2(q^2) \left[ \epsilon^*_\mu
  (m_B^2-m_{V}^2) - (\epsilon^*\cdot P_1) \,(P_1+P_2)_\mu \right]
  \nonumber\\
& & {} + T_3(q^2)
(\epsilon^*\cdot P_1) \left[ q_\mu - \frac{q^2}{m_B^2-m_{V}^2}\,
(P_1+P_2)_\mu\right]\;,
\label{eq:T}
\end{eqnarray}
respectively, where $\epsilon$ denotes the polarization of the vector
meson $V$. I simply quote the results in \cite{BallBraun}
as shown in Fig.~\ref{tBud}.
LCSR has been also applied to the $B\to \rho \gamma$ weak annihilation 
\cite{KSW,AB}, the penguin form factor in the $B\to\eta$ transition
\cite{AS03}, and the $B\to \mu \nu \gamma$ width \cite{KSW} 
employing the photon distribution amplitude. 


\subsection{\it Lattice QCD \label{sec:lat}}

The $B$ meson decay constant and transition form factors, defined as
hadronic matrix elements in the previous subsection, can be calculated
directly on the lattice. For recent reviews on the application of
lattice QCD to exclusive $B$ meson decays, refer to \cite{DB02,LL02}.
Many results have been obtained by different
groups using different heavy quark methods, each of which has different
systematic errors. For example, the UKQCD and APE groups used an
$O(a)$-improved Sheikholeslami-Wohlert (SW)
action, $a$ being the lattice spacing, which is defined at the scales of
the $c$ quark mass. Outcomes are then extrapolated to the $b$ quark mass
following the evolution governed by HQET.
The Fermilab group (FNAL) \cite{FnalHQ,ZS-fullFNAL} identified and 
correctly renormalized nonrelativistic operators in the SW action,
such that discretization errors reduce from $O(am_Q)$ to
$O(a\Lambda_{\rm QCD})$. JLQCD adopted the action derived from
non-relativistic QCD (NRQCD). 

\begin{figure}[t!]
\begin{center}
 \epsfig{file=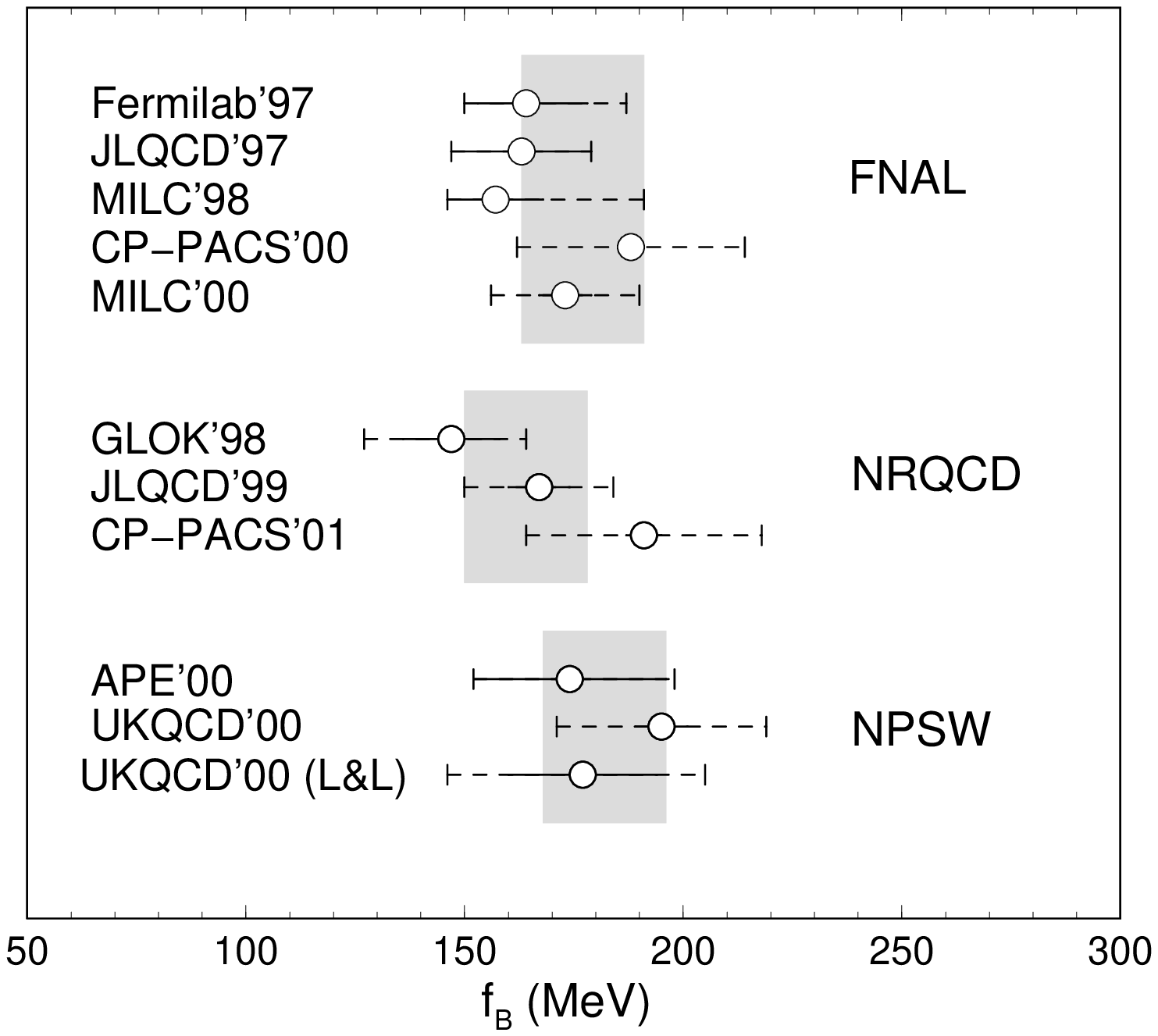,height=2.5in} 
 \hspace{1.0cm}
 \epsfig{file=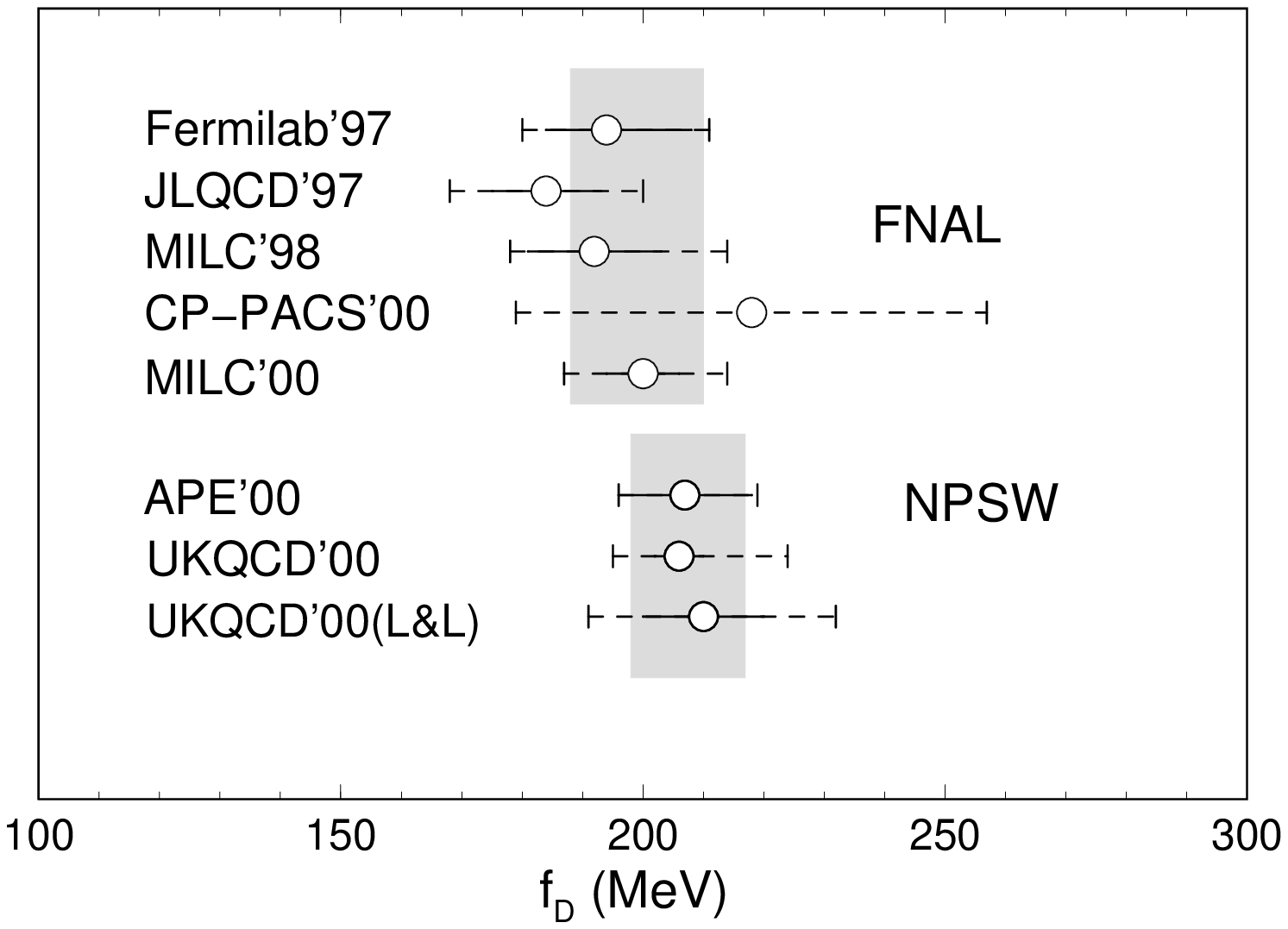,height=2.5in}
\caption{Recent determinations of $f_B$ and $f_D$ in the quenched 
approximation.}
\label{fig8}
\end{center}
\end{figure}

Recent determinations of $f_B$ and $f_D$ in the quenched approximation
are summarized in
Fig.~\ref{fig8} \cite{SRl} with the references, from top to bottom,
\cite{FNAL:fB},\cite{JLQCD:fB}, \cite{MILC98:fB}, \cite{CPPACS:fBKKM}, 
\cite{MILClat00}, \cite{GLOK98:fb}, \cite{JLQCD99:fB}, 
\cite{CPPACS:fBNRQCD}, \cite{APE00:fB}, \cite{UKQCD:fB}, and
\cite{UKQCD:fBLL} for $f_B$, and \cite{FNAL:fB},\cite{JLQCD:fB}, 
\cite{MILC98:fB}, \cite{CPPACS:fBKKM}, \cite{MILClat00}, 
\cite{APE00:fB}, and \cite{UKQCD:fB}, and \cite{UKQCD:fBLL} for $f_D$. 
Values of $f_B$ derived for a given heavy quark action
are consistent. The solid bands, representing the
average for a particular heavy quark action, are in agreement 
with each other. The values of $f_B$ and $f_D$ in Fig.~\ref{fig8} are
also consistent with the LCSR results in \cite{KRWWY00}.

A large source of uncertainty comes from the extrapolation from the
scales of the $c$ quark mass to those of the $b$ quark mass.
There are other subtle
issues, such as the scaling violation from discretization, and the
determination of $f_B$ at rest and at non-zero momentum \cite{davies01}
and from the temporal $A_0$ and spatial $A_k$ currents in the matrix
element $\langle 0|A_\mu |B(P)\rangle = f_BP_\mu$.
More detailed discussion on the above topics can be found \cite{SRl}.

	\begin{table*}[htb]
\caption{$B$ and $D$ meson decay constants with $N_f=2$.}
	\label{Tab:fBNf2}
	\begin{tabular}{@{}|l|cccc|}
	\hline
        Group & $f_B$ (MeV)
        & $\displaystyle{\frac{f_B^{N_f=2}}{f_B^{N_f=0}}}$
        &$f_D$ (MeV) & $\displaystyle{\frac{f_D^{N_f=2}}{f_D^{N_f=0}}}$\\
	\hline
Collins99~\cite{NRQCDfBNf2} &$186(5)(25)(^{+50}_{-0})$  &$\simeq 1.26$ &  &\\
MILC'00~\cite{MILClat00}    &$191(6)(^{+24}_{-18})(^{+11}_{-0})$   
		       &$\simeq 1.10$ 
                       &$215(5)(^{+17}_{-13})(^{+8}_{-0})$ 
		       &$\simeq 1.08$ \\
MILC'01 ($N_f=2+1$)~\cite{bernard01}    
		       &  & $1.23(3)(11)$  &   &\\
CP-PACS'00(FNAL)~\cite{CPPACS:fBKKM}     
		       &$208(10)(29)$ & 1.11(6) & 225(14)(40) & 1.03(6)\\
CP-PACS'00(NR)~\cite{CPPACS:fBNRQCD} 
		       & $204(8)(29)(^{+44}_{-0})$ 
		       & $1.10(5)$ &  &\\     
JLQCD~\cite{yamada-talk}  
		       & $190(14)(7)$  &$\simeq 1.14$ &  &\\
	\hline 
	\end{tabular}\vskip -.15truein
	\end{table*}

Calculations of decay constants with dynamical quarks have been available,
whose results are listed in Table~\ref{Tab:fBNf2} \cite{SRl}.
It is observed that $f_B$ is larger in the unquenched theory.
The difference between the quenched and unquenched predictions depends on
which type of the valence chiral extrapolation (linear, quadratic or
logarithmic \cite{sharpe-zhang}) is used.
%
%
Note that $f_D$ may in fact be smaller
than the reported value due to discretization effects.
It is also observed from Table~\ref{Tab:fBNf2} that 
the dynamical effect for $D$ mesons is smaller than 
for $B$ mesons. 


\begin{figure}[t!]
\begin{center}
 \epsfig{file=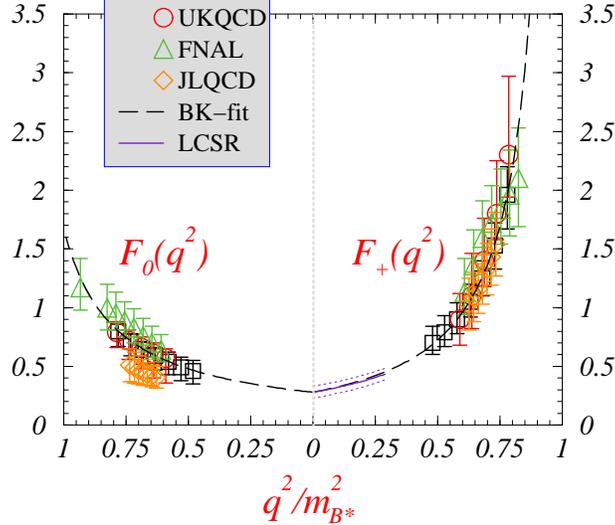,height=3.0in} 
\caption{Recent lattice results for $F_+(q^2)$ and $F_0(q^2)$ in the
quenched approximation
with $F_+(0)=F_0(0)$. Prediction obtained from LCSR \cite{KRWWY00}
are also shown.}
\label{fig9}
\end{center}
\end{figure}

The $B\to\pi$ transition form factors $F_+(q^2)$ and $F_0(q^2)$
have been calculated in lattice QCD recently
\cite{ukqcd-bpi,ape-bpi,fnal-bpi,jlqcd-bpi,shigemitsu-bpi}, and the 
results are presented in Fig.~\ref{fig9} \cite{DB02}. 
The data show general agreement among different groups, within the
quoted uncertainties. The agreement is especially good for the form 
factor $F_+(q^2)$. Note that the quenching effects may be significant, 
particularly for the form factor $F_0(q^2)$ \cite{bpi-bpz}. 
The lattice results are available only for 
large $q^2$ (smail recoil). Since the bulk of the experimental data
is located at small $q^2$, one needs to extend the calculation 
to this region at currently accessible lattice spacings to avoid a 
model-dependent extrapolation from large $q^2$.
The new value for the $B\to K^*$ form factor at maximal recoil,
$T_1(0)=T_2(0)\equiv T(0)$ defined in Eq.~(\ref{eq:T}), is given by
\cite{DB02}
\bea
T^{BK^\ast}(0) = 0.24(5)\left(^{+1}_{-2}\right)\;,
\eea
which is smaller than the LCSR ones 
\cite{BallBraun,qsr-bkstar} shown in Fig.~\ref{tBud}.


\subsection{\it Hard-scattering Picture \label{sec:pqcd}}

In the PQCD approach hard dynamics is assumed to dominate in the
$B$ meson transition form factors. Soft contribution, 
though indeed playing a role, is less important because of 
suppression from the Sudakov
mechanism. Unlike QCD sum rules, soft contribution can not be included
into the PQCD formalism in a consistent way: if there is no hard gluon 
exchange to provide a large characteristic scale, twist expansion does not 
hold. Therefore, soft contribution can not be estimated using the same
meson distribution amplitudes resulting from
twist expansion. If it has to be added, it must be introduced as 
an independent input, similar to the treatment in the QCDF approach.
The values of these inputs usually come from QCD sum-rule or lattice 
calculations, which, in principle, can contain perturbative contributions.
Then a double counting of the perturbative contribution, which exists 
already in the one-gluon-exchange diagrams, may not be avoidable.

It has been explained that the internal $\bar b$ quark 
involved in the hard kernel becomes on-shell as the momentum fraction 
$x$ of the $d$ quark vanishes in Fig.~\ref{aa2} \cite{L5}. The 
contributions to the $B\to \pi$ form factor $F^{B\pi}$ are then 
logarithmically divergent at twist 2 and linearly divergent at twist 3 
in collinear factorization theorem. It has been argued that as the 
end-point region is important, the corresponding large double logarithms
$\alpha_s\ln^2 x$ need to be organized into a jet function $J(x)$ as
a consequence of threshold resummation \cite{L5}. This jet function,
diminishing as $x\to 0, 1$, modifies the end-point behavior of
meson distribution amplitudes effectively.
In \cite{TLS} the following approximate form has been proposed
for convenience,
\begin{eqnarray}
J(x)=\frac{2^{1+2c}\Gamma(3/2+c)}{\sqrt{\pi}\Gamma(1+c)} [x(1-x)]^c\;,
\label{trs}
\end{eqnarray}
where the parameter $c\approx 0.3$ is determined from the best fit to 
Eq.~(\ref{mjx}). The above expression is normalized 
to unity.

Similarly, the inclusion of $k_T$ also regulates the end-point
singularity, and large double logarithms $\alpha_s\ln^2 k_T$
are then produced from higher-order corrections. These double logarithms
should be organized to all orders, leading to $k_T$
resummation \cite{BS,CS}. The resultant Sudakov form factor, 
constructed in Sec.~\ref{sec:kt} \cite{LY1},
controls the magnitude of $k_T^2$ to be roughly
$O(\Lambda m_B)$ by suppressing the region with
$k_T^2\sim O(\Lambda^2)$. The coupling constant
$\alpha_s(\sqrt{\Lambda m_B})/\pi \sim 0.13$ is then small enough to
justify the PQCD evaluation of heavy-to-light form factors \cite{KLS}.
Note that either threshold or $k_T$ resummation smears the end-point
singularity. However, to suppress the soft contribution sufficiently,
both resummations are required, such that the reasonable values of the 
$B\to\pi$ form factors can be obtained.

The $B\to\pi$ form factors $F_+$ and $F_0$ are written as,
\begin{eqnarray}
F_+&=&\frac{1}{2}(f_1+f_2)\;,
\nonumber\\
F_0&=& \frac{1}{2}\left[\left(1+ \frac{q^2}{m_B^2}\right)f_1
+ \left(1 - \frac{q^2}{m_B^2}\right)f_2\right]\;,
\label{f12}
\end{eqnarray}
with 
\begin{eqnarray}
f_1&=&16\pi m_B^2C_Fr_\pi\int dx_1dx_2\int b_1db_1 b_2db_2\Phi_+(x_1,b_1)
[\phi_\pi^p(x_2)-\phi_\pi^t(x_2)] E(t^{(1)})h(x_1,x_2,b_1,b_2)\;,
\label{f1}\\
f_2&=&16\pi m_B^2C_F\int dx_1dx_2\int b_1db_1 b_2db_2\Phi_+(x_1,b_1)
\nonumber\\
&\times& \Biggl\{\left[\phi_\pi(x_2)(1+x_2\eta)
+2r_\pi\left((\frac{1}{\eta} -x_2 )\phi_\pi^t(x_2) -x_2\phi_\pi^p(x_2)
 \right)\right]E(t^{(1)})h(x_1,x_2,b_1,b_2)
\nonumber\\
& &+ 2r_\pi\phi_\pi^p E(t^{(2})h(x_2,x_1,b_2,b_1)\Biggr\}\;.
\label{fpi}
\end{eqnarray}
The variable $\eta$ is the pion energy fraction, and the evolution 
factor given by,
\begin{eqnarray}
E(t)=\alpha_s(t)e^{-S_B(t)-S_\pi(t)}\;.
\label{evol}
\end{eqnarray}
The explicit expressions of the Sudakov exponents $S_B$ and $S_\pi$
are referred to \cite{LY1}. The ratio $r_\pi$ is defined as 
$r_\pi=m_0/m_B$, where the mass scale $m_0$, related to the chiral
symmetry breaking, comes from the normalization of the two-parton twist-3
distribution amplitudes $\phi_\pi^{p,t}$. The contributions from 
$\phi_\pi^{p,t}$ are of leading power \cite{TLS}, and need to be taken 
into account as mentioned in Sec.~\ref{sec:b}. The first (second) terms 
in Eq.~(\ref{fpi}) correspond to Fig.~\ref{aa2}(a) [Fig.~\ref{aa2}(b)].

The hard function is written as
\begin{eqnarray}
h(x_1,x_2,b_1,b_2)&=&J(x_2)K_{0}\left(\sqrt{x_1x_2\eta}m_Bb_1\right)
\nonumber \\
& &\times \left[\theta(b_1-b_2)K_0\left(\sqrt{x_2\eta}m_B
b_1\right)I_0\left(\sqrt{x_2\eta}m_Bb_3\right)\right.
\nonumber \\
& &\left.+\theta(b_2-b_1)K_0\left(\sqrt{x_2\eta}m_Bb_2\right)
I_0\left(\sqrt{x_2\eta}m_Bb_1\right)\right]\;.
\label{dh}
\end{eqnarray}
The jet function $J(x)$ suppresses the end-point behavior of the pion
distribution amplitudes, especially of the twist-3 ones.
The hard scales $t$ are defined as
\begin{eqnarray}
t^{(1)}&=&{\rm max}(\sqrt{x_2\eta}m_B,1/b_1,1/b_2)\;,
\nonumber\\
t^{(2)}&=&{\rm max}(\sqrt{x_1\eta}m_B,1/b_1,1/b_2)\;.
\end{eqnarray}
It is obvious that turning off threshold and $k_T$ resummations with
$\alpha_s$ fixed, Eqs.~(\ref{f1}) and (\ref{fpi}) are infrared divergent.

As stated in Sec.~\ref{sec:lat}, lattice calculations become more 
difficult in the large recoil region of the light meson. However, this 
region is the one where PQCD is applicable \cite{LY1,DJK}, indicating that 
the PQCD and lattice approaches complement each other. In LCSR 
\cite{PB3,KR98} dynamics of the $B\to\pi$ form factors have been assumed 
to be dominated by a scale larger than $O(\sqrt{\Lambda m_B})$. This is 
the reason the twist expansion into Fock states in powers of $1/m_B$ 
applies to the pion bound state. If this assumption is valid, PQCD should 
be also applicable to the $B\to\pi$ form factors. 
I emphasize that the ``soft" contributions have different
meanings in LCSR and in PQCD. The soft contribution defined in the former
has been multiplied by the perturbative Sudakov factor in the latter,
such that the soft contribution is large in the former, but
small in the latter. A good explanation has been provided in \cite{SR}.
The definitions of the ``hard" contributions are also different,
since the twist expansion has been employed for the $B$ meson bound state
in PQCD, but not in LCSR. Briefly speaking, the contributions of
various orders and powers have been organized in different ways in
LCSR and in PQCD (also different in LFQCD discussed below).
Hence, the soft dominance concluded in LCSR does not apply to PQCD 
\cite{KMU}, and there is no conflict between the basic assumptions
in the two approaches. For PQCD to be a self-consistent theory, it is 
only necessary to examine the converegnce of subleading contributions.

For the $B$ meson wave function, the model \cite{KLS}
\begin{eqnarray}
\Phi_+(x,b)=N_Bx^2(1-x)^2
\exp\left[-\frac{1}{2}\left(\frac{xm_B}{\omega_B}\right)^2
-\frac{\omega_B^2 b^2}{2}\right]\;,
\label{os}
\end{eqnarray}
has been adopted in \cite{TLS}. The shape parameter $\omega_B\sim 0.4$ GeV 
has been fixed from the fit to the $B\to\pi$ form factors derived from 
lattice QCD \cite{UKQCD} and from LCSR \cite{PB3}. The normalization 
constant $N_B$ is related to the decay constant $f_B=190$ MeV through the 
relation
\begin{eqnarray}
\int_0^1 dx\phi_+(x)=\int_0^1 \Phi(x,b=0)=\frac{f_B}{2\sqrt{2N_c}}\;.
\end{eqnarray}
It is easy to find that Eq.~(\ref{os}) has a maximum at 
$x\sim \Lambda/m_B$. 
The models for the pion distribution amplitudes
can be found in \cite{PB1}.

\begin{figure}[t!]
\begin{center}
\epsfig{file=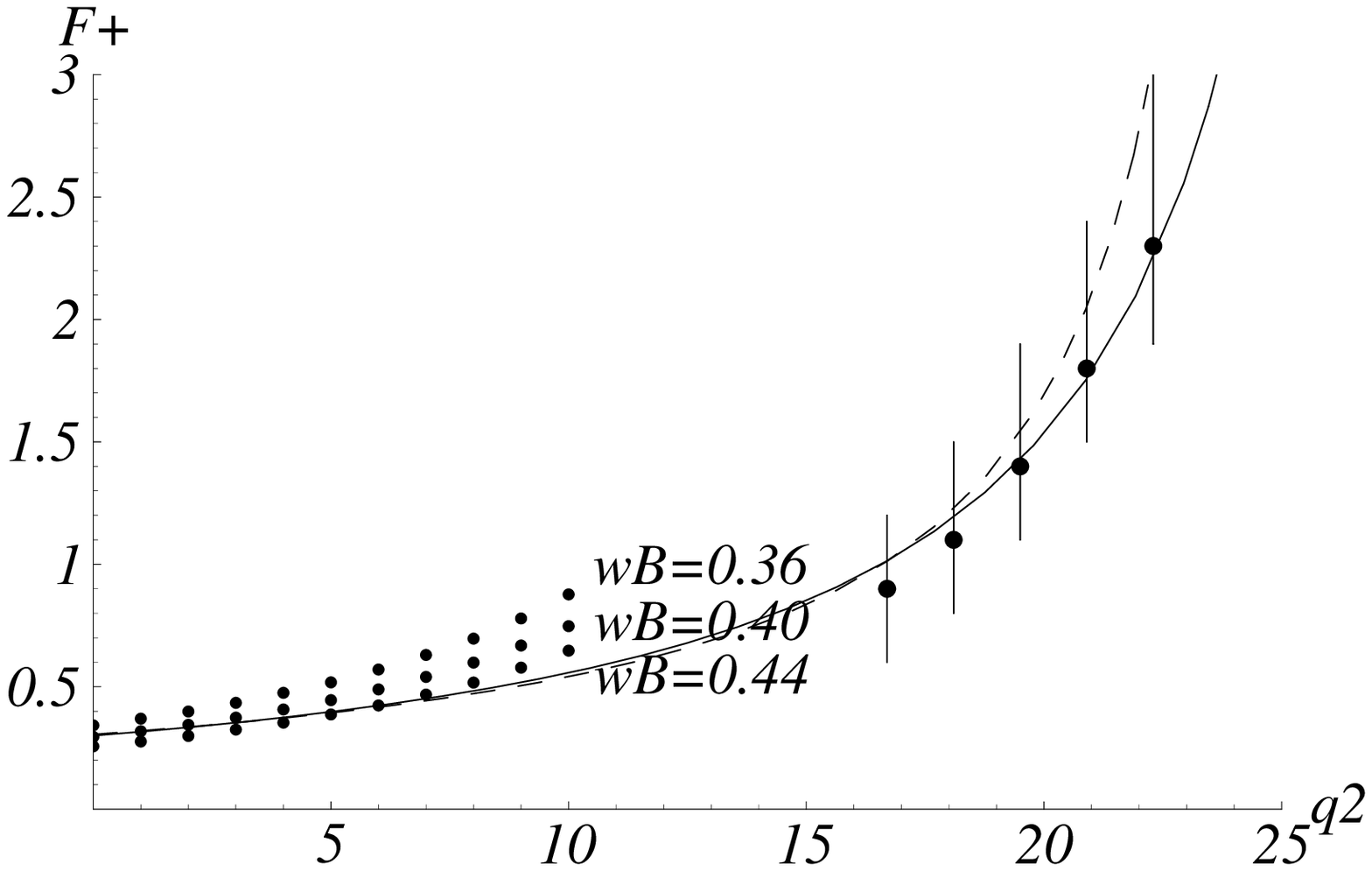,height=2.0in}
\epsfig{file=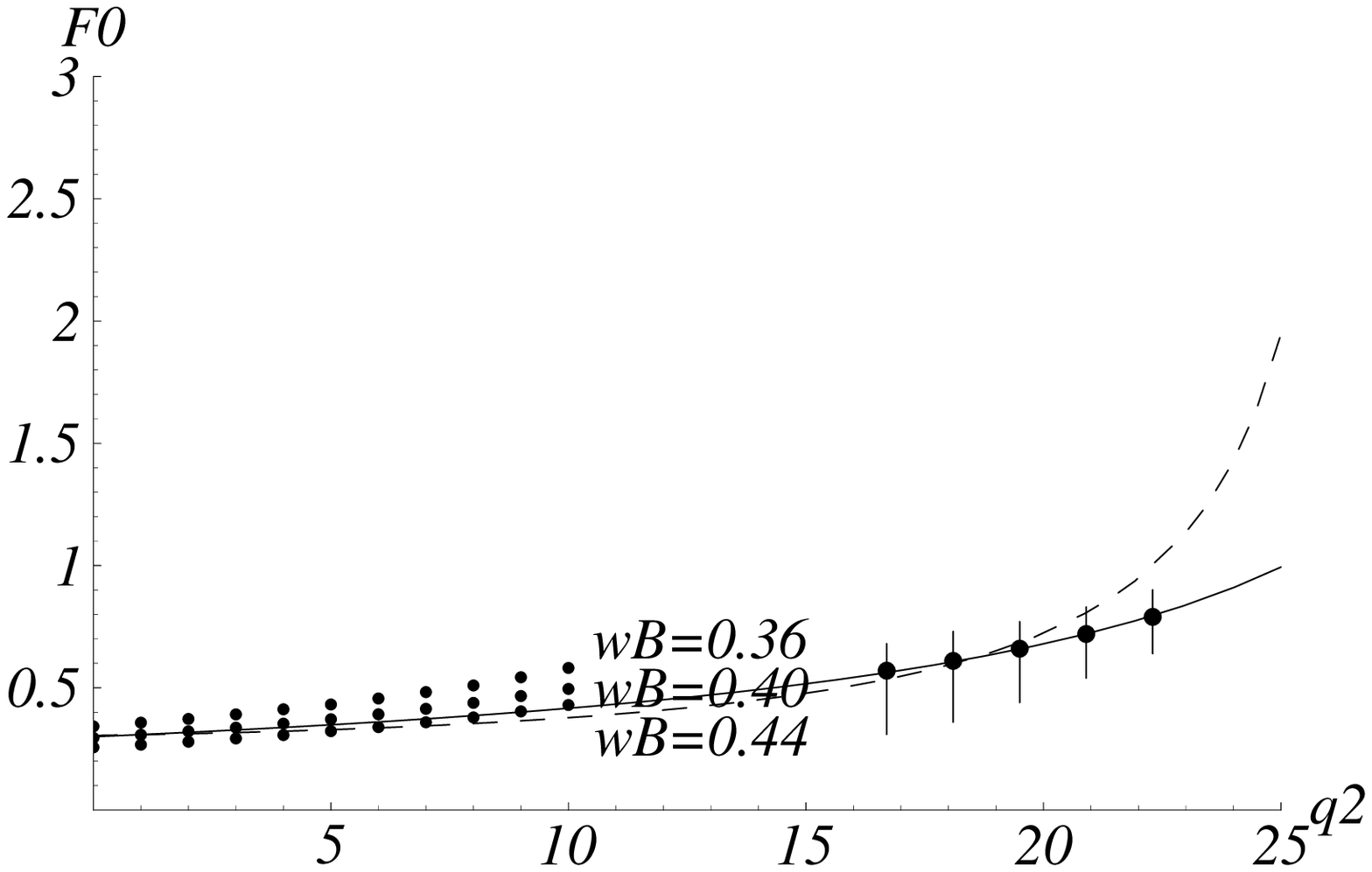,height=2.0in}
\end{center}
\caption{The $B\to \pi$ form factors $f_+$ and $f_0$ as
functions of $q^2$ (GeV$^2$). PQCD results for $\omega_B=0.36$,
0.40, and 0.44 GeV are shown in dots.
The solid lines correspond to fits to the lattice QCD results 
\cite{UKQCD} with errors. The dashed lines come from LCSR 
\cite{PB3}.}
\label{bpi}
\end{figure}

The relative importance of the twist-2 and twist-3 contributions to 
$F_+(q^2)$ has been investigated, and the results are listed in
Table~\ref{tbl-twist}. It is found that the latter are comparable
to the former, consistent with the argument that the twist-3
contributions are not power-suppressed. 
The approximately equal weights of the twist-2 and higher-twist 
contributions to $F_+$ have been also observed in LCSR \cite{KRWY}. 
We compare the PQCD results of 
$F_+(q^2)$ and $F_0(q^2)$ for $q^2 = 0 \sim 10$ GeV$^2$ with those 
derived from lattice QCD \cite{UKQCD} and from LCSR
\cite{PB3} in Fig.~\ref{bpi}, where lattice results have been 
extrapolated to the small $q^2$ region. Different extrapolation 
methods cause uncertainty only of about 5\% \cite{DB}.  
The agreement at large recoil indicates that 
$\omega_B\sim 0.4$ GeV is a good choice. The fast rise of the
PQCD curves at slow recoil indicates that perturbative calculation
becomes unreliable gradually.  
The range $\omega_B = 0.40\pm 0.04$ GeV, corresponding to 
$F_+(0)=0.30\pm 0.04$, has been taken as one of the inputs
of the PQCD approach to two-body nonleptonic $B$ meson decays. 

\begin{table}
\begin{center}
\begin{tabular}{|c|ccccccccccc|}
\hline
$q^2$ (GeV${}^2$)&0.0&1.0&2.0&3.0&4.0&5.0&6.0&7.0&8.0&9.0&10.0\\\hline
twist 2 &0.120&0.128&0.138&0.148&0.159&0.172&0.188&0.204&0.223&0.243&0.270\\
twist 3 &0.177&0.193&0.210&0.230&0.253&0.279&0.308&0.344&0.385&0.432&0.487\\\hline
total&0.297&0.321&0.348&0.378&0.412&0.451&0.496&0.548&0.608&0.675&0.757\\
\hline
\end{tabular}
\end{center}
\caption{
Contributions to $f_+(q^2)$ from the twist-2 and two-parton twist-3
pion distribution amplitudes.}
\label{tbl-twist}
\end{table}

\begin{figure}[t!]
\begin{center}
\epsfig{file=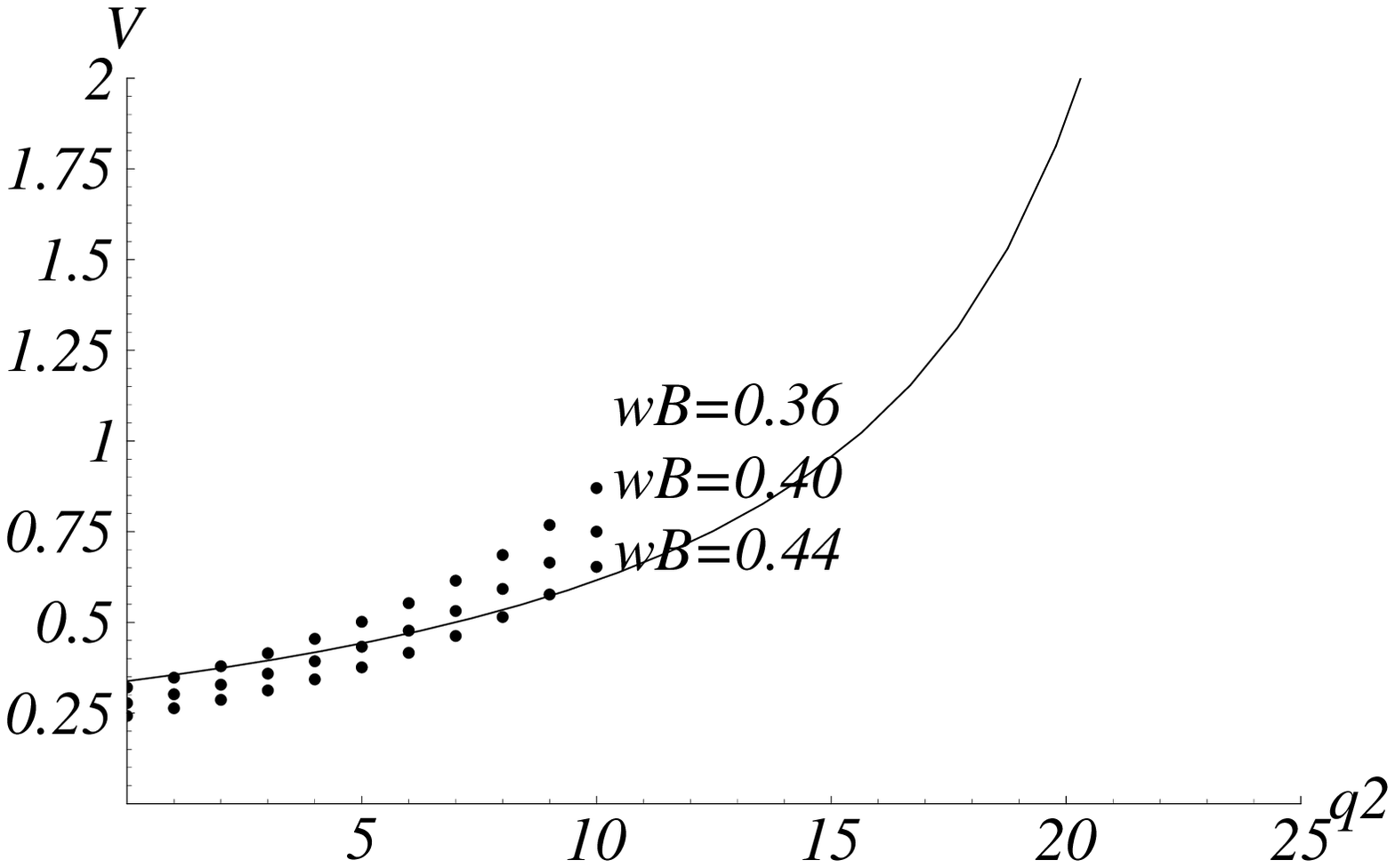,height=5.3cm}
\epsfig{file=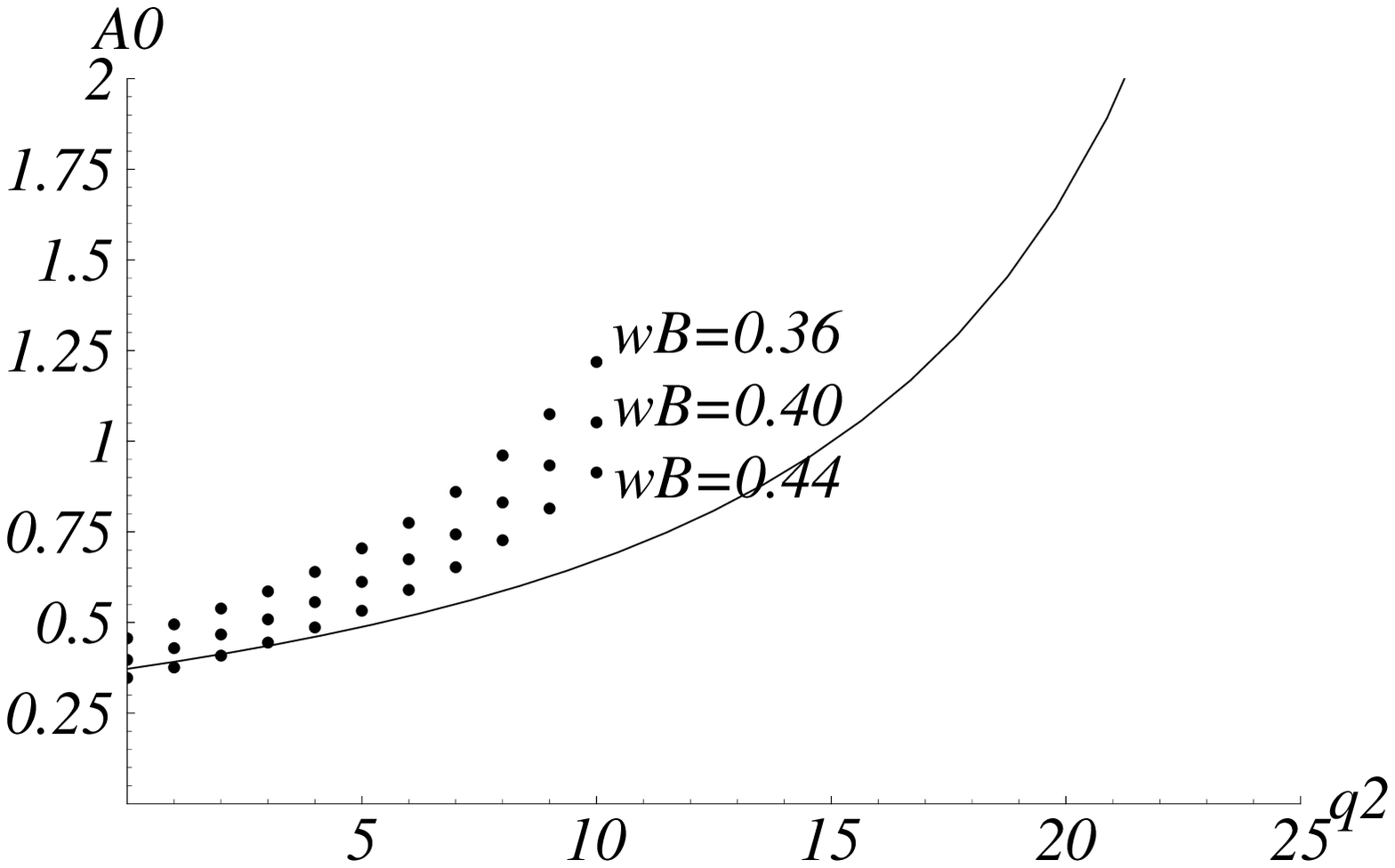,height=5.3cm}
\epsfig{file=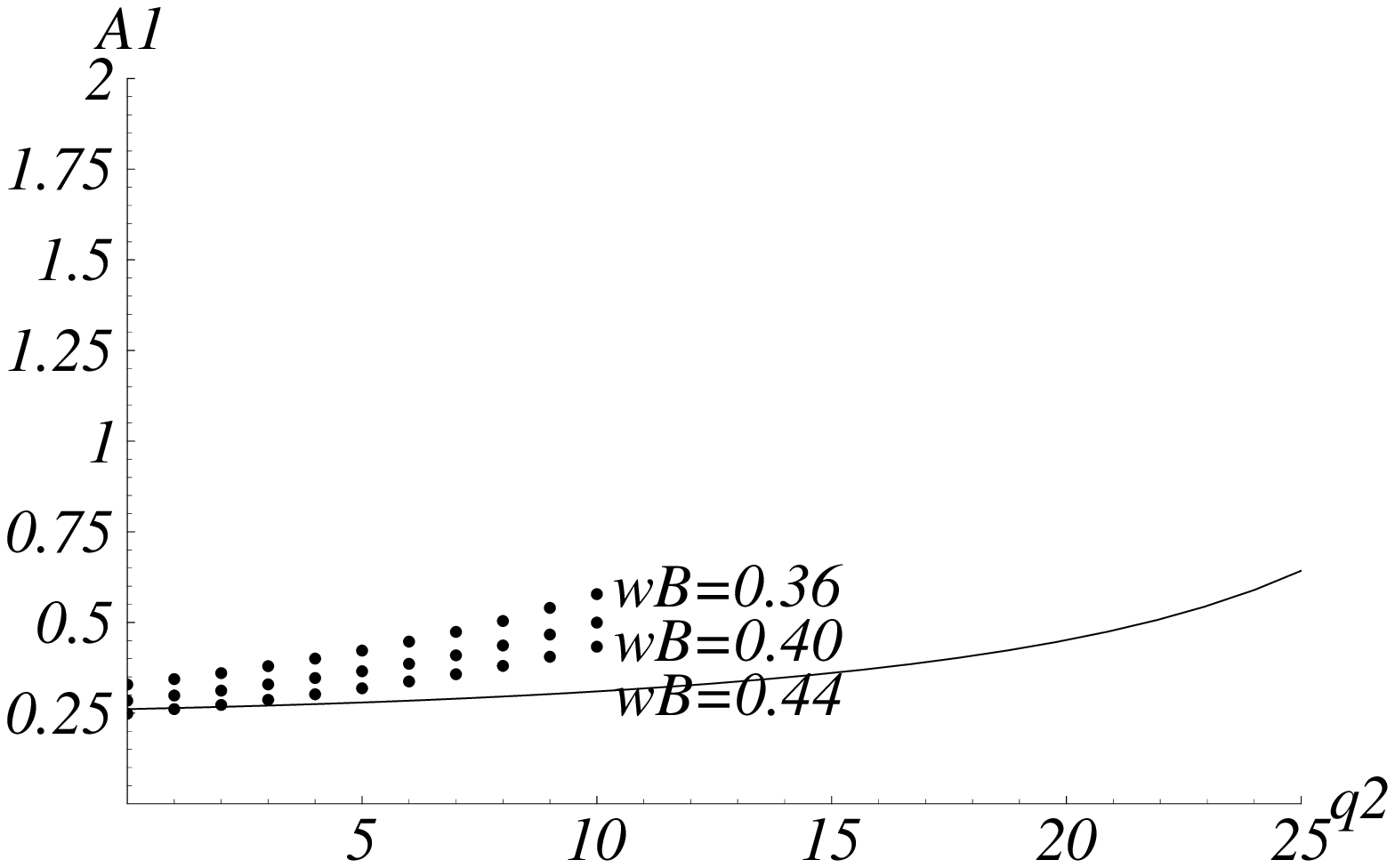,height=5.3cm}
\epsfig{file=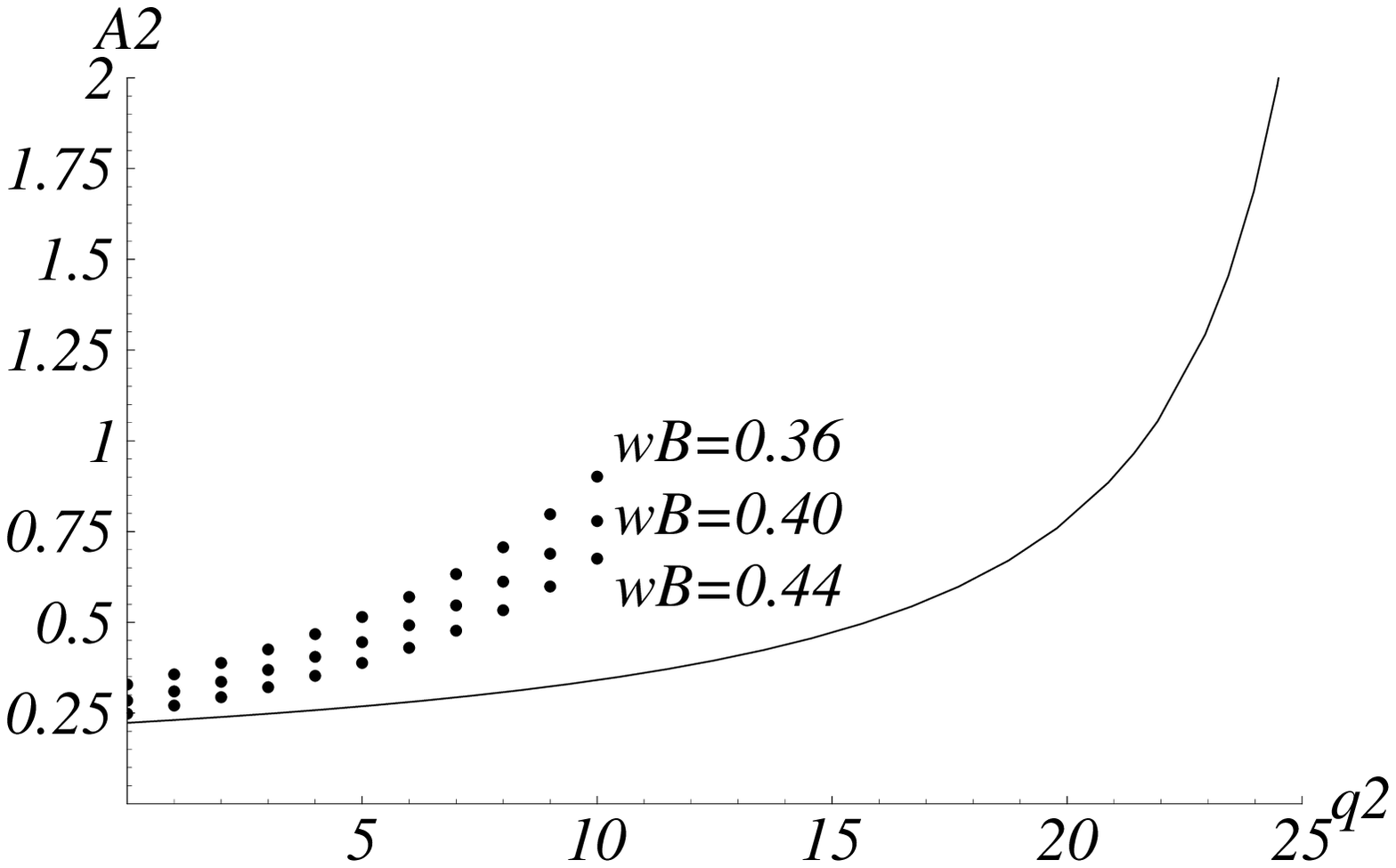,height=5.3cm}
\end{center}
\caption{%
The $B\to \rho$ form factors $V$, $A_0$, $A_1$ and $A_2$ as
functions of $q^2$. PQCD results are given in dots.
The solid lines come from light-cone sum rules.
}
\label{Brho}
\end{figure}

The same range of $\omega_B$ has been adopted in the evaluation of the
$B\to\rho$ transition form factors.
The results, displayed in Fig.~\ref{Brho},
are also consistent with those from LCSR \cite{sumrho} at small $q^2$. 
It is found that the symmetry relation in Eq.~(\ref{helicity}) below holds
very well: $A_1$ is larger than $V$ only by 2\% in the large recoil 
region, even after considering the pre-asymptotic forms of
the $\rho$ meson distribution amplitudes \cite{PB1}.
Taking the fast recoil limit with $\eta\to 1$ and assuming the asymptotic
behavior $\phi_\rho^v=\phi_\rho^a$, the above form factors 
are found to obey the symmetry relations \cite{BF,Charles:1998dr},
\begin{eqnarray}
V=A_1\;,\;\;\; A_2=A_1-2r_\rho A_0\;.
\label{sr}
\end{eqnarray}
Note that the form factors, treated as 
nonperturbative objects, are not calculated in \cite{BF}. Instead, the 
diagrams in Fig.~\ref{aa2} under an infrared regularization scheme
are regarded as perturbative corrections 
to the relations in Eq.~(\ref{sr}).
For the application of the PQCD approach to the radiative decay
$B\to K^*\gamma$, refer to \cite{LL99}.

\subsection{\it Power Corrections \label{sec:so2}}

SCET provides a systematic framework for discussing power corrections
to heavy-to-light transitions at large recoil.
Here I review the heavy quark expansion of the $B$ meson transition
form factors in SCET. It will be observed
that SCET is a powerful tool of deriving the relations among
various form factors in the large energy limit.
There are three independent form factors associated with
decays into pseudoscalars, and seven independent form
factors with decays into vector mesons. 
In the former case except for the form factors $F_{+,0}$ introduced
in Eq.~(\ref{f+0}), another one $F_T$ is defined via the matrix element, 
\begin{equation}
\langle P(P_2)|\bar q \, \sigma^{\mu\nu} q_\nu b|\bar{B}(P_1) \rangle =
\frac{i F_T(q^2)}{m_B+m_P}\left[q^2(P_1^\mu+P_2^{\mu})-
(m_B^2-m_P^2)\,q^\mu\right]\;.
\label{ftensor}
\end{equation}

At leading power, the number of independent form factors reduces.
Assume the energy $E\sim O(m_B)$ of the final state meson, which is 
related to the momentum transfer $q^2$ by 
$E=(m_B^2+m_P^2-q^2)/(2 m_B)$. The universal
functions $A_P^{(0)}(E,v,\bar n)$ and $A_V^{(0)}(E,v,\bar n,\epsilon)$,
defined via the matrix elements, 
$\langle P(P_2) | \xi \, \Gamma \, W \, h_v |\bar B(P_1)\rangle$ and 
$\langle V(P_2,\epsilon) | \bar \xi \, \Gamma \, W \, h_v |\bar B(P_1)
\rangle$, respectively, can be
decomposed into their most general independent Dirac structures
allowed by Lorentz invariance and parity
\cite{BF,Charles:1998dr},
\begin{eqnarray}
   A_P^{(0)}(E,v,\bar n) &=& 2  E \, \xi_P^{(0)}(E)\;, 
     \nonumber\\
   A_V^{(0)}(E,v,\bar n,\epsilon) &=& - 2 E \,  
\not\epsilon_T^\ast \, \gamma_5 \, \xi_T^{(0)}(E)
     - 2 E \, (v \cdot \epsilon^\ast) \, 
                   \gamma_5 \, \xi_\parallel^{(0)}(E)\;.
\label{APVdef}
\end{eqnarray}
That is, only one universal form factor $\xi_P^{(0)}(E)$ is left for
the former, and two form factors, $\xi_T^{(0)}(E) $ and
$\xi_\parallel^{(0)}(E)$, corresponding to transversely and
longitudinally polarized light vector mesons, respectively, for the latter.

It has been argued \cite{Beneke:2002ph} that the small expansion 
parameter in SCET should be taken as $\lambda\sim \sqrt{\Lambda/E}$
for the heavy-to-light transition form factors. The pion emitted in a
heavy-to-light decay at large recoil carries momentum scaling like 
$p_\pi\sim(\Lambda^2/E,E,\Lambda)$. This pion is produced in a highly 
asymmetric state, composing of a soft quark with momentum 
$p_s\sim(\Lambda,\Lambda,\Lambda)$ and a 
collinear jet. This jet must have an invariant mass squared 
$p_c^2=(p_\pi-p_s)^2\sim E\Lambda=\lambda^2 E^2$. Hence, one has the
expansion parameter $\lambda\sim\sqrt{\Lambda/E}$ 
based on the above kinematical consideration. 
In this case soft fields carrying momenta of order 
$\Lambda$ scale like $E(\lambda^2,\lambda^2,\lambda^2)$
and are assigned as being usoft as stated in Sec.~\ref{sec:b}. 

One decomposes the
matrix element of the heavy-to-light current in full QCD as
\begin{eqnarray}
\label{ff:general}
\langle P(P_2) | \bar q \, \Gamma \, Q |\bar B(P_1)\rangle &=& 
{\rm tr}\left[ A_P^{(++)}\, \frac{\not n \not{\bar n}}{4}  \, \Gamma
     \, \frac{1+\not v}{2}  \right] 
 + {\rm tr}\left[ A_P^{(+-)} \, \frac{\not n \not{\bar n}}{4}  \, \Gamma
     \, \frac{1-\not v}{2}  \right]  
\nonumber \\[0.15em]
  && \hspace*{-2cm}+ \,{\rm tr}\left[ A_P^{(-+)} \, 
\frac{\not{\bar n} \not n}{4}  \, \Gamma
     \, \frac{1+\not v}{2}  \right]  
 + {\rm tr}\left[ A_P^{(--)} \, \frac{\not{\bar n} \not n}{4}  \, \Gamma
     \, \frac{1-\not v}{2}  \right] \;,
\end{eqnarray}
using the projectors $(1\pm \not v)/2$, $\not n \not{\bar n}/4$ 
and $\not{\bar n}\not n/4$ on
the large and small components of the quark fields, labelled by $+$
and $-$, respectively. The definition for decays into vector mesons is
analogous. The functions $A_{P}^{(kl)}$ and $A_{V}^{(kl)}$ with $k,l=+,-$ 
can again be decomposed as
\begin{eqnarray}
   A_P^{(kl)}(E,v,\bar n) &=& 2  E \, f_P^{(kl)}(E)\;, 
 \nonumber\\[0.15em]
   A_V^{(kl)}(E,v,\bar n,\epsilon) &=& - 2 E \,  \not
   \epsilon_T^\ast \, \gamma_5 \, f_T^{(kl)}(E)
     - 2 E \, (v \cdot \epsilon^\ast) \, 
                   \gamma_5 \, f_\parallel^{(kl)}(E)\;.
\label{APV-full-def}
\end{eqnarray}
Among the $4+8$ form factors $f_{P}^{(kl)}$ and
$f_{T,\,\parallel}^{(kl)}$ only $3+7$ are independent due to 
the equations of motion for light and heavy quarks in QCD and
translational invariance,
\begin{eqnarray}
 q^\mu \langle P| \bar q \, \gamma_\mu \, b |B\rangle 
&=& (m_b-m_q) \, \langle P| \bar q \, b |B\rangle\;,
\nonumber \\
 q^\mu \langle V| \bar q \, \gamma_\mu\gamma_5 \, b |B\rangle 
&=& - (m_b+m_q) \, \langle V| \bar q \, \gamma_5 \, b |B\rangle \;,
\label{seq}
\end{eqnarray}
with $q=P_1-P_2$. 

With the SCET expansion of the heavy-to-light currents, it is easy to
identity the scaling of the above form factors:
\bea
& &f_i^{(++)}(E) = \xi_i^{(0)}(E) \, 
             \Big( 1 + O(\alpha_s,\lambda) \Big)\;,
\nn\\
& &f_i^{(+-)}(E) \sim  \lambda\, f_i^{(++)}(E)\; , \qquad
f_i^{(-+)}(E) \sim  \lambda\, f_i^{(++)}(E)\; , \qquad
f_i^{(--)}(E) \sim  \lambda^2 f_i^{(++)}(E)\;.
\eea
At $O(\lambda)$, the neglect of $f_i^{(--)}(E)$ leaves 3+6 form factors 
for pseudoscalar and vector mesons, respectively. The equations of motion 
in Eq.~(\ref{seq}) give two more constraints at this order. Therefore, 
one has 2+5 independent form factors, implying 1+2 form factor relations.

At $O(\lambda)$, the three form factor relations are written as
\cite{Beneke:2002ph}
\begin{eqnarray} 
R_P &=& \frac{F_+-F_0}{F_T} = \frac{q^2}{m_B \, (m_B+m_P)} 
        \left(1+ O(\alpha_s,\lambda^2)\right)\;,
\nonumber\\[0.5cm]
R_T &=& 
\frac{\displaystyle \left(1-\frac{q^2}{m_B^2}\right)  T_1 - T_2 +
   \frac{q^2}{m_B^2} \left(1+\frac{m_V}{m_B}\right) A_1}
{\displaystyle \left(1-\frac{q^2}{m_B^2}\right)  V} = 
\frac{q^2}{m_B \, (m_B+m_V)}
 \left(1 +
 O(\alpha_s,\lambda^2) \right)\;,
\nonumber\\[0.3cm]
R_\parallel &=& 
\frac{\displaystyle \left(1+\frac{m_V}{m_B}\right)  A_1 -
  \left(1-\frac{m_V}{m_B}\right) \left(1-\frac{q^2}{m_B^2}\right) A_2 - 2
  \, \frac{m_V}{m_B}
  \left(1-\frac{q^2}{m_B^2}\right)  A_0}
{\displaystyle T_2 -
  \left(1-\frac{q^2}{m_B^2}\right) 
   T_3}\;,
\nonumber\\
& =& \frac{q^2}{m_B^2} \left(1 +
 O(\alpha_s,\lambda^2)\right)\;,
\label{rel3}
\end{eqnarray}
for the decays into light pseudoscalar, 
transversely and longitudinally polarized mesons, respectively.
As $q^2 \to 0$, the left-hand sides of the above relations vanish exactly.
Other form factor relations, which receive $O(\lambda)$ corrections, 
have been also derived in \cite{Beneke:2002ph}:
\begin{equation}
  \left(1-\frac{q^2}{m_B^2}\right)  V 
- \left(1+\frac{m_V}{m_B}\right)^2 A_1
= O(\lambda)\;, 
\qquad
\left(1-\frac{q^2}{m_B^2}\right) T_1 - T_2
= \frac{q^2}{m_B^2}\, O(\lambda)\;.
\label{helicity}
\end{equation}
The above result differs from that in \cite{Chay:2002vy}, where
the first equation in (\ref{helicity}) does not receive an $O(\lambda)$
correction. 

\begin{figure}[t!]
\begin{center}
 \epsfig{file=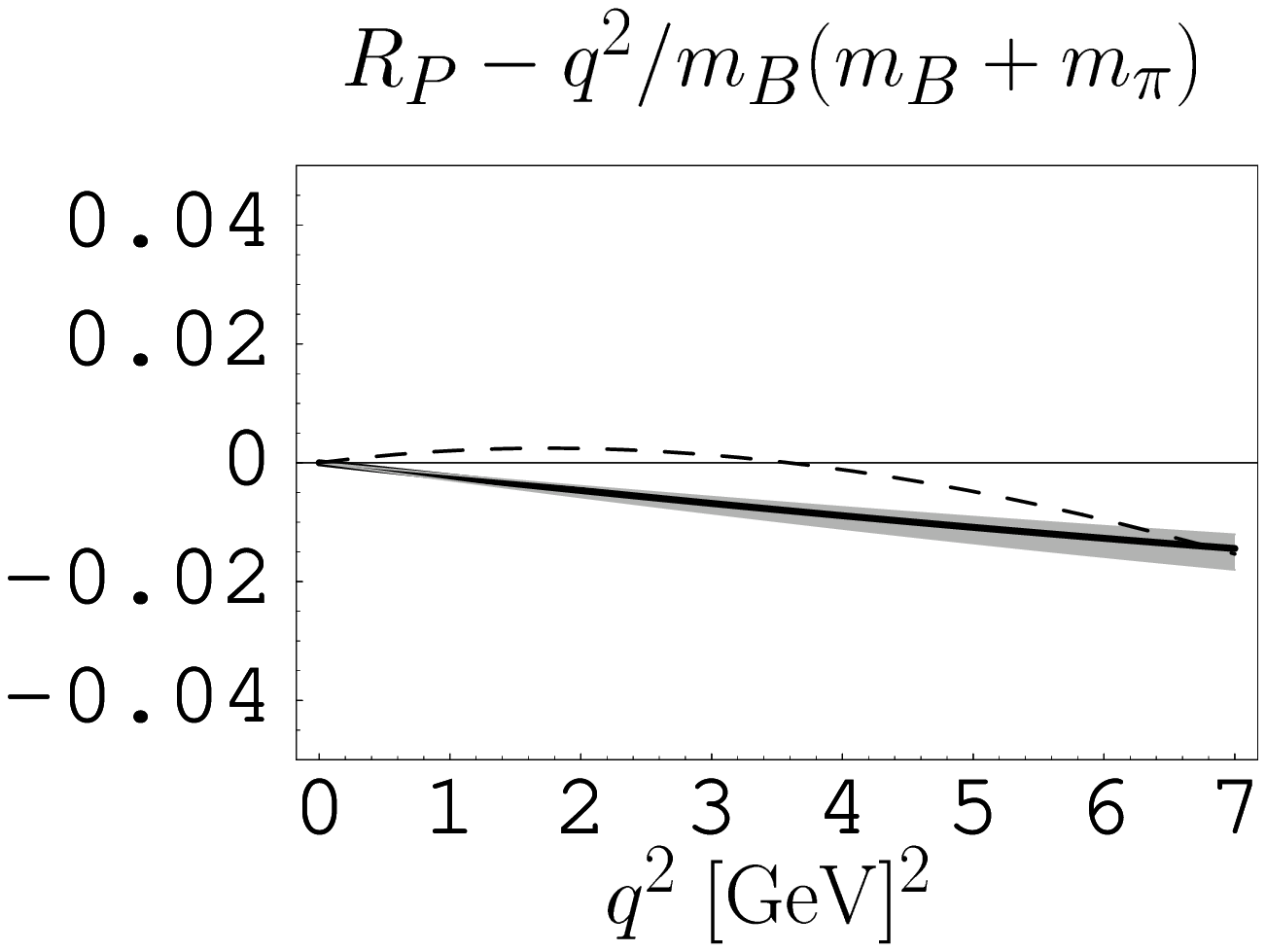,height=2.0in} 
 \hspace{1.3cm}
 \epsfig{file=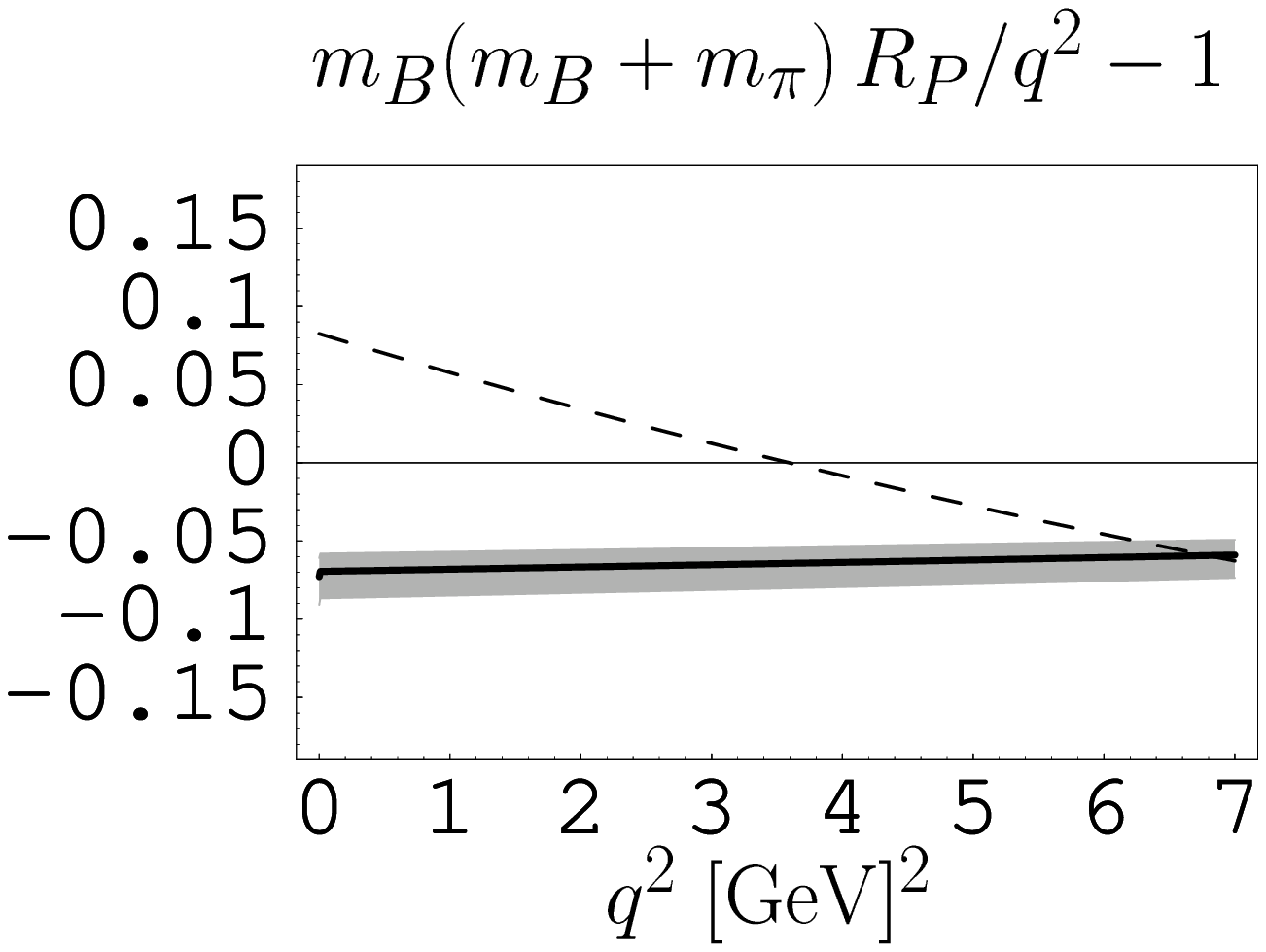,height=2.0in}
\vskip 1.0cm
 \epsfig{file=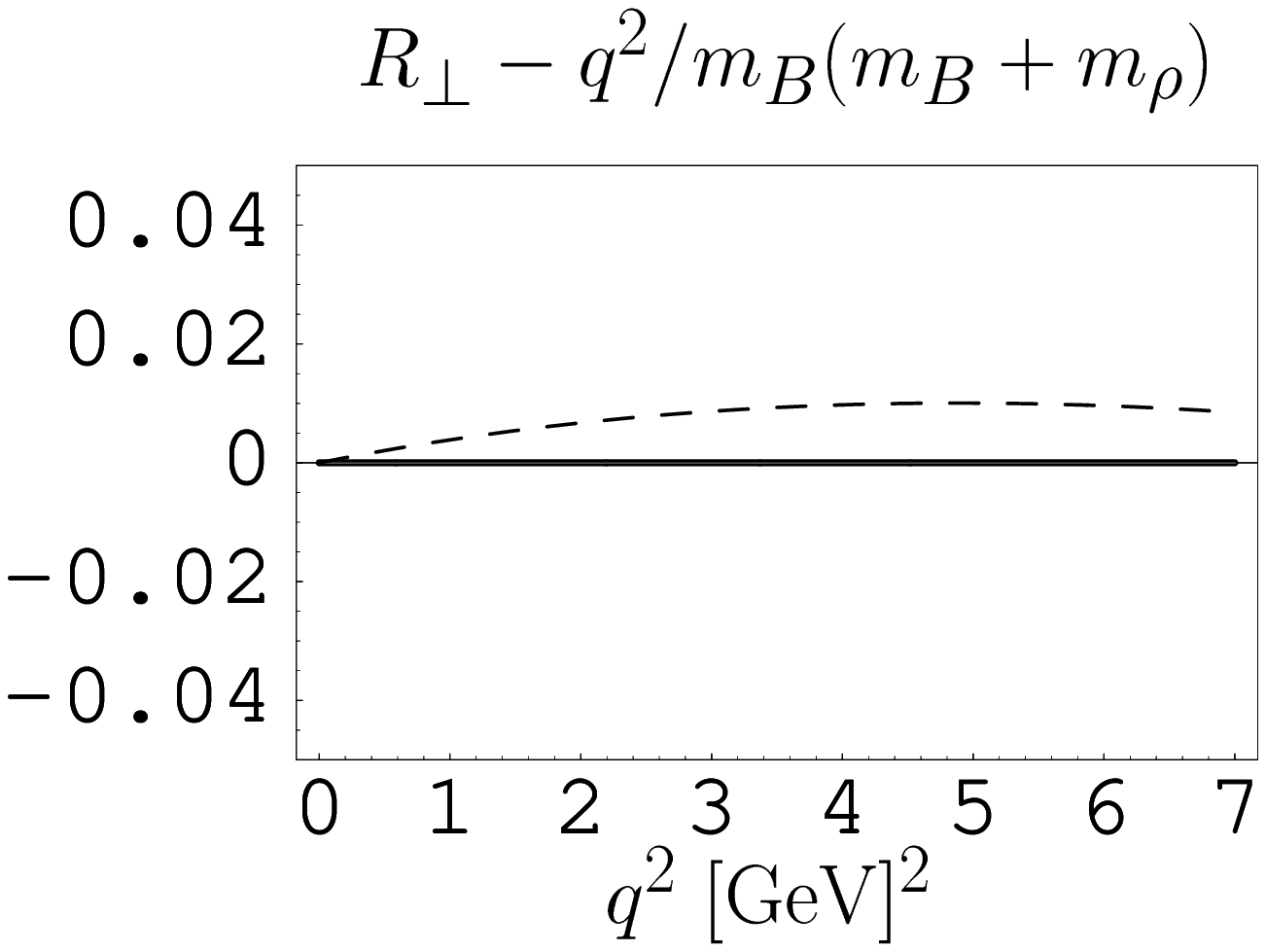,height=2.0in} 
 \hspace{1.3cm}
 \epsfig{file=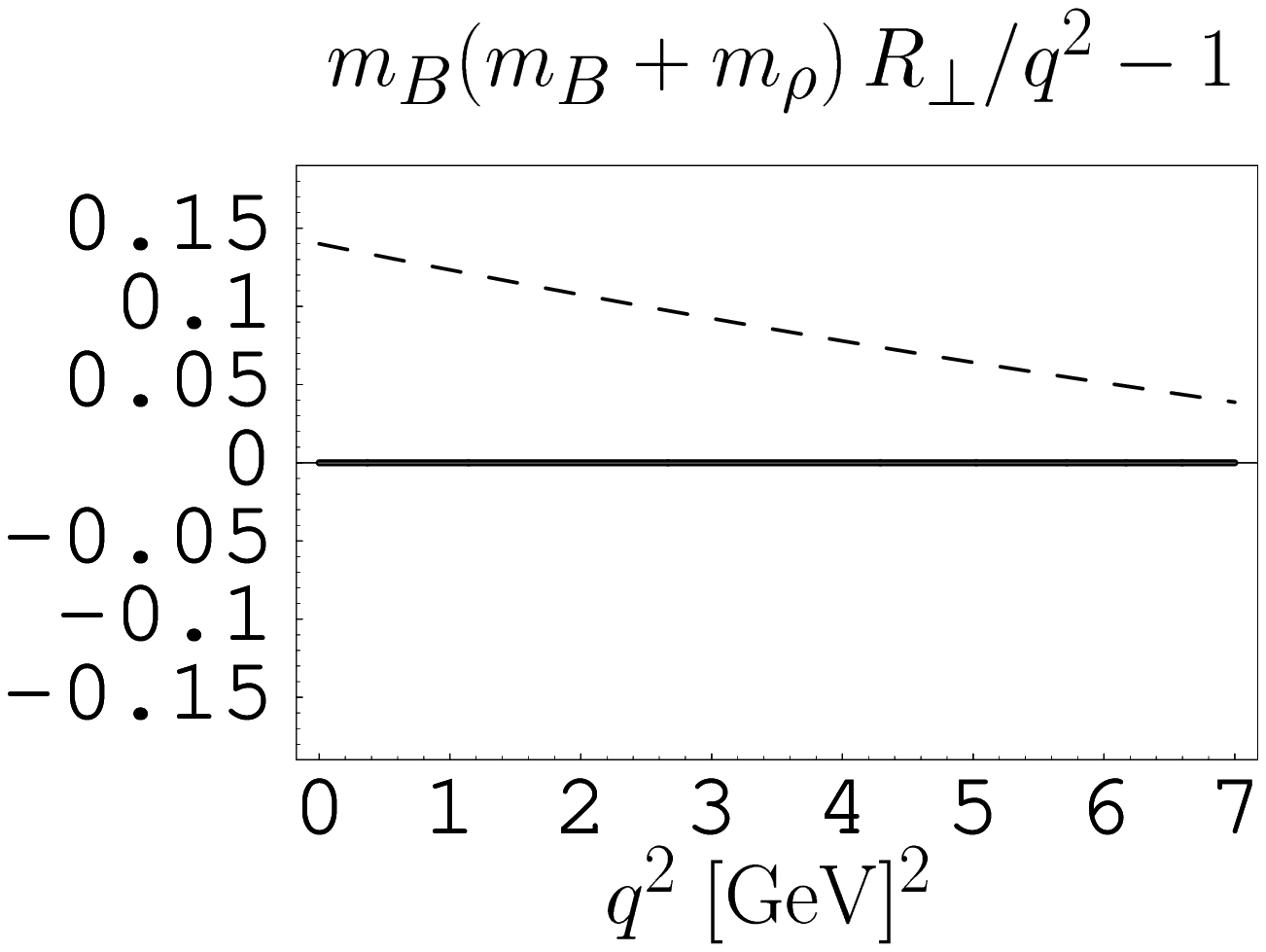,height=2.0in}
\vskip 1.0cm
 \epsfig{file=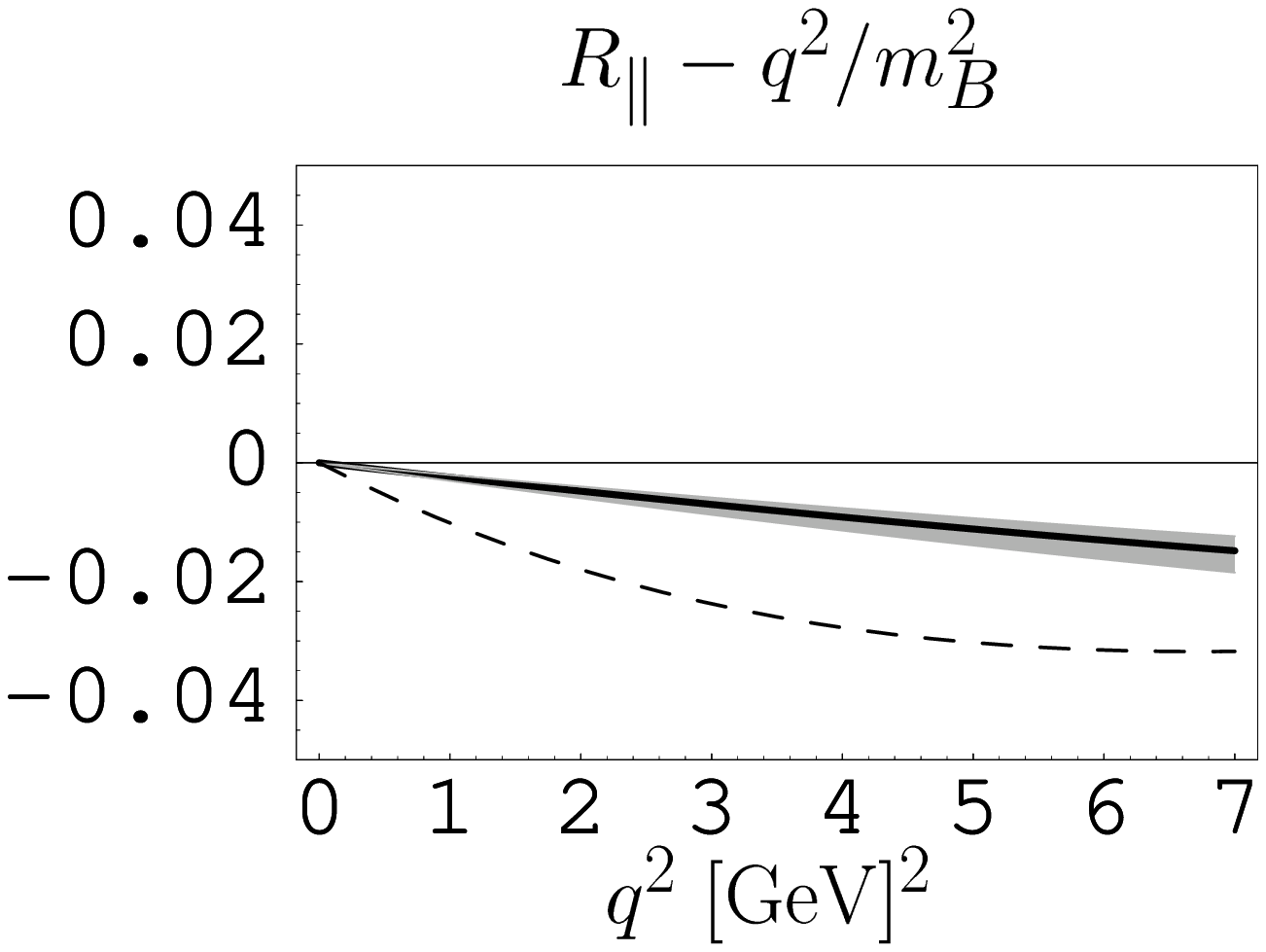,height=2.0in} 
 \hspace{1.3cm}
 \epsfig{file=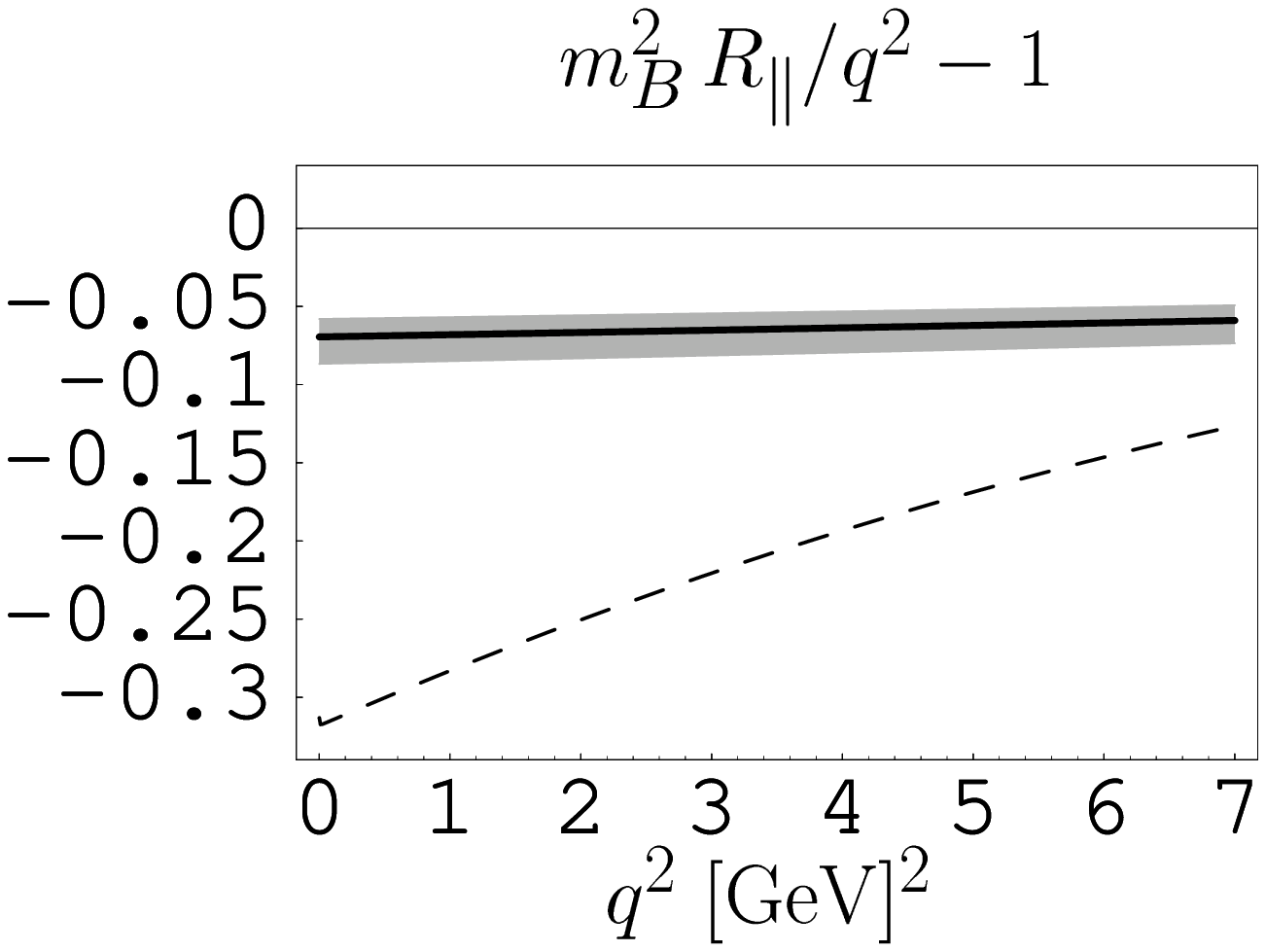,height=2.0in}
\caption{Test of the relations in Eq.~(\ref{rel3}) 
for $B\to \pi$ and $B \to \rho$ form factors.
The dashed line corresponds to the form factors predicted from LCSR
\cite{PB3,BallBraun}.
The solid line is the $O(\lambda^0)$ result including the
$\alpha_s$ corrections \cite{BF}.}
\label{LCQCD}
\end{center}
\end{figure}

To test the form factor relations in Eq.~(\ref{rel3}), one substitutes
the form factors derived from LCSR
\cite{PB3,BallBraun} into the left-hand sides and compare the
outcomes with the right-hand sides. The $O(\alpha_s)$ corrections to the
heavy quark limit have been included on the right-hand sides. 
The results are displayed in Fig.~\ref{LCQCD} \cite{Beneke:2002ph}. The
ratios $R_{P,T,\parallel}$ minus the corresponding values in the
symmetry limit are shown on the left-hand side.
The tensor form factors are evaluated at the scale $m_b$.
The grey error band reflects the theoretical uncertainty from varying the
scale of $\alpha_s$ from $m_b/2$ to $2 m_b$.
The difference between the dashed and solid curves is an
estimate of $O(\lambda^2)$ or $O(\alpha_s\lambda)$ corrections,
which are at most 3\% deviations from zero for $q^2$ up to 7 GeV$^2$.
$R_{P,T,\parallel}$ divided by their symmetry limits minus 1 are
shown on the right-hand side of Fig.~\ref{LCQCD}. It is expected that
deviations from 0 are of $O(\lambda^2)$ and/or $O(\alpha_s)$, since
the $q^2$ suppression has been divided out.


Note that the scaling behavior of the quark fields and of the
meson states with $\lambda$ is not sufficient to determine the
scaling behavior of form factors. Considering soft contribution to the 
form factor, the $u$ quark created in the decay of the $b$ quark 
carries almost all the energy of the light meson, while the spectator 
quark is soft. To leading order in 
$\Lambda/E$, the interaction of energetic quarks with soft
gluons is described by the eikonal Lagrangian 
\cite{Charles:1998dr,leet}
\be
\label{leet}
  {\cal L}_{\rm eik} = 
    \bar q_n \, \frac{\not n}{2} \, (i \, \bar n \cdot D) \, q_n
   + O(1/E_q),
\ee
where $q_n(x)=
e^{i E_q \bar n \cdot x} \not{\bar n} \not {n} q(x)/4$
are the large components of the light quark spinor field.  
Even for light-cone dominated 
processes this is an atypical configuration (the preferred one having 
nearly equal momenta of the quark and antiquark).
For this reason, although the interaction in Eq.~(\ref{leet}) 
is spin-symmetric, the symmetry is not realized in the hadronic 
spectrum, and there exists no relation among the soft contributions 
to the form factors of pseudoscalar and vector mesons.
Furthermore, 
the probability that such an asymmetric parton configuration
hadronises into a light meson depends on the energy of the meson. 
Hence, the soft contributions to the form factors are energy-dependent
functions, whose absolute normalization is not known. 

In this respect SCET applied to heavy-to-light decays at large recoil 
differs from HQET for $B \to D^{(*)}$ form factors. In the 
case of heavy-heavy form factors, spin symmetry relates pseudoscalar and 
vector mesons, and the Isgur-Wise form factor $\xi(v\cdot v')$ is 
independent of the heavy quark mass.
One obtains non-trivial form factor relations beyond the
leading order in $\lambda$, because the heavy quark flavor symmetry
also relates the initial and final hadronic states.


Recently, an expansion parameter $\lambda\sim \Lambda/E$ for
heavy-to-light form factors has been proposed \cite{HN02}, which differs
from $\lambda\sim\sqrt{\Lambda/E}$ discussed above 
\cite{Chay:2002vy,Beneke:2002ph}. For $\lambda\sim\sqrt{\Lambda/E}$,
the external pion, whose 
momentum scales like $p_\pi\sim E(1,\lambda^4,\lambda^2)$,  
cannot be built up from the combination of a generic usoft momentum
$p_s\sim E(\lambda^2,\lambda^2,\lambda^2)$ with a generic collinear 
momentum $p_c\sim E(1,\lambda^2,\lambda)$. It implies that 
the soft mechanism is strongly suppressed in this picture.
For $\lambda\sim\Lambda/E$, the pion momentum scales
like a collinear momentum. In order to make a light meson out of
collinear particles and soft particles, one has to require the
minus component of the total soft momentum, which would scale 
like $E\lambda$, to be accidentally small, of order $E\lambda^2$.
However, this implies a phase-space suppression of $O(\Lambda/E)$
as explained in \cite{TLS}. It has been expected \cite{HN02} that
under the different choices of the expansion parameters, the
violations of heavy quark symmetry relations between form factors 
may start at different power of $\Lambda/E$. From the view point of
the PQCD approach, the spectator on the pion side is as energetic
as the collinear particles. That is, all the momenta scale like
a collinear momentum, and there is no phase-space suppression.

\subsection{\it Radiative Corrections \label{sec:fac}}


Form factors for heavy-to-light transitions are presumably dominated 
by nonperturbative QCD dynamics at small momentum transfer  
and not computable in perturbation theory. Charles {\em et al.}
have shown that certain symmetries apply to this soft contribution,
when the momentum of the final light meson is large \cite{Charles:1998dr}. 
These symmetries reduce the number of independent form factors from ten to 
three as shown in Sec.~\ref{sec:so2}. The corresponding symmetry
relations for the form factors are broken by power corrections
discussed above and by radiative corrections. In this subsection 
I review the evaluation of the symmetry-breaking corrections
at first order in the strong coupling constant $\alpha_s$, which
are dominated by short-distance contributions. The formalism 
adopted below is referred to as the QCDF approach \cite{BF}.

\begin{figure}[t!]
\begin{center}
\epsfig{file=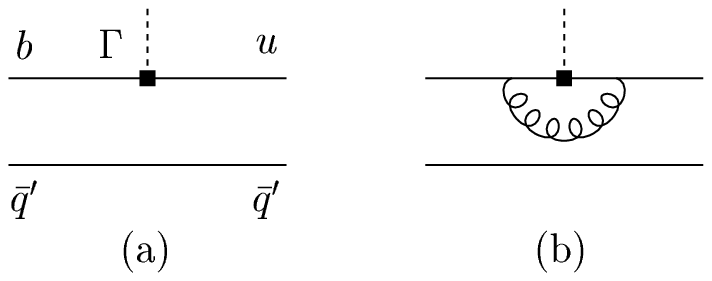,height=1.5in}
\vskip 0.5cm
\epsfig{file=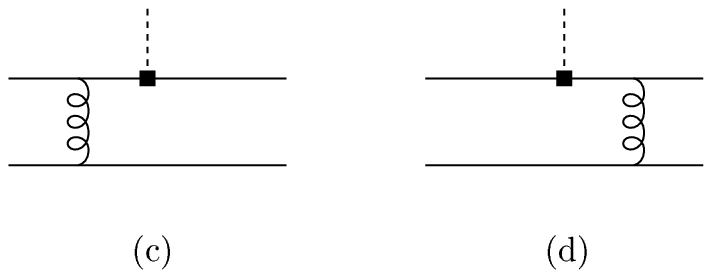,height=1.5in}
\end{center}
\caption{Different contributions to the 
$B\to P(V)$ transition. (a) Soft contribution (soft interactions with 
the spectator antiquark $\bar{q}'$ are not drawn). (b) Hard vertex 
renormalization. (c,d) Hard spectator interaction.}
\label{alp}
\end{figure}

In the absence of a hard spectator interaction shown in Fig.~\ref{alp}(a), 
the light meson is produced in a parton configuration, in which the $u$ 
quark carries all the momentum of the meson, up to an amount of 
$O(\Lambda)$ in the $B$ meson rest frame. The hard part of the vertex 
correction in Fig.~\ref{alp}(b) does not respect the symmetry relations, 
but can be accounted for in perturbation theory by multiplicatively
renormalizing the current $[u_n\Gamma b_v]_{\rm eff}$ in the effective
theory. A hard interaction with the 
spectator quark shown in Figs.~\ref{alp}(c) and \ref{alp}(d)
allows the meson to be formed in a preferred
configuration, in which the momentum is distributed nearly equally
between the two quarks. 

The soft contribution scales like \cite{Chernyak:1990ag}
\begin{equation}
\label{softbpi}
F_{+,0,T}^{\rm soft}(q^2\approx 0) \sim \xi_P(E\approx m_B/2) 
\sim \sqrt{\frac{m_B}{E}}\left(\frac{\Lambda}{E}\right)^{\!3/2}
\sim \left(\frac{\Lambda}{M}\right)^{\!3/2}\;.
\end{equation}
For the hard contribution, both quarks that form the light 
meson have momenta of $O(m_B)$, and the gluon in Figs.~\ref{alp}(c) and 
\ref{alp}(d) has virtuality of order $\Lambda m_B$. The resulting scaling 
behaviour for the pseudoscalar meson form factors is 
\begin{equation}
\label{hardbpi}
F_{+,0,T}^{\rm hard}(q^2\approx 0) 
\sim \alpha_s(\sqrt{\Lambda m_B})\,
\left(\frac{\Lambda}{m_B}\right)^{\!3/2}\;.
\end{equation}
Therefore, the hard spectator interaction is suppressed by one power of 
$\alpha_s$ relative to the soft contribution.

The QCDF formula for a heavy-light form factor at large recoil at leading 
power in $1/m_B$ is then written as \cite{BF}
\be
\label{fff}
F_i(q^2)  = C_i \, \xi_P(E)  + 
\phi_B \otimes T_i \otimes \phi_P\;.
\ee
The soft form factor $\xi_P(E)$, defined in Eq.~(\ref{APVdef}) and 
represented by Fig.~\ref{alp}(a), obeys the symmetries discussed above. 
The hard-scattering kernel $T_i$  
from Figs.~\ref{alp}(c) and \ref{alp}(d) is
convoluted with the light-cone distribution amplitudes of the 
$B$ meson and of the light pseudoscalar meson, for which the endpoint 
divergence together with some finite contribution have been
absorbed into the leading soft term. The coefficient $C_i=1+O(\alpha_s)$ 
is the hard vertex renormalization from Fig.~\ref{alp}(b). 
The correction at $O(\alpha_s)$ from the hard vertex renormalization and 
from the hard spectator interaction have been obtained in
\cite{bfps,BF}.

To absorb the end-point singularities, the factorization scheme has been 
defined by imposing the condition,
\bea
 && F_+ \equiv  \xi_P, 
  \qquad \ 
 V   \equiv  \frac{m_B+m_V}{m_B} \, \xi_\perp, 
  \qquad
  A_0 \equiv  \frac{E}{m_V} \, \xi_\parallel, 
\label{constraint}
\eea
exactly to all orders in perturbation theory, similar to the
DIS scheme for inclusive processes. Having fixed the factorization 
scheme, all other form factors can be expressed, for example, as
\bea
  F_0 &=& \frac{2E}{m_B} \, \xi_P \, \left[1 +
    \frac{\alpha_s \, C_F}{4\pi} \,
    \left( 2 - 2 \, L \right) \right]
  + \frac{\alpha_s \, C_F}{4\pi} \, \Delta F_0\;, 
\label{add1}
 \\[0.5em]
  F_T &=& \frac{M+m_P}{M} \, \xi_P \, \left[1 +
    \frac{\alpha_s \, C_F}{4\pi} \,
    \left( \ln \frac{m_b^2}{\mu^2} + 2 \, L \right)
  \right]
+\frac{\alpha_s \, C_F}{4\pi} \, \Delta F_T\;,
\label{vertexcorr1}
\eea
with the notation,
\be
  L = - \frac{2E}{m_B-2E} \, \ln\frac{2E}{m_B}\;.
\label{Labbrev}
\ee
Note $L \to 1$ for $E \to m_B/2$ ($q^2 \to 0$).

The $O(\alpha_s)$ corrections to the form factors are given by
\be
  \Delta F_0  =  \frac{m_B-2 E}{2 E} \, \Delta F_P\;,  \qquad
  \Delta F_T  = - \frac{m_B+m_P}{2 E}  \, \Delta F_P \; ,
\label{eq:pi_final}
\ee
with the quantity,
\bea
\Delta F_P= \frac{8 \pi^2  f_B f_P}{N_C m_B} 
\int dk^+ \frac{\phi_+(k^+)}{k^+}
\int du \frac{\phi(u)}{\bar u}\;.
\label{pihard}
\eea
The theoretical uncertainties in the computation of the hard-scattering 
correction from the moments of the meson distribution amplitudes 
\cite {BK03} and
from the $B$ meson decay constant are all contained $\Delta F_P$. 

Equation (\ref{fff}) has been further elucidated in the framework
of SCET \cite{PS02}. At leading power in $1/m_b$ and all orders in $\alpha_s$, a $B\to P$ transition form factor $F$ can be split into 
factorizable and nonfactorizable components,
\begin{eqnarray}
F(E) &=& f^{\rm F}(E) +f^{\rm NF}(E)\;,
\nn\\
f^{\rm F}(E) &=& \frac{f_B f_P}{E^2} \int_0^1\!\!\!\! dz\! 
    \int_0^1\!\!\!\! dx\! \int_0^\infty\!\!\!\!\! dk^+ \,
    T(z,E,\mu_{\rm 0}) 
\nn\\
 &&
\quad \times J(z,x,k^+,E,\mu_{\rm 0},\mu) \phi_P(x,\mu) 
\phi_{+}(k^+,\mu) \,,
\label{fF}\\
f^{\rm NF}(E) &=&C_i(E,\mu)\: \xi_i(E,\mu) \,.
\label{fNF}
\end{eqnarray}
Compared to Eq.~(\ref{fff}), the hard-scattering kernel has been further
factorized into a function $T$ characterized by the scale $m_b$ and
a jet function $J$ characterized by the scale 
$\mu_{\rm 0}\simeq \sqrt{m_b\Lambda}$. Hence,
the hard coefficients $C_k$ and $T$ are calculated in an
expansion in $\alpha_s(m_b)$. The jet function $J$ is calculable in
terms of a matrix element involving $\alpha_s(\sqrt{\Lambda m_b})$. 
That is, the contributions characterized by
$m_b$ and $\sqrt{m_b\Lambda}$ have been clearly separated.
Endpoint singularities arise only in the 
soft, nonperturbative form factors $\xi_i(E,\mu)$. The 
convolution integrals in the factorizable terms are infrared finite
in collinear factorization theorem.

I explain the difference between the QCDF formulas based on collinear
factorization and those in the PQCD approach based on $k_T$ 
factorization. In the former the piece with an end-point singularity in 
collinear factorization theorem has been regularized and absorbed into the 
soft term $f^{\rm NF}$. In $k_T$ factorization theorem the end-point 
singularity is absent, and both $f^{\rm NF}$ and $f^{\rm F}$ can
be formulated into the factorization formulas. Since $f^{\rm NF}$
remains in the formulas, the form factor symmetry relations at
large recoil are still respected in the PQCD approach, which are then 
modified by the less important term $f^{\rm F}$. This has been shown
explicitly in Eq.~(\ref{sr}), contrary to the criticism in \cite{BF}.
It is then realized that the definition of soft contributions is in fact
ambiguous, depending on the theoretical framework that is adopted.
In QCDF the soft contribution refers to the one with 
the end-point singularity in collinear factorization theorem (plus an 
arbitrary infrared-finite piece related to a factorization scheme). In 
PQCD it refers to the one from a large (but arbitrary) coupling
constant. Therefore, the hard-scattering terms in both approaches
[in Eq.~(\ref{fff}) and in Eq.~(\ref{fpi})] also collect different contributions.  

To be more specific, I compare the explicit expression for the form 
factor $F_+$ derived based on Eq.~(\ref{fNF}) \cite{PS02},
\begin{eqnarray}
F_+(E) &=&  N_0 \int dxdk^+ \left[\frac{2E-m_B}{m_B} C_a(E,\mu_{\rm 0})
+\frac{2E}{m_b} C_b(E,\mu_{\rm 0})\right]
\nn\\
& &\times \frac{\alpha_s(\mu_{\rm 0})}{x \, k^+} \phi_\pi(x) \phi_+(k^+) 
+ C(E,\mu) \, \zeta(E,\mu)\,,
\label{F+}
\end{eqnarray} 
with the constant, 
\begin{eqnarray}
N_0=\frac{\pi C_F}{4}\frac{f_B f_\pi m_B}{N_c E^2}\;,
\end{eqnarray}
to Eq.~(\ref{f12}). The Wilson coefficients 
satisfy $C_a = C_b = 1$ at the tree level. Removing all the Sudakov 
factors and dropping the twist-3 contributions, $F_+$ in Eq.~(\ref{f12})
from the PQCD approach reduces to
\begin{eqnarray}
F_+(E)&=&N_0\alpha_s m_B\int dx_1dx_2
\left(\frac{\eta}{x_1x_2}+\frac{\eta+1}{x_1x_2}+\frac{1}{x_1x_2^2}
\right)\phi_\pi(x_2)\phi_+(x_1)\;.
\label{f+}
\end{eqnarray}
Using the variable change $\eta=2E/m_B-1$, it is easy to identify 
the three terms in Eq.~(\ref{f+}) as the three terms in Eq.~(\ref{F+})
in sequence. The third term in Eq.~(\ref{f+}) with the end-point 
singularity comes from the term 1
in the coefficient $1+x_2\eta$ of $\phi_\pi$ in Eq.~(\ref{fpi}).
This piece obeys the large-energy symmetry mentioned above.
The term $x_2\eta$, whose $x_2$ cancels a power of $x_2$ in the 
denominator, corresponds to the hard-scattering piece in Eq.~(\ref{F+}).

A study of the relative importance of soft and hard dynamics has been
done in the framework of QCD sum rules \cite{SR}.
The soft contribution without Sudakov suppression was estimated to be
0.22 (corresponding to $f_B\sim 130$ MeV). The soft contribution to
$f_BF^{B\pi}$ obtained in \cite{KRWY} is consistent with the above
value. It was then shown that the Sudakov effect decreases the soft
contribution by a factor 0.4-0.7, depending on infrared cutoffs for
loop corrections to the weak decay vertex. Therefore, the soft
contribution turns out to be about 0.09-0.15, and smaller than
the perturbative contribution about 0.19. It is then possible
that the $B\to\pi$ form factors receive significant 
perturbative contributions, in spite of the large theoretical
uncertainty in sum rules, for example, from the variation of the
Borel mass \cite{KRWY}. 

\begin{figure}[t!]
\begin{center}
 \epsfig{file=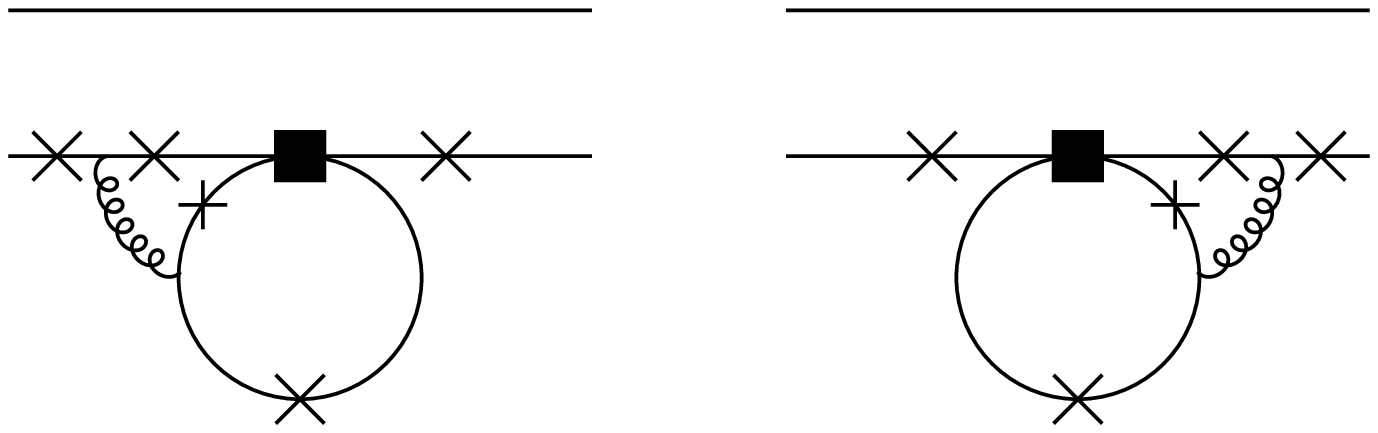,height=1.2in}
\caption{Typical $O(\alpha_s)$ contribution to the hard-scattering
kernel $T^{I}_i$. The crosses represent the photon vertices.}
\label{qit1}
\end{center}
\end{figure}

\begin{figure}[t!]
\begin{center}
 \epsfig{file=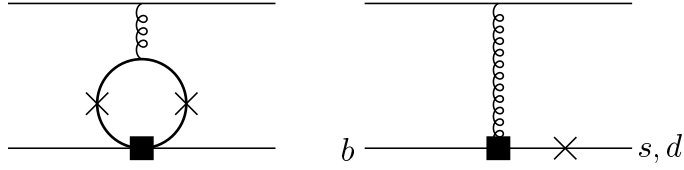,height=1.0in}
\caption{Typical $O(\alpha_s)$ contribution to the hard-scattering
kernel $T^{II}_i$.}
\label{qit2}
\end{center}
\end{figure}

The QCDF formalism has been applied to the study of the forward-backward 
asymmetry in the rare decay $B\to V\ell^+\ell^-$, where $V$ is a vector 
meson \cite{Ali00}. Below I discuss the simpler modes $B\to V\gamma$
\cite{Beneke:2001at,BVgam}. The hadronic matrix elements are written 
as,
\begin{eqnarray}
\langle V\gamma(\epsilon)|O_i|B\rangle
  =\Big[ F^{BV} T^I_{i}
  + \!\int^1_0 \!\!d\xi\, dv \, T^{II}_i(\xi,v) \phi_B(\xi) \phi_V(v)\Big]
  \!\cdot\epsilon\;,
\label{fform}
\end{eqnarray}
where $\epsilon$ is the photon polarization vector and the operators
$O_i$ come from the effective weak Hamiltonian. The soft form factor 
$F^{BV}$ for the $B\to V$ transition obeys the symmetry
relations in the large energy limit. In QCDF the next-to-leading-order 
hard corrections to the weak decay vertex in Fig.~\ref{qit1} 
contribute to $T_i^I$ \cite{GHWBCMU}. 
These contributions are dominated by scales of
$O(m_b)$ and infrared finite. $T^{II}_i$ involves the hard scattering of
the spectator shown in Fig.~\ref{qit2}. 

\begin{figure}[t!]
\begin{center}
 \epsfig{file=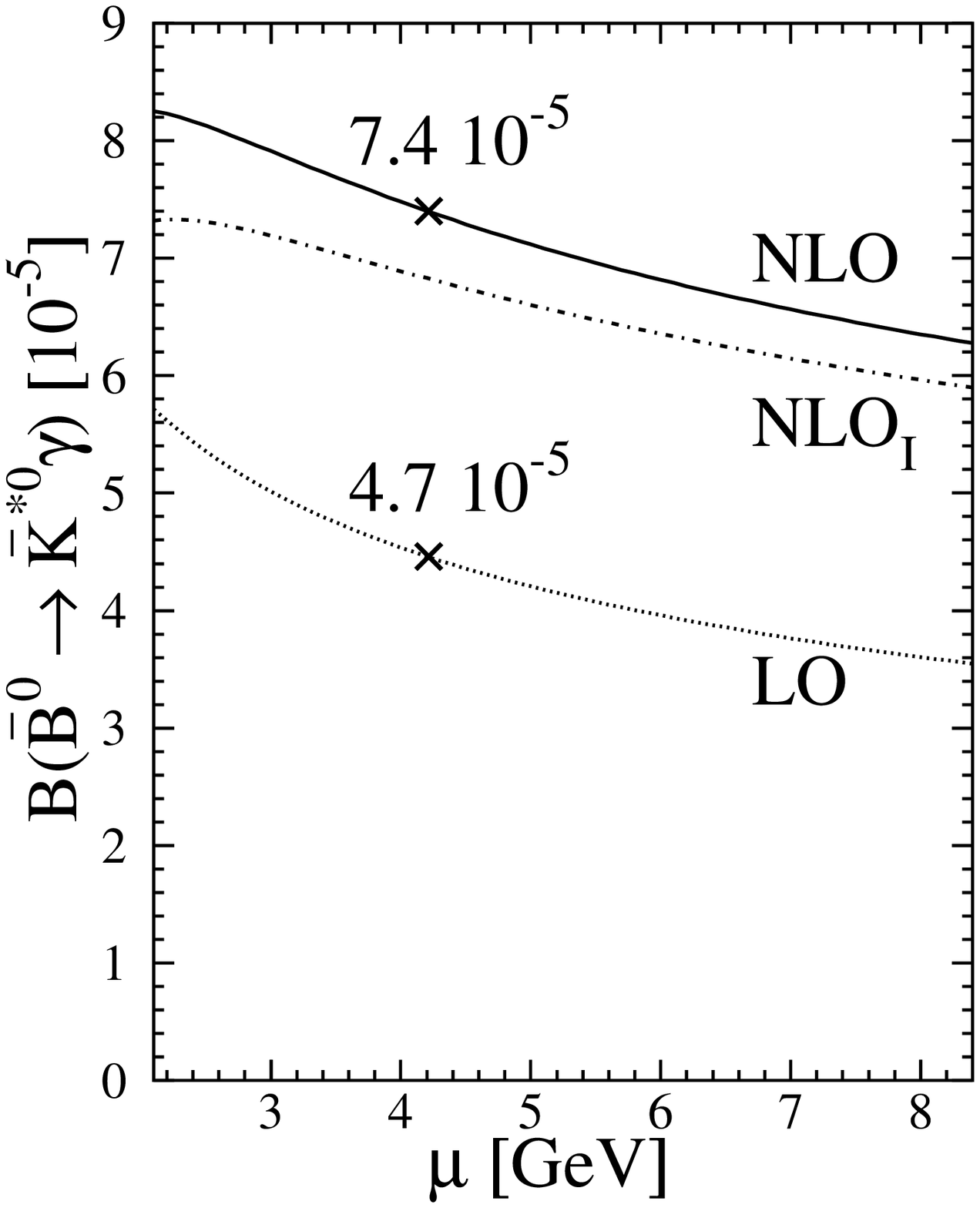,height=3.0in}
 \hspace{1.3cm}
 \epsfig{file=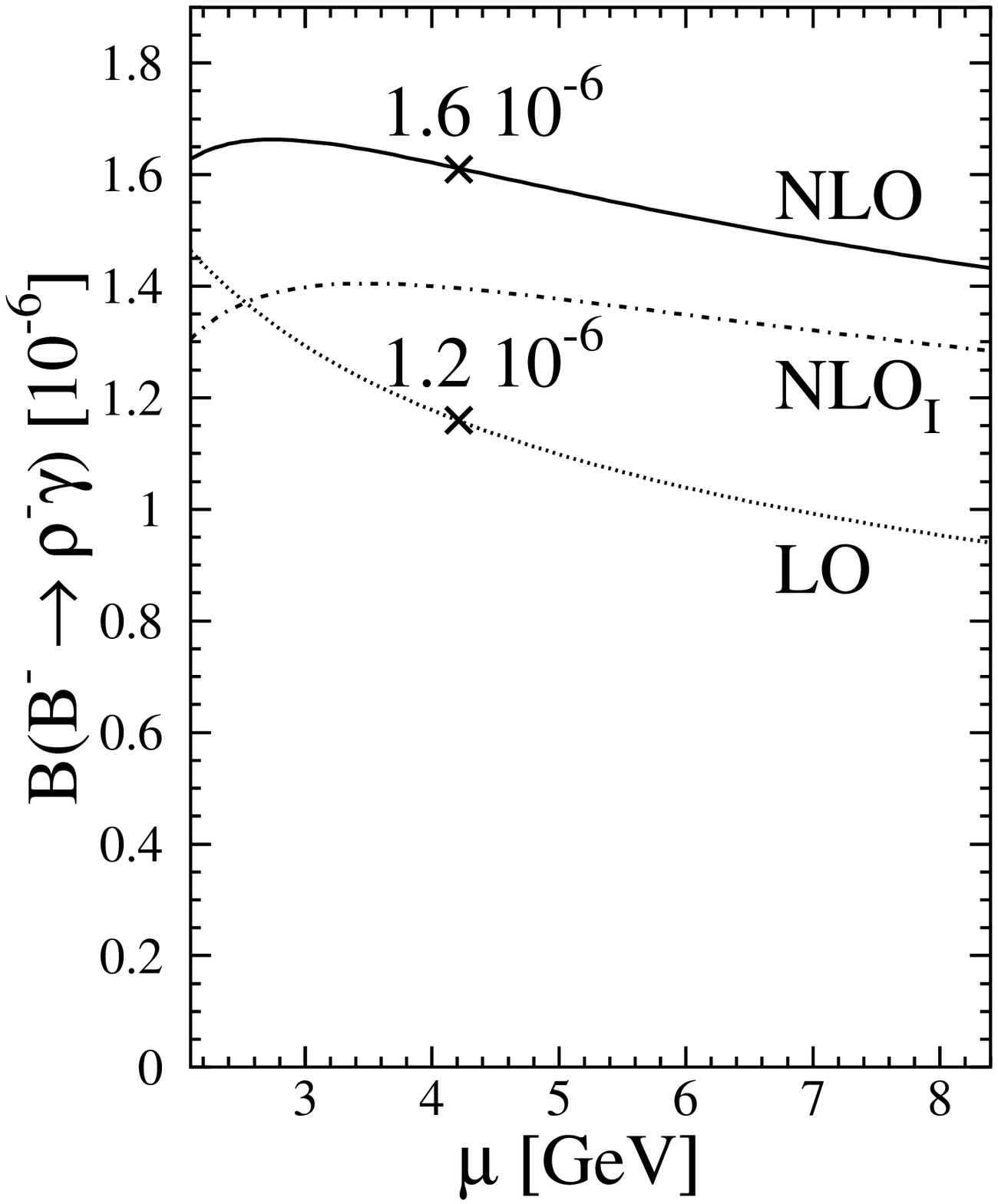,height=3.0in}
\caption{Dependence of the branching ratios
$B(\bar{B}^0 \to \bar{K}^{*0} \gamma)$ and $B(B^- \to \rho^- \gamma)$ on
the renormalization scale $\mu$, where the leading-order contributions
(dotted lines), the next-to-leading-order contributions including only
type-I corrections (dash-dotted lines), and the complete
next-to-leading-order contributions (solid lines) are explicitly shown.}
\label{bkrhomu}
\end{center}
\end{figure}

For the $B\to V\gamma$ decay,
both the type I and type II contributions can be expressed
in terms of the matrix element $\langle O_7\rangle$:
\be
A(B\to V\gamma)=\frac{G_F}{\sqrt{2}}\left[\lambda_u^{(s)}a^u_7
+\lambda_c^{(s)} a^c_7\right]\langle V\gamma|O_7| B\rangle\;,
\ee
where $\lambda_{u,c}^{(s)}$ are the products of the
CKM matrix elements, the coefficients $a_7^{(u,c)}$ consist of the
Wilson coefficient $C_7$ and the contributions from the type-I and
type-II hard-scattering corrections. It has been observed that
the leading-order value is enhanced by the $T^I$-type correction.
The net enhancement of $a_7$ at the next-to-leading order increases
the branching ratios as illustrated in Fig.~\ref{bkrhomu} \cite{BBgamgam}.
The residual scale dependence for
$B(\bar{B}\to \bar{K}^{*0}\gamma)$ and $B(B^-\to\rho^-\gamma)$ at
leading and next-to-leading orders is also exhibited.

The branching ratios of the exclusive radiative decays have been
measured to be
$B(B^0\!\to K^{*0}\gamma)=(4.44\pm 0.35)\times 10^{-5}$ and
$B(B^+\!\to K^{*+}\gamma)=(3.82\pm 0.47)\times 10^{-5}$ \cite{meas}. For
the $B\to\rho\gamma$ decay, only upper bound 
exists. The leading-order results from QCDF have
more or less saturated the experimental data, and the inclusion of
the next-to-leading-order contributions overshoot the data.
Note that the transition form factor $F^{BV}$ and the distribution
amplitudes $\phi_B$ and $\phi_V$ are both the independent inputs in QCDF.
The large predicted branching ratio about
$B(B^0\!\to K^{*0}\gamma)=7\times 10^{-5}$ could
indicate a double counting of hard contributions between the two terms
in Eq.~(\ref{fform}). Therefore, the basic assumption of QCDF, in which
the transition form factor is a completely soft object, requires a more
careful examination.

\subsection{\it Light-Front QCD \label{sec:lig}}

In the non-relativistic quark model,
wave functions best resemble meson states in the rest frame, or
where the meson velocities are small. Therefore, the form factors
calculated in this model are reliable only at small recoil.
At large recoil, relativistic effects must be taken into account.
A consistent and fully relativistic treatment \cite{Iva}
of quark spins and the center-of-mass motion can be carried out
in LFQCD \cite{Ter,Chung,Zhang}.
This method has several advantages: the light-front (LF)
wave function is manifestly Lorentz invariant 
in terms of the momentum fraction variables (in ``+" components), which
is in analogy with parton distributions in the infinite momentum
frame. Hadron spin can be correctly constructed
using the Melosh rotation. The kinematic subgroup of the
LF formalism has the maximal number of interaction-free
generators, including the boost operator which describes the
center-of-mass motion of the bound state. 

LFQCD has been applied
to the heavy-to-heavy and heavy-to-light transition form
factors \cite{Jaus,Don94,CJK02,GHLZ}. However, the form factors were
calculated only for $q^2\leq 0$, whereas physical
decays occur in the time-like region with $0\leq q^2\leq (m_i-m_f)^2$,
$m_{i,f}$ being the initial and final meson masses. Hence,
extra assumptions are needed to extrapolate the form factors from the
space-like region to the time-like region \cite{Jaus96,Mel}.
Recently, the $P\to P$ transition form factors were
calculated in the entire range of $q^2$ \cite{Sima,Cheung2},
such that the additional extrapolation is
no longer required. This is based on the observation \cite{Dubin}
that in the frame where the momentum transfer is purely longitudinal,
i.e., $q_T=0$, the invariant $q^2=q^+q^-$ covers the entire
range of momentum transfer. The price to pay is that, besides the
conventional valence-quark contribution, one must also consider the
nonvalence configuration (or the so-called Z graph) in order to maintain
covariance. The nonvalence contribution vanishes at $q^+=0$, but is
expected to be more important for heavy-to-light transitions near zero
recoil \cite{Jaus,Jaus96,Dubin,Saw}. Prescriptions for treating
this nonvalence configuration have been proposed. For example, the
authors of \cite{Cheung2} considered the effective higher Fock state, and
calculated the contribution in chiral perturbation theory.
For a relevant discussion of covariance in the LFQCD framework,
refer to \cite{Ja1}.

\begin{figure}[t!]
\begin{center}
 \epsfig{file=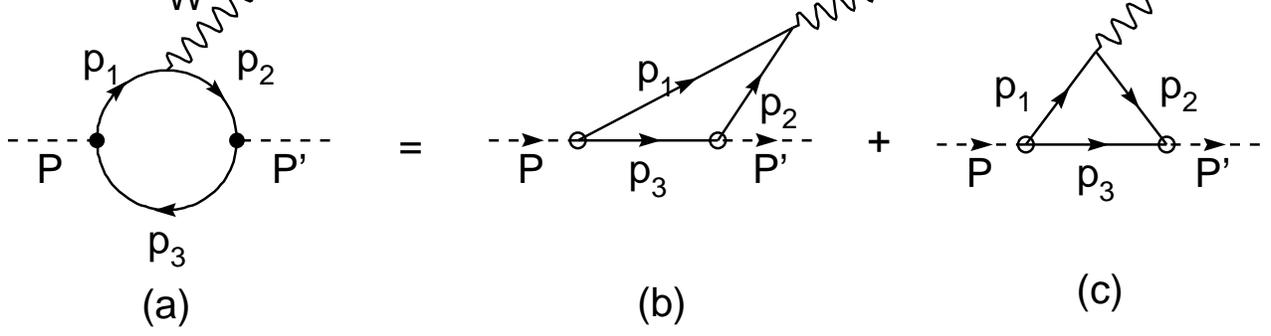,height=2.0in} 
\caption{(a) The Feynman triangle
diagram. (b) corresponds to the LF nonvalence configuration and
diagram (c) to the valence one. Filled and empty circles indicate
vertex functions and LF wave functions, respectively.}
\label{fig10}
\end{center}
\end{figure}

Below I mention the other two prescriptions \cite{Hwang,JC00}.
Start with the matrix element,
\bea
\la P'(P_2)|\bar {Q'} \gamma^\mu Q|P(P_1)
\ra=F_+(q^2)(P_1+P_2)^\mu+F_-(q^2)~q^\mu\;, \label{PPform}
\eea
where $q=P_1-P_2$ is the momentum transfer. The form factor $F_0$
is related to $F_\pm(q^2)$ by
\bea
F_0(q^2)=F_+(q^2)+{q^2\over{m_P^2-m_{P'}^2}}F_-(q^2)\;. \label{f0}
\eea
Assume a vertex function $\Lambda_P$ \cite{Jaus,Don94}, related to
bound state $Q\bar q$ of the meson $P$. The quark-meson diagram in
Fig.~\ref{fig10}(a) gives
\bea
\la P'(P_2)|\bar {Q'} \gamma^\mu Q|P(P_1) \ra=-\int {d^4 p_1 \over{(2
\pi)^4}} \Lambda_P \Lambda_{P'}tr\Bigg[\gamma_5
        {i(\not{\! p_3}+m_3)\over{p_3^2-m^2_3+i\epsilon}} \gamma_5
       {i(\not{\!
        p_2}+m_2)\over{p_2^2-m^2_2+i\epsilon}}\gamma^\mu
        {i(\not{\! p_1}+m_1)\over{p_1^2-m^2_1+i\epsilon}}\Bigg]\;,
        \label{amp}
\eea
with $p_2=p_1-q$ and $p_3=p_1-P_1$. Consider the poles in
denominators and perform the integration over the ``energy" $p_1^-$ in Eq.
(\ref{amp}). One derives
\bea
\la P'(P_2)|\bar {Q'} \gamma^\mu Q|P(P_1) \ra &=& \int^q_0 [d^3 p_1]
          \Bigg({\Lambda_P \over{S_1+S_3}}I^\mu
                {\Lambda_{P'} \over{S_2+S_3}}\Bigg)
                \Bigg |_{S_{1}=0}\;, \nn \\
&+&\int^P_q [d^3 p_1]
          \Bigg({\Lambda_P \over{S_1+S_3}}I^\mu
                {\Lambda_{P'} \over{S_2+S_3}}\Bigg)
                \Bigg |_{S_{3}=0}\;,  \label{SSS}
\eea
with the definitions,
\bea
&&[d^3 p_1]=dp^+_1 d^2p_{1T}/(64\pi^3 \prod_{i=1}^3 p^+_i)\;,
\nn \\
&&I^\mu=tr[\gamma_5(\not{\!
p_3}+m_3)\gamma_5(\not{\! p_2}+m_2)\gamma^\mu(\not{\! p_1}+m_1)]\;,
\nn \\
& &S_i \equiv p^-_i-p^-_{i{\rm on}}\;, \;\;\;\;i=1,2,3\;,
\nn\\
&&p^-_{1(3)}=P^-_{1\rm on}-p^-_{3(1){\rm on}}\;,
\nn \\
&&p^-_{i{\rm on}}=(m_i^2+p^2_{iT})/(2p_i^+)\;.
\eea
$p^-_2$ is equal to $p^-_{3{\rm on}}-P^-_{2\rm on}$
($P^-_{2\rm on}-p^-_{3{\rm on}}$) in the first (second) term of
Eq.~(\ref{SSS}). The non-pointlike vertex structures $\Lambda_P$
are necessary in order to smear the divergences in the integrals.

Associate 
\bea
R^{\rm v}_{1,2}&=&{\sqrt{P_1^+p^+_1 p^+_2}
              \over{2\sqrt{p_{1{\rm on}}\cdot p_{2{\rm on}}+m_1 m_2}}}\;,
\nn\\
R^{\rm n}_{2,3}&=&{\sqrt{P_2^+p^+_2 p^+_3}
              \over{2\sqrt{p_{2{\rm on}}\cdot p_{3{\rm on}}-m_2 m_3}}}\;,
\eea
with the valence and nonvalence configurations of the Melosh 
transformation \cite{Melosh} $R^{S,S_z}_{\lambda_1,\lambda_2}$, which
creats a state of definite spin ($S,S_z$) out of LF helicity
($\lambda_1,\lambda_2$) eigenstates \cite{Cheung2,cheng,hwcw2}.
Both of them come from the internal structure $\Lambda_P$.
One then further makes the substitution,
\bea
{\Lambda_P \over{S_1+S_3}}\Bigg|_{S_{1}=0}
\longrightarrow R^{\rm v}_{1,3}~\phi^{\rm v}_P\;, \;\;\;\; 
{\Lambda_{P'}\over{S_2+S_3}} \Bigg|_{S_{1}=0}
\longrightarrow R^{\rm n}_{2,3}~\phi^{\rm n}_{P'}\;.
\eea
The wave funciton $\phi^{v(n)}$, normalized to unity, describes the
momentum distribution of the constituents in the bound state.

At last, Eq.~(\ref{SSS}) becomes
\bea
& &\la P'(P_2)|\bar {Q'} \gamma^+ Q |P(P_1) \ra=2P_1^+H(r)\;,
\nn\\
& &H(r)= \int {d^2k_T\over 2(2\pi)^3}
\Bigg\{\int^{r}_0dx~ \phi^{\rm v}_P(x,k_T) \phi^{\rm
v}_{P'}(x',k_T)\,{{\cal A}{\cal A}'+k^2_T\over\sqrt{{\cal
A}^2 +k_T^2}\sqrt{{\cal A}'^2+k_T^2}}
\nn \\
&&\hspace{1.5cm}+\int^{1}_rdx~ \phi^{\rm v}_P(x,k_T) \phi^{\rm
n}_{P'}(x',k_T) \,{{\cal A}{\cal A}'+k^2_T\over\sqrt{{\cal
A}^2 +k_T^2}\sqrt{{\cal A}'^2+k_T^2}} \Bigg\}\;,
\label{Hi}
\eea
with the ratio $r=P_2^+/P_1^+$ and the variables,
\bea
{\cal A}=m_1x+m_3(1-x)\;, \;\;\;\;
{\cal A}'=m_2x'+m_3(1-x')\;,
\eea
$x~(x'=x/r)$ being the momentum fraction carried by the spectator antiquark
in the initial (final) state in the first term of Eq.~(\ref{Hi}). However,
$x' \geq 1$ in the second term of Eq.~(\ref{Hi}) indicates that the
momentum $p^+_3$ of the spectator quark is larger than $P'^+$
of the final meson. Therefore, the wave function
$\phi^{\rm n}_{P'}$ plays the role of a fragmentation function in
inclusive QCD processes.

The form factors are then given, in terms of $H(r)$,  
by
\bea
F_{\pm}(q^2) &=&\pm {(1 \mp r_-)H(r_+)-(1 \mp r_+)H(r_-)\over
r_+-r_-}\,, \label{fpm}
\eea
where the ratios,
\bea
r_{\pm}&=&{1\over {2m^2_P}}\big[m_P^2+m_{P'}^2-q^2\pm
2m_P{\cal Q}(q^2)~\big]\;,
\nn\\
{\cal Q}(q^2)&=&\sqrt{(m^2_P+m^2_{P'}-q^2)^2-4 m^2_P
m^2_{P'}}/2m_P\;,
\label{y12}
\eea
are the solutions of $q^2=(1-r)(m_P^2-m_{P'}^2/r)$.

\begin{figure}[t!]
\begin{center}
 \epsfig{file=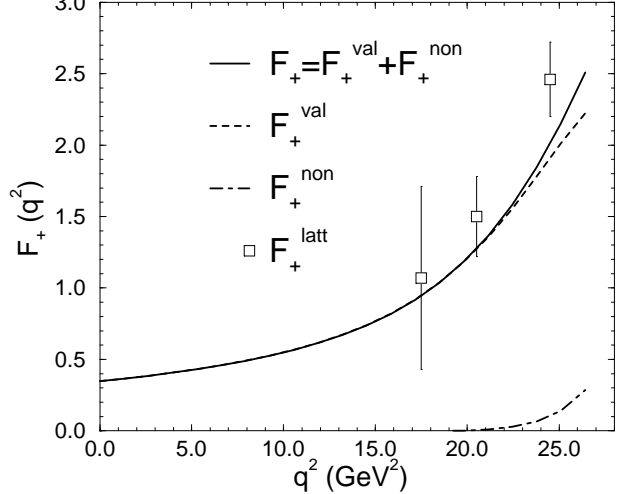,height=3.0in} 
\caption{The $B\to\pi$ form factor $F_+$
compared with the lattice QCD results \cite{Bpilat}.}
\label{fig11}
\end{center}
\end{figure}


Assume some model wave functions, whose parameters can be fixed from 
the quark masses, decay constants, and other experimental 
data \cite{Hwang}. The numerical results of the $B\to\pi$ form factor 
$F_+$ are ploted in Fig.~\ref{fig11}. The form factor values are
consistent with those obtained in the $q^{+}=0$ frame followed by
an anatylic continuation to the time-like region \cite{CJ2}.
It is found that
the nonvalence contribution to heavy-to-light transitions
is negligible in the whole region of $q^2$ except
near zero recoil ($q^2\sim q^2_{\rm max}$). 
In addition, for the same final
meson, the nonvalence contributions are smaller when the inital mesons
are heavier. This conclusion is consistent with those drawn
in \cite{Jaus,Jaus96,Dubin,Saw}.

\begin{figure}[t!]
\begin{center}
 \epsfig{file=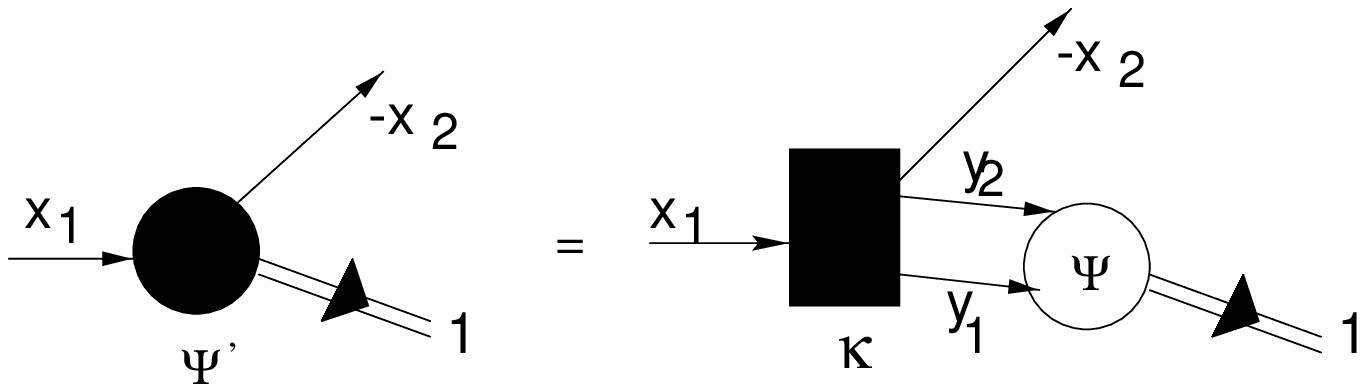,height=1.0in} 
\caption{Non-wave-function vertex(black blob) linked to an ordinary
LF wave function(white blob).}
\label{figSD}
\end{center}
\end{figure}

Another treatment of the nonvalence state is to adopt the Schwinger-Dyson
equation, which connects the embedded state (the black blob
in Fig.~\ref{figSD}) to the ordinary light-cone wave function (white blob
in Fig.~\ref{figSD}) \cite{JC00}. This connection from one-body to 
three-body sector can be achieved by introducing an operator ${\cal K}$, 
which in general depends on the involved momenta.
It is easy to see that the following link between the 
non-wave-function vertex (black blob) and the 
ordinary LF wave function (white blob)
naturally arises,
\bea\label{eq:SD}
(M^2-M'^{2}_0)\Psi'(x_i,{\bf k}_{T i})
=\int[dy][d^2{\bf l}_{T}]
{\cal K}(x_i,{\bf k}_{T i};y_j,{\bf l}_{T j})
\Psi(y_j,{\bf l}_{T j})\;,
\label{SD}
\eea 
where $M$ is the mass of outgoing meson and $M'^{2}_0=(m^2_1+{\bf 
k}^2_{T 1})/x_1 - (m^2_2+{\bf k}^2_{T 2})/(-x_2)$ with
$x_1 = 1-x_2 > 1$ due to the kinematics of the non-wave-function vertex.
Note that Eq.~(\ref{SD}) essentially takes the same form as 
the LF bound-state equation.

\begin{figure}[t!]
\begin{center}
 \epsfig{file=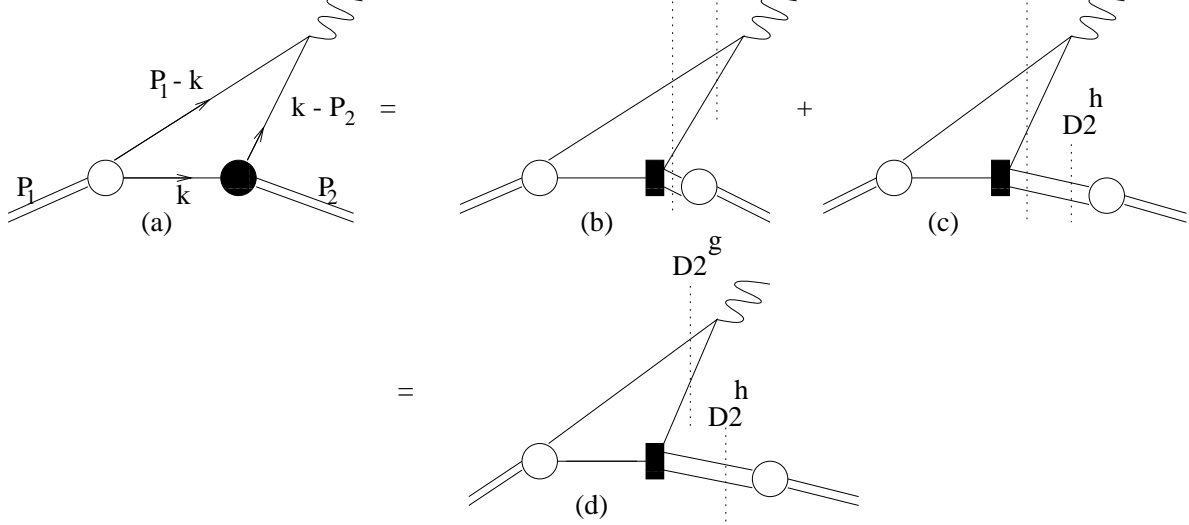,height=3.0in} 
\caption{Effective calculation of the embedded state (black blob) in 
terms of the usual LC wave fucntion (white blob).}
\label{fig12}
\end{center}
\end{figure}

Next step is to remove the four-body energy 
denomenator $D_4$ using the identity, 
\bea
\frac{1}{D_4D^g_2} + \frac{1}{D_4D^h_2}=\frac{1}{D^g_2D^h_2}\;.
\eea
One then obtains the amplitude identical to the nonvalence
contribution in terms of ordinary LF wave functions of gauge boson  
and hadron (white blob) as shown in Fig.~\ref{fig12}(d). 
Hence, the valence and nonvalence contributions can be
calculated by menas of the ordinary LF wave functions with the latter
involving an unknown operator $\cal K$. It has been argued that  
the right-hand side of Eq.~(\ref{SD}) can be approximated as a
constant for heavy meson decays in the region with small momentum
transfer \cite{JC00}. In contact interaction case, it has been verified 
that the prescription of a constant operator in Fig.~\ref{fig12}(d) is 
an exact solution of  Fig.~\ref{fig12}(a). The above formalism has been 
applied to the $D\to K$ transition form factors \cite{JC00}.

\section{Two-Body Nonleptonic Decays}

In this section I review progress on the understanding 
of QCD dynamics in two-body charmless nonleptonic $B$ meson decays. 
Topics related to charmed decays will be arranged in the next section.
Intuitively, decay products from the heavy $b$ quark move fast without 
further interaction between them. This simple picture, supported by the 
color-transparency argument \cite{transparency}, leads to 
the naive factorization. 
Although the factorization assumption (FA) \cite{BSW} gives predictions 
in relatively good agreement with data (apart from
the color-suppressed modes), it provides no insight into dynamics. 
Moreover, FA suffers serious theoretical drawbacks as explained
below. To improve FA, several frameworks, based on different assumptions 
of the dominant dynamics in exclusive $B$ meson decays, have been 
developed. Amomg these, I will discuss the QCDF, PQCD and LCSR
approaches.

\subsection{\it Factorization Assumption \label{sec:nai}}

To explain the idea of FA, I take the decay 
$\bar B^0\to D^+\pi^-$ as an example.
The relevant effective weak Hamiltonian is given by
\begin{eqnarray}
{\cal H}_{\rm eff} = {G_F\over\sqrt{2}}\, V_{cb}V_{ud}^*
\Big[C_1(\mu)O_1(\mu)+C_2(\mu)O_2(\mu)\Big]\;,
\end{eqnarray}
with the four-fermion operatos,
\begin{eqnarray}
O_1 = (\bar db)_\vma(\bar cu)_\vma\;,\qquad\qquad  
O_2= (\bar cb)_\vma(\bar du)_\vma,
\end{eqnarray}
and the definition $(\bar q_1q_2)_{_{V\pm A}}\equiv
\bar q_1\gamma_\mu(1\pm \gamma_5)q_2$. 
To ensure the renormalization-scale and -scheme independences
of physical amplitudes, the matrix elements of four-fermion
operators have to be evaluated in the same renormalization scheme
as that for Wilson coefficients and renormalized at the same scale $\mu$.
Under FA, the matrix element 
$\langle O(\mu)\rangle$ is factorized into
the product of two matrix elements of single currents, governed by decay 
constants and form factors. The naive
factorization was first proved in the framework of large
energy effective theory \cite{leet}, and justified
in the large $N_c$ limit \cite{largeN}.
For nice reviews, refer to \cite{NRSX}.

In spite of its simplicity, FA encounters three principal difficulties. 
First, the hadronic matrix element under FA is independent of the 
renormalization scale $\mu$, as the vector or axial-vector current is 
partially conserved. 
Consequently, the amplitude $C_i(\mu)\langle O\rangle_{\rm fact}$ is not 
truly physical as the scale dependence of Wilson coefficients does not 
get compensation from the matrix elements. Second, nonfactorizable 
contributions have been neglected. This may be the reason FA fails to 
describe the color-suppressed decays, such as $B\to J/\psi K^{(*)}$. 
Strong phases are essential for predicting CP asymmetries in
exclusive $B$ meson decays. These phases,
arising from the Bander-Silverman-Soni (BSS) mechanism \cite{BSS},
are ambiguous in FA. In this mechanism the $c$ quark loop
contributes an imaginary piece proportional to,
\begin{equation}
C_2(t)\alpha_s(t)\int du u(1-u)\theta(q^2 u(1-u)-m_c^2)\;,
\label{stp}
\end{equation}
where $q^2$ is the invariant mass of the gluon emitted from the penguin.
Since $q^2$ is not clearly defined in FA, one can not obtain definite 
information of the strong phase from Eq.~(\ref{stp}).

The scale problem in FA
can be circumvented in two prescriptions.
First, one incorporates nonfactorizable effects into the
effective coefficients \cite{Cheng94,Cheng96,Soares}:
\begin{eqnarray}
a_1^{\rm eff}& =& C_2(\mu) + C_1(\mu) \left[{1\over N_c}
+\chi_1(\mu)\right]\,, 
\nn\\
a_2^{\rm eff}& =& C_1(\mu) + C_2(\mu)\left[{1\over N_c} 
+ \chi_2(\mu)\right]\,,
\end{eqnarray}
where nonfactorizable terms are characterized by the parameters $\chi_i$.
The $\mu$ dependence of Wilson coefficients is assumed to be exactly
compensated by that of $\chi_i(\mu)$ \cite{NRSX}. 
However, the renormalized four-fermion operator by itself
still depends on $\mu$, though the scale dependence of 
$\langle O(\mu)\rangle$ is lost in FA. To
next-to-leading order, the Wilson coefficients depend on the
choice of the renormalization scheme, and it is not clear if $\chi_i(\mu)$
can restore the scheme independence of the matrix element.

In the second prescription, $\langle O(\mu)\rangle$ is
related to the tree-level hadronic matrix element via the relation
$\langle O(\mu)\rangle=g(\mu)\langle O\rangle_{\rm tree}$. The factor
$g(\mu)$ is obtained by calculating loop corrections to the
weak decay vertices. Then schematically one writes
\begin{eqnarray}
C(\mu)\langle O(\mu)\rangle=
C(\mu)g(\mu)\langle O\rangle_{\rm tree}
\equiv C^{\rm eff}\langle O\rangle_{\rm tree}.
\label{ude}
\end{eqnarray}
FA is applied afterwards to the hadronic 
matrix element of the operator $O$ at the tree level. Since the 
tree-level matrix element $\langle O\rangle_{\rm tree}$ is renormalization
scheme and scale independent, so are the effective Wilson coefficients
$C_i^{\rm eff}$ and the effective parameters $a_i^{\rm eff}$ expressed by
\cite{Ali,CT98}
\begin{eqnarray} 
a_1^{\rm eff} &=& C_1^{\rm eff} + C_2^{\rm eff} \left({1\over N_c}
+\chi_1\right)\,, 
\nn\\
a_2^{\rm eff} &=& C_2^{\rm eff} + C_1^{\rm eff}\left({1\over N_c} +
\chi_2\right)\,.
\label{aeff}
\end{eqnarray}

Unfortunately, the extraction of $g(\mu)$ from the matrix
element is infrared divergent.
The divergences are usually regularized by considering off-shell 
momenta $p$ for the external quark lines with $p^2 <0$.
What one has achieved is actually
\begin{eqnarray}
C^{\rm eff}=C(\mu)g(\mu,-p^2,\lambda)\;,
\label{gfc}
\end{eqnarray}
with $p^2$ being the infrared cutoff, and $\lambda$ a gauge parameter.
Obviously, $c^{\rm eff}$ is subject to the ambiguities of the
infrared cutoff and the gauge dependence. As stressed in
\cite{Buras98}, the gauge and infrared dependences always appear
as long as the matrix elements are evaluated between
quark states. The reason has been implicitly pointed out in \cite{ACMP} 
that ``off-shell renormalized vertices of gauge-invariant operators are 
in general gauge dependent".
Also, the nonfactorizable contributions are included by introducing
more free parameters as shown in Eq.~(\ref{aeff}). These parameters,
being process-dependent, then make FA even less predictive. The 
difficulty in predicting strong phases in FA also remains.

\subsection{\it QCD Factorization \label{sec:fa2}}

An important step towards a rigorous framework for
two-body nonleptonic $B$ meson decays in the heavy quark limit
has been made \cite{BBNS1,BBNS2,BBNS3}. 
The infrared divergences appearing in the loop corrections to the
weak deacy vertices are absorbed into a $B$ meson transition form
factor, such that $g(\mu)$ can be evaluated in terms of on-shell
external quarks. In this way, the infrared divergences are regularized
without breaking the gauge invariance. The nonfactorizable
contribution is calculated in the framework of collinear factorization
theorem, since it is not suffered by the end-point singularity at least 
at leading twist due to the color-transparency argument. The gluon 
invariant mass $q^2$ in the BSS mechanism can be 
clearly defined and related to parton momentum fractions in collinear
factorization theorem. Therefore, the theoretical difficulties 
in FA are resolved in principle in this QCDF approach.

\begin{figure}[t!]
\begin{center}
\epsfig{file=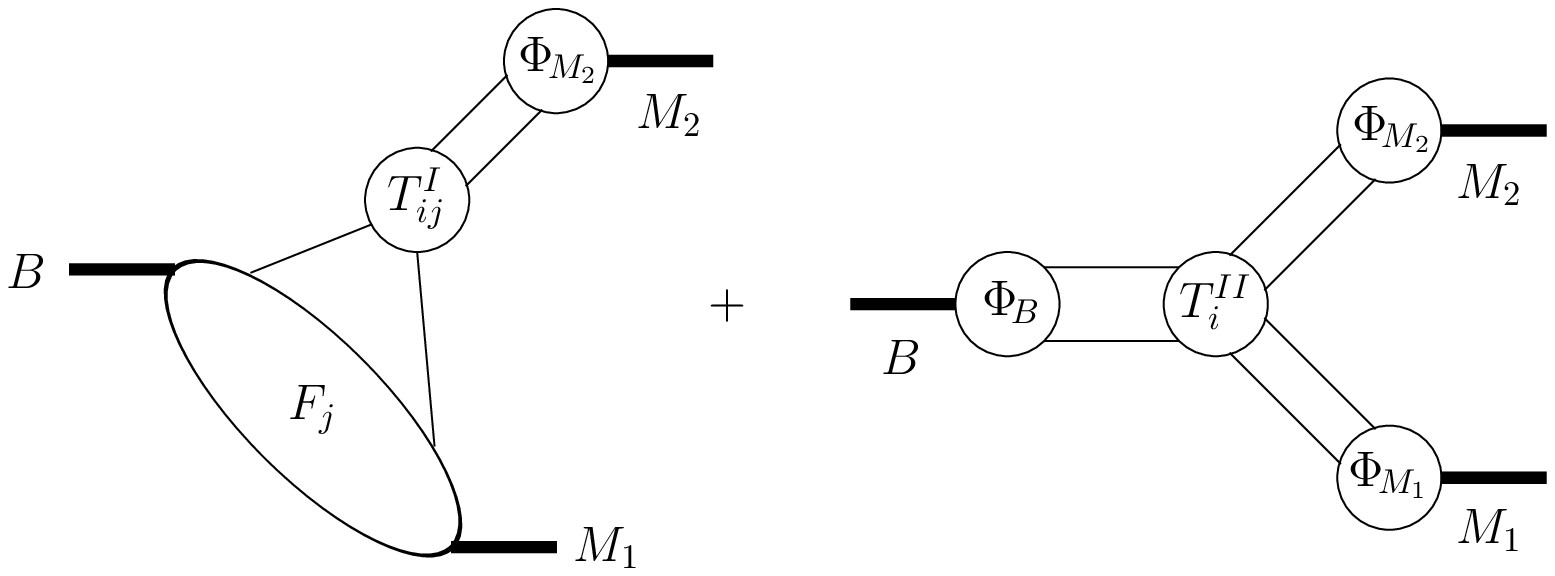,height=2.0in}
\end{center}
\caption{Graphic representation of QCD factorization}
\label{qcdff}
\end{figure}

The resulting factorization formula for the decay amplitudes 
incorporates elements both of FA
sketched above and of the hard-scattering picture. Consider a decay 
$B\to M_1M_2$, where $M_1$ picks up the spectator quark. 
If $M_1$ is either a heavy ($D$) or a light ($\pi$, $K$) meson, and 
$M_2$ a light ($\pi$, $K$) meson, QCDF gives the following structure,
\begin{equation}
A(B\to M_1M_2)=\left[\mbox{``naive factorization''}\right]
\times\left[1+O(\alpha_s)+O(\Lambda_{\rm QCD}/m_b)\right]\;.
\label{qcdfe}
\end{equation}
The $O(\alpha_s)$ term contains the finite piece of the loop 
corrections to the decay vertices and the nonfactorizable contributions.
The $O(\Lambda_{\rm QCD}/m_B)$ term collects the power corrections
to FA, such as those from the annihilation topology. 
Both terms are supposed to be calculable in a systematic way in QCDF. 
Equation (\ref{qcdfe}), without the chirally enhanced twist-3 
contributions, has been proved to all orders in $\alpha_s$
in the framework of SCET \cite{CK03}.

The leading term in $1/m_b$ is then
written, according to Eq.~(\ref{qcdfe}), as
\begin{eqnarray}
\label{fact}
\langle M_1 M_2|O_i|\bar B\rangle 
&\hspace*{-0.2cm}=&\hspace*{-0.2cm} 
F^{BM_1}(0)\int_0^1 \!\!du\,T^I(u)
\phi_{M_2}(u) \nonumber\\[0.0cm]
&&\hspace*{-2.8cm}
+\!\int \!d\xi du dv \,T^{II}(\xi,u,v)\,\phi_B(\xi)\phi_{M_1}(v) 
\phi_{M_2}(u)\;,
\end{eqnarray}
which is graphically described in Fig.~\ref{qcdff}. 
$F^{BM_1}$ is a nonperturbative form factor, and
$\phi_{M_i}$ and $\phi_B$ the light-cone distribution 
amplitudes. The hard kernel $T^{I}$ absorbs the finite
part of the loop corrections to the decay vertices. The hard kernel
$T^{II}$ corresponds to the nonfactorizable contributions. It is easy
to find that Eq.~(\ref{fact}) is similar to the leading-power
QCDF expression for a $B$ meson transition form factor in 
Eq.~(\ref{fff}).

For QCDF, the universal nonperturbative inputs include not only hadron
distribution amplitudes, but also $B$ meson transition form factors.
It has been found that the end-point singularities exist in the
twist-3 factorization formula for the nonfactorizable amplitudes,
and in power-suppressed annihilation amplitudes.
Because of these end-point singularities, the $O(\alpha_s)$ 
and $O(\Lambda_{\rm QCD}/m_B)$ terms in Eq.~(\ref{qcdfe}) turn out
to be uncalculable, and their contributions have been parametrized
into
\begin{eqnarray}
\ln\frac{m_B}{\Lambda}\left(1+\rho_He^{i\delta_H}\right)\;,\;\;\;\;
\ln\frac{m_B}{\Lambda}\left(1+\rho_Ae^{i\delta_A}\right)\;,
\label{rhoa}
\end{eqnarray}
respectively. In fact, the singularities signal the breakdown 
of factorization. QCDF then contains the non-universal and
uncontrollable parameters $\rho_{H,A}$ and $\delta_{H,A}$.
These arbitrary parameters can be determined, together with the
unitarity angles, from the best fit to experimental data
\cite{BBNS3,DZ}. To make
predictions, QCDF usually presents large theoretical uncertainty
due to the arbitrary parameters.
Another source of theoretical uncertainty comes from the scalar 
currents, which are proportional to the chiral symmetry breaking scale
$m_0=m_\pi^2/(m_u+m_d)$ \cite{BBNS1}.

\begin{figure}[t!]
\begin{center}
\epsfig{file=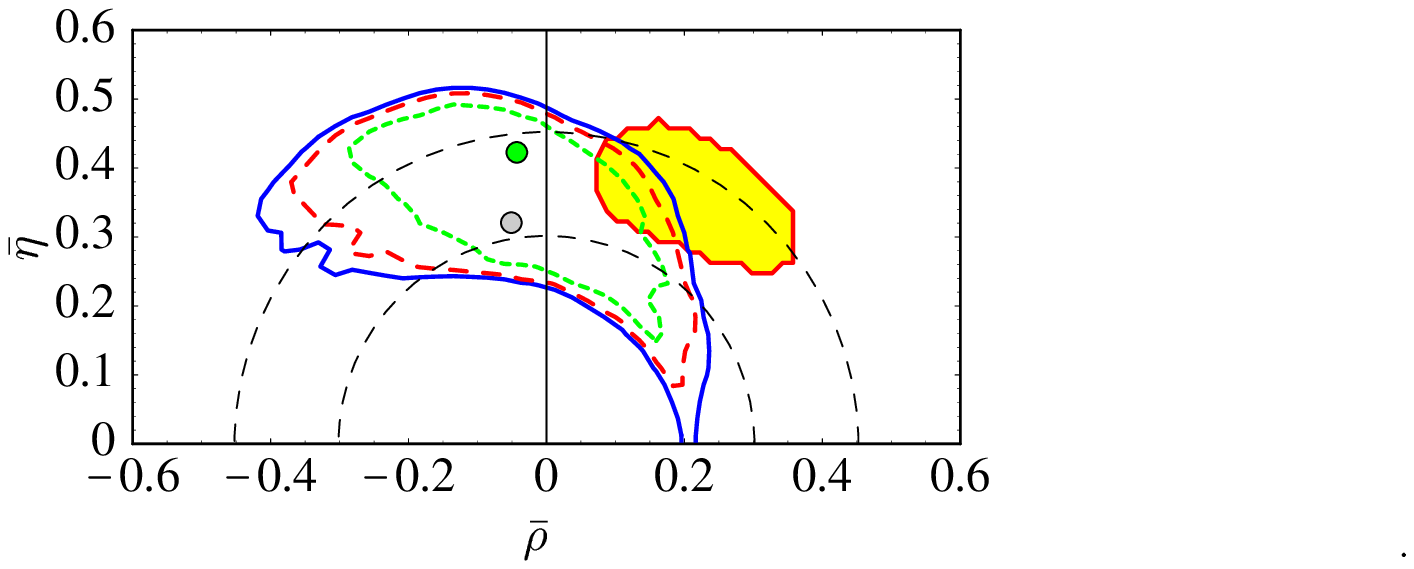,height=2.2in}
\end{center}
\caption{
95\% (solid), 90\% (dashed) and 68\% (short-dashed) confidence level 
contours in the $(\bar\rho,\bar\eta)$ plane obtained from a global 
fit to the CP averaged $B\to \pi\pi,K\pi$ branching ratios.
The darker dot shows the overall best fit, whereas the lighter dot
indicates the best fit for the default hadronic parameter set.
The light-shaded region indicates the region 
preferred by the standard global fit \cite{Hocker:2001xe}, 
including the direct measurement of $\sin(2\phi_1)$.}
\label{pik}
\end{figure}

It is possible to extract the unitarity angle $\phi_3$
from the data of the CP-averaged $B\to\pi\pi$ and $B\to K\pi$ branching 
ratios. The global fit of the Wolfenstein parameters $(\bar\rho,\bar\eta)$ 
to the six measured $B\to\pi\pi, K\pi$ world-averaged branching
ratios is displayed in Fig.~\ref{pik} \cite{Be02}. The best fits
with theory parameters in the allowed ranges \cite{BBNS3}
have $\chi^2\approx 0.5$.
The ranges of the strange quark mass and of the $B$ meson decay
constant are $[75,125]\,\mbox{MeV}$
and $[170,230]\,\mbox{MeV}$,
respectively. Since a wide range of $\phi_3$ is allowed,
the result is consistent with the standard fit based on meson mixing and
$|V_{ub}|$. Figure~\ref{pik} shows a preference for $\phi_3$ slightly
greater than $90^\circ$. This result is similar to that from a fit
based on naive FA \cite{WS}.

\begin{table}
\begin{center}  
\begin{tabular}{|l|c|c|c|}
\hline
 Decay Mode & Exp. Average\hspace{0.15cm}  & 
\hspace*{0.1cm}  Default fit \hspace*{0.1cm} & \hspace*{0.4cm}  Fit2 
\hspace*{0.4cm}\\
\hline
$B^0\to\pi^+\pi^-$& $5.15\pm 0.61$  & 5.12 & 5.24\\
$B^\pm\to\pi^\pm\pi^0$ &$4.88\pm 1.06$  & 5.00 & 4.57 \\
$B^0\to\pi^0\pi^0$ &$-$  & 0.78 & 0.94\\
\hline
$B^0\to\ K^\pm \pi^\mp$ & $18.56\pm 1.08$  & 17.99 & 18.47 \\
$B^\pm\to K^\pm\pi^0$ & $11.49\pm 1.26$  & 12.07 & 11.83 \\
$B^\pm\to K^0\pi^\pm$ & $17.93\pm 1.70$  & 15.65 & 17.88 \\
$B^0\to K^0\pi^0 $     & $8.82\pm 2.20$   & 5.55  & 6.87 \\
\hline
\end{tabular}
\caption{\label{tab1} CP-averaged $B\to\pi\pi, K\pi$ branching
ratios (in units of $10^{-6}$): 
data vs. results from the fits. The default fit to
$(\bar\rho,\bar\eta)$ (returning $|V_{ub}/V_{cb}|=0.085$, 
$\phi_3=116^\circ$ with $\chi^2 =4.5$) refers to the 
default theory parameter set used in \cite{BBNS3}. 
``Fit2'' (returning $|V_{ub}/V_{cb}|=0.079$, 
$\phi_3=97^\circ$, $\chi^2 =1.0$) refers to a fit without
annihilation contributions and chirally enhanced spectator corrections.
}
\end{center}
\end{table}

The experimental data and QCDF fits are presented in Table~\ref{tab1}
\cite{Be02}. The last two columns come from
the fitted branching ratios for the default theory parameter set 
in \cite{BBNS3} and the central values of the above ranges for $m_s$ and 
$f_B$, and for a second set, where 
all annihilation effects and chirally enhanced spectator interactions 
are removed. The second set also leads to a good fit 
without these uncertain power-suppressed effects.
The normalization of the $B\to K\pi$ modes are
sensitive to weak annihilation and to the strange quark mass through the
scalar penguin amplitude. If $\phi_3$ is assumed to take
values around $55^\circ$ as favored by indirect constraints,
the agreement becomes worse for the branching ratios with significant
tree and penguin interference. Note that the inclusion of the 
annihilation contribution weakens the constraint from nonleptonic $B$ 
meson decays on a global fit of $\phi_3$ 
\cite{Ciuchini:2000de,CFMP}.


The direct and mixing-induced CP asymmetries in the $B_d^0\to\pi^+\pi^-$ 
decay are also the important quantities for extracting the unitarity
angles. The time-dependent asymmetry is defined as
\bea
A(t) &\equiv& {\Gamma(\bar{B}^0(t) \to \pi^{+}\pi^{-}) - 
\Gamma(B^0(t) \to \pi^{+}\pi^{-}) \over \Gamma(\bar{B}^0(t) \to
\pi^{+}\pi^{-}) +
\Gamma(B^0(t) \to \pi^{+}\pi^{-})}\;,
\nonumber\\
&=& S_{\pi\pi} \,\sin(\Delta m t) - C_{\pi\pi}\,\cos(\Delta m t)\;,
\label{cppp}
\eea
where the direct and mixing-induced asymmetries,
\be
C_{\pi\pi}={1-|\lambda_{\pi\pi}|^2 \over 1+|\lambda_{\pi\pi}|^2}\;,
\hspace{20mm}
S_{\pi\pi}={2 \,Im(\lambda_{\pi\pi}) \over 1+|\lambda_{\pi\pi}|^2}\;,
\ee
respectively, satisfy the relation $C_{\pi\pi}^2 + S_{\pi\pi}^2 \leq 1$.
The factor $\lambda_{\pi\pi}$ is given by
\be
\lambda_{\pi\pi} = |\lambda_{\pi\pi}|\, e^{2i(\phi_2 + \Delta\phi_2)}
=e^{2i\phi_2} \left[{1+R_c e^{i\delta} \,e^{i\phi_3} \over 
1+R_c e^{i\delta} \,e^{-i\phi_3} } \right]\;,
\ee
with the penguin-over-tree ratio $R_c=|P/T|$ and the strong phase difference between penguin and tree amplitudes, $\delta=\delta_P-\delta_T$.
We have $S_{\pi\pi} = \sin(2\phi_2)$, if the penguin amplitude is zero.

\begin{figure}[t!]
\begin{center}
\epsfig{file=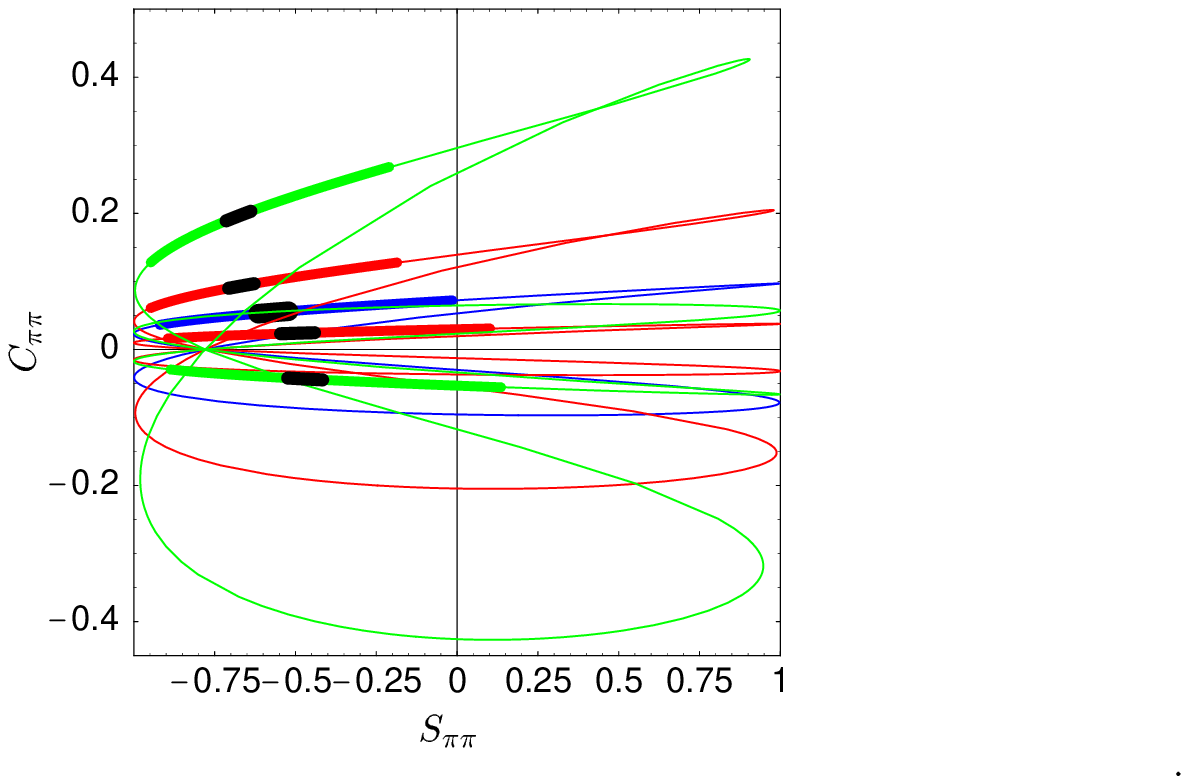,height=3.5in}
\end{center}
\caption{
Predicted correlation between the mixing-induced and direct CP 
asymmetries in the $B_d\to\pi^+\pi^-$ decay. See text for explanation 
of the different curves.}
\label{pipi}
\end{figure}

The predicted correlation between $S_{\pi\pi}$ and $C_{\pi\pi}$ from QCDF
is shown in Fig.~\ref{pipi} \cite{Be02}, where the $B\bar B$ mixing phase has been fixed to $\sin(2\phi_1)=0.78$. Each closed curve is generated by 
specifying the theory input and varying $\phi_3$ from 0 to $360^\circ$. 
The central (dark) curve refers to the calculation of $P/T$ with the 
default theory parameter set, the two neighboring
(lighter) curves refer to $P/T$ plus-minus its theoretical error without
the error from weak annihilation (but including the one from $|V_{ub}|$),
and the final (lightest) curves also include the error from weak
annihilation. The black part on each curve marks the point
$\phi_3=60^\circ$; the fat line segment marks the range
$[40^\circ,80^\circ]$ favored by the standard unitarity triangle fit
with larger $\phi_3$ to the right of the black part.
Note that the Belle data \cite{alpha2} are close to the boundary
of the physical region.

The recent BaBar measurement \cite{babar} with $90\%$ C.L. interval,
taking into account the systematic errors, are
\bea
&S_{\pi\pi}=0.02\pm0.34\pm0.05\;, 
& [-0.54,\hspace{5mm} +0.58]\;,
\nn\\
&C_{\pi\pi}=-0.30\pm0.25\pm0.04 \;,
& [-0.72,\hspace{5mm} +0.12]\;.
\label{scpp}
\eea
It is observed that the QCDF predictions prefer a small $C_{\pi\pi}$,
which are close to the upper bound of the BaBar data.
The predictions for $S_{\pi\pi}$ depend on $\phi_2$. 
The determination of the CKM matrix elements from the measurement
of $S_{\pi\pi}$ in the QCDF framework has been performed in \cite{GB03}.

\begin{table}
\begin{center}
\begin{tabular}{|l|c|c|c|c|}
\hline
Mode & Default & $m_s=80$\,MeV & $F_2=0.1$ & Experiment \\
\hline
$B^-\to K^-\eta'$ & $42_{\,-12\,-11}^{\,+16\,+27}$
 & $59_{\,-16\,-17}^{\,+22\,+41}$
 & $56_{\,-14\,-13}^{\,+19\,+31}$
 & $72.2\pm 5.3$ \\
$\bar B^0\to\bar K^0\eta'$ & $41_{\,-11\,-11}^{\,+15\,+26}$
 & $57_{\,-15\,-16}^{\,+21\,+39}$
 & $56_{\,-13\,-13}^{\,+18\,+30}$
& $54.8\pm 10.1$ \\
$B^-\to K^-\eta$ & $1.7_{\,-1.5\,-0.5}^{\,+2.0\,+1.3}$
 & $2.2_{\,-2.0\,-0.8}^{\,+2.7\,+1.9}$
 & $1.4_{\,-1.2\,-0.5}^{\,+1.8\,+1.1}$
 & $<6.9$ \\
$\bar B^0\to\bar K^0\eta$ & $1.0_{\,-1.2\,-0.4}^{\,+1.7\,+1.1}$
 & $1.4_{\,-1.7\,-0.6}^{\,+2.4\,+1.6}$
 & $0.7_{\,-0.9\,-0.4}^{\,+1.5\,+0.9}$
 & $<9.3$ \\
$B^-\to K^-\pi^0$ & $9.4_{\,-2.9\,-2.4}^{\,+3.2\,+5.6}$
 & $12.6_{\,-3.8\,-3.5}^{\,+4.3+8.2}$
 & $9.4_{\,-2.9\,-2.4}^{\,+3.2\,+5.6}$
 & $12.7\pm 1.2$ \\
$\bar B^0\to\bar K^0\pi^0$ & $5.9_{\,-2.3\,-1.9}^{\,+2.7\,+4.5}$
 & $8.5_{\,-3.1\,-2.8}^{\,+3.7\,+6.8}$
 & $5.9_{\,-2.3\,-1.9}^{\,+2.7\,+4.5}$
 & $10.2\pm 1.5$ \\
\hline
$B^-\to K^{*-}\eta'$ & $3.5_{\,-3.7\,-1.7}^{\,+4.4\,+4.7}$
 & $7.7_{\,-6.7\,-3.2}^{\,+7.6\,+8.0}$
 & $2.7_{\,-2.6\,-1.3}^{\,+3.5\,+3.9}$
 & $<35$ \\
$\bar B^0\to\bar K^{*0}\eta'$ & $2.5_{\,-3.1\,-1.5}^{\,+3.8\,+4.3}$
 & $6.3_{\,-5.8\,-2.9}^{\,+6.8\,+7.4}$
 & $1.2_{\,-1.8\,-0.9}^{\,+2.7\,+3.2}$
 & $<13$ \\
$B^-\to K^{*-}\eta$ & $8.6_{\,-2.6\,-\phantom{1}4.4}^{\,+3.0\,+14.0}$
 & $13.8_{\,-4.2\,-\phantom{1}6.7}^{\,+4.8\,+19.8}$
 & $9.1_{\,-2.7\,-\phantom{1}4.6}^{\,+3.1\,+14.3}$
 & $26.5\pm 6.1$ \\
$\bar B^0\to\bar K^{*0}\eta$
 & $8.7_{\,-2.6\,-\phantom{1}4.5}^{\,+2.9\,+14.0}$
 & $13.9_{\,-4.1\,-\phantom{1}6.7}^{\,+4.6\,+19.5}$
 & $9.2_{\,-2.7\,-\phantom{1}4.7}^{\,+3.0\,+14.2}$
 & $16.4\pm 3.0$ \\
$B^-\to K^{*-}\pi^0$ & $3.2_{\,-1.1\,-1.3}^{\,+1.2\,+4.0}$
 & $3.3_{\,-1.2\,-1.5}^{\,+1.3\,+4.8}$
 & $3.2_{\,-1.1\,-1.3}^{\,+1.2\,+4.0}$
 & $<31$ \\
$\bar B^0\to\bar K^{*0}\pi^0$ & $0.7_{\,-0.5\,-0.6}^{\,+0.6\,+2.4}$
 & $0.7_{\,-0.5\,-0.6}^{\,+0.6\,+3.0}$
 & $0.7_{\,-0.5\,-0.6}^{\,+0.6\,+2.4}$
 & $<3.6$ \\
\hline
\end{tabular}
\caption{\label{pred}
Predictions for the CP-averaged branching ratios (in units of $10^{-6}$), 
assuming $\phi_3=70^\circ$, $|V_{cb}|=0.041$ and $|V_{ub}/V_{cb}|=0.09$. 
The first error is due to parameter variations, and the second one 
shows the estimate of the uncertainty due to weak annihilation. The 
column labeled ``default'' refers to 
$m_s=100$\,MeV and $F_2=0$.}
\end{center}
\end{table}

The QCDF formalism has been extended to the study of $B\to VP$ modes
\cite{HMW,CY,DZ02,AGMP}. The $B\to\phi K$ branching ratios were predicted
to be around $4\times 10^{-6}$ \cite{HMW,CY}, which seems to be smaller
than the experimental data (see Table~7). The reason is that the same
set of free parameters in Eq.~(\ref{rhoa}) has been adopted for the 
$B\to PP$ and $VP$ decay amplitudes. The annihilation contribution is
then constrained by the $B\to K\pi$ branching ratios, and can not help
to increase the $B\to\phi K$ branching ratios. In a global fit performed 
in \cite{DZ02} these parameters have been assumed to be different for the 
$B\to PP$ and $VP$ modes. Introducing two independent sets of free parameters, the $B\to\phi K$ branching ratios can be fit (due to a
larger annihilation contribution) without increasing the $B\to K\pi$ ones.
However, the data for the $B\to K^*\pi$ modes were not included into the
global fit. It has been noticed \cite{AGMP} that once the $B\to K^*\pi$ 
modes are included, the confidence level of the best fit drops to below
0.1\%. The possible large direct CP asymmetry in the $B\to\rho\pi$
decays \cite{GT00} also deteriorate the fit.

QCDF has been also applied to flavor-singlet $B$ meson decays, such as 
$B\to K^{(*)}\eta^{(\prime)}$ \cite{BN02}. It is difficult to account 
for the branching ratios of these modes in FA
\cite{Ali:1997ex}. The scheme for the $\eta$-$\eta^\prime$ 
mixing, with a single mixing angle advocated in \cite{FKS}, was assumed. 
The contributions from the $b\to c\bar cs$ and $b\to sgg$ 
transitions through the gluon content of singlet mesons were analyzed 
carefully. Also, a singlet annihilation amplitude, where two gluons 
radiating from the spectator quark form an $\eta^{(\prime)}$ meson,
contributes at leading power. The unknown form factors 
$F_0^{B\eta^{(\prime)}}(0)$ were parametrized as
\begin{equation}
   F_0^{B P}(0) = F_1\,\frac{f_P^q}{f_\pi}
   + F_2\,\frac{\sqrt2 f_P^q+f_P^s}{\sqrt3 f_\pi} \,,
\end{equation}
with $P=\eta$ or $\eta'$ and $q=u$, $d$. The decay constants $f_P^{q}$ 
and $f_P^{s}$ are defined through the quark currents. In \cite{BN02} 
$F_1=F_0^{B\pi}(0)$ and $F_2=0$ or 0.1 were adopted. Combining the 
above effects, the predictions for the CP-averaged 
$B\to K^{(*)} (\eta^{(\prime)},\pi^0)$ 
branching ratios from QCDF are summarized in Table~\ref{pred}.

\subsection{\it Perturbative QCD \label{sec:pqc2}}

As stated before, the PQCD approach to two-body nonleptonic 
$B$ meson decays is based on $k_T$ factorization. Therefore, the 
theoretical difficulties in FA can also be resolved but in a 
way different from that of QCDF. The infrared divergences appearing in 
the loop corrections to the
weak deacy vertices are absorbed into meson wave functions, such that the 
infrared divergences are regularized without breaking the gauge 
invariance. The factorizable, nonfactorizable and power-suppressed
annihilation contributions are calculated in the framework of $k_T$ 
factorization theorem without the end-point singularities. The arbitrary cutoffs introduced in QCDF \cite{BBNS2,BBNS3} are not necessary, and
PQCD involves only universal and controllable inputs. 
The gluon invariant mass $q^2$ in the BSS 
mechanism can also be clearly defined and related to parton momentum 
fractions.

\begin{figure}[t!]
\begin{center}
\epsfig{file=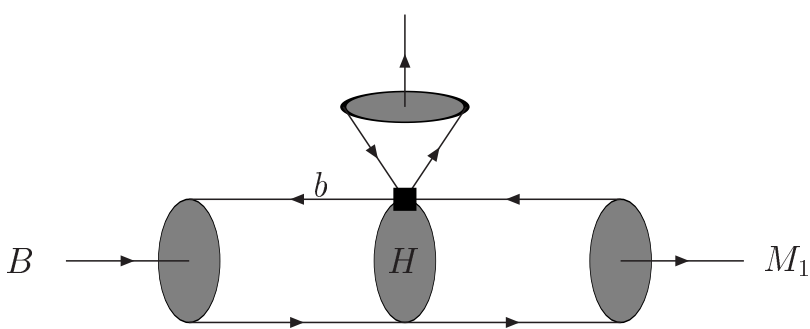,height=2.0in}
\end{center}
\caption{Factorization of two-body nonleptonic $B$ meson decays
in the PQCD approach.}
\label{pqcdf}
\end{figure}

The amplitude for the $B\to M_1M_2$ decay is then factorized 
into the convolution of the six-quark hard kernel, the Wilson 
coefficient, the jet function and the Sudakov factor 
with the bound-state wave functions 
as shown in Fig.~\ref{pqcdf} \cite{CL,YL,CLY,L365},
\begin{eqnarray}
A=\phi_B\otimes H^{(6)}\otimes J\otimes S \otimes\phi_{M_1}\otimes 
\phi_{M_2}\;,
\label{six}
\end{eqnarray}
all of which are well-defined and gauge-invariant.  $J$ denotes the jet
function from threshold resummation
discussed in Sec.~\ref{sec:thr}, and $S$ denotes the Sudakov factor
from $k_T$ resummation discussed in Sec.~\ref{sec:kt}. $J$ ($S$), 
organizing the double logarithms in the hard kernel (meson wave functions), 
is hidden in $H$ (the three meson states)
in Fig.~\ref{pqcdf}. The partition of nonperturbative and
perturbative contributions is quite arbitrary. Different
partitions correspond to different factorization schemes. However,
the decay amplitude, as the convolution of the above factors, is
independent of factorization schemes as it should be.

\begin{figure}[t!]
\begin{center}
\epsfig{file=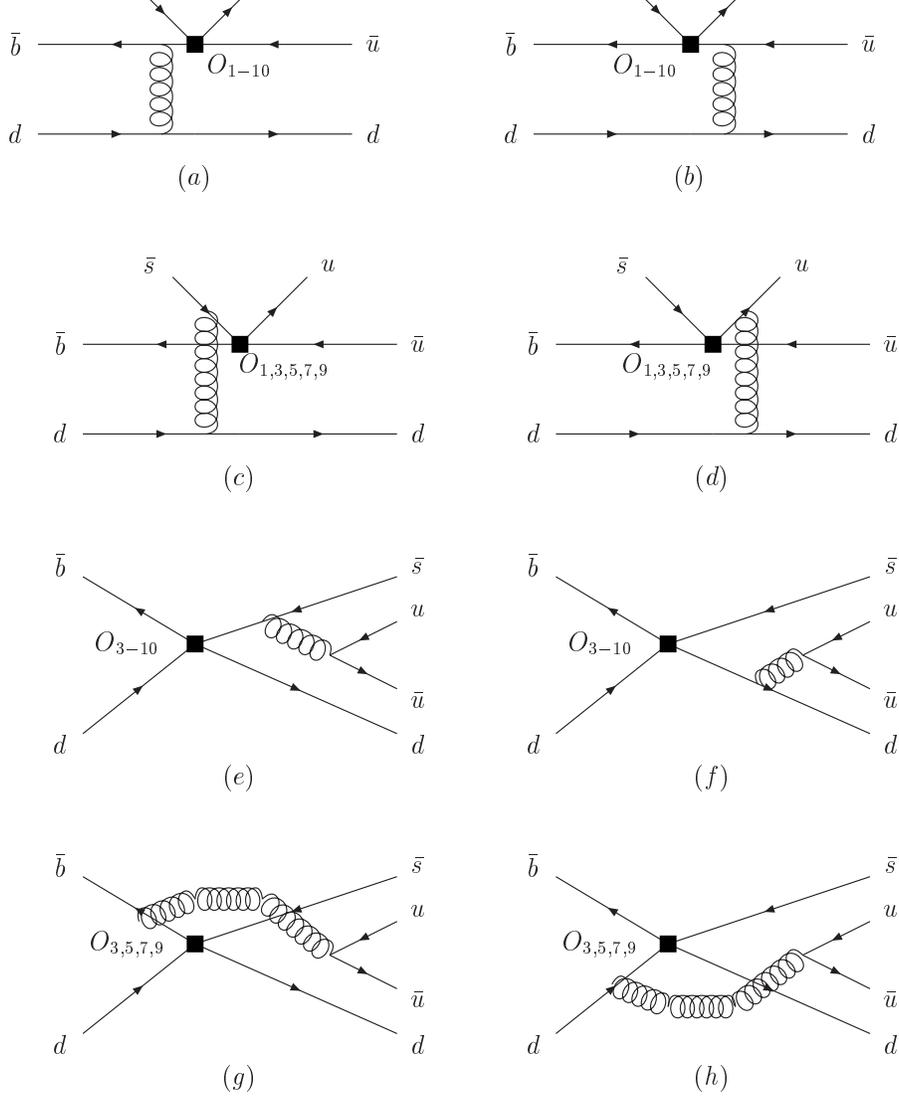,height=6.0in}
\end{center}
\caption{Leading-order diagrams for the six-quark hard kernel
in the PQCD approach.}
\label{sixh}
\end{figure}

The six-quark hard kernel $H^{(6)}$ consists of the diagrams with at 
least one hard gluon \cite{WYL}. The complete set of leading-order 
diagrams for the $B\to K\pi$ decays is displayed in Fig.~\ref{sixh}. 
Figures~\ref{sixh}(a) and \ref{sixh}(b), referred to as the factorizable 
emission, correspond to the leading contribution in QCDF [the left-hand 
diagram in Fig.~\ref{qcdff}]. Figures~\ref{sixh}(c) and \ref{sixh}(d), 
referred to as the nonfactorizable emission, correspond to the 
next-to-leading-order contribution in QCDF [the right-hand diagram in 
Fig.~\ref{qcdff}]. Figures~\ref{sixh}(e) and 
\ref{sixh}(f), and Figs.~\ref{sixh}(g) and \ref{sixh}(h) are referred to as 
the factorizable annihilation and the nonfactorizable annihilation, 
respectively. They are explicitly power-suppressed. However, 
for the physical mass $m_B\sim 5$ GeV, the scalar contribution from
the penguin operators, proportional to $m_0/m_B$, is not really 
negligible.

\begin{figure}[t!]
\begin{center}
\epsfig{file=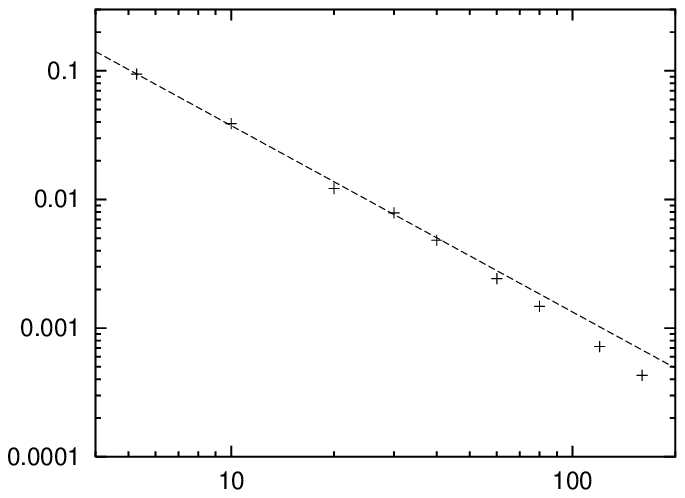,height=2.0in}
\hspace{1.0cm}
\epsfig{file=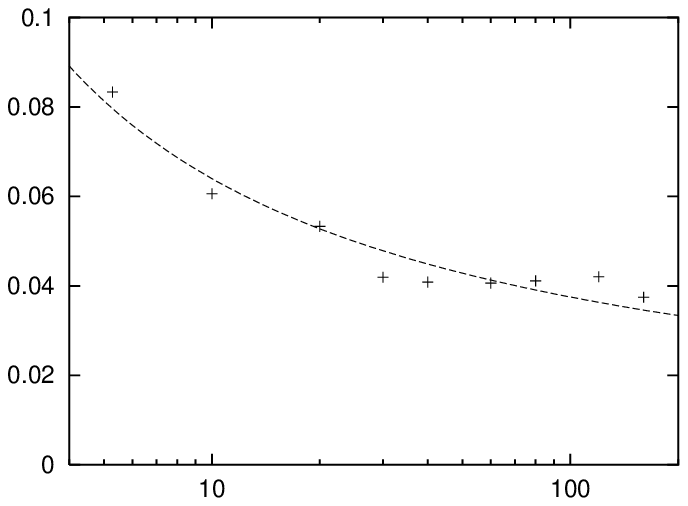,height=2.0in}
\end{center}
\caption{(a) The factorizable emission amplitude as a function of 
$m_B$ in unit of GeV. (b) The ratio $r$ of the nonfactorizable emission 
amplitude over the factorizable one as a function of $m_B$.}
\label{fig:limit}
\end{figure}

The factorization limit of the PQCD approach at large $m_B$, which is not
as obvious as in QCDF, has been examined \cite{LU02}. It was found that
the factorizable emission amplitude in Figs.~\ref{sixh}(a) and 
\ref{sixh}(b) decreases like $m_B^{-3/2}$
as displayed in Fig.~\ref{fig:limit}(a), if the $B$
meson decay constant $f_B$ scales like $f_B\propto m_B^{-1/2}$. This
power-law behavior is consistent with that obtained in 
\cite{BBNS2,Chernyak:1990ag}.
Define $r$ as the ratio of the magnitude of the 
nonfactorizable emission amplitude [from Figs.~\ref{sixh}(c) and 
\ref{sixh}(d)] over the factorizable one.
Figure~\ref{fig:limit}(b), exhibiting $r$ as a function of 
$m_B$, indicates that the curve actually descends with $m_B$ despite 
of small oscillation. If parametrizing the ratio as
\begin{eqnarray}
r\equiv \frac{|\rm Nonfact.|}{\rm Fact.}\propto
\frac{1}{\ln^\alpha(m_B/\Lambda)}\;,
\label{r}
\end{eqnarray}
the best fit to the curve gives the power $\alpha\sim 1.0$ for
$\Lambda\sim 0.4$ GeV \cite{LU02}. This logarithmic decrease has been 
confirmed up to $m_B=300$ GeV. It implies that the PQCD formalism 
approaches FA logarithmically. 

Surprisingly, the behavior of the ratio $r$ with $m_B$ in PQCD is 
close to that in QCDF. However, the reasonings for achieving the same 
power counting are quite different. In QCDF the 
factorizable contribution is assumed to be dominated by soft dynamics, 
and identified as being of $O(\alpha_s^0)$. The nonfactorizable 
contribution, being calculable, starts from $O(\alpha_s)$. Because of 
the soft cancellation at $x_3\sim O(\Lambda/m_B)$, the
nonfactorizable emission amplitude is dominated by the contribution from
the region with $x_3\sim O(1)$. In this region there is no further power
suppression, and one has the ratio,
\begin{eqnarray}
r_{\rm QCDF}\sim \alpha_s(m_B)\propto 
\frac{1}{\ln(m_B/\Lambda_{\rm QCD})}\;.
\end{eqnarray}
In PQCD based on $k_T$ factorization theorem \cite{BS,LS},
both the factorizable and nonfactorizable contributions, being
calculable, start from $O(\alpha_s)$. However, the Sudakov factor
modifies the factorization formulas in the way that
a pair of nonfactorizable diagrams exhibits a stronger cancellation as
$m_B$ increases \cite{CKL}. It turns out that the ratio $r$ also
vanishes logarithmically as shown in Eq.~(\ref{r}).

\begin{figure}[t!]
\begin{center}
\epsfig{file=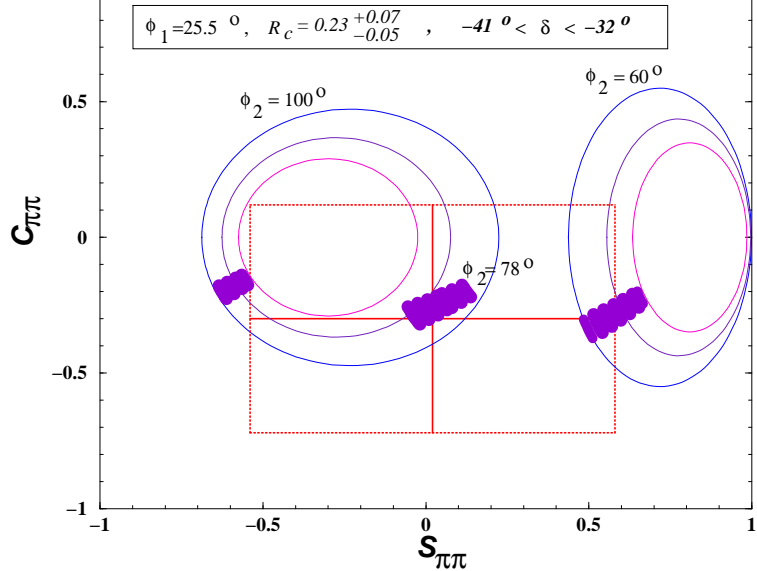,height=4.0in, angle=-90}
\end{center}
\caption{Plot of $C_{\pi\pi}$ vs $S_{\pi\pi}$  for various values
of $\phi_2$ with $\phi_1=25.5^o$, $0.18 < R_c < 0.30$ and $-41^o <
\delta < -32^o$ in PQCD. The allowed experimental ranges from BaBar 
measurment within $90\%$ C.L. was considered.
Dark areas are the regions allowed by PQCD for different $\phi_2$.}
\label{phi2}
\end{figure}

I then discuss the applications of PQCD to two-body nonleptonic
$B$ meson decays. An alternative way to determine $\phi_2$ is to
use the time-dependent CP asymmetry in the $B_d^0(t) \to \pi^{+}\pi^{-}$ 
decay, which provides two constraints from
$C_{\pi\pi}$ and $S_{\pi\pi}$ for three unknown variables 
$R_c,\delta$ and $\phi_2$ in Eq.~(\ref{cppp}).
If one knows $R_c$ and $\delta$, $\phi_2$ can be extracted 
from the experimental data of $C_{\pi\pi}$ vs $S_{\pi\pi}$. 
Since PQCD gives $R_c=0.23^{+0.07}_{-0.05}$ and $-41^o
<\delta<-32^o$, the allowed range of $\phi_2$ at present stage has
been fixed to be $55^o <\phi_2< 100^o$ as shown in Fig.~\ref{phi2}
\cite{Keum}. 
Because the strong phase in PQCD is relatively large compared to that 
in QCDF as explained below, a significant direct CP 
asymmetry $C_{\pi\pi} = -(23\pm7) \%$ was predicted, which could 
be tested by more precise experimental measurement in the near future
\cite{Nir,Ros}. The central point of the BaBar data in Eq.~(\ref{scpp})
\cite{babar} then corresponds to $\phi_2 = 78^o$. 
Denote $\Delta \phi_2$ as the uncertainty of $\phi_2$ due to the 
penguin contribution. For the allowed region of $\phi_2=(55 \sim 100)^o$, 
one obtains $\Delta \phi_2 =(8\sim 16)^o$, implying sizable
penguin contributions in the $B_d^0 \to \pi^{+}\pi^{-}$ decay. 
The main uncertainty comes from the value of $|V_{ub}|$. 

Here I give a simple explanation for the different phenomenological 
consequences of the CP asymmetries in two-body nonleptonic $B$ meson
decays derived from QCDF and from PQCD.
According to the QCDF power counting rules
\cite{BBNS1,BBNS2} based on collinear factorization, the factorizable
emission diagram gives the leading contribution of $O(\alpha_s^0)$, since
the $B\to\pi$ form factor is not calculable. Because
the leading contribution is real, the strong phase arises from the 
factorizable annihilation diagram, being of $O(\alpha_s m_0/m_B)$, and 
from the vertex correction to the leading diagram, being of $O(\alpha_s)$. 
For $m_0/m_B$ slightly smaller than unity, the vertex correction 
is the leading source of strong phases. In $k_T$ factorization the
power counting rules change \cite{CKL}. The factorizable emission diagram
is calculable and of $O(\alpha_s)$. The factorizable annihilation diagram
has the same power counting as in QCDF. The vertex correction becomes of
$O(\alpha_s^2)$. Therefore, the annihilation diagram contributes the
leading strong phase. This is the reason the strong phase derived from
PQCD and from QCDF could be opposite in sign, and the former has a large 
magnitude. As a consequence of the different power counting rules, QCDF 
prefers a small and positive CP asymmetry $C_{\pi\pi}$ \cite{Be02}, while
PQCD prefers a large and negative $C_{\pi\pi}$
\cite{LUY,Keum,KL,LMY,US}. 
Significant CP asymmetries are also expected in the 
$B\to K\pi$ \cite{KLS}, $B\to KK$
\cite{CHL} and $B\to \rho K$, $\omega K$ \cite{CHC} decays. The last two
modes are especially sensititve to the annihilation contributions. 

\begin{figure}[t!]
\begin{center}
\epsfig{file=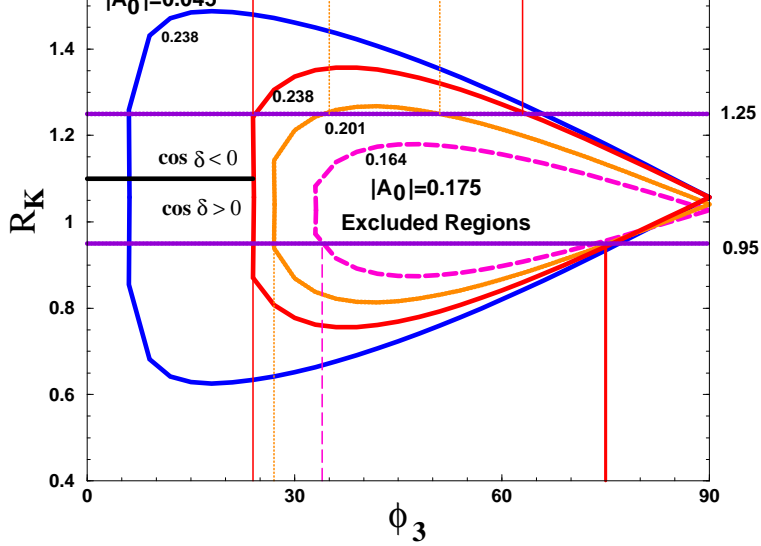,height=4.0in, angle=-90}
\end{center}
\caption{Plot of $R_K$ vs $\phi_3$ with $r_K=0.164,0.201$ and $0.238$.}
\label{Rkpi}
\end{figure}

The CP-averaged $B\to K\pi$ branching ratios may lead to nontrivial 
constaints on the angle $\phi_3$ \cite{FM,NR}. Introduce the 
observables,
\bea
R_K &=&{{Br}(B^0\to K^{\pm}\pi^{\mp}) \,\, \tau_{+} \over
{Br}(B^\pm\to K^{0}\pi^{\pm}) \,\, \tau_{0} }\;,
\nn\\
A_0 &=& A_{CP}(B^0 \to K^{+}\pi^{-}) \,\, R_K \;,
\eea
with the tree-over-penguin ratio $r_K = |T^{'}/P^{'}|$ for
the $B \to K\pi$ decays, and 
the strong phase difference between the tree and penguin amplitides,
$\delta = \delta_{T'} -\delta_{P'}$.
One has \cite{Keum}
\be
R_K = 1 + r_K^2 \pm \sqrt{4 r_K^2 \cos^2\phi_3 -A_0^2 \cot^2\phi_3}\;,
\ee
with $r_K = 0.201\pm 0.037$ from PQCD \cite{KLS},
and $A_0=-0.110\pm 0.065$ from the recent BaBar measurement 
$A_{CP}(B_d^0\to K^+\pi^-)=-10.2\pm5.0\pm1.6 \%$ \cite{babar}
and the present world-averaged value of $R_K=1.10\pm 0.15$
\cite{rk}. 

With the values $\delta_{P'} = 157^o$, $\delta_{T'} = 1.4^o$ and the 
negative $\cos\delta= -0.91$ derived in \cite{Keum},
one constrains the allowed range of 
$\phi_3$ within $1\,\sigma$ as displayed in Fig.~\ref{Rkpi},
\begin{itemize}
\item[]For $r_K=0.164$: exclude
$0^o \leq \phi_3 \leq 6^o$. 
\item[]For $r_K=0.201$: exclude
$0^o \leq \phi_3 \leq 6^o$ and $ 35^o \leq \phi_3 \leq 51^o$. 
\item[]For $r_K=0.238$: exclude
$0^o \leq \phi_3 \leq 6^o$ and $ 24^o \leq \phi_3 \leq 62^o$.
\end{itemize}
Taking the central value $r_K=0.201$,
$\phi_3$ is allowed in the range of $51^o \leq \phi_3 \leq
129^o$, because of the symmetric property between $R_K$ vs $\cos\phi_3$. 
This range is consistent with that from the model-independent
CKM-fit in \cite{Hocker:2001xe}.
The PQCD predictions for the CP-averaged $B\to\pi\pi$
and $K\pi$ branching ratios are listed in Table~6.

\begin{table}
\begin{center}  
\begin{tabular}{|l|c|c|}
\hline
 Decay Mode & Exp. Average\hspace{0.15cm}  & 
\hspace*{0.1cm}  PQCD \hspace*{0.1cm} \\
\hline
$B^0\to\pi^+\pi^-$& $5.15\pm 0.61$  & $7.0^{+2.0}_{-1.5}$\\
$B^\pm\to\pi^\pm\pi^0$ &$4.88\pm 1.06$  & $3.7^{+1.3}_{-1.1}$ \\
$B^0\to\pi^0\pi^0$ &$-$  & $0.3 \pm 0.1$\\
\hline
$B^0\to\ K^\pm \pi^\mp$ & $18.56\pm 1.08$  & $15.5^{+3.1}_{-2.5}$ \\
$B^\pm\to K^\pm\pi^0$ & $11.49\pm 1.26$  & $9.1^{+1.9}_{-1.5}$ \\
$B^\pm\to K^0\pi^\pm$ & $17.93\pm 1.70$  & $17.3 \pm 2.7$ \\
$B^0\to K^0\pi^0 $     & $8.82\pm 2.20$   & $8.6 \pm 0.3$ \\
\hline
\end{tabular}
\label{cpav}
\caption{PQCD predictions for the
$B \to \pi \pi $ and $K \pi $ branching ratios in unit of $10^{-6}$
with $\phi_3=80^0$, $R_b=\sqrt{\rho^2+\eta^2}=0.38$. 
}
\end{center}
\end{table}

The leading factorizable contributions involve four-quark hard
kernels in QCDF, but six-quark hard kernels in PQCD. This
distinction also implies different characteristic scales in the two
approaches: the former is characterized by $m_B$, while the latter is
characterized by the virtuality of internal particles of order
$\sqrt{\Lambda m_B}\sim 1.5$ GeV \cite{KLS,LUY,KL}. It has been known
that to accommodate the $B\to K\pi$ and $\pi\pi$ data, penguin
contributions must be large enough. In QCDF one relies on the chiral
enhancement by increasng the chiral symmetry breaking scale to a large
value $m_0\sim 3$-4 GeV \cite{WS}. Because of the renormalization-group
evolution effect of the Wilson coefficients, the lower hard scale leads 
to the dynamical penguin enhancement in PQCD.
The dynamical enhancement of penguin contributions in the PQCD approach
also appears in the study of $B\to VP$ modes \cite{CKL,L6,M,Me}. The
predictions are listed in Table~7.
For a vector meson, the mass $m_0$ is replaced by the
physical mass $m_V\sim 1$ GeV, and the chiral enhancement does not
exist. Therefore, the ways to account for the $B\to VP$ branching ratios
in PQCD and in QCDF are different. As stated in the previous
subsection, the infrared cutoffs in Eq.~(\ref{rhoa}) for the $B\to VP$ 
modes have been assumed to differ from those for the $B\to PP$ modes
\cite{DZ02}. A larger annihilation contribution can then help to
enhance the $B\to\phi K$ branching ratios without 
increasing the $B\to K\pi$ ones.

At last, the $B\to K eta^{(\prime)}$ decays have been analyzed in the
PQCD approach \cite{KS00,AP00}. However, the analysis is not yet complete.
\begin{table}
\begin{center}
\begin{tabular}{|c|ccc|c|} \hline
Decay Channel & CLEO & BELLE & BABAR & PQCD   \\
\hline  
$\phi K^{\pm}$ & 
 $5.5^{+2.1}_{-1.8}\pm 0.6$ &
 $11.2^{+2.2}_{-2.0} \pm 0.14$ &  
 $7.7^{+1.6}_{-1.4}\pm 0.8$ & 
 $10.2^{+3.9}_{-2.1}$  \\
$\phi K^{0}$ & 
 $ < 12.3 $ &
 $8.9^{+3.4}_{-2.7}\pm 1.0$ &  
 $8.1^{+3.1}_{-2.5}\pm 0.8 $ &
 $9.6^{+3.7}_{-2.0}$    \\ 
\hline
$K^{*\pm}\pi^{\mp}$ & $16^{+6.3}_{-5.4}$  &
$26.0\pm 9.0$ & --- & $9.1^{+4.9+0.3}_{-3.9-0.2}$ \\ \hline
$K^{*0}\pi^{\pm}$   & $< 16$ &
$16.2^{+4.8}_{-4.5}$ &
$15.5\pm 3.8$ & $10.0^{+5.3}_{-3.5}\pm 0.0$ \\ \hline
$K^{*\pm}\pi^{0}$ & --- & --- & --- & $3.2^{+1.9+0.6}_{-1.2-0.2}$ \\ \hline
$K^{*0}\pi^{0}$   & --- & --- & --- & $2.8^{+1.6}_{-1.0}\pm0.0$ \\
\hline 
\end{tabular}
\end{center}
\label{phikg}
\caption{PQCD predictions and experimental data for the
$B \to \phi K^{(*)}$ and $B \to K^{*}\pi$ branching ratios in unit of 
$10^{-6}$ for $\phi_3=80^0$, $R_b=0.38$. 
}
\end{table} 

\subsection{\it Light-Cone QCD Sum Rules \label{sec:su2}}

\begin{figure}[t!]
\begin{center}
\epsfig{file=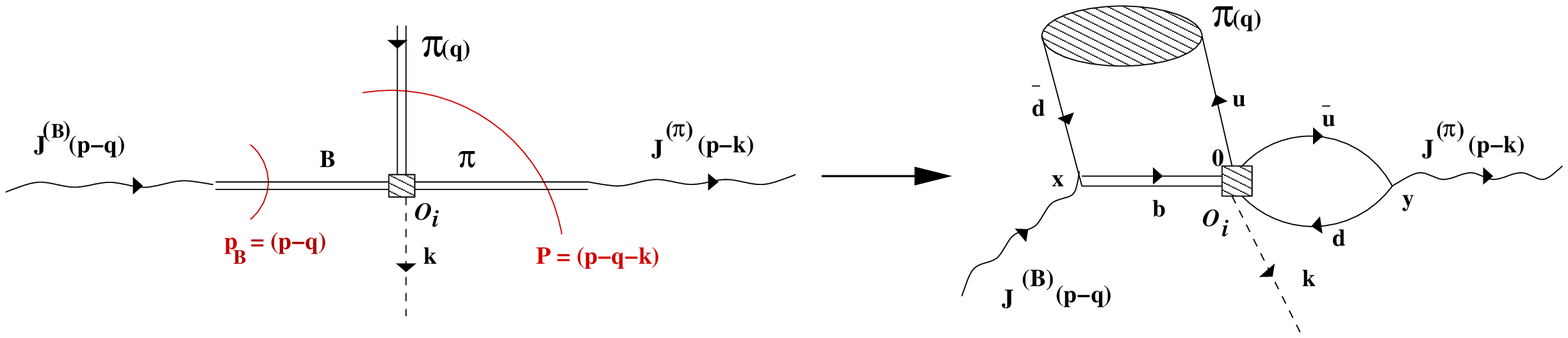,height=1.5in}
\end{center}
\caption{
Sum rule calculation of the $B \rightarrow \pi\pi$ decay. 
The shaded oval region denotes nonperturbative input, the pion
distribution amplitude. The other pion and the $B$ meson 
are represented by the currents $J^{(\pi)}(p-k)$ and $J^{(B)}(p-q)$, 
respectively. The square represents the four-quark operators 
$O_i$.}
\label{defP}
\end{figure}

LCSR has been applied to the $B \rightarrow \pi\pi$ decays \cite{Khodja}
and the $B \rightarrow \pi\pi$ decays \cite{LWH} recently. The 
fundamental concept, such as the quark-hadron
duality, has been briefly explained in Sec.~\ref{sec:sum}.
Start with the correlation function,
\begin{equation}
F_{\nu}(p,q,k) = \int d^4 x e^{-i(p-q)\cdot x} 
\int d^4 y e^{i(p-k)\cdot y} \langle 0 |
T [ J_{\nu 5}^{(\pi)}(y) O(0) J_5^{(B)}(x) ] |\pi(q)  
\rangle \,,
\label{corr}
\end{equation}
where the external pion state has been specified,
the two interpolating currents,
\be
J_{\nu 5}^{(\pi)}= \overline{u} \gamma_{\nu} \gamma_5 d\;,\;\;\;\;
J_5^{(B)} = m_b \overline{b} i\gamma_{5}  d\;,
\ee
are for the pion and for the $B$ meson, respectively, and
the relevant operators are
\be
O_1 = (\overline{d}\Gamma_{\mu}u)(\overline{u}\Gamma^{\mu}b)\;,
\;\;\;\;
\tilde{O}_1 = \left(\overline{d}\Gamma_{\mu}\frac{\lambda^a}{2} u\right)
\left(\overline{u}\Gamma^{\mu} \frac{\lambda^a}{2} b\right)\;.
\ee
The configuration is illustrated in Fig.~\ref{defP} with an unphysical 
momentum $k$ coming out of the weak vertex. This momentum
was introduced to prevent the $B$ meson four-momenta from being the same
before ($p_B =p-q$) and after ($p_B=P$) the decay. Then
the continuum of light states will not enter the dispersion relation 
of the $B$ meson channel. 

Take $p^2 = k^2= q^2= 0$, and consider the region of large spacelike 
momenta,
\begin{equation}
|(p-k)^2| \sim |(p-q)^2| \sim |P^2| \gg \Lambda_{\rm QCD}^2 \, , 
\label{kin}
\end{equation}
in which the correlation function is explicitly calculable by means of
OPE. The decomposition of the correlation function in Eq.~(\ref{corr})
in the independent momentum structures contains four
invariant amplitudes,
\be
F_\nu^{(O)}= 
(p-k)_\nu F^{(O)} 
+ q_\nu \widetilde{F}_1 ^{(O)} 
+ k_\nu\widetilde{F}_2^{(O)} 
+ \epsilon_{\nu\beta\lambda\rho}q^\beta p^\lambda k^\rho
\widetilde{F}_3^{(O)}\,,
\label{decompos}
\ee
for the operators $O=O_1,\widetilde{O}_1$,
where only the amplitude $F^{(O)}$ is relevant. 
The procedure to derive a double dispersion relation is as follows
\cite{Melic02}. One first makes a dispersion 
relation in a pion channel of momentum $(p-k)^2$ and applies the 
quark-hadron duality for this channel. Thereafter, to extract the physical 
$B$ meson state, one performs an analytical continuation of the invariant 
mass $P^2$ to its positive value, $P^2 = m_B^2$. This procedure is 
analogous to the one in the 
transition from spacelike to timelike form factors. Finally, a 
dispersion relation in the $B$ meson channel of 
momentum $(p-q)^2$ is derived, together with the application of the 
quark-hadron duality \cite{Khodja}.


At the diagrammatical level, there are four topologically 
different contributions to  the correlation function in Eq.~(\ref{corr}),
corresponding to four possible
combinations of $\bar{u}$ and $d$ fields in the 
pion distribution amplitude 
$\langle 0 \mid \bar{u}_\alpha(z_1)d_\beta(z_2)\mid \pi^-\rangle$,
with $z_{1}=0$ or $y$, 
and $z_2=x$ or $y$, $ \alpha, \beta$ being the spinor indices. 
Drawing the quark diagrams, one finds that 
each contribution yields a $B\to \pi\pi$ 
matrix element with a certain quark topology: 
emission ($z_1=0$, $z_2=x$), 
annihilation ($z_1=0$, $z_2=y$), penguin ($z_1=y$, $z_2=x$)
and penguin annihilation ($z_1=z_2=y$). So far, only the emission
topology $F_{\nu E}^{(O)}$ has been calculated \cite{Khodja}.

\begin{figure}[t!]
\begin{center}
\epsfig{file=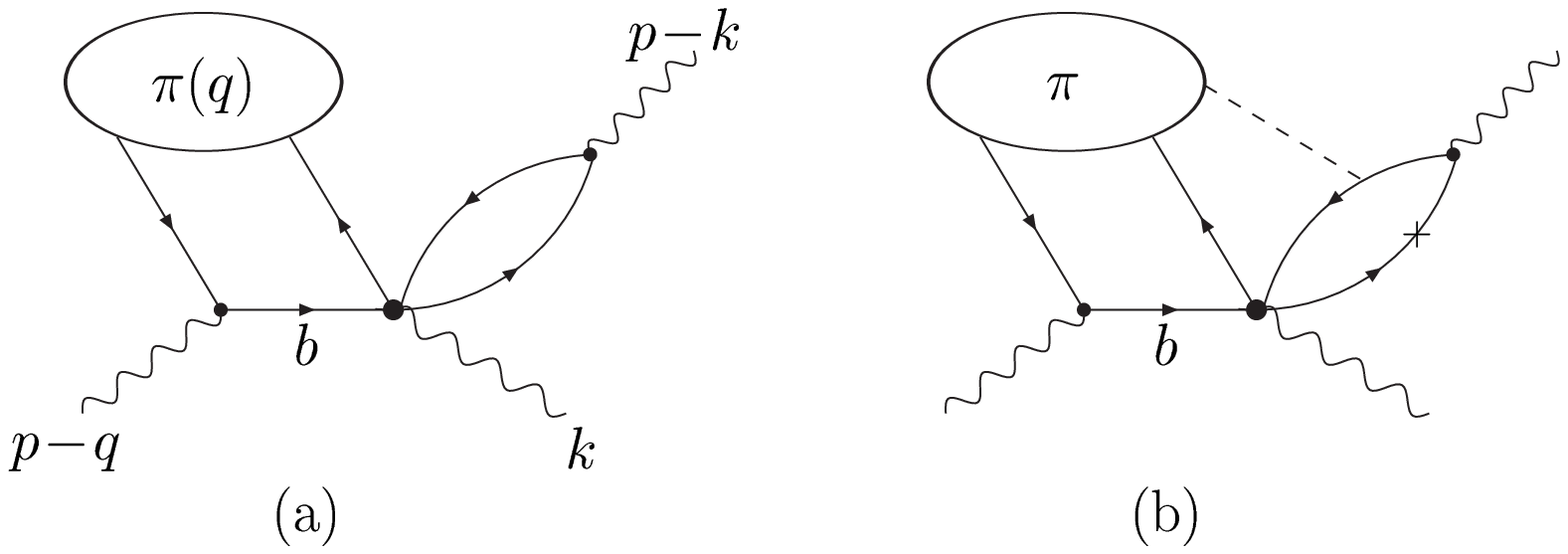,height=1.5in}
\end{center}
\caption{Diagrams corresponding (a) to the leading-order
of the correlation function in Eq.~(\ref{corr}) for $O=O_1$;
(b) to the higher-twist soft-gluon nonfactorizable contribution
for $O=\widetilde{O}_1$. 
Solid, dashed and wavy lines represent quarks, gluons, and 
external momenta, respectively. Thick points denote the 
weak interaction vertices, and ovals the pion distribution amplitudes. 
The cross represents another attachment of the 
gluon.}
\label{pp1}
\end{figure}

For the matrix elements of $O_1$,
the factorizable diagrams are those, in which the quarks 
of the heavy-light currents do not interact with the quarks of the 
light-quark currents. A typical example is shown in Fig.~\ref{pp1}(a).
To calculate the factorizable contribution, 
one inserts an intermediate vacuum state between the weak currents 
of the operator $O_1$. Equation~(\ref{corr}) is then converted 
into a product of two disconnected 
two-point correlation functions,
\bea
F^{(O_1)}_{\nu E} (p,q,k)=
\Bigg(i\!\int d^4y\; e^{i(p-k)\cdot y}
\langle\, 0 \mid T[j^{(\pi)}_{\nu 5}(y)\bar{d}(0)
\gamma_\mu\gamma_5 u(0) ]\mid 0\rangle
\Bigg)
\nonumber\\
\times \Bigg(i\!\int d^4x\; e^{i(p-q)\cdot x} 
\langle\, 0 \mid T[\bar{u}(0)\gamma^\mu b(0)
j^{(B)}_5(x)]\mid \pi^-(q)\rangle
\Bigg )\,.
\label{corrE}
\eea
The analysis then reduces to that of the $B\to\pi$ transition 
form factors.

\begin{figure}[t!]
\begin{center}
\epsfig{file=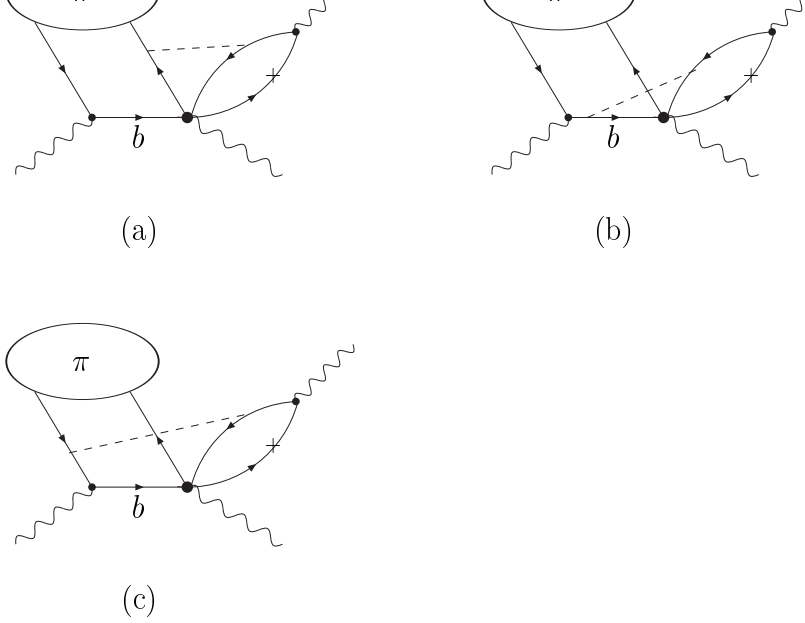,height=3.5in}
\end{center}
\caption{Diagrams corresponding to the 
$O(\alpha_s)$ nonfactorizable contributions 
to the correlation function in Eq.~(\ref{corr})
for $O=\widetilde{O}_1$.}
\label{pp2}
\end{figure}

For the operator $O_1$, nonfactorizable corrections to Eq.~(\ref{corr}), 
appearing at a two-gluon level, are negligible. In the case of 
$\widetilde{O}_1$, nonfactorizable effects start at the one-gluon 
level. The relevant correlation
function $F_\nu^{(\widetilde{O}_1)}$ receives contributions of
hard gluon exchanges, whose $O(\alpha_s)$ examples are shown
in Fig.~\ref{pp2}. These two-loop diagrams, not yet calculated 
because of their complexity, are very important: they give the scale
dependence of the matrix element, which partially compensates the 
scale dependence of the Wilson coefficients $C_{1,2}$ in the 
effective weak Hamiltonian. Moreover, the analytic continuation of these 
two-loop contributions in $P^2$ generates an imaginary part, 
which is essential for predicting the CP asymmetries in the
$B\to\pi\pi$ decays. The diagrams in Figs.~\ref{pp2}(a) and \ref{pp2}(b)
correspond to the corrections to the weak decay vertex in the literature.
Figure~\ref{pp2}(c) corresponds to the hard spectator contribution.  
However, as explained before, the soft and perturbative contributions
in the QCDF, PQCD and LCSR approaches all have different definitions.

There is another type of nonfactorizable effects, which comes from
the diagram with on-shell gluons being emitted from the quarks of the 
pion current and absorbed into the pion 
distribution amplitude, as shown in Fig.~\ref{pp1}(b). 
In terms of the light-cone expansion these contributions are of  
higher twist, starting from twist 3. It has been argued that
the higher-twist nonfactorizable effects 
are suppressed by a power of $1/m_B$ compared to the 
twist-2 factorizable amplitude \cite{Khodja}.
To quantify the magnitude of the nonfactorizable  
effect from the twist-3 pion distribution amplitude, 
the ratio has been introduced, 
\be
\frac{\lambda_{E}(\bar{B}^0_d\to\pi^+\pi^-)}{m_B}
\equiv \frac{A^{(\widetilde{O}_1)}_{E}(
\bar{B}^0_d\to\pi^+\pi^-)}{A^{(O_1)}_{E}( \bar{B}^0_d\to\pi^+\pi^-)}\;.
\label{lambdaE}
\ee
$\lambda_{E}$ was estimated to be
\be 
\lambda_{E}(\bar{B}^0_d\to\pi^+\pi^-)= 0.05 \div 0.15 ~\mbox{GeV}\,,
\label{lambdaEnum}
\ee
indicating that
nonfactorizable soft corrections from the twsit-3 pion distribution 
amplitude appeared to be numerically small ($\sim 1\%$). 

Recently, a piece of higher-order contribution to the
$B\to\pi\pi$ decays from the chromomagnetic dipole operator $O_{8g}$
(gluonic penguin) has been evaluated in LCSR \cite{KMU}. 
Similarly, consider the ratio
\be
r^{(O_{8g})}( \bar{B}^0_d \to \pi^+ \pi^-)= 
\frac{A^{(O_{8g})}( \bar{B}^0_d \to \pi^+ \pi^-)}
{A^{(O_1)}_E( \bar{B}^0_d \to \pi^+ \pi^-)}\,,
\label{ratio}
\ee
which determines (up to the known Wilson coefficient $C_{8g}$) 
the gluonic-penguin correction to the factorizable $B\to \pi\pi$ decay
amplitude. The result
\be
r^{(O_{8g})}(\bar{B}^0_d\to \pi^+\pi^-)= 0.035 \pm 0.015\;,
\label{number}
\ee
is of the same order of magnitude as $\lambda_{E}$ in
Eq.~(\ref{lambdaEnum}).
Though the impact of gluonic penguins on the $\bar{B}^0_d\to \pi^+\pi^-$ 
mode is very small, it might be noticeable for the
$\bar{B}^0_d\to \pi^0\pi^0$ mode \cite{KMU}.

\section{Charmed Decays}

\subsection{\it Transition Form Factors \label{sec:tra}}

$B$ and $D$ meson hadronic matrix elements play a crucial role in 
the determination of the CKM matrix elements and in overconstraining the 
unitarity triangle of the Standard Model. 
The $B\to D^{(*)}$ transitions are defined by the matrix elements,
\begin{eqnarray}
\langle D (P_2)|{\bar b}(0)\gamma_\mu c(0)|B(P_1)\rangle
&=&\sqrt{m_Bm_D}\left[\xi_+(\eta)(v_1+v_2)_\mu+
\xi_-(\eta)(v_1-v_2)_\mu\right]\;,
\nonumber\\
\langle D^*(P_2,\epsilon)|{\bar b}(0)\gamma_\mu\gamma_5 c(0)|B(P_1)
\rangle&=&\sqrt{m_Bm_{D^*}}
\left[{\xi_{A1}}(\eta)(\eta+1)\epsilon^*_\mu-{\xi_{A2}}(\eta)
\epsilon^*\cdot v_1v_{1\mu}\right.
\nn\\
&&\left.-{\xi_{A3}}(\eta)\epsilon^*\cdot v_1 v_{2\mu}\right]\;,
\nonumber\\
\langle D^*(P_2,\epsilon)|{\bar b}(0)\gamma_\mu c(0)|B(P_1)\rangle&=&
i\sqrt{m_Bm_{D^*}}\xi_V(\eta)\epsilon^{\mu\nu\alpha\beta}
\epsilon^*_\nu v_{2\alpha}v_{1\beta}\;,
\label{ff}
\end{eqnarray}
with the velocity transfer $\eta=v_1\cdot v_2$, $v_1=P_1/m_B$ and 
$v_2=P_2/m_{D^{(*)}}$. 
The form factors $\xi_+$, $\xi_-$, $\xi_{A_1}$, $\xi_{A_2}$, $\xi_{A_3}$,
and $\xi_V$ satisfy the relations in the heavy quark limit,
\begin{equation}
\xi_+=\xi_V=\xi_{A_1}=\xi_{A_3}=\xi,\;\;\;\;  \xi_-=\xi_{A_2}=0\;,
\label{iwr}
\end{equation}
where $\xi$ is the Isgur-Wise (IW) function \cite{IW}.

The $O(1/m_c)$ corrections introduce four new functions \cite{Luke}, and
the $O(1/m_b)$ corrections do not \cite{NRSX}. Taking the matrix elements
of the vector current as an example, one has
\begin{eqnarray}
\xi_+ &=& \xi \left[ 1 + \left(\frac1{m_c}+\frac1{m_b}\right)\rho_1 \right],
\nonumber\\
\xi_- &=& \xi \left(\frac1{m_c}-\frac1{m_b}\right)
\left(-\frac{\varepsilon}{2}+\rho_4\right),
\label{xi}\\
\xi_V &=& \xi \left[ 1
+ \left(\frac1{m_c}+\frac1{m_b}\right)\frac{\varepsilon}{2}
+ \frac{\rho_2}{m_c} + \frac{\rho_1-\rho_4}{m_b} \right],
\nonumber
\end{eqnarray}
with the mass difference $\varepsilon=m_B-m_b=m_D-m_c=m_{D^*}-m_c$, all 
of which are equal up to $1/m$ corrections.
The Luke's theorem \cite{NRSX,Luke,proof} 
leads to the values of the subleading form factors at zero recoil,
$\rho_1(1)=0$, $\rho_2(1)=0$.
It is essential to examine these $1/m$ corrections, especially those
from $1/m_c$. If they are modest, HQET will be self-consistent and
useful.

The form factors in Eq.~(\ref{ff}) have been calculated using
QCD sum rules at finite $m_b$, $m_c$ in \cite{BG1,OS}.
HQET sum rules for the IW function were derived in \cite{R,N},
which coincide with the limit $m_{b,c}\to\infty$ of the QCD sum-rule
calculation. The above results for finite and infinite masses were 
compared in \cite{Ball}. There are two alternative ways to obtain sum 
rules for the subleading form factors $\rho_i$. One can either expand the 
known finite-mass QCD results to the first order in $1/m$, or start 
from the HQET Lagrangian and currents in the first power in $1/m$.
It has been noticed that the slope of $\xi(\eta)$ near $\eta=1$ depends on 
how to model the contribution from the continuum state on the hadron 
side \cite{R,N,BG99}.

\begin{figure}[t!]
\begin{center}
\epsfig{file=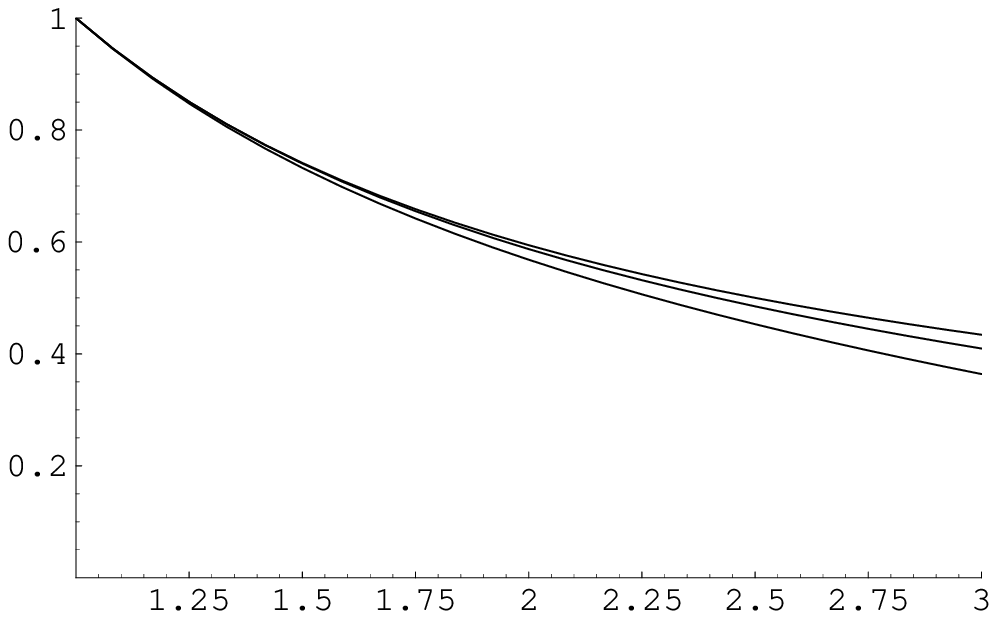, height=2.0in}
\end{center}
\caption{Isgur-Wise form factor $\xi(\eta)$ for different Borel 
parameters. }
\label{F2}
\end{figure}

\begin{figure}[t!]
\begin{center}
\epsfig{file=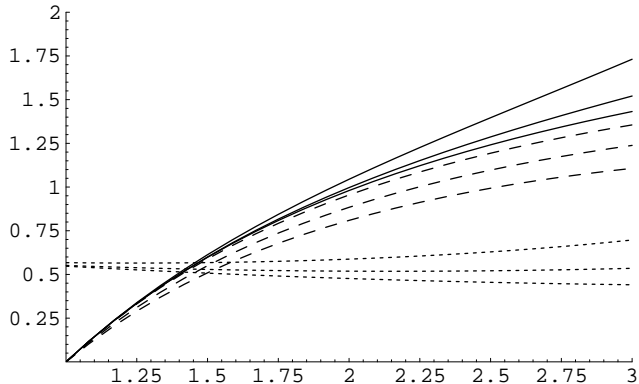, height=2.0in}
\end{center}
\caption{$1/m$ correction form factors $\rho_1(\eta)$ 
(solid curves), $\rho_2(\eta)$ (dashed curves),
and $\rho_4(\eta)$ (dashed-dotted curves) in the unit of
100 MeV.}
\label{F3}
\end{figure}

The sum rule for $\xi(\eta)$ is well known \cite{R,N}, whose results
for several Borel parameters are shown in Fig.~\ref{F2}. 
The predictions for the $1/m$-correction form factors $\rho_{1,2,4}(\eta)$ 
are presented in Fig.~\ref{F3} \cite{BG99}. The curves of
$\rho_{1,2}(\eta)$ are pinned at the origin by the Luke's theorem,
and increase for $\eta>1$. $\rho_4(\eta)$ is small and nearly constant.
Variation of the results with the Borel parameter can be employed to 
estimate their accuracy. Hence, it is expected that the $O(1/m)$ 
corrections to the $B\to D^{(*)}$ transition form factors are at 
most 10\%. The above analysis provides the behavior of the
$B\to D$ transition form factors with the velocity transfer, but
the values at zero recoil, which are specially important for the 
extraction of the CKM matrix element $|V_{cb}|$, remain intact.

The IW function near the zero recoil is parametrized as 
\begin{eqnarray}
\xi(\eta)=F_{B\to D^{(*)}}(1)[1-{\hat\rho}^2_{D^{(*)}}(\eta-1) 
+ {\hat c}_{D^{(*)}}(\eta-1)^2
+O((\eta-1)^3)]\;.
\end{eqnarray}
Including the $1/m$ and radiative corrections, the normalizations
$F_{B\to D}(1)=0.98\pm 0.07$ and $F_{B\to D^*}(1)=0.91\pm 0.03$ have been
derived in \cite{CLN}.  These normalizations can also be studied in 
lattice QCD, and the results are 
\bea
F_{B\rightarrow D}(1)& =& 
1.058\pm 0.016\pm 0.003^{+0.014}_{-0.005}\;\;\cite{SH1}\;,
\nn\\
F_{B\rightarrow D^\ast}(1)& = &
0.9130^{+0.0238}_{-0.0173}{}^{ +0.0171}_{ -0.0302}\;\;\cite{SH2}\;.
\label{niw}
\eea
The errors come from fitting, matching lattice gauge theory and HQET 
to QCD, lattice spacing dependence, light quark mass dependence and the 
quenched approximation. The above values agree with those from other 
methods, such as non-relativistic quark models \cite{b2Dstar:neubert} 
and a zero-recoil sum rule~\cite{B2Dstar:BSU,B2Dstar:SUV}. 
For a QCD sum-rule analysis of the subleading form factors  
involved in the semileptonic decays $B\to D_1 l\bar\nu$ and 
$B\to D_2^* l\bar\nu$, refer to \cite{HD01}.

The LFQCD formalism has been applied to the $B\to D^*$ form factors,
whose results can be found in \cite{BCJ}.

The PQCD formalism for $B\to D^{(*)}$ transitions has been developed
recently \cite{TLS2}, which applies under the hierachy,
\begin{eqnarray}
m_B\gg m_{D^{(*)}}\gg \Lambda\;,
\label{ll}
\end{eqnarray}
with $m_{D^{(*)}}$ being the $D^{(*)}$ meson mass. The relation
$m_B\gg m_{D^{(*)}}$ justifies perturbative evaluation of the
$B\to D^{(*)}$ form factors at large recoil and the definition of
light-cone $D^{(*)}$ meson wave functions. The relation  
$m_{D^{(*)}}\gg\Lambda$ justifies the power expansion in the 
parameter $\Lambda/m_{D^{(*)}}$. Equation~(\ref{ll}),
corresponding to the heavy quark and large recoil limits,
may not be realistic. Nevertheless, an attempt to construct a 
self-consistent theory under this hierachy is worthwhile.  

It has been argued that the wave function for an 
energetic $D^{(*)}$ meson absorbs collinear dynamics, but with the $c$ 
quark line being eikonalized. That is, its definition is a mixture of 
those for a $B$ meson dominated by soft dynamics and for a pion dominated 
by collinear dynamics. The behavior of the heavy meson wave 
functions under Eq.~(\ref{ll}) has been examined. For 
$\Lambda/m_B$, $\Lambda/m_{D^{(*)}}\ll 1$, only a single $B$ meson
wave function $\phi_+(x)$ and a single $D^{(*)}$ meson wave function 
$\phi_{D^{(*)}}(x)$ are involved in the $B\to D^{(*)}$ form factors, $x$
being the momentum fraction associated with the light spectator quark. 
Equations of motion for the relevant nonlocal matrix elements imply that 
$\phi_+(x)$ and $\phi_{D^{(*)}}(x)$ exhibit maxima at 
$x\sim \Lambda/m_B$ and at $x\sim \Lambda/m_{D^{(*)}}$, respectively. 
To proceed a numerical analysis, the simple model \cite{TLS2},
\begin{eqnarray}
\phi_{D^{(*)}}(x)=\frac{3}{\sqrt{2N_c}}f_{D^{(*)}} 
x(1-x)[1+C_{D^{(*)}}(1-2x)]\;,
\label{phid}
\end{eqnarray}
has been adopted. The free shape parameter $C_{D^{(*)}}$ is expected to 
take a value, such that $\phi_{D^{(*)}}$ has a maximum at 
$x\sim \Lambda/m_{D^{(*)}}$.
The intrinsic $b$ dependence of the $D$ meson wave 
function was not included.

\begin{figure}[t!]
\begin{center}
\epsfig{file=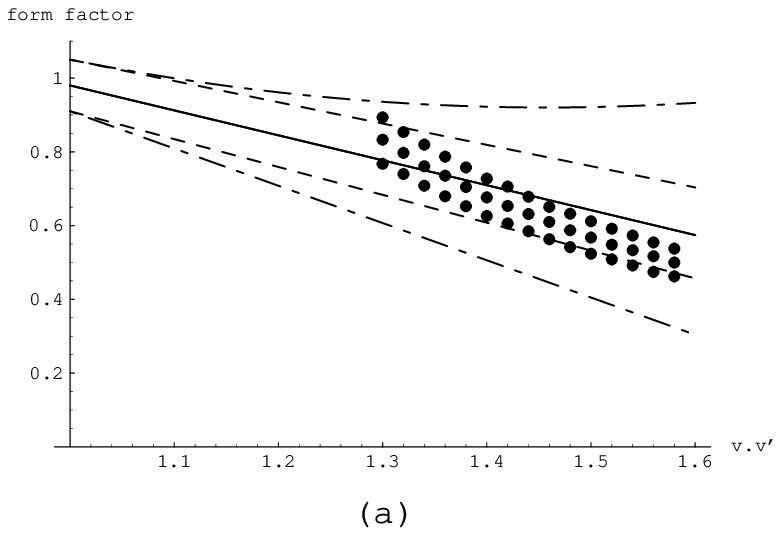, height=3.0in}
\epsfig{file=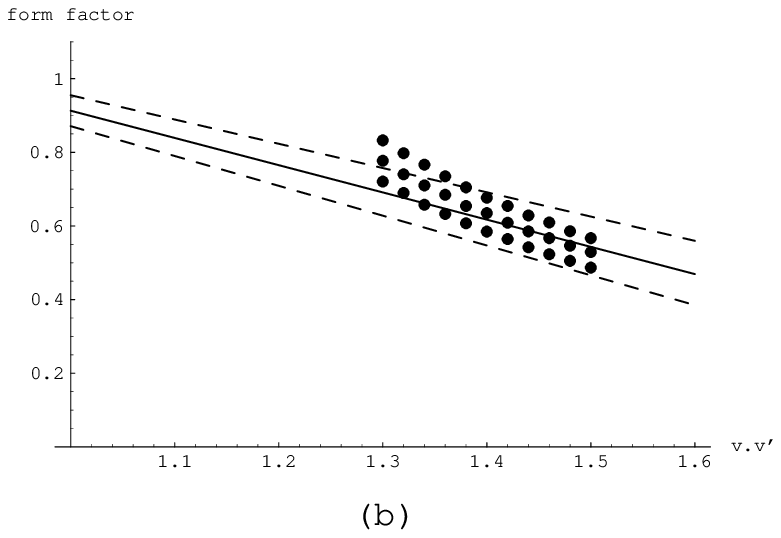, height=3.0in}
\end{center}
\caption{(a) [(b)] $\xi$ as a function of the velocity transfer from the
$B\to D^{(*)}l\nu$ decay.
The solid lines represent the central values, the dashed
(dot-dashed) lines give the bounds from the linear (quadratic) fits.
The circles correspond to $C_{D^{(*)}}=0.5$, 0.7, and 0.9 from
bottom to top.}
\label{dlv}
\end{figure}

The free parameters $C_{D^{(*)}}$ can be fixed by fitting
the leading-power PQCD predictions to the measured decay spectra
\cite{Beld,Belds,CLEOs}. With the normalization given in Eq.~(\ref{niw}),
the linear and quadratic fits to the data give \cite{Beld,Belds}
\begin{eqnarray}
& &{\hat\rho}^2_{D}=0.69\pm 0.14\;,\;\;\;\;{\hat c}_{D}=0\;,\;\;\;\;
{\hat\rho}^2_{D^*}=0.81\pm 0.12\;,\;\;\;\;{\hat c}_{D^*}=0\;,
\nonumber\\
& &{\hat\rho}^2_{D}=0.69^{+0.42}_{-0.15}\;,\;\;\;\;
{\hat c}_{D}=0.00^{+0.59}_{-0.00}\;,
\end{eqnarray}
respectively. Choosing the decay constants $f_B=190$ MeV and 
$f_D=f_{D^*}=240$ MeV, it has been found that
$C_D\sim C_{D^*}=0.7\pm 0.2$ leads to an excellent agreement with the 
data as exhibited in Fig.~\ref{dlv}. For these values, the corresponding 
$D^{(*)}$ meson
distribution amplitude shows a maximum at $x\sim 0.36$, consistent
with the expectation. The rough equality of $C_D$ and $C_{D^*}$ hints
that the heavy quark symmetry holds well. It has been shown that the 
leading PQCD factorization formulas $B\to D^{(*)}$ transitions indeed
respect the heavy quark symmetry.

\subsection{\it $B\to D\pi$ \label{sec:two}}

The recent measurement of the $\bar B^0\to D^0\pi^0$ branching ratio
reveals interesting QCD dynamics. In naive (or generalized)
FA, nonfactorizable effects are parameterized through the
phenomenological coefficients $a_i^{\rm eff}$ in Eq.~(\ref{aeff}), which 
depend on the color and Dirac structure of the operators, but otherwise 
are postulated to be universal \cite{BSW,NRSX,Dean93}.  
Class-1 and class-2 decay topologies refer to the cases, where a charged 
and a neutral final-state meson are produced from the four-quark 
operators, respectively. For instance, the decay $\bar B^0\to D^+\pi^-$ 
is a class-1 process, in which the charged pion is generated at the weak 
vertex, whereas $\bar B^0\to D^0\pi^0$ is a class-2 process, in which the 
$D^0$ meson is directly produced. The corresponding amplitudes are then 
expressed as 
\begin{eqnarray}\label{ampl}
   A(\bar B^0\to D^+\pi^-) &=& i\,\frac{G_F}{\sqrt 2}\,
    V_{cb} V_{ud}^*\,(m_B^2-m_D^2)\,f_\pi\,F_0^{B D}(m_\pi^2)\,
    a_1(D\pi) \,, \nonumber\\
   \sqrt 2\,A(\bar B^0\to D^0\pi^0) &=& i\,\frac{G_F}{\sqrt 2}\,
    V_{cb} V_{ud}^*\,(m_B^2-m_\pi^2)\,f_D\,F_0^{B\pi}(m_D^2)\,
    a_2(D\pi) \,, 
\end{eqnarray}
where the coefficients have the orders of magnitude $a_1(D\pi)\sim O(1)$ 
and $a_2(D\pi)\sim O(1/N_c)$. The isospin symmetry then implies 
\begin{equation}
   A(B^-\to D^0\pi^-) = A(\bar B^0\to D^+\pi^-)
   + \sqrt 2\,A(\bar B^0\to D^0\pi^0) \,.
\end{equation} 

\begin{table}
\begin{center}
\begin{tabular}{|l| l| c| }
\hline
{Decay mode} &
 Belle \cite{BelleC} & CLEO \cite{CLEOC} \\ \hline
 $\bar B^0\to D^0\pi^0$ &  $3.1\pm0.4\pm0.5$ &
 $2.74^{+0.36}_{-0.32}\pm 0.55$ \\
 $\bar B^0\to D^{*0}\pi^0$ &  $2.7^{+0.8+0.5}_{-0.7-0.6}$ &
 $2.20^{+0.59}_{-0.52}\pm0.79$ \\
 $\bar B^0\to D^0\eta$ & $1.4^{+0.5}_{-0.4}\pm 0.3$ & \\
 $\bar B^0\to D^{*0}\eta$ & $2.0^{+0.9}_{-0.8}\pm0.4$ & \\
 $\bar B^0\to D^0\omega$ &  $1.8\pm0.5^{+0.4}_{-0.3}$ & \\
 $\bar B^0\to D^{*0}\omega$ &  $3.1^{+1.3}_{-1.1}\pm0.8$ &
 \\\hline
\end{tabular}
\caption{Data (in units of $10^{-4}$) of the
$\bar B^0\to D^{(*)0}M^0$ $(M=\pi,\eta,\omega)$ branching ratios.}
\label{dpda}
\end{center}
\end{table}

Within errors, the class-1 decays $\bar B^0\to D^{(*)+} M^-$ with 
$M=\pi,\rho,a_1,D_s,D_s^*$ can be described using a universal value 
$|a_1|\approx 1.1\pm 0.1$, whereas the class-2 decays 
$\bar B\to\bar K^{(*)} M$ with $M=J/\psi,\psi(2S)$ suggest a nearly 
universal value $|a_2|\approx 0.2$--0.3 \cite{a1a2}. The wide range of 
$|a_2|$ is due to the uncertainty in the $B\to K^{(*)}$ form factors. 
Moreover, the class-3 decays $B^-\to D^{(*)0} M^-$ with $M=\pi,\rho$, 
which are sensitive to the interference of the two decay topologies, 
could be explained by a real, positive ratio $a_2/a_1\approx 0.2$--0.3, 
which seemed to agree with the determinations of $|a_1|$ and $|a_2|$ from 
other modes. The observed branching ratios of the color-suppressed modes
are listed in Table~\ref{dpda} \cite{BelleC,CLEOC}.
The parameter $a_2$ extracted from Table~\ref{dpda} falls into the
range of $|a_2(D\pi)|\sim 0.35-0.60$ and $|a_2(D^*\pi)|\sim
0.25-0.50$ \cite{C02}. The phases of $a_2/a_1$ are $59^\circ$ for the
$D\pi$ system and $63^\circ$ for $D^*\pi$ \cite{C02}, implying sizeable 
relative strong-interaction phases between class-1 and class-2 
$\bar B\to D^{(*)}\pi$ decay amplitudes \cite{NPe,CR02}. These results 
can be regarded as a failure of naive FA: the parameters $a_2$ in 
different types of decays such as $\bar B\to D^{(*)}\,\pi$ and 
$\bar B\to\bar K^{(*)} J/\psi$ differ by almost a factor 2 in magnitude, 
indicating a strong nonuniversality of nonfactorizable effects.

If the $c$ quark is treated as a massive quark,  
QCDF does not apply to the class-2 
decays $\bar B^0\to D^{(*)0} M^0$, because of the uncancelled
end-point singularities. Therefore, the magnitude and phase of the 
$a_2(D^{(*)}M)$ parameters are not calculable in QCDF.
However, these decays are calculable in the PQCD approach
based on $k_T$ factorization theorem, in which the end-point
singularity does not exist. From the power counting rules 
proposed in Eq.~(\ref{ll}) \cite{TLS2}, it has been shown that the 
relative importance of the different topologies
of diagrams for the $B\to D\pi$ decays is roughly
\begin{eqnarray}
{\rm emission} : {\rm nonfactorizable} 
\sim 1 : \frac{m_D}{m_B}\;,
\end{eqnarray}
which approaches $1:\Lambda/m_B$ as the $D$ meson mass $m_D$ reduces 
to the pion mass of $O(\Lambda)$. Since the factorizable and
nonfactorizable diagrams contribute to the parameters $a_1$ and $a_2$ in
PQCD, respectively, the ratio $|a_2|/a_1\sim 0.5$ is obtained.
Moreover, the imaginary nonfactorizable amplitudes determine the
relative phase of the factorizable and nonfactorizable contributions,
which is about $-57^o$. 

To obtain the above results, the $D^{(*)}$ meson wave function determined
from the semileptonic $B\to D^{(*)}l\nu$ decay in the previous subsection
has been adopted. Therefore, there is no free parameter in the above
calculation. The PQCD predictions for the $B\to D\pi$ branching ratios 
\cite{KLL},
\begin{eqnarray}
& &B(B^-\to D^{0}\pi^-)\sim 5.5\times 10^{-3}\;,
\nonumber\\
& &B({\bar B}^0\to D^{+}\pi^-)\sim 2.8\times 10^{-3}\;,
\nonumber\\
& &B({\bar B}^0\to D^{0}\pi^0)\sim 2.6\times 10^{-4}\;,
\end{eqnarray}
are consistent with the experimental data \cite{BelleC,CLEOC,Bab}. 
The conclusion that the $B\to D^0\pi^0$ data hint large final-state 
interaction was drawn from the analyses based on FA 
\cite{C02,NPe,X,CHY}. Hence, it is more reasonable to conclude that 
experimental data just hint large strong phases, but do not tell which 
mechanism generates these phases \cite{JPL}. From the viewpoint of
PQCD, these strong phases are of short distance, and produced from
hard kernels. For the application of the PQCD approach to other
charmed nonleptonic decays, refer to \cite{LU003}.

\subsection{\it Decays into Charmonia \label{sec:cha}}

The $B$ meson decays into charmonia, such as $B \to J/\psi K$, 
are theoretically notorious. As defined above, this mode 
belongs to the color-suppressed class-2 decays, in which one expects 
large nonfactorizable contributions. Experimentally, the measurement
of the branching ratios \cite{exp},
\begin{eqnarray}
B(B^+ \rightarrow J/\psi K^+) &=& (10.1 \pm 0.3 \pm 0.5) 
\times 10^{-4} \, , 
\nonumber \\
B(B^0 \rightarrow J/\psi K^0) &=& (8.3 \pm 0.4 \pm 0.5)
\times 10^{-4} \, , 
\label{eq:BRexp}
\end{eqnarray}
provides the information of the coefficient $a_2(J/\psi K)$,
being of order $0.20-0.30$ \cite{a1a2}. However,
it is very difficult to understand such a large $a_2(J/\psi K)$, and also 
those for other similar decays into charmonia:  the observed 
branching ratio differs from the naive FA prediction by at least a 
factor of 3. In this subsection I discuss the attempts made in the LCSR 
and QCDF approaches, and explain why they fail.

The QCDF method is usually not applicable, if the emitted meson is heavy. 
Take the $\bar B^0\to D^0\pi^0$ decay as an example. Since the
$D^0$ meson is not a compact object with small transverse extension, 
it will strongly interact with the $(B\pi)$ system, such that the
factorization breaks down. The parameter $a_2(\pi D)$ has been roughly 
estimated in \cite{BBNS2} by treating the charmed meson as a light meson, 
which is certainly a dubious approximation. Fortunately, QCDF
is applicable to the $B\to J/\psi K$ decay, because
the transverse size of $J/\psi$ becomes small in the heavy quark
limit. However, a recent study \cite{chay} indicates that the 
leading-twist (twist-2) contributions from QCDF are too small 
to explain the data.

\begin{figure}[t!]
\begin{center}
\epsfig{file=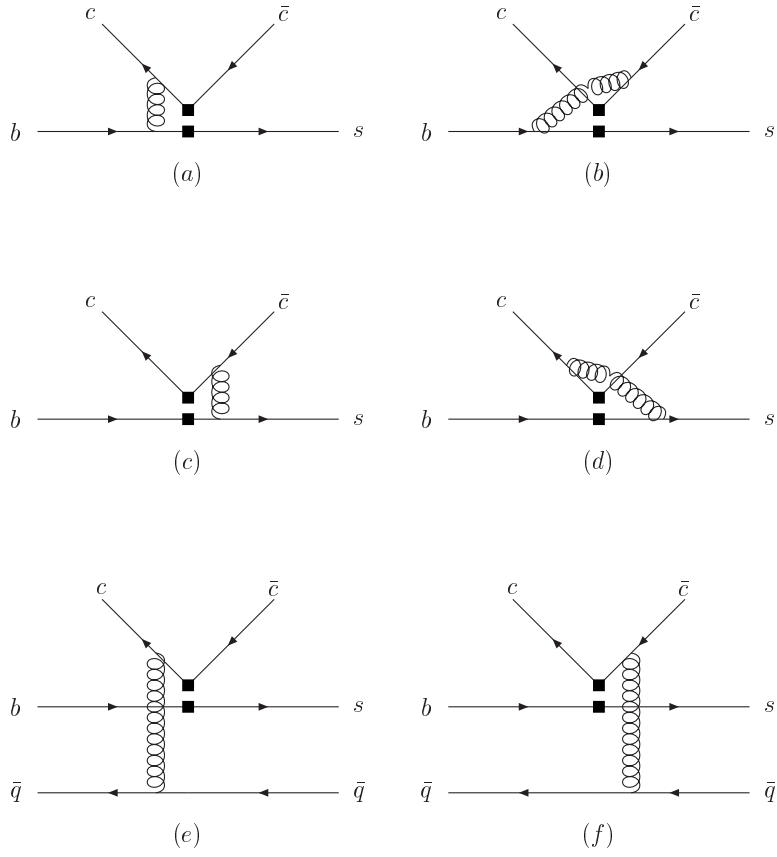, height=5.0in}
\end{center}
\caption{Feynman diagrams for nonfactorizable corrections to
$\overline{B} \rightarrow J/\psi K$.} 
\label{fvs}
\end{figure}

The authors of \cite{Cheng} then calculated twist-3 contributions from
the diagrams in Fig.~\ref{fvs}. The result for 
Figs.~\ref{fvs}(a)-\ref{fvs}(d) is free of the end-point singularity.
The contribution from Fig.~\ref{fvs}(e) and \ref{fvs}(f) is written,
up to twist 3, as
\be
f_{II}=f_{II}^{(2)}+f_{II}^{(3)}+\cdots,
\ee
where the superscript denotes the twist of the kaon distribution
amplitude. The expressions for $f_{II}^{(2)}$ and $f_{II}^{(3)}$ are
\bea
f^{(2)}_{II} &=& {4\pi^2\over N_c}\,{f_Kf_B\over
F_+^{BK}(m^2_{J/\psi})m^2_B}\,{1\over 1-r}\int^1_0 dx_1\,
{\phi_+(x_1)\over x_1}\int^1_0 dx_2 \,{\phi_{J/\psi}(x_2)\over
x_2}\int^1_0 dx_3\, {\phi_K(x_3)\over x_3}\;, 
\label{fII2'}\\
f^{(3)}_{II}& =&\left({2m_0\over m_B}\right)\,{4\pi^2\over
N_c}\,{f_K f_B\over F_+^{BK}(m^2_{J/\psi})m_B^2} \int^1_0
{dx_1\over x_1}\,\phi_+(x_1)\int^1_0 {dx_2\over
x_2}\,\phi_{J/\psi}(x_2)\int^1_0 {dx_3\over x_3^2}\,
{\phi_K^\sigma(x_3)\over 6(1-r)^3}\;, \label{fII3}
\eea
with the mass ratio $r\equiv m_{J/\psi}^2/m_B^2$ and 
the $B \rightarrow K$ transition form factor 
$F_+^{BK}(m_{J/\psi}^2)$ being evaluated at the $J/\psi$ meson mass. 
The twist-2 and one of the two-parton twist-3 distribution 
amplitudes are given, in their asymptotic form, by
\bea
\phi_K(x)=6x(1-x)\;,\;\;\;\;
\phi_K^\sigma(x) = 6x(1-x)\;,
\eea
respectively. It is observed that
$f^{(2)}_{II}$ is finite, because the potential logarithmic divergence
cancels between Fig.~\ref{fvs}(e) and \ref{fvs}(f).
$f^3_{II}$ is singular,
since only the potential linear divergence cancels, leaving
the logarithmic one.

\begin{figure}[t!]
\begin{center}
\epsfig{file=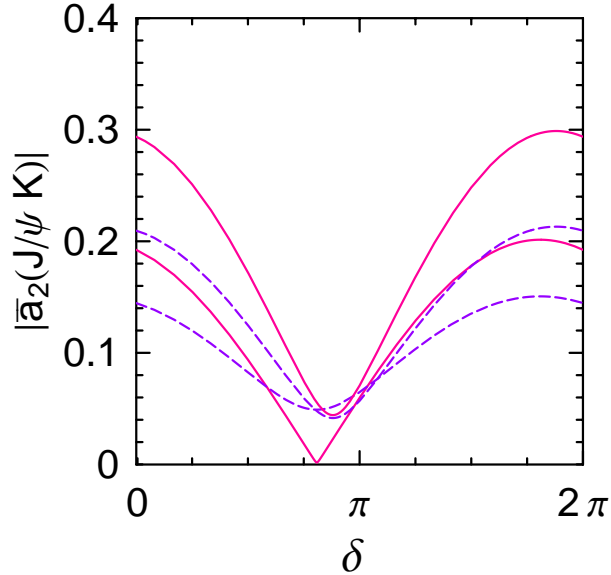, height=3.0in}
\end{center}
\caption{The coefficient $|\bar a_2(J/\psi K)|$ vs. the
    phase of the parameter $r$. Solid and dashed curves are for
    $|\rho|=6$ and 3, respectively.
    The upper and lower solid curves are for $r=m_{J/\psi}^2/m_b^2$ and
    $m_{J/\psi}^2/m_B^2$,
    respectively, and likewise for the dashed curves.} 
\label{jpsi}
\end{figure}

To estimate the twist-3 effect, the divergent integral has been 
parametrized as \cite{Cheng}
\be
X\equiv\int^1_0 {dx_3\over
x_3}=\ln\left({m_B\over\Lambda_{\rm QCD}}\right)+\rho, 
\label{logdiv}
\ee
as in Eq.~(\ref{rhoa}),
with $\rho=|\rho|\exp(i\delta)$ being a complex random number.
The variation of $|a_2(J/\psi K)|$ with the arbitrary parameter
$\rho$ is exhibited in Fig.~\ref{jpsi} \cite{Cheng}, which implies
that $|a_2(J/\psi K)|$ can fit the data, only when $|\rho|$ is almost
as large as 6. For such a huge subleading contribution, the 
self-consistency of the QCDF analysis is in doubt.

A worse situation has been observed: QCDF breaks down for the
$B\to \chi_{c1}K$ deacy even at twist 2 \cite{Chao2}. Similarly, the function $f_{II}$ is obtained 
by computing the two spectator corrections in Figs.~\ref{fvs}(e) 
and \ref{fvs}(f), whose contribution is given by
 \bea
  f_{II}
= \frac{4\pi^2}{N_c} \frac{f_K f_B}{F_+^{BK} (m_{\chi_{c1}}^2) m_B^2}
\frac{1}{1-r} \int_0^1 dx_1 \frac{\phi_+ (x_1)}{x_1} \int_0^1 dx_3
\frac{\phi_K (x_3)}{x_3} \int_0^1 dx_2
{\phi_{\chi_{c1}}(x_2)}\left[\frac{1}{x_2}+\frac{2r}{x_3(1-r)}\right]\;, 
\label{fIIc}
\eea
with the mass ratio $r=m_{\chi_{c1}}/m_B$. Obviously,
the integral over $x_3$ in Eq.~(\ref {fIIc}) gives
logarithmic divergence. Therefore, QCDF breaks down
even at leading twist. This is different from the $B \rightarrow
J/\psi K$ decay, which does not have logarithmic divergence at
leading twist \cite{chay,Cheng}. The reason is that the logarithmic
divergences arising from the contribution of the vector and tensor
currents are cancelled out in the $B \rightarrow J/\psi K$ decay,
while there is no such cancellation for the 
$B \rightarrow\chi_{c1} K$ decay. 

The logarithmically divergent integral has been parametrized 
in the same way as in Eq.~(\ref{logdiv}). In \cite{Chao2}, the 
parameter $X$ is chosen as $X\approx 2.4$ to make a rough 
estimate. For $\phi_{\chi_{c1}}(x) = \delta (x-1/2)$,
$B (\overline{B} \rightarrow \chi_{c1} K) = 0.16
\times 10^{-4}$ was obtained \cite{Chao2}. The measured branching ratio
\cite{babarc}
  \be
B(B^0 \rightarrow \chi_{c1} K^0) = (5.4 \pm 1.4)
\times 10^{-4},
  \ee
is about thirty times larger than the theoretical prediction.
Choosing $X\approx 2.4$, instead of around 6 \cite{Cheng},
the QCDF prediction for the $B\to J/\psi K$
branching ratio is also too small.

The end-point singularity becomes more serious in the
$B^+ \rightarrow \chi_{c0} K^+$ mode. In the previous calculations, 
the contribution of the four vertex diagrams in Fig.~\ref {fvs} is
infrared safe. However, for the $B \rightarrow \chi_{c0} K$ decay 
\cite{Chao2}, these four diagrams produce infrared divergences.
The QCDF predictions for the $B\to\eta_c^{(')}K$ branching ratios,
thought infrared safe, are also too small \cite{Chao1}.
The above analyses indicate that it is difficult to apply QCDF to
$B$ meson decays into charmonia.

I now turn to the LCSR approach. For the 
$B \rightarrow J/\psi K$ decay, the relevant operators are
\be
O_1 = (\overline{c} \Gamma_{\mu} c)(\overline{s} \Gamma^{\mu} b)\;,
\;\;\;\;
\tilde{O}_1 = \left(\overline{c} \Gamma_{\mu} 
\frac{\lambda_a}{2} c\right)
\left(\overline{s} \Gamma^{\mu} \frac{\lambda_a}{2} b\right)\;.
\ee 
In the factorization limit, the matrix element of 
$\tilde{O}_1$ vanishes, and the matrix element of the
operator $O_1$ can be factorized into
\begin{eqnarray}
\langle  J/\psi(p,\epsilon) K(q) | O_1 | B(p+q) \rangle &=& 
\langle J/\psi(p,\epsilon) | \overline{c} \Gamma_{\mu} c| 0 \rangle
\langle K(q)| \overline{s} \Gamma^{\mu} b | B(p+q) \rangle 
\nonumber \\
&=& 2 \epsilon \cdot q \, m_{J/\psi}  f_{J/\psi} 
F_+^{BK}(m_{J/\psi}^2) \, ,
\label{eq:fac}
\end{eqnarray}
with $f_{J/\psi}$ being the $J/\psi$ meson decay constant.
The $B \rightarrow K$ transition form factor 
$F_+^{BK}(m_{J/\psi}^2)= 0.55 \pm 0.05$ \cite{Melic02} was obtained
using LCSR in a way the same as for the $B \rightarrow \pi$ form factor. 
Evaluating the $B \rightarrow J/\psi K$ 
branching ratio with the next-to-leading-order Wilson coefficients, 
one arrives at \cite{Melic02}
\begin{equation}
B(B \rightarrow J/\psi K)^{fact} = 3.3 \times 10^{-4}\,.
\label{eq:BRnf}
\end{equation}
This value, representing a prediction from FA, is too small
compared to the experimental data in Eq.~(\ref{eq:BRexp}).

Similarly, the nonfactorizable contribution associated with the
operator ${\tilde O}_1$ starts at the one-gluon level, and that with
$O_1$ starts at the two-gluon level. One then considers only 
the matrix element of ${\tilde O}_1$, which is expressed as
\begin{equation}
\langle J/\psi K | {\tilde O}_1(\mu) | B \rangle = 
2 \epsilon \cdot q \,
m_{J/\psi} f_{J/\psi} \tilde{F}^{BK}_+(\mu^2)\;.
\end{equation}
To calculate the factor $\tilde{F}^{BK}_+$, one studies the
correlation function,
\begin{equation}
F_{\nu}(p,q,k) =
i^2 \int d^4 x e^{-i(p+q)\cdot x} \int d^4 y e^{i(p-k)\cdot y} 
\langle K(q) |
T [ J_{\nu}^{(J/\psi)}(y) O(0) J_5^{(B)}(x) ] | 0 \rangle\;,
\end{equation}
with the interpolating currents $J_{\nu}^{(J/\Psi)} =
\overline{c} \gamma_{\nu} c$
and $J_5^{(B)} = m_b \overline{b} i \gamma_5 u$, 
and follows the derivation 
for the $B \rightarrow \pi\pi$ decays in Sec.~\ref{sec:su2}.

The results are $\tilde{F}_+^{BK(3)}(\mu_b) = 0.003 - 0.0055$ from the
twist-3 kaon distribution amplitudes and
$\tilde{F}_+^{BK(4)}(\mu_b) = 0.006 - 0.0012$ from the
twist-4 kaon distribution amplitudes with the scale
$\mu_b  \simeq m_b/2$. 
The wide range for $\tilde{F}_+^{BK}$ is attributed to the
variation of sum-rule parameters.
Combined with the factorizable
contribution, one derives the effective parameter $a_2$,
\begin{equation}
a_2 \sim 0.14 -0.17\, |_{\mu = \mu_b} \, ,
\label{eq:a2calc}
\end{equation}
which is still too small.

Since the PQCD approach based on $k_T$ factorization is free from
end-point singularities, it is applicable to $B$ meson decays into
charmonia. The formalism for the color-suppressed nonfactorizable
amplitude is the same as that for the $\bar B\to D^0\pi^0$ decay
in the previous subsection. The $B\to J/\psi K^{(*)}$ decays have been 
analyzed recently \cite{SSU,Chen03}, and the predicted branching ratios, 
together with the experimental data, are listed in Table~\ref{bjpk}.
Briefly speaking, the measured $B^0\to J/\Psi K^{0}$ branching ratio 
\cite{babarc} is employed to determine the unknown $J/\psi$ meson wave 
function. This wave function is then used to predict the
$B^0\to J/\Psi K^{*0}$ branching ratio, whose
consistency with the data is obvious.

\begin{table}[h]
\begin{center}
\begin{tabular}{|c|c|c|c|}
\hline
Mode &   Belle \cite{BelleBr} & BaBar \cite{babarc} & PQCD \\
\hline 
$J/\Psi K^{0}$ & $7.9\pm 0.4 \pm 0.9$  & $8.3\pm 0.4 \pm 0.5$ & $8.3$ \\ 
\hline        
$J/\Psi K^{*0}$ & $12.9\pm0.5 \pm 1.3$  & $12.4\pm 0.5 \pm 0.9$&  $13.37$
\\\hline
\end{tabular}
\caption{PQCD results of the $B\to J/\psi K^{(*)}$ branching ratios
in unit of $10^{-4}$ with the $J/\psi$ meson decay constant 
$f_{J/\Psi}=0.405$ GeV.} 
\label{bjpk}
\end{center}
\end{table}

\section{Other Topics}

\subsection{\it Final-State Interaction \label{sec:fin}}

A strong phase can come from short-distance dynamics and from 
long-distance dynamics. The former may be calculable, but the latter
is completely nonperturbative. The strong phases for two-body
nonleptonic $B$ meson decays, derived in the QCDF or PQCD approaches,
are all of the short-distance type. That is, the long-distance phases
from final-state interaction (FSI) have been neglected in both
approaches. It is then essential to look for any means to justify
this assumption. In this subsection I review the recent study
on this subtle and complicated topic.

Most estimates of FSI effects in
the literature \cite{FSI} suffer ambiguities or difficulties. 
Some opinions favor small FSI effects in two-body nonleptonic
$B$ meson decays. For example, the smallness of FSI effects has been put
forward based on the color-transparency argument \cite{transparency}. 
The renormalization-group analysis of soft gluon
exchanges among initial- and final-state mesons \cite{LT98} has 
indicated that FSI effects are not important in two-body $B$ meson decays.
For a limited number of decay channels, one 
can extract the FSI phases $\Delta\delta$ directly from experiments.
The phases are large for $D$ meson decays \cite{Bis} and small for $B$ 
meson decays \cite{MS1}: 
\begin{eqnarray}
       \Delta\delta &=& 80^{\circ}\pm 7^{\circ}; \;\; D\rightarrow
                                         K^-\pi^+/\overline{K}^0\pi^+, \\
       \Delta\delta &<& \left\{ \begin{array}{ll}
                          11^{\circ};& B\rightarrow
                                         D^+\pi^-/D^0\pi^-\\ 
                          16^{\circ};& B\rightarrow
                                         D^+\rho^-/D^0\rho^-\\
                          29^{\circ};& B\rightarrow
                                         D^{*+}\pi^-/D^{*0}\pi^- .\\
                        \end{array} \right.
\end{eqnarray}
implying that FSI phases diminish as the initial mass 
increases \cite{S00}. 

However, an opposite opinion was raised in \cite{S00}.
Take the decay $B^+(\overline{b}u)\rightarrow
K^0(\overline{s}d)\pi^+(\overline{d}u)$ as an example. 
The gluons exchanged between $\overline{d}$ and $\overline{s}$ 
(or $d$) are hard, but the gluons between the soft spectator $u$ and 
$\overline{s}$ (or $d$) are not so hard. By simple kinematics,  the
invariant mass
$m_{\overline{d}u} \approx (\Lambda_{\rm QCD}m_B)^{1/2}\simeq$ 1.2 GeV 
for $E_u\simeq\Lambda_{\rm QCD}$ is in the middle of the 
resonance region of the ${\overline{s}d}$ channel. Then long-distance 
interactions cannot be ignored between $K^0$ and $\pi^+$. 

Some attempts have been made to estimate FSI effects. 
Applying the time-reversal operation on the decay amplitude
$A_n=\langle n^{out}|O|B\rangle$, one obtains
\begin{equation}
A_n \stackrel{T}{\rightarrow}\langle B|T O
T^{-1}|n^{in}\rangle\;.
\label{tss}
\end{equation}
Insert a complete set of {\it out} states, and employ the
symmetry propery $S_{nn'} = S_{n'n}$,  
$S_{nn'} = \langle n^{out}|n'^{in}\rangle$. From the time reversal 
invariance of strong interaction, Eq.~(\ref{tss}) becomes 
\begin{equation}
      A_n = \sum_{n'} S_{nn'}A_{n'}^* \;,  \label{time-reversal}
\end{equation}
for a T-even decay operator $O$. Subtracting the complex conjugate of 
$A$ from both sides in Eq.(\ref{time-reversal}) and dividing it by $i$, 
the familiar relation emerges 
\begin{equation}
   2{\rm Im}A_n = \sum_{n'}t_{nn'}A_{n'}^*\;, \label{reduction} 
\end{equation}
with $t = (S-1)/i$.
The above expression states that the strong phase
is the sum of the contributions over all intermediate states $n'$.

If approximating the intermediate states $n'$ by two-body states, 
those connected to the final state $n$ by Pomeron exchange dominate in 
the sum, and $t_{nn'}$ will be almost imaginary.
The decay amplitude $A_n$ will be also almost 
imaginary, no matter what operator is responsible for the decay.
Certainly, this approximation may not be justified.
Without the two-body-state approximation, one can take advantage of the
presence of many states. A statistical approach or a random approximation 
then helps \cite{CS97}. Note that since $A_{n'}$ and $S_{nn'}$ come from
two different sources, weak and strong interactions, the phase of 
product $S_{n'n}A_{n'}^*$ for $n'\neq n$ takes equally likely a positive 
or a negative value as $n'$ is varied with $n$ fixed. While $A_{n'}$ is 
related to $A_{n}$ ($n'\neq n$) by rescattering, there exist so many states 
that the influence of $n$ on $n'$ can be disregarded. Therefore, the phase 
of $S_{nn'}A_{n'}^*$ takes random values as $n'$ varies.  
Under this approximation, a typical FSI phase in the
$B\to\pi\pi$ and $K\pi$ decays has been estimated
to be about $20^o$ from the meson-meson scattering at 5 GeV \cite{SW}. 
The limitation of the random approximation is that one can only compute 
the statistically likely values of FSI phases as their standard 
deviations from zero, instead of predicting values of individual phases.

\begin{figure}[t!]
\begin{center}
\epsfig{file=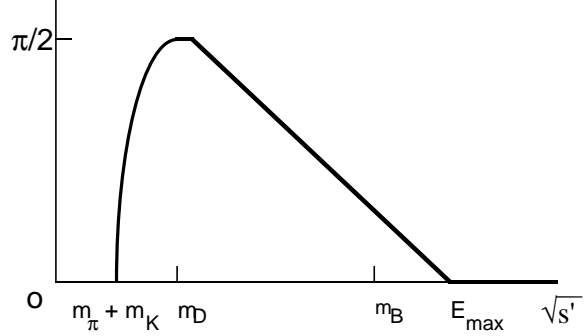, height=2.0in}
\caption{A model FSI phase.}
\label{FSI1}
\end{center}
\end{figure}

\begin{figure}[t!]
\begin{center}
\epsfig{file=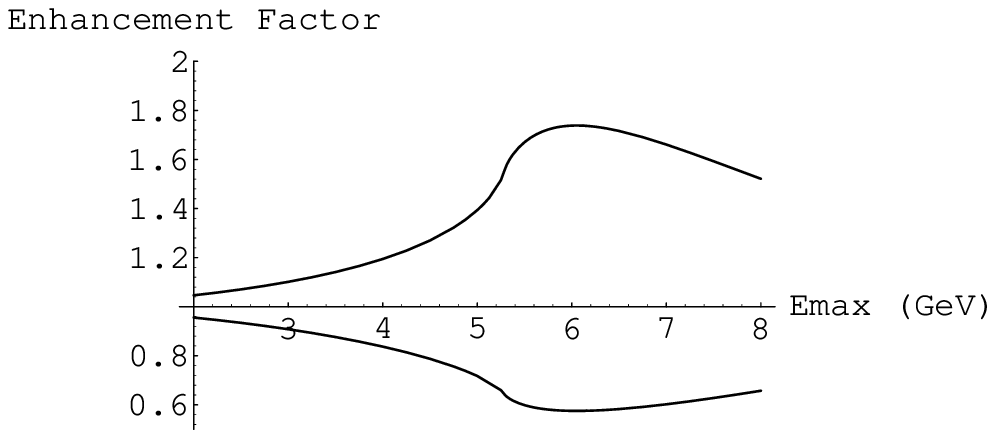, height=2.0in}
\caption{The enhancement (suppression) factor 
$E$ vs $E_{max}$. The lower curve ($E<1$) for $\delta(s')>0$ shown here,
and the upper curve for $\delta(s')\rightarrow -\delta(s')$.}
\label{FSI2}
\end{center}
\end{figure}

FSI not only generates a strong phase but also changes a
magnitude of amplitude. It has been argued that the decay amplitude 
could be enhanced by a FSI factor \cite{Jacob},
\begin{equation}
  E = \exp\biggl(\frac{{\cal P}}{\pi}\int^{\infty}_{s_{\rm th}}
    \frac{\delta(s')}{s'-m_B^2}ds'\biggr)\;.  
\label{enhance}
\end{equation}
In the approximation of two-body 
intermediate states, the FSI phase approaches $\pm 90^{\circ}$ at high 
energies. The enhancement or suppression effect is then very large due to
a constructive integration over $s'$. In the random approximation, the 
sign of $\delta(s')$ may fluctuate with $s'$. In this case, the effect 
of enhancement and suppression would be much smaller.
An estimate for the $B\rightarrow K\pi$, $\pi\pi$ decays 
with one model FSI phase motivated by experiments has been performed. 
Choose the FSI phase as shown in Fig.~\ref{FSI1} \cite{S00}:
$\delta(s')$ rises to large values ($\sim 90^{\circ}$) around 
2 GeV and falls linearly to zero at $E_{\rm max}= O(m_B)$.
This $\delta(s')$ suppresses the decay amplitude, as indicated by
the lower curve in Fig.~\ref{FSI2}, since the support 
of the phase integral is mostly below $m_B$, where  
$1/(s'-m_B^2)$ is negative. A 10$\sim$40\%
correction to the decay amplitude is possible \cite{S00}. 

An application of Eq.~(\ref{time-reversal}) to the study of the 
$B\to D\pi$ decays was provided in \cite{CHY}.
The matrix equation can be formally solved to give 
\begin{equation}
       A = S^{1/2}A^0\;, \label{solution}
\end{equation}
where $A^0$ is an arbitrary {\it real} vector of the same dimension as 
$A$. One may consider the vector $A^0$ as representing the decay 
amplitude in the absence of the final state phases
from strong interaction. Assume elastic final state rescattering in the 
$\bar B\to D\pi$ modes. Assigning the real factorization amplitudes 
in FA to $A^0$, one obtains
\begin{equation}
\left(
\begin{array}{l}
A_{D^0\pi^-}\\
A_{D^+\pi^-}\\
A_{D^0\pi^0}
\end{array}
\right)
=S^{1/2}\,
\left(
\begin{array}{l}
A^f_{D^0\pi^-}\\
A^f_{D^+\pi^-}\\
A^f_{D^0\pi^0}
\end{array}
\right)\;,
\label{eq:FSI}
\end{equation}
where the superscript $f$ denotes FA. 


\begin{figure}[t!]
\begin{center}
\epsfig{file=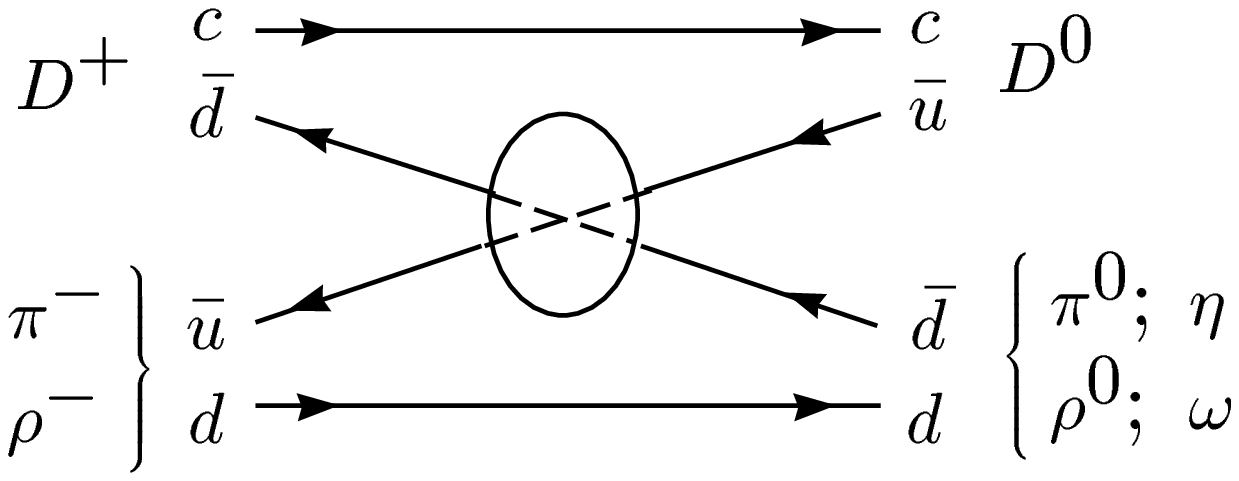, height=1.2in}\vskip 0.5cm
\epsfig{file=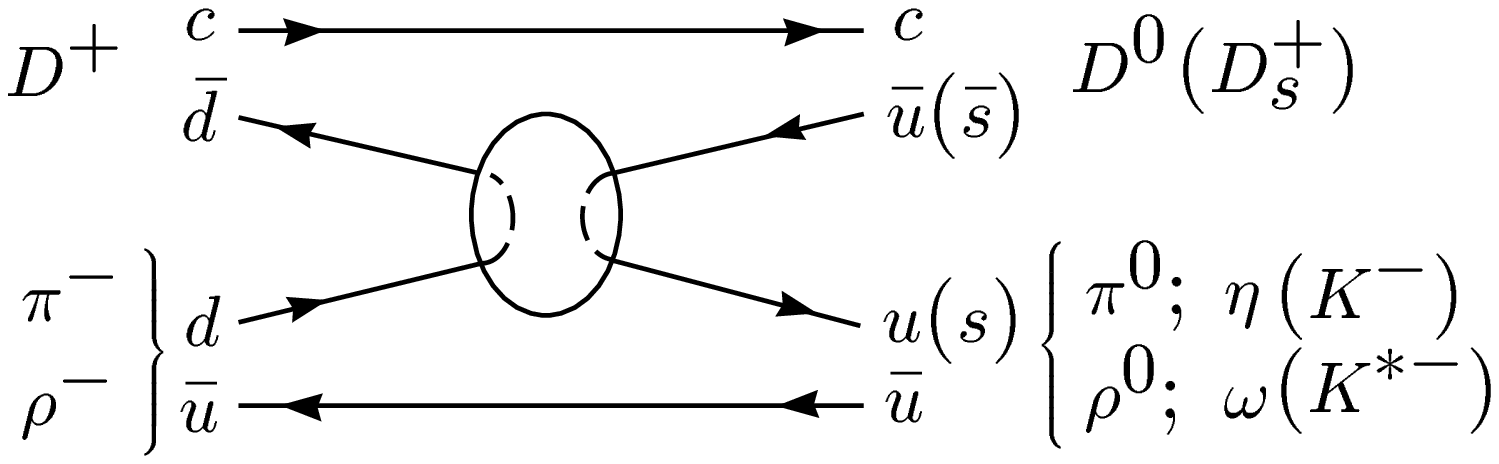, height=1.2in}\vskip 0.5cm
\epsfig{file=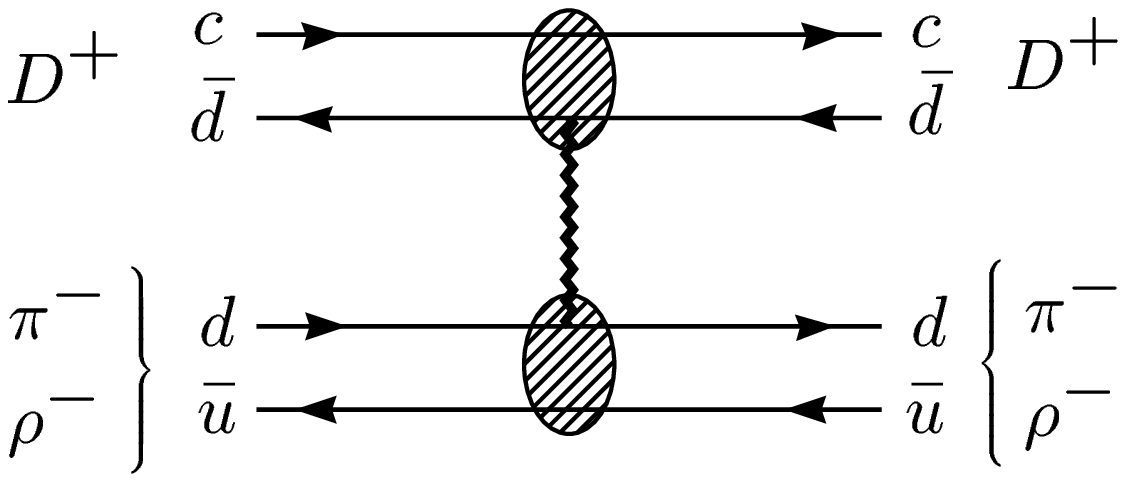, height=1.2in}
\caption{Pictorial representation of
  (a) $r_e$ (charge exchange), (b) $r_a$ (annihilation) and (c) $r_0$ (singlet exchange).}
\label{FSI3}
\end{center}
\end{figure}

To obtain $S^{1/2}$, one derives the scattering matrix $t$ through 
the optical theorem. This way makes transparent the mechanism involved 
in final-state rescattering \cite{CHY}.
By means of Eq.~(\ref{reduction}) and the various scattering 
mechanism displayed in Fig.~\ref{FSI3}, one has 
\begin{equation}
t=\left(
\begin{array}{ccc}
r_0^*+r^*_e &0 &0\\
0 &r^*_0+r^*_a &{1\over \sqrt2}(r^*_a-r^*_e)\\
0 &{1\over \sqrt2}(r^*_a-r^*_e) &r^*_0+{1\over2}(r^*_a+r^*_e)\\
\end{array}
\right)\;,
\label{eq:t}
\end{equation}
where $r_e$, $r_a$, and $r_0$ parametrize the rescattering effects from
charge exchange, annihilation, and flavor singlet (Pomeron) exchange, 
respectively, as defined in Fig.~\ref{FSI3}.
The definition $S=1+it$ then gives the constraints,
\begin{eqnarray}
r_0+r_e&=&2\sin\delta_{3/2}\,e^{i\delta_{3/2}},
\nonumber\\
r_0+{1\over2}(3r_a-r_e)&=&2\sin\delta_{1/2}\,e^{i\delta_{1/2}}\;,
\label{eq:r1r2}
\end{eqnarray}
where $\delta_{1/2}$ and $\delta_{3/2}$ are the phases associated with 
the corresponding isospin amplitudes.

To perform the global fit, the data are put into the left-hand side
of Eq.~(\ref{eq:FSI}). The amplitudes in FA with $a_2/a_1\sim 0.25$
(input from the $B\to J/\psi K$ decay)
and some model form factors are put into the right-hand side. The 
numerical analysis leads to \cite{CHY}
\bea
& &(1+i r_0) e^{-2i\delta_{3/2}}= 0.45+0.50i\;,
\nn\\
& &i r_e e^{-2i\delta_{3/2}}=0.55-0.50i\;,
\nn\\
& &i r_a e^{-2i\delta_{3/2}}=0.14+0.04i\;.
\label{rea}
\eea
One then realizes the relative importance of the various rescattering
mechanism: the Pomeron exchange and the charge exchange give roughly 
similar effect, larger than that from the annihilation
mechanism. It was then concluded that FSI is crucial for enhancing
$a_2$ from 0.25 for the $B\to J/\psi K$ decay to 0.5 for the
$\bar B^0\to D^0\pi^0$ decay \cite{CHY}. Similar formalism has been 
applied to two-body charmless modes, and large FSI phases were also
postulated. 

I emphasize that the major assumption in the above
analysis is the absence of short-distance phases in two-body nonleptonic
$B$ meson decays. If short-distance phases exist, the decay 
amplitudes in Fig.~\ref{FSI3} have carried phases already
before rescattering. Therefore, the absence of short-distance phases 
must be assumed, so that the amplitudes before rescattering can be 
identified as the real amplitudes on the right-hand side of 
Eq.~(\ref{eq:FSI}). If short-distance phases do exist, the FSI effects
could be small, and the conclusion on the relative
importance of different mechanism might be altered. For example,
the $B\to D\pi$ branching ratios can be accounted for in the PQCD
approach without resort to FSI as shown in Sec.~\ref{sec:two}.

A stringent test of the small FSI assumption is provided
by measuring the $B\to KK$ decays.
In particular, large observed $B_d^0\to K^\pm K^\mp$ 
branching ratios and CP asymmetry in the $B_d^0\to K^0 {\bar K}^0$ modes 
will imply large FSI effects. So far, the PQCD predictions for the 
branching ratios \cite{CL00},
\begin{eqnarray}
& &B(B^+\to K^+K^0)=1.47\times 10^{-6}\;,
\nonumber\\
& &B(B^-\to K^-K^0)=1.84\times 10^{-6}\;,
\nonumber\\
& &B(B_d^0\to K^+K^-)=3.27\times 10^{-8} \;,
\nonumber\\
& &B({\bar B}_d^0\to K^-K^+)=5.90\times 10^{-8}\;,
\nonumber\\
& &B(B_d^0\to K^0{\bar K}^0)=1.75\times 10^{-6} \;,
\nonumber\\
& &B({\bar B}_d^0\to K^0{\bar K}^0)=1.75\times 10^{-6}\;.
\end{eqnarray}
are still below the experimental bounds.

\subsection{\it Intrinsic Charm and Charming Penguin \label{sec:int}}

One of the higher-power contributions comes from the Fock states of 
arbitrarily many particles. For example, a $B$ meson bound state contains  
\bea
|B^- \rangle =
\psi_{b \bar u} |b \bar u \rangle+
\psi_{b \bar u g} |b \bar u g \rangle +
\psi_{b \bar u d \bar d} |b \bar u d \bar d \rangle 
+ \psi_{b \bar u c \bar c} |b \bar u c \bar c \rangle + \cdots.
\eea
The Fock state decomposition is usually performed at equal light-cone
time using light-cone quantization in the gauge $A^+ =0$ 
\cite{Dirac:1949cp,Brodsky:1989pv}. The non-valence partons in the higher 
Fock states are generated by QCD splitting mechanism. All the partons in 
a Fock component are almost on mass shell with long lifetimes, and 
interact with each other through multiple infrared gluon exchanges.
This is the reason they are {\it intrinsic} to the hadron structure. 
The intrinsic heavy quarks are part of the hadron bound state 
\cite{Brodsky:1980pb}. Due to the hierachical
structure of the CKM matrix elements, the weak transition 
$b\to s c{\overline c}$ is doubly Cabibbo enhanced with respect to the 
$b\to s u{\overline u}$ transition. This is the argument
for the potential importance of the intrinsic charm (IC) \cite{sjbsvg}. 
In contrast, a perturbative correction to the weak transition 
matrix element can produce a $c\bar c$ pair through gluon splitting. The
quark pair is generally not multiply connected to the partons of the 
bound state, and {\it extrinsic} to the hadron structure.
Generally, ``intrinsic'' contributions in $B$ meson decays
are of higher twist, whereas ``extrinsic'' contributions
are of higher order in $\alpha_s$. 

Some estimates based on phenomenological hints suggest that the IC 
probability in the $B$ meson could be as large as a few percent 
\cite{Chang:2001iy}. The slight excess in the inclusive 
$B\to J/\psi X$ yield at low $J/\psi$ momentum \cite{CLEOV} implies  
the presence of the $\bar B \to J/\psi D^{(*)}$ channel, which occurs 
through IC, though such an effect could also be generated by the
$\bar B\to J/\psi \Lambda \bar n$ decay \cite{Brodsky:1997yr}. 
IC could help to understand the large 
$B\to \eta^\prime K\;,\eta^\prime X$ branching ratios. 
A valence $c\bar c$ component in the $\eta^\prime$ meson has
been introduced to resolve this discrepancy \cite{Halperin:1997as}, but
the decay constant $f_{\eta^\prime}^{(c)}\sim -2$ MeV, defined via
$\langle 0 | {\bar c}\gamma_\mu \gamma_5 c| \eta^\prime(p)\rangle
\equiv i f_{\eta^\prime}^{(c)} p_\mu$, 
is too small \cite{Franz:2000}. Discussions on the above subjects
can be found in \cite{SG02}.
It was proposed recently \cite{GQV} that IC in the $B$ meson 
could be probed by measuring the $B\to J/\psi e\nu X$ decay.

\begin{figure}[t!]
\begin{center}
\epsfig{file=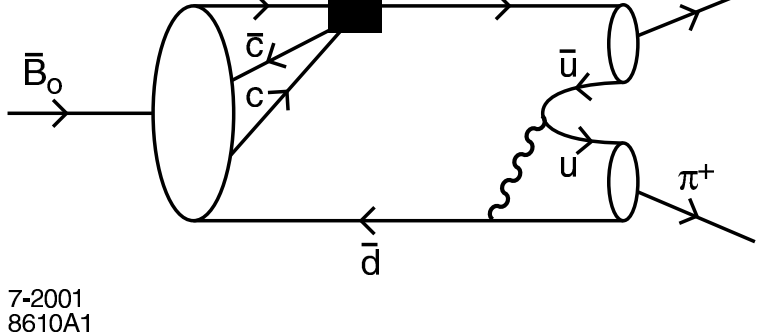,height=2.0in} 
\end{center}
\caption{Intrinsic charm in the B-meson can mediate the decay to a
strange, charmless final state via the weak transition
$b\to s c {\overline c}$.}
\label{fig:icdiag}
\end{figure}

As emphasized in \cite{sjbsvg}, IC plays an important role in
the $B\to K\pi$ decays. Considering only the valence contribution,
the decay amplitude $A(B^0 \to K^+\pi^-)$ is written as \cite{Buras:2000ra}
\bea
A(B_d^0 \to K^+\pi^-) =V_{us}V_{ub}^\ast(
E_1  - P_1^{\rm GIM})- V_{ts}V_{tb}^\ast P_1 \;.
\label{burasparam}
\eea
The parameter $E_1$ denotes the contribution from the
$W$ emission topologies, $P_1$ denotes the penguin topologies, and 
$P_1^{\rm GIM}$ contains penguin contributions, which
vanish in the $m_c=m_u$ limit. Beyond the valence approximation,
the additional contribution of IC through Fig.~\ref{fig:icdiag}
arises, which is expressed as $V_{cs} V_{cb}^\ast A_1^{\rm IC}$. 
There are no hard gluon 
exchanges across the weak vertex in Fig.~\ref{fig:icdiag}, so that the 
computation of the hard 
scattering amplitude factorizes. One portion of the weak vertex
mediates the annihilation of the ${\bar b} c$ quarks, and the
other describes the amplitude for the $(c\bar q)$ state
to emerge with the parton content of the $K\pi$ final state,
namely ${\bar s} q' q {\bar q}$. Note that
the above two pieces of the hard scattering kernels are in fact convoluted
together over the momentum fraction of the light-cone wave function
for the $(b\bar c)(c\bar q)$ state, which is still unknown.

\begin{figure}[t!]
\begin{center}
\vskip 1.0cm
\epsfig{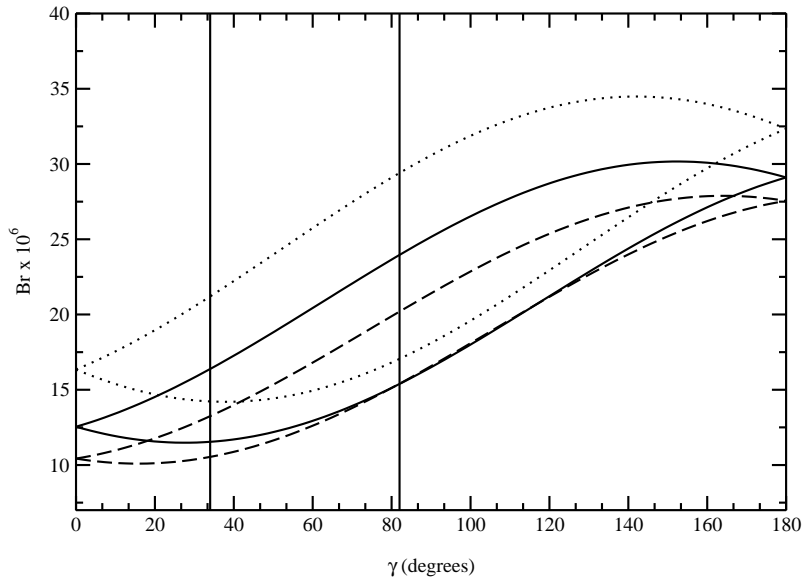} 
\end{center}
\caption{Intrinsic charm in the B-meson can mediate the decay to a
strange, charmless final state via the weak transition
$b\to s c {\overline c}$.}
\label{fig:keum}
\end{figure}

The amplitude for the $(c\bar q)$ state to emerge as
${\bar s} q' q {\bar q}$ cosmetically resembles Fig.~\ref{sixh}(f).
Hence, the IC contribution has been parametrized as \cite{sjbsvg}
\begin{equation}
A_1^{\rm IC}(s,q,B,K,\pi) \sim
f_{B_c^\ast} F_a^{P} \frac{a_1(m_b/2)}{a_6(m_b/2)}B \;,
\label{ICestimate}
\end{equation}
where $f_{B_c^\ast}\sim 0.317$ GeV \cite{Colangelo:1993cx} arises
from the annihilation of the $\bar b c$ pair, and the remaining
factors come from an estimate of the lower half of the diagram
in Fig.~\ref{fig:icdiag}. The factorizable annihilation
amplitude $F_a^P$ has been calculated in the PQCD approach \cite{KLS}.
It is easy to get ${a_1(m_b/2)}/{a_6(m_b/2)}\sim -20$. The
parameter $B$ reflects the probability amplitude to find the
B meson in an IC configuration, as well as an adjustment for the
$\sim$ 50\% penguin enhancement \cite{KLS}. Hence, one has
$B\sim 2(0.02)/3$. The impact of IC on the $B\to K\pi$ decays
has been investigated in \cite{sjbsvg} based on 
Eq.~(\ref{ICestimate}), and the results are exhibited
in Fig.~\ref{fig:keum}. Note that IC can act to either enhance or 
decrease the CP asymmetry. The 
IC contribution $|A_1^{\rm IC}|/|P_1| \sim O(10)\%$ 
is a nontrivial fraction of the penguin parameter
$P_1$, which reflects the accuracy we can reach in calculating
the effective value of $P_1$. 

``Charming penguins" refer to the nonperturbative piece of the extrinsic
$c$ quark loop, whose contributions are parametrized into a term 
$\tilde P_{1}$. Theoretically, their importance is not clear.
However, a recent analysis of the $B\to K \pi$, $\pi \pi$,
$K \eta$, and $K \eta'$ decays indicates that they could be important. 
The parameter $\tilde P_{1}$ includes not only the charming penguin 
contributions, 
but also annihilation and penguin contractions of penguin operators. It 
does not include leading emission amplitudes of penguin operators
($O_3$--$O_6$), which have been explicitly evaluated in FA.  
In this respect, it is a general parameterization of all the perturbative 
and nonperturbative $O(\Lambda_{\rm QCD}/m_{b})$ contributions of the 
operators $O_{5}$ and $O_{6}$. $\tilde P_{1}$ has
the same quantum numbers and physical effects
as the original charming penguins proposed in \cite{charming}.

The charming penguin contribution in $\tilde P_{1}$ can be, taking the 
$B_d^0 \to K^+\pi^-$ decay as an example, included in the term $P_1$ in 
Eq.~(\ref{burasparam}). Another $O(\Lambda_{\rm QCD}/m_b)$ contribution
from the GIM-penguin $\tilde P_1^{GIM}$ has been neglected 
for simplicity \cite{CFMP}. The $B\to K\pi$ data do not constrain this 
parameter very effectively, since its contribution is doubly Cabibbo 
suppressed with respect to $\tilde P_1$. The remaining $\pi^+\pi^-$ mode
alone is not sufficient to fully determine the complex parameter 
$\tilde P_1^{GIM}$. However, the 
GIM-penguin contribution may be able to enhance the 
${\cal B}(B\to\pi^0\pi^0)$ up to few $\times 10^{-6}$
\cite{Ciuchini:2001gv}.

\begin{table}
\begin{center}
\begin{tabular}{|c|c|c|c|c|}
\hline
$\vert V_{cb}\vert\!\times\! 10^{3}$ & $\vert V_{ub}\vert\!\times\! 10^{3}$ & $\hat B_K$ & $f_{B_d}\sqrt{B_d}$ (MeV) & $\xi$\\
$40.9\!\pm\! 1.0$&$3.70\!\pm\! 0.42$ & $0.86\!\pm\! 0.06\!\pm\!  0.14$& $230\!\pm\! 30\!\pm\! 15$ &
$1.16\!\pm\!0.03\!\pm\!0.04$\\[4pt]
\hline
$F_K(M_K^2)$ & $B(K^+\pi^-)\!\times\! 10^6  $ &  $B(K^+\pi^0)\!\times\! 10^6  $ & $B(K^0\pi^+)\!\times\! 10^6  $&
 $B(K^0\pi^0)\!\times\! 10^6  $\\
$0.32\pm0.12$ & $18.6 \pm 1.1$  & $11.5 \pm 1.3$  & $17.9 \pm 1.7$ & $8.9 \pm 2.3$\\[4pt]
$F_\pi(m_\pi^2)$ & $B(\pi^+\pi^-)\!\times\! 10^6 $ &  $B(\pi^+\pi^0)\!\times\! 10^6  $ & $B(\pi^0\pi^0)\!\times\!
 10^6 $ & \\
\hline
$0.27\pm0.08$&$5.2 \pm 0.6$ &$4.9\pm 1.1$ & $<\! 3.4\,$ BaBar &\\[4pt]
\hline
\end{tabular}
\caption{Values of the input parameters used in the global fit. }
\label{tab:expbr}
\end{center}
\end{table}

\begin{table}
\begin{center}
\begin{tabular}{|c|cc|cc|}
\hline
  Mode  & \multicolumn{2}{c}{UTA} & \multicolumn{2}{c}{$\vert V_{ub}/V_{cb}\vert$} \\
        & $B$ ($10^{-6}$) & $\vert A_{CP}\vert$ &  $B$ ($10^{-6}$) & $\vert A_{CP}\vert$\\
\hline
 $\pi^+\pi^-$ & $8.9 \pm 3.3$ & $0.37\pm0.17$    & $8.7 \pm 3.6$ & $0.39\pm0.20$    \\
 $\pi^+\pi^0$ & $5.4 \pm 2.1$ & -- & $5.5 \pm 2.2$ & --    \\
 $\pi^0\pi^0$ & $0.44\pm 0.13$ & $0.61\pm0.26$   & $0.69 \pm 0.27$ & $0.45\pm0.27$  \\
\hline
 $K^+\pi^-$   & $18.4 \pm 1.0$ & $0.21\pm0.10$   & $18.8 \pm 1.0$ & $0.21\pm0.12$   \\
 $K^+\pi^0$   & $10.3 \pm 0.9$ & $0.22\pm0.11$   & $10.7 \pm 1.0$ & $0.22\pm0.13$   \\
 $K^0\pi^+$   & $19.3 \pm 1.2$ & $0.00\pm0.00$   & $18.1 \pm 1.5$ & $0.00\pm0.00$   \\
 $K^0\pi^0$   &  $8.7 \pm 0.8$ & $0.04\pm0.02$   &  $8.2 \pm 1.2$ & $0.04\pm0.03$   \\
\hline
\end{tabular}
\caption{Predictions for CP-averaged branching ratios and
absolute values of the CP asymmetries $\vert A_{CP}\vert$. The left
(right) columns show results obtained using constraints on the CKM
parameters $\rho$ and $\eta$ obtained from the 
unitarity triangle analysis (UTA) \cite{Ciuchini:2000de} (the measurement 
of $\vert V_{ub}/V_{cb}\vert$). The last four channels are those used for
fitting the charming penguin
parameter $\tilde P_1$.}
\label{tab:results}
\end{center}
\end{table}

Using the inputs collected in Table~\ref{tab:expbr} \cite{CFMP},
the complex parameter
$\tilde P_1=(0.13\pm 0.02)\, e^{\pm i (114\pm 35)^o}$
in units of $f_\pi F^{B\pi}(m_\pi^2)$ has been 
determined from the data fitting, which has the expected size of 
$O(\Lambda_{\rm QCD}/m_b)$. Note that the sign of the phase is practically 
not constrained by the data. 
Table~\ref{tab:results} \cite{CFMP} shows the predicted CP-averaged
branching ratios and the absolute value of the CP-asymmetries
$\vert A_{CP}\vert$ for the $B\to K\pi$ and $\pi\pi$ modes.
The angle $\phi_3$ is determined through the effect of interefence
terms in the $B\to K\pi$ branching ratios.
The nonperturbative parameter $\tilde P_1$ with an additional phase in 
the amplitudes makes the extraction of $\phi_3$ difficult. 
It has been checked using the
$\vert V_{ub}/V_{cb}\vert $-constrained fit that almost any value of
$\phi_3$ is allowed, given the uncertainty on $\tilde P_1$. 

I stress that the intrinsic penguins and charming penguins represent
distinct dynamics. However, both contributions are Cabibbo-enhanced and 
contain an $O(1)$ Wilson coefficient. Indeed, the parameter 
$A_1^{\rm IC}$ can also be absorbed into $P_1$ and $P_1^{\rm GIM}$. 
Hence, a global fit can not distinguish them. Note that the intrinsic
penguin contribution in \cite{sjbsvg} was parametrized to be proportional
to the annihilation penguin amplitude calculated in the PQCD approach,
such that its phase, including the sign, can be fixed.

\section{Conclusion}
 
In this article I have reviewed the two fundamental tools in QCD 
perturbation theory, collinear and $k_T$ factorization theorems,
in which soft dynamics and hard dynamics of a process are factorized
into hadron wave functions and hard kernels, respectively.
Both factorization theorems can be constructed in a gauge-invariant
way up to all orders, such that infrared-finite and gauge-invariant
predictions are obtained. $k_T$ factorization is more appropriate,
when the end-point region of parton momentum fractions becomes
important. To improve perturbative expansion, threshold resummation
for collinear factorization and $k_T$ resummation for $k_T$
factorization can be performed. 

I then discussed the various QCD approaches to exclusive $B$ meson decays 
available in the literature, which were developed based on the above
fundamental concepts for perturbaton theory. I reviewed
the recent progress made in these approaches, emphasizing the basic
ideas behind and the comparison of their phenomenological implications.
The competition among different approaches and the confrontation
between theoretical predictions and experimental data have 
stimulated tremendous progress. In this section I briefly summarize the 
advantage of each method and the important issues, whcih require further 
investigation in order for a more solid and complete framework.

The advantage of LCSR is that both soft and hard contributions can
be analyzed in the same framework, and that it is easy to examine
the self-consistency of expansions in $\alpha_s$ and in $1/m_b$.
Recently, an essential step toward the evaluation of nonfactorizable
contributions (from twist-3 hadron distribution amplitudes) has been 
made. However, the analysis is not yet complete. The urgent subject is 
to include the nonfactorizable contributions
associated with the twist-2 distribution amplitudes, i.e.,
from the diagrams in Fig.~\ref{pp2}. These diagrams can not only
moderate the scale dependence of predictions, but introduce strong
phases, which are necessary for generating direct CP asymmetries.
Since the strong phases come from higher-order contributions,
compared to the leading soft transition form factors,
they are expected to be small, the same as those from the QCDF approach.

For QCDF, the advantage is its explicit factorization picture in the
heavy quark limit. Talking about the treatment of leading contribtuions, 
it is most complete among all approaches, since the scale and scheme
dependences have been greatly reduced. However, the end-point
singularity appearing at subleading level (twist-3 nonfactorizable
amplitudes and annihilation amplitudes) makes QCDF less predictive.
This is also the reason the experimental constraint on the unitarity 
angles derived from QCDF is not very strong. Except for the end-point 
singularity, another challenge comes from the explanation of the
possible large direct CP asymmetries in the two-body nonleptonic 
$B$ meson decays.

The PQCD approach, based on $k_T$ factorization theorem, is free of
the end-point singularities. Therefore, the nonperturbative inputs are
only universal hadron distribution amplitudes, without the transition
form factors and the arbitrary infrared cutoffs due to the
singularities. So far, the PQCD predictions are in agreement with data,
and phenomenologically successful.
However, it is still under debate whether the crucial Sudakov
effect is strong enough to suppress the end-point contribution
for the physcial mass $m_B\sim 5$ GeV. The urgent subject is then to
calculate higher-order corrections, and examine whether they converge.
Before proving this, PQCD is not yet a self-consistent theory.
The calculation can also verify the argument about
the characteristic hard scale $\sqrt{\Lambda m_B}$ for exclusive
$B$ meson decays, which was made based on the hard-scattering picture.
This characteristic scale is the key for PQCD to explain the 
$B\to VP$ data.

SCET provides the most systematic framework for constructing collinear
factorization formulas of exclusive $B$ meson decays at large recoil.
Its advantage is that contributions characterized by different scales can 
be separated easily. A progress has been made recently in deriving the 
symmetry relations among various transition form factors, and the
power corrections and radiative corrections to these relations.
However, to make explicit predictions, which can be compared with
data, it is necessary to calculate the Wilson coefficients
of effective operators from the matching of the effective theory to the
full one. The calculation is as complicated as in other approaches.
 
The advantage of LFQCD is its simplicity. At leading power, the
formula for a transition form factor involves only an overlap
integral of the initial- and final-state hadron wave fuctions.
Both soft dynamics and hard dynamics have been absorbed into the
wave functions. The inclusion of nonvalence contributions
has been worked out, which is crucial for guaranteeing the
covariance of predictions. The next step is to extend the LFQCD to two-body
nonleptonic decays, especially to nonfactorizable amplitudes.
This extention will be a challenge.
 
Though some study has been done, the behavior of the $B$ meson 
bound state is still not clear. This subject is important, since
the $B$ meson wave functions, including those associated with higher
Fock states (such as intrinsic charms), are the input of all QCD
approaches.
There is still no reliable method to analyze long-distance FSI effects 
in exclusive $B$ meson decays. Even an ansatz does not exist, in 
which the importance of FSI can be determined from data fitting. The 
ansatz based on the factorization assumption, i.e., on the absence of 
short-distance phases, is not convincing as explained in 
Sec.~\ref{sec:two} and in Sec.~\ref{sec:fin}. Another crucial
contribution from charming penguins also requires investigation.
More clever ideas are necessary for understanding the above 
nonperturbative dynamics. 

As demonstrated in this article,
exclusive $B$ meson decays exhibit exciting QCD dynamics, which
could be, fortunately, studied in a self-consistent way due to the 
large $b$ quark mass. Some plausible mechanism 
has been explored. For example, annihilation contribution may not be  
as small as we thought with the hint from the possible large
CP asymmetry in the $B_d^0\to\pi^+\pi^-$ decay. However, more challenging 
topics are waiting for our effort.

\vskip 1.0cm

I thank Profs. S. Brodsky and A.I. Sanda for their encouragement and 
useful discussions. 
I also thank my collaborators and competitors for their constructive
opinions. This work was supported by the National Science Council of
R.O.C. under the Grant No. NSC-91-2112-M-001-053.

\end{document}